\documentclass[11pt]{article}
\newcommand{\ignore}[1]{}
\usepackage[T1]{fontenc}
\usepackage[utf8]{inputenc}
\usepackage[english]{babel}

\usepackage[font=small, labelfont = bf]{caption}
\usepackage{graphicx}
\usepackage{float}
\floatstyle{ruled}
\newfloat{algorithm}{htbp}{loa}
\floatname{algorithm}{Algorithm}
\usepackage{color}

\usepackage{amsmath,amsfonts}
\usepackage{mathtools}
\usepackage{amsbsy}
\usepackage{amssymb}
\usepackage{amsthm}
\newtheorem{remark}{Remark}
\usepackage{bm}
\usepackage{natbib}
\usepackage[hidelinks]{hyperref}
\usepackage{booktabs}
\usepackage{multirow}
\usepackage{makecell}
\usepackage{enumitem}
\usepackage{lineno}
\usepackage{comment}
\usepackage{xcolor}

\usepackage{authblk}

\usepackage[normalem]{ulem}
\newif\ifshowdeletions
\showdeletionstrue

\usepackage{longtable}
\usepackage{pdflscape}
\usepackage{subcaption}
\DeclareGraphicsExtensions{.pdf,.png}
\graphicspath{{numerical_results/}}
\usepackage[section]{placeins}
\newcommand{\defeq}{\mathrel{\vcentcolon=}}   

\numberwithin{equation}{section}
\newtheorem{theorem}{Theorem}[section]
\newtheorem{lemma}{Lemma}[section]

\newtheorem{proposition}{Proposition}[section]

\newtheorem{corollary}{Corollary}[section]
\newtheorem{example}{Example}[section]

\DeclareMathOperator{\vecop}{vec}

\DeclareMathOperator{\diag}{diag}
\providecommand{\ip}[2]{\langle #1,#2\rangle}
\providecommand{\tr}{\operatorname{tr}}

\providecommand{\E}{\mathbb{E}}

\newtheorem{definition}{Definition}

\newcommand{\one}{\mathbf{1}}
\newcommand{\beq}{\begin{equation}}
\newcommand{\eeq}{\end{equation}}
\newcommand{\beas}{\begin{eqnarray*}}
\newcommand{\eeas}{\end{eqnarray*}}
\newcommand{\bea}{\begin{eqnarray}}
\newcommand{\eea}{\end{eqnarray}}
\newcommand{\bei}{\begin{itemize}}
\newcommand{\eei}{\end{itemize}}
\newcommand{\ben}{\begin{enumerate}}
\newcommand{\een}{\end{enumerate}}
\newcommand{\bet}{\begin{theorem}}
\newcommand{\eet}{\end{theorem}}
\newcommand{\bel}{\begin{lemma}}
\newcommand{\eel}{\end{lemma}}
\newcommand{\bep}{\begin{proposition}}
\newcommand{\eep}{\end{proposition}}
\newcommand{\bed}{\begin{definition}}
\newcommand{\eed}{\end{definition}}
\newcommand{\bec}{\begin{corollary}}
\newcommand{\eec}{\end{corollary}}
\newcommand{\bex}{\begin{example}}
\newcommand{\eex}{\end{example}}

\newcommand{\IR}{\mathbb{R}}
\newcommand{\IP}{\mathbb{P}}

\def\limsup{\mathop{\overline{\rm lim}}}
\def\liminf{\mathop{\underline{\rm lim}}}

\makeatletter
\def\saveenum{\xdef\@savedenum{\the\c@enumi\relax}}
\def\resetenum{\global\c@enumi\@savedenum}
\makeatother

\newcommand{\Cov}{\mathop{\rm Cov}}
\newcommand{\Var}{\mathop{\rm Var}}

\newtheorem{assumption}{Assumption}[section]

\title{Localizing Mean Shifts in High-Dimensional Vector Autoregressive Models}

\author[1]{Hongqiang Sun}
\author[1]{Sayar Karmakar}
\author[2,*]{George Michailidis}

\affil[1]{Department of Statistics, University of Florida, Gainesville, FL, USA}
\affil[2]{Department of Statistics and Data Science, University of California, Los Angeles, Los Angeles, CA, USA \thanks{Corresponding author: \texttt{gmichail@ucla.edu}}}

\date{\today}
\date{}

\begin{document}
\maketitle

\textsc{Abstract:}
We study the localization of mean changes in high-dimensional vector autoregressions (VARs) when the autoregressive dynamics remain \textit{common across regimes}. Temporal dependence can substantially affect the accuracy with which a change in the mean is localized, raising the question of whether exploiting and estimating the dependence structure can improve localization. We develop a three-step procedure that first obtains preliminary change-point estimates
from the raw series ignoring the dependence structure, then estimates a common sparse VAR using observations away from the estimated changes, and finally filters the series to refine the change point locations.

For a single change point, we show that the preliminary localization rate depends directly on the strength of temporal dependence, whereas filtering removes this dependence penalty from the oracle rate. With an estimated filter, the refined rate additionally reflects the error in estimating the VAR dynamics and the relation between the change in the stationary mean and the corresponding change in the VAR intercept. We characterize these rates explicitly and establish conditions under which filtering improves localization. Because filtering near a change point uses lags from adjacent regimes, it also creates a short-lived distortion in the filtered mean; we characterize when this boundary effect is innocuous and develop a transition-corrected
alternative.

We extend the procedure to an unknown number of change points. Overlapping local windows are used to generate candidate locations, an information criterion removes redundant and spurious candidates and estimates the number of changes, and the retained locations are subsequently refined using the estimated VAR filter. We establish consistency for the estimated number of
changes and simultaneous localization rates for their locations. Numerical experiments corroborate the localization gains and show robustness to moderate
departures from common VAR dynamics. An application to a large panel of U.S.\ macroeconomic indicators identifies economically interpretable changes.


\section{Introduction}\label{sec:intro}

Vector autoregressions (VARs) provide a flexible framework for modeling dynamic interactions among economic variables and tracing the propagation of shocks through the underlying system. They are widely used in macroeconomics and finance (see, e.g., \citep{sims1980macroeconomics,stock2001vector,stock2016dynamic,kilian2017structural,zivot2006vector}). Classical VAR theory considers a fixed number of variables \citep{lutkepohl2005introduction}, while more recent work accommodates settings in which the dimension is comparable to, or exceeds, the sample size by imposing structural restrictions on the VAR dynamics such as (group) sparsity or low rankness \citep{basu2015regularized,kock2015oracle,medeiros2016l1,lin2017regularized,basu2019low}. Bayesian methods offer an alternative approach to estimation in high-dimensional VARs while naturally providing uncertainty quantification \citep{ghosh2019high,ghosh2021strong}.

We study changes in the means of a high-dimensional time series whose autoregressive dynamics remain \textit{common} across regimes. This restriction isolates the role of the dependence structure in mean-change localization, allowing us to quantify the accuracy gains from estimating common autoregressive dynamics. Our setting is distinct from that of changes in the VAR dynamics, as considered by \citet{safikhani2022joint,bai2023multiple,safikhani2022fast,wang2019localizing} and, in a factor-adjusted framework, by \citet{cho2024high}. It is also related to the literature on mean-change detection and localization \citep{WangSamworth2018,cho2015multiple,cho2016change,bai2010common,HorvathLiuRiceZhao2022}. While several of these methods allow for temporal dependence, our approach explicitly estimates the common VAR dynamics and filters out the resulting autoregressive dependence before refining the location of each mean change.

Let $p$ denote the time series dimension, $d$ the VAR lag order, $n$ the sample size, and $\tau \in \{1, \dots, n-1\}$ the true change-point location. To model a \textit{single mean change}, we define two regime-specific stationary processes $y_t^{(j)} \in \mathbb{R}^p$ for $j \in \{1, 2\}$, given by
\begin{equation}\label{eq:varmodelwithmean}
y_t^{(j)} = \mu_j + \sum_{i=1}^d A_i y_{t-i}^{(j)} + \varepsilon_t, \qquad t \in \mathbb{Z}, \quad j = 1, 2,
\end{equation}
where $\mu_j \in \mathbb{R}^p$ is the regime-specific intercept, $\varepsilon_t \in \mathbb{R}^p$ is the innovation vector, and $A_1, \ldots, A_d \in \mathbb{R}^{p \times p}$ are autoregressive coefficient matrices common to both regimes.

The observed time series $y_1, \ldots, y_n$ switches from $y_t^{(1)}$ to $y_t^{(2)}$ at the change point $\tau$:
\begin{equation}\label{eq:single-cp}
y_t = y_t^{(1)} \one\{t \le \tau\} + y_t^{(2)} \one\{t > \tau\}, \qquad t = 1, \ldots, n.
\end{equation} 
To accommodate high-dimensional settings where the dimension $p$ may exceed the sample size $n$, we assume that the common autoregressive coefficient matrices $A_1, \ldots, A_d$ are \textit{sparse}. Let $\eta_j = \Big(\mathbf{I}_p - \sum_{i=1}^d A_i\Big)^{-1} \mu_j$ denote the stationary mean of regime $j \in \{1, 2\}$. Under this specification, both regimes share the same zero-mean stationary process $e_t$:
\begin{equation}\label{eq:common-centered}
y_t^{(j)} - \eta_j = e_t := \sum_{\ell=0}^{\infty} \Psi_\ell \varepsilon_{t-\ell}, \qquad \Psi_0 = \mathbf{I}_p, \quad \Psi_\ell = \sum_{i=1}^{\min(d,\ell)} A_i \Psi_{\ell-i}.
\end{equation}
The recursion holds for $\ell \ge 1$, with $\Psi_\ell = 0$ for $\ell < 0$. Under this formulation, the observed series reduces to a deterministic mean shift superimposed on a stationary error process. Preserving a common dependence structure across regimes isolates the change point as a pure level shift, allowing the VAR dynamics to be estimated from \textit{stationary observations} within each regime. Similar additive mean-shift models appear in the panel change-point literature: \citet{bai2010common} studies common changes in panel means and variances, whereas \citet{bhattacharjee2019change} consider mean shifts under temporally and cross-sectionally dependent linear processes, including VAR errors. Section~\ref{sec:multiple} extends this model to accommodate multiple mean changes.

By holding the propagation mechanism fixed, the assumption of common VAR dynamics isolates structural shifts in the mean. While evidence of instability in autoregressive relations \citep{stock1996evidence} and time-varying parameter VARs \citep{cogley2005drifts,primiceri2005time} emphasizes settings where dynamics evolve, our framework explicitly isolates mean changes to evaluate whether estimating and filtering out autoregressive dependence \textit{improves change-point localization}. This perspective is particularly relevant to episodes where macroeconomic fluctuations are driven by unusually large shocks rather than changing propagation mechanisms \citep[see, e.g.,][]{stock2012disentangling}.

Our main methodological contribution is a three-step estimator that leverages this common VAR structure to refine an initial change-point location. First, we fit piecewise-constant means to the raw series using ridge regularization. Second, excluding a neighborhood around the preliminary change point, we estimate a common sparse VAR from the remaining observations while profiling out regime-specific intercepts. Third, we filter the series using the estimated VAR dynamics and re-estimate the change point.

Let $\Delta_\eta := \eta_2 - \eta_1$, $\Delta_\mu := \mu_2 - \mu_1$, and $\gamma_p := \sum_{\ell \ge 0} \|\Psi_\ell\|_2$, where the filter-estimation scale $b_n$ is formally defined in \eqref{eq:single-ell2-rho-alpha}. For a single change point, the preliminary and refined estimators achieve localization rates of
\[
\frac{\gamma_p^2}{\|\Delta_\eta\|_2^2} \qquad\text{and}\qquad \frac{1+b_n}{\|\Delta_\mu\|_2^2},
\]
respectively, wherein $\gamma_p$ quantifies the strength of temporal dependence, while $b_n$ reflects the estimation error from fitting the common VAR dynamics. Oracle filtering ($b_n = 0$) completely eliminates the dependence penalty $\gamma_p^2$, and the feasible estimator approaches the oracle rate as $b_n \to 0$. Our theoretical results show that filtering sharpens change-point localization whenever this gain dominates $b_n$, a trade-off that also depends on the relationship between the stationary-mean jump $\Delta_\eta$ and the corresponding VAR intercept jump $\Delta_\mu$.

Establishing these results also requires new technical arguments. First, the blocks used to estimate the common VAR dynamics depend on the preliminary change-point estimate and are therefore \textit{data dependent}. We address this by
establishing probability bounds uniformly over the admissible retained sets of observations. Second, filtering removes the stochastic temporal dependence, but induces a deterministic transition layer of at most $d$ observations after the change point, which in turn affects the comparison between the true change-point and nearby candidates. We provide an alignment condition (Assumption~\ref{assump:transition-alignment}) under which the filtered two-mean criterion attains the stated localization rate. In Appendix~\ref{sec:remove-transition-alignment}, we also provide examples of stable VAR models that violate this condition and develop a transition-corrected criterion that explicitly accounts for the boundary observations; its population criterion separates the true change-point without the alignment condition. More broadly, our analysis develops concentration and
localization arguments for high-dimensional dependent observations that may be useful in other time-series change-point problems.

Our second methodological contribution extends the procedure to multiple change points, going beyond the framework of \citet{bai2010common} and \citet{bhattacharjee2019change}. Because multiple change points introduce distinct structural challenges (Section~\ref{sec:multiple}), we adopt a localized screening strategy. Building on the rolling-window approach of \citet{bai2023multiple}, we generate candidate change points within overlapping local windows, prune redundant candidates using an information criterion, and refine the retained locations via the estimated common VAR dynamics. While our theoretical guarantees establish properties for the global minimizer of the information criterion, practical implementation uses a backward-elimination heuristic. Under the conditions in Remark~\ref{rem:complexity}, this implementation runs in linear time $\mathcal{O}(n)$ with respect to the sample size. Extensive numerical experiments corroborate the theoretical localization gains and demonstrate robustness to departures from common dynamics, including regime-dependent VAR parameters. Finally, an empirical application to a large panel of US macroeconomic indicators demonstrates the practical utility of the methodology.

The remainder of the paper is organized as follows. Section~\ref{sec:single} develops the single-change estimator and its theoretical properties, while Section~\ref{sec:multiple} extends the methodology to multiple change points. Sections~\ref{sec:simulations} and \ref{sec:fred_md_empirical} present the simulation results and a macroeconomics inidcators application, respectively. The Appendix contains all mathematical proofs, details on the transition-corrected refinement procedure, supplementary numerical results and implementation specifics, and documentation of the macroeconomic indicators data set and the corresponding variable transformations.

\paragraph{Notation:}\label{sec:notation}
We write $\mathbb Z$ and $\mathbb R$ for the integers and real numbers. For a set $S$, $|S|$ denotes its cardinality and $S^c$ its complement. We use $\mathbf1\{\cdot\}$ for the indicator function, $\operatorname{supp}(v)$ for the support of a vector, and $v_S$ for its restriction to $S$. For vectors, $\|\cdot\|_q$ denotes the $\ell_q$ norm for $q\ge1$; for matrices, $\|\cdot\|_2$ denotes the operator norm and $\|\cdot\|_F$ the Frobenius norm. We write $\|\cdot\|_{\psi_1}$ for the $\psi_1$-Orlicz norm. The symbols $H^\top$ and $H^*$ denote transpose and conjugate transpose, while $\operatorname{vec}$, $\otimes$, and $\langle\cdot,\cdot\rangle$ denote columnwise vectorization, Kronecker product, and the Euclidean inner product, respectively. We write $\mathbf{I}_r$ for the $r\times r$ identity matrix. An $\mathbf{I}$ without a subscript has the dimension required by context; $\mathbf1_r$ and $\mathbf0_r$ denote the all-ones and zero vectors. For symmetric or Hermitian matrices, $\lambda_{\min}$ and $\lambda_{\max}$ denote the smallest and largest eigenvalues, and $\rho(H)$ denotes the spectral radius of a square matrix $H$. We use $O(\cdot)$, $o(\cdot)$, $O_{\mathbb P}(\cdot)$, and $o_{\mathbb P}(\cdot)$ in their usual senses. Finally, $\lesssim$, $\gtrsim$, and $\asymp$ denote comparisons up to constant factors, while $\ll$ denotes little-o order and $\gg$ its reverse. For real numbers, $a\vee b:=\max\{a,b\}$, while $\lfloor\cdot\rfloor$ and $\lceil\cdot\rceil$ denote the floor and ceiling functions.

\section{Single change point detection}\label{sec:single}

We consider the single-change spliced model \eqref{eq:varmodelwithmean}--\eqref{eq:single-cp}
under the following innovation and stability conditions, assumed throughout the paper. 

\begin{assumption}\label{ass:innov}
The innovations $\{\varepsilon_t\}_{t\in\mathbb Z}$ are independent and
identically distributed Gaussian $N(0,\sigma^2 \mathbf{I}_p)$ random variables, with fixed $0<\sigma^2<\infty$.
The VAR polynomial $A(z):=\mathbf{I}_p-\sum_{i=1}^d A_i z^i$ satisfies
$\det A(z)\ne0$ for every $|z|\le1$. In particular, $A(1)=\mathbf{I}_p-\sum_{i=1}^d A_i$ is a nonsingular
$p\times p$ matrix. Each stationary segment in
\eqref{eq:varmodelwithmean} is the causal stationary solution driven by the same two-sided innovation sequence. 
\end{assumption}

Let $\mathcal A\in\mathbb R^{dp\times dp}$ denote the VAR companion matrix \citep{lutkepohl2005introduction}, whose  first block row is $(A_1,\ldots,A_d)$, its first block subdiagonal consists of $\mathbf{I}_p$, and all remaining blocks are zero. Note that for $d=1$, $\mathcal A\equiv A_1$.

\noindent 
Assumption~\ref{ass:innov} ensures that both stationary segments are defined for all $t\in\mathbb Z$, and share the centered process in \eqref{eq:common-centered}, with stationary means $\eta_j=A(1)^{-1}\mu_j$, $j=1,2$. The filtering identity \eqref{eq:intro-filtered-mean} shows that the first $d$ post-change observations form a transition layer requiring separate treatment in the refinement step. We measure the strength of temporal dependence by $\gamma_p:=\sum_{\ell\ge0}\|\Psi_\ell\|_2$. Since $A(1)^{-1}=\sum_{\ell\ge0}\Psi_\ell$, it follows that $\|A(1)^{-1}\|_2\le\gamma_p$ and $\|\eta_2-\eta_1\|_2\le\gamma_p\|\mu_2-\mu_1\|_2$.

\begin{figure}[t]
    \centering
    \includegraphics[width=\textwidth]
    {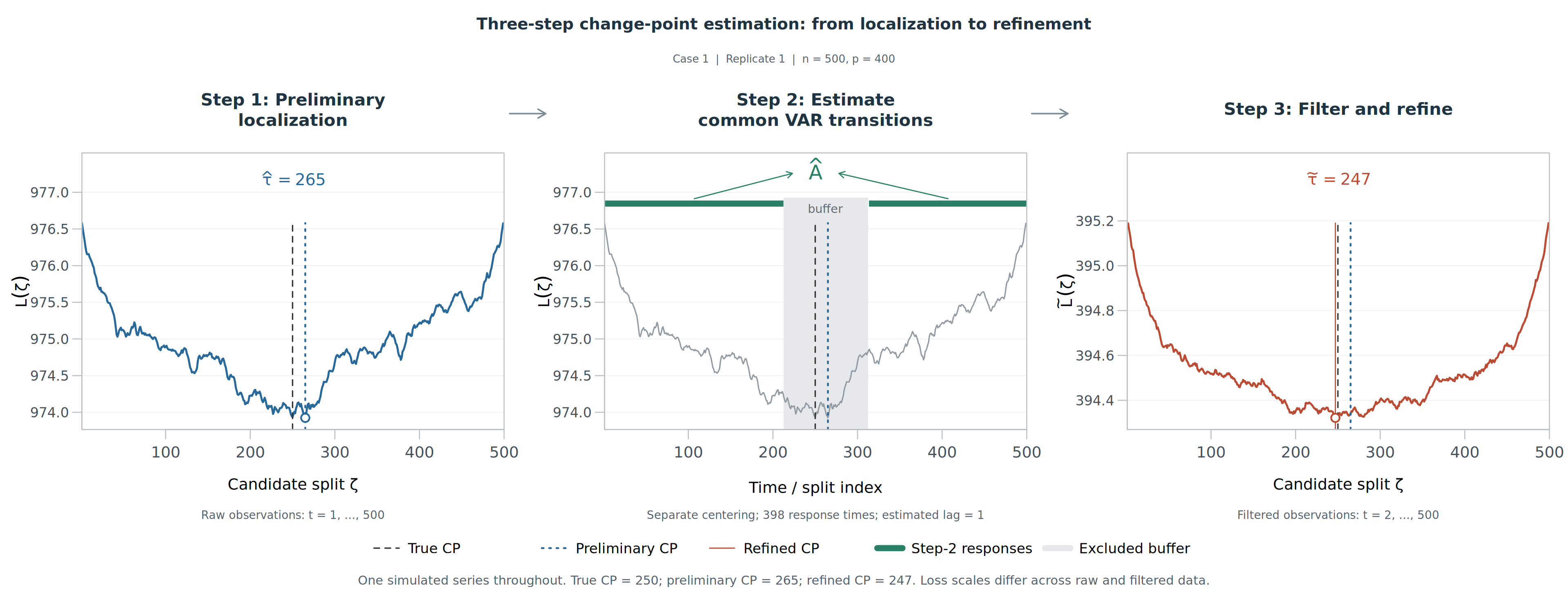}
    \caption{Illustration of the three-step estimation procedure for a simulated series with $n=500$, $p=400$, and true change point $\tau=250$. Left panel: The raw-data profile loss $L(\zeta)$ yields the preliminary estimate $\widehat\tau=265$.
Middle panel: Observations near $\widehat\tau$ are excluded (gray shading), and the retained observations (green bars) are centered separately within each segment to estimate the common VAR transition matrices. Right panel: The filtered profile loss $\widetilde L(\zeta)$ yields the refined estimate $\widehat\tau^{\,r}=247$.
Vertical lines mark the true and estimated change points; the raw and
filtered losses have different vertical scales.}
    \label{fig:single-cp-workflow}
\end{figure}

Our estimation procedure consists of three steps. Step~1 obtains a preliminary change-point estimate from the raw observations. Step~2 removes a buffer around this estimate and estimates the common sparse VAR transition matrices from the remaining observations. Step~3 filters the series using the estimated
transition matrices and refines the change-point estimate by applying a two-mean segmentation criterion to the filtered observations. Figure~\ref{fig:single-cp-workflow} illustrates how preliminary
localization, estimation of the common VAR coefficients, and
filtering followed by refinement are combined for a single
simulated series.

\subsection{A 3-step estimator for a single change point}\label{ssc:singlecptau}

\paragraph{Step 1: initial localization.}

Let $\lambda_1\ge0$ be the Step~1 ridge parameter.
For $\zeta\in\{1,\ldots,n-1\}$, estimate the two raw segment means by minimizing over $\eta_1,\eta_2\in\mathbb R^p$:
\begin{equation}\label{eq:Q-def}
Q(\eta_1,\eta_2;\zeta)=\frac1n\left[\sum_{t=1}^{\zeta}\|y_t-\eta_1\|_2^2+\sum_{t=\zeta+1}^{n}\|y_t-\eta_2\|_2^2\right]+\lambda_1\|\eta_1-\eta_2\|_2^2.
\end{equation}
Let $(\widehat\eta_1(\zeta),\widehat\eta_2(\zeta))$ denote a minimizer. The ridge penalty shrinks the estimated mean contrast toward zero, reducing the variability induced by estimating high-dimensional segment means, particularly when $p$ is large relative to $n$. This regularization aids in improving the stability of change-point localization, as supported by our numerical results (Appendix~\ref{app:single-cp-tables}). The penalty is used only to estimate the segment means, while candidate splits are compared using the unpenalized profiled loss 
\begin{equation}\label{eq:L-def}
L(\zeta)=\frac1n\left[\sum_{t=1}^{\zeta}\|y_t-\widehat\eta_1(\zeta)\|_2^2+\sum_{t=\zeta+1}^{n}\|y_t-\widehat\eta_2(\zeta)\|_2^2\right].
\end{equation}
Fix a trimming constant $c^\ast\in(0,1/2)$.  The preliminary estimator is given by
\begin{equation}\label{eq:tau0-def}
\widehat\tau\in\arg\min_{\zeta\in\mathcal T_n}L(\zeta),\qquad
\mathcal T_n:=\{\zeta\in\mathbb Z:c^\ast n\le \zeta\le (1-c^\ast)n\},
\end{equation}
with ties broken arbitrarily.

The buffer used in Step~2 is determined by the localization rate of $\widehat\tau$. Let $K_n\to\infty$ satisfy $K_n\frac{\gamma_p^2}{\|\eta_1-\eta_2\|_2^2}=o(n)$, and define
\begin{equation}\label{eq:single-buffer}
w:=\left\lceil K_n\frac{\gamma_p^2}{\|\eta_1-\eta_2\|_2^2}\right\rceil,
\qquad
\widehat\tau^-:=\widehat\tau-w,
\qquad
\widehat\tau^+:=\widehat\tau+w.
\end{equation}
By construction, the buffer is asymptotically larger than the Step~1 localization error, while remaining $o(n)$. In practice, a conservative data-driven choice may be used instead.

\paragraph{Step 2: Buffered estimation of the common VAR dynamics.}
Define the retained response-time sets
\begin{equation}\label{eq:single-retained-sets}
I_1:=\{d+1,\ldots,\widehat\tau^-\},\qquad
I_2:=\{\widehat\tau^+ + d+1,\ldots,n\},\qquad I:=I_1\cup I_2.
\end{equation}
On the event $|\widehat\tau-\tau|\le w$, the additional $d$ observations excluded after the right buffer ensure that every retained response and its $d$ lagged covariates belong to the same true regime. Let
\[
n_1:=|I_1|=\widehat\tau^- -d,
\qquad n_2:=|I_2|=n-\widehat\tau^+-d,
\qquad n_{\rm obs}:=n_1+n_2.
\]
For $t=d+1,\ldots,n$, define the lag vector
$x_t:=(y_{t-1}^\top,\ldots,y_{t-d}^\top)^\top\in\IR^{dp}$. Let $Y_I=(y_t^\top)_{t\in I}$, $X_I=(x_t^\top)_{t\in I}$, and $E_I=(\varepsilon_t^\top)_{t\in I}$, wherein the subscript $I$ restricts rows to the retained observations. 

Define
\[
B:=\begin{pmatrix}A_1^\top\\ \vdots\\ A_d^\top\end{pmatrix}\in\IR^{dp\times p},\qquad
\beta^\star:=\vecop(B),
\]
and let $S:=\operatorname{supp}(\beta^\star)$. Throughout, $s\ge1$ is a deterministic sparsity upper bound satisfying $|S|\le s$; thus the curvature and rate expressions are also defined when $S$ is empty.
 On the event $|\widehat\tau-\tau|\le w$, the retained observations satisfy
\begin{equation}\label{eq:single-integrated-regression}
Y_I=X_IB+u_1\mu_1^\top+u_2\mu_2^\top+E_I,
\end{equation}
where
\[
u_1:=\begin{pmatrix}\mathbf1_{n_1}\\\mathbf0_{n_2}\end{pmatrix},\qquad
u_2:=\begin{pmatrix}\mathbf0_{n_1}\\\mathbf1_{n_2}\end{pmatrix}.
\]
We profile out the regime-specific intercepts before estimating the sparse transition matrices $A_j, j=1,\cdots,d$. Define
\[
X^\mu:=[\mathbf{I}_p\otimes u_1,\ \mathbf{I}_p\otimes u_2],\qquad
P_{\rm blk}:=\begin{pmatrix}n_1^{-1}J_{n_1}&0\\0&n_2^{-1}J_{n_2}\end{pmatrix},
\]
where $J_m=\mathbf1_m\mathbf1_m^\top$, and let
\[
H_0:=X^\mu\{(X^\mu)^\top X^\mu\}^{-1}(X^\mu)^\top=\mathbf{I}_p\otimes P_{\rm blk},\qquad
G_{\rm blk}:=\mathbf{I}_{n_{\rm obs}}-P_{\rm blk}.
\]
Set $\widetilde{\mathcal Y}_I:=(\mathbf{I}-H_0)\vecop(Y_I)=\vecop(G_{\rm blk}Y_I)$ and $
\widetilde Z_I:=(\mathbf{I}-H_0)(\mathbf{I}_p\otimes X_I)=\mathbf{I}_p\otimes(G_{\rm blk}X_I)$.
Let $\lambda_n\ge0$ denote the Step~2 Lasso parameter. We estimate the common sparse VAR coefficients by solving the following optimization problem:
\begin{equation}\label{eq:single-lasso}
\widehat\beta\in\arg\min_{\beta\in\IR^{dp^2}}\left\{\frac1{n_{\rm obs}}\|\widetilde{\mathcal Y}_I-\widetilde Z_I\beta\|_2^2+\lambda_n\|\beta\|_1\right\}.
\end{equation}
Reshaping $\widehat\beta$, we obtain
$\widehat B=\bigl(\widehat A_1^\top \cdots \widehat A_d^\top\bigr)^\top$.

\paragraph{Step 3: filtered two-mean refinement.}
Using the Step~2 estimates, define the observed and oracle filtered series,
\begin{equation}\label{eq:single-filtered-series}
\widehat Y_t:=y_t-\sum_{i=1}^d\widehat A_i y_{t-i},\qquad
Y_t:=y_t-\sum_{i=1}^d A_i y_{t-i},\qquad t=d+1,\ldots,n.
\end{equation}
Let
\[
\mathcal T_n^{(3)}:=\{\zeta\in\mathbb Z:c_{\rm tr}n\le \zeta\le (1-c_{\rm tr})n\},\qquad 0<c_{\rm tr}<c^\ast .
\]
The requirement \(c_{\rm tr}<c^\ast\) guarantees, via
Assumption~\ref{assump:tau-domain}, that \(\tau\) is an interior point of
\(\mathcal T_n^{(3)}\) with a margin of at least \((c^\ast-c_{\rm tr})n\); the
refinement search therefore covers a full neighborhood of \(\tau\).
Let $\nu_1,\nu_2\in\mathbb R^p$ be candidate filtered means and $\lambda_3\ge0$ the Step~3 ridge parameter.
For each $\zeta\in\mathcal T_n^{(3)}$, define
\begin{equation}\label{eq:tildeQ-def}
\widetilde Q(\nu_1,\nu_2;\zeta)=\frac1{n-d}\left[\sum_{t=d+1}^{\zeta}\|\widehat Y_t-\nu_1\|_2^2+\sum_{t=\zeta+1}^{n}\|\widehat Y_t-\nu_2\|_2^2\right]+\lambda_3\|\nu_1-\nu_2\|_2^2.
\end{equation}
Let $(\widehat\nu_1(\zeta),\widehat\nu_2(\zeta))$ be a minimizer of \eqref{eq:tildeQ-def}.  As in Step~1, candidate splits are compared using the
unpenalized profiled loss
\begin{equation}\label{eq:tildeL-def}
\widetilde L(\zeta)=\frac1{n-d}\left[\sum_{t=d+1}^{\zeta}\|\widehat Y_t-\widehat\nu_1(\zeta)\|_2^2+\sum_{t=\zeta+1}^{n}\|\widehat Y_t-\widehat\nu_2(\zeta)\|_2^2\right],
\end{equation}
and define the refined estimator 
\begin{equation}\label{eq:tau-hat}
\widehat\tau^{\,r}\in\arg\min_{\zeta\in\mathcal T_n^{(3)}}\widetilde L(\zeta).
\end{equation}
This criterion mirrors its Step~1 counterpart, but its analysis is different, because the oracle filtered series contains a deterministic transition layer: its mean is $\mu_1$ before the change and $\mu_2$ after the first $d$ post-change observations, with the intervening means determined by the VAR dynamics.

\begin{remark}[Role of ridge regularization]\label{rem:ridge-mean-jump}
Our framework allows for dense mean changes, making ridge regularization a natural choice for stabilizing estimation of the mean contrast without imposing sparsity on the affected coordinates. This is particularly useful when the mean shift is small relative to estimation noise. The numerical results in Appendix~\ref{app:single-cp-tables} support this motivation: removing ridge regularization from Steps~1 and~3 substantially worsens localization accuracy. In this comparison, the preliminary estimate and the Step~2 VAR filter are re-estimated without ridge regularization in Step~1, while the Step~2 Lasso penalty is retained. Thus, the observed gains provide finite-sample support for regularizing potentially dense mean changes.
\end{remark}

\subsection{Theoretical guarantees for the refined estimator \texorpdfstring{\(\widehat{\tau}^r\)}{tau}}

Throughout this subsection, let
$\eta_j=(\mathbf{I}_p-\sum_{\ell=1}^d A_\ell)^{-1}\mu_j$, $j=1,2$, and define
$\Delta_\eta:=\eta_2-\eta_1$ and $\Delta_\mu:=\mu_2-\mu_1$. Recall that
$\gamma_p:=\sum_{\ell\ge0}\|\Psi_\ell\|_2$ measures the strength of temporal dependence.

\subsubsection{Preliminary change-point localization (Step 1)}
\label{sec:theory-step1}

We first establish the localization rate of the preliminary estimator $\widehat\tau$ defined in \eqref{eq:tau0-def} under the following assumptions.
\begin{assumption}\label{H1}
It is assumed that 
\[
\frac{n}{p\gamma_p^2}\|\Delta_\eta\|_2^2
=
\frac{n}{p\gamma_p^2}
\left\|
\left(\mathbf{I}_p-\sum_{\ell=1}^d A_\ell\right)^{-1}
(\mu_2-\mu_1)
\right\|_2^2
\longrightarrow\infty .
\]
\end{assumption}
\begin{assumption}\label{assump:tau-domain}
For some constant \(c^\ast\in(0,1/2)\),
$c^\ast n\le \tau\le (1-c^\ast)n$.
\end{assumption}

\begin{remark}
Assumption~\ref{H1} is the signal-to-noise ratio (SNR) condition for the preliminary estimator. Since Step~1 operates directly on $y_t$, its signal is the stationary-mean shift $\Delta_\eta$. The condition requires the average squared signal $\|\Delta_\eta\|_2^2/p$ to dominate $\gamma_p^2/n$, thereby accounting jointly for dimension and temporal dependence. When $\gamma_p=\mathcal{O}(1)$, it reduces to $n\|\Delta_\eta\|_2^2/p\to\infty$, the usual form of a high-dimensional mean-shift detectability condition. Related dependence conditions are discussed by \citet[Example~6.2]{bhattacharjee2019change}. Compared with the conditions in \citet{bai2010common}, Assumption~\ref{H1} combines the signal,
dimension, and dependence requirements into a single condition.
Assumption~\ref{assump:tau-domain} ensures that both regimes contain order $n$ observations.
\end{remark}

\begin{proposition}\label{prop:tau0}
Suppose Assumptions~\ref{ass:innov}, \ref{H1}, and \ref{assump:tau-domain} hold and $0\le\lambda_1\le\bar\lambda$ for a fixed $\bar\lambda<\infty$. Then,
\begin{equation}\label{eq:step1-rate}
|\widehat\tau-\tau|
=O_{\IP}\left(\frac{\gamma_p^2}{\|\Delta_\eta\|_2^2}\right).
\end{equation}
\end{proposition}

The proof of Proposition~\ref{prop:tau0} leverages the empirical-process machinery underlying Theorem~3.2.5 of \citet{van1996weak}. Specifically,
we establish the requisite modulus bound \eqref{asf:eq:vdvw_fluct} and quadratic drift condition \eqref{asf:eq:vdvw_drift}, whose balance
yields the localization rate in \eqref{eq:step1-rate}. We provide a self-contained argument that explicitly accounts for the effects of serial dependence on both stochastic fluctuations and population drift. The resulting proof techniques may be of independent interest for change-point problems involving high-dimensional, temporally dependent observations.

The proof proceeds in three steps. First, Lemma~\ref{asf:lem:profile} reduces the profiled criterion to a weighted contrast, with ridge
regularization entering only through deterministic weights. Second, Propositions~\ref{asf:prop:fluctuation} and \ref{asf:prop:drift} establish uniform fluctuation bounds and quadratic population separation,
respectively. The centered fluctuations decompose into a Gaussian linear component, controlled by convolution variance bounds and Gaussian maximal
inequalities, and a quadratic component in the second Gaussian Wiener chaos, controlled through an exact variance identity for Gaussian bilinear forms and hypercontractivity. The drift analysis separately controls the dependence-induced noise contribution, which is dominated by the signal under Assumption~\ref{H1}. Third, a dyadic-shell argument combines these bounds, using the Borell-Tsirelson-Ibragimov-Sudakov inequality and a $\psi_1$ maximal bound, to establish the exponential tail inequality in ~\ref{asf:eq:tau0-tail}. This yields the stated localization rate and provides the tail control needed for simultaneous localization of multiple change points in Section~\ref{sec:multiple}. Details are given in Appendix~\ref{app:step1-proof}.

\subsubsection{Estimation of the common VAR transition matrices (Step 2)}
\label{sec:theory-step2}

Step~2 estimates these matrices from observations buffered away from the preliminary change point estimate $\widehat{\tau}$. Define the ``purity" event $\mathcal E_{\mathrm{pure}}:=\{|\widehat\tau-\tau|\le w\}$. By Proposition~\ref{prop:tau0} and the choice of $K_n\to\infty$ in \eqref{eq:single-buffer},
\[
P(\mathcal E_{\mathrm{pure}})\to1,
\qquad
w=o(n).
\]
On $\mathcal E_{\mathrm{pure}}$, the retained blocks $I_1$ and $I_2$ defined in \eqref{eq:single-retained-sets} contain observations from a single regime; the additional offset by $d$ ensures the same for all lagged covariates.

After profiling out the regime-specific intercepts by block demeaning, the Lasso estimator \eqref{eq:single-lasso} can equivalently be written as
\begin{equation}\label{eq:quad-argmin}
\widehat\beta
\in\arg\min_{\beta}\left\{-2\beta^\top\widehat\gamma
+\beta^\top\widehat\Gamma\beta
+\lambda_n\|\beta\|_1\right\},
\end{equation}
where, with $G_{\mathrm{blk}}=\diag(G_{n_1},G_{n_2})$ and
$G_r=\mathbf{I}_r-r^{-1}\mathbf1_r\mathbf1_r^\top$,
\[
Y_I^G:=G_{\mathrm{blk}}Y_I,\qquad
X_I^G:=G_{\mathrm{blk}}X_I,\qquad
E_I^G:=G_{\mathrm{blk}}E_I,
\]
and
\begin{equation}\label{eq:gamma-hat}
\widehat\gamma=\frac{1}{n_{\mathrm{obs}}}
\vecop\!\bigl((X_I^G)^\top Y_I^G\bigr), \qquad
\widehat\Gamma=\frac{1}{n_{\mathrm{obs}}}
\mathbf{I}_p\otimes\bigl((X_I^G)^\top X_I^G\bigr).
\end{equation}
On $\mathcal E_{\mathrm{pure}}$,
\begin{equation}\label{eq:gamma-minus-Gamma-beta-IG}
\widehat\gamma-\widehat\Gamma\beta^\star=
\frac{1}{n_{\mathrm{obs}}}
\vecop\!\bigl((X_I^G)^\top E_I^G\bigr).
\end{equation}

We use the following restricted strong convexity (RSC) condition throughout.
\begin{definition}\label{def:RSC}
A symmetric matrix $\Gamma$ satisfies the restricted strong convexity condition with curvature $\alpha > 0$ and tolerance $\kappa \ge 0$, denoted $\Gamma \sim \mathrm{RSC}(\alpha, \kappa)$, if
\[
\theta^\top \Gamma \theta \ge \alpha \|\theta\|_2^2 - \kappa \|\theta\|_1^2 \qquad \text{for all } \theta.
\]
\end{definition}

For the centered lag process
$\widetilde x_t=(e_{t-1}^\top,\ldots,e_{t-d}^\top)^\top$, let $f_x$ denote its spectral density and define
\[
M_x:=\sup_{\omega\in[-\pi,\pi]}\lambda_{\max}\{f_x(\omega)\},
\qquad L_x:=2\pi M_x.
\]
These quantities may depend on $(n,p)$. Their finiteness follows from Assumption~\ref{ass:innov}, as verified below; no uniform bound in $(n,p)$ is imposed.

Let
\[
\mu_{\min}(A)=\min_{|z|=1}\lambda_{\min}\{A(z)^*A(z)\},
\qquad
\mu_{\max}(A)=\max_{|z|=1}\lambda_{\max}\{A(z)^*A(z)\}.
\]
and define 
\begin{equation}\label{eq:single-step2-constants}
\alpha=\frac{\sigma^2}{4\mu_{\max}(A)},\qquad
\chi_x=\frac{L_x}{\alpha},\qquad \kappa=\frac{\alpha}{64s}.
\end{equation}
Here $\alpha$ provides a lower curvature bound, while $\chi_x$ measures the spectral cost of concentration. We select $\kappa$ so that the Lasso cone retains sufficient curvature. Since the lag-stacked spectral density satisfies $\lambda_{\max}\{f_x(\omega)\}\le d\sigma^2/\{2\pi\mu_{\min}(A)\}$, we have $L_x\le d\sigma^2/\mu_{\min}(A)$ and $\chi_x\le4d\,\mu_{\max}(A)/\mu_{\min}(A)$; in particular, $M_x<\infty$ under Assumption~\ref{ass:innov}.

We also require the following deviation condition. Let $(\gamma,\Gamma)$ denote a generic pair consisting of the normalized design--response cross-product vector and Gram matrix of a profiled regression (corresponding to $(\widehat\gamma,\widehat\Gamma)$ in the single-change setting and $(\widehat\gamma_m,\widehat\Gamma_m)$ in the multiple-change setting).

\begin{definition}\label{def:deviation}
For a sample size $N$ and logarithmic factor $\varpi$, the pair $(\gamma, \Gamma)$ satisfies the \textit{deviation condition} relative to parameter $\beta^\star$ if
\[
\|\gamma - \Gamma \beta^\star\|_\infty \le Q_{\mathrm{dev}} \sqrt{\frac{\varpi}{N}},
\]
where
\begin{equation}\label{eq:Qdev-explicit}
Q_{\mathrm{dev}} = c_0 \left[ \sigma^2 \left( 6 + 4 \frac{\mu_{\max}(A)}{\mu_{\min}(A)} \right) + C_\Psi \sigma \sqrt{L_x} \right],
\end{equation}
for sufficiently large universal constants $c_0, C_\Psi > 0$.
\end{definition}

\begin{remark}\label{rem:deviation-constants}
In \eqref{eq:Qdev-explicit}, the first term matches the deviation bound of \citet{basu2015regularized}. However, only the second term, $Q_{\mathrm{dev}} \gtrsim \sigma \sqrt{L_x}$, is strictly needed in the proof of Proposition~\ref{prop:uniform-block-concentration}; the first term can be dropped without affecting any theoretical results.
\end{remark}

Define
\begin{equation}\label{eq:Qbar-def}
\bar Q:=Q_{\mathrm{dev}}/\alpha,
\qquad
\ell_{2,n}:=1+\log(dp^2)+4\log n\asymp\log(ndp^2).
\end{equation}
The quantity $\bar Q$ captures the spectral dependence of the VAR
estimation bounds and is allowed to diverge.

\begin{proposition}\label{thm:S2}
Suppose the conditions of Proposition~\ref{prop:tau0} hold, and use the buffer in \eqref{eq:single-buffer}, with $K_n\to\infty$ and
$K_n\gamma_p^2/\|\Delta_\eta\|_2^2=o(n)$. Let $d$ be fixed and $\beta^\star$ be $s$-sparse, with $s\ge1$. Thus, $w+d=o(n)$ and $n_{\mathrm{obs}}=n-2w-2d\asymp n$. Suppose, for a sufficiently large universal constant $C_{\mathrm{S2}}$,
\begin{equation}\label{eq:n-regime}
n_{\mathrm{obs}}\ge C_{\mathrm{S2}}
\left[\chi_x^2\{s\{1+\log(dp)\}+\log n\}+\chi_x+\ell_{2,n}\right].
\end{equation}
Select
$\lambda_n=C_\lambda Q_{\mathrm{dev}}\sqrt{\ell_{2,n}/n_{\mathrm{obs}}}$,
{with a fixed $C_\lambda\ge4$.}
Define $\mathcal E_{\mathrm{RSC}}:=\{\widehat\Gamma\sim\mathrm{RSC}(\alpha,\kappa)\}$,
$\mathcal E_{\mathrm{dev}}:=\{\|\widehat\gamma-\widehat\Gamma\beta^\star\|_\infty
\le Q_{\mathrm{dev}}\sqrt{\ell_{2,n}/n_{\mathrm{obs}}}\}$, and
$\mathcal E_{\mathrm{S2}}:=\mathcal E_{\mathrm{pure}}\cap\mathcal E_{\mathrm{RSC}}\cap\mathcal E_{\mathrm{dev}}$.
Then
\begin{equation}\label{eq:single-S2-probability}
P(\mathcal E_{\mathrm{S2}}^c)
\le P(\mathcal E_{\mathrm{pure}}^c)+Cn^{-2}\longrightarrow0.
\end{equation}
On $\mathcal E_{\mathrm{S2}}$, every minimizer satisfies
\begin{equation}\label{eq:single-S2-oracle}
\begin{gathered}
\|\widehat\beta-\beta^\star\|_1
\le\frac{12s\lambda_n}{\alpha}, \qquad
\|\widehat B-B\|_F=
\|\widehat\beta-\beta^\star\|_2
\le\frac{3\sqrt{s}\lambda_n}{\alpha},\\
(\widehat\beta-\beta^\star)^\top\widehat\Gamma
(\widehat\beta-\beta^\star)
\le\frac{9s\lambda_n^2}{2\alpha}, \qquad
\widehat\beta-\beta^\star\in\mathcal C(S),
\end{gathered}
\end{equation}
where $S=\operatorname{supp}(\beta^\star)$ and
$\mathcal C(S):=\{v:\|v_{S^c}\|_1\le3\|v_S\|_1\}$. {Moreover, on the same event,}
\[
\frac{\|G_{\mathrm{blk}}X_I(\widehat B-B)\|_F^2}{n_{\mathrm{obs}}}
\le\frac{9s\lambda_n^2}{2\alpha}.
\]
Consequently,
\[
\|\widehat B-B\|_F
=O_{\IP}\left(\bar Q\sqrt{\frac{s\ell_{2,n}}{n_{\mathrm{obs}}}}\right).
\]
\end{proposition}

The proof of Proposition~\ref{thm:S2} follows the framework of \citet[Section~4.2]{basu2015regularized}, which combines probabilistic
control of the Gram matrix and score with deterministic Lasso inequalities.
Our main technical contribution is to extend this framework to data-dependent blocks with separately profiled intercepts. Lemma~\ref{asf:lem:S2-profile} identifies the resulting Gram matrix
and score, while Proposition~\ref{prop:uniform-block-concentration} establishes uniform RSC and score bounds over all admissible block pairs.
These bounds use Gaussian quadratic-form concentration and sparse nets, while accounting for the loss of curvature and nonzero score mean induced by profiling, as well as the additional complexity of selecting block endpoints. Proposition~\ref{asf:prop:S2-eventwise} then applies the deterministic Lasso argument to obtain the stated estimation bounds. Details are given in Appendix~\ref{asf:sec:S2-proof}.

The prediction bound applies to the retained, block-demeaned observations, whereas Step~3 relies on the cone and Frobenius bounds to control the estimated filter over the full sample. As in \citet{basu2015regularized}, the key ingredients are RSC and score bounds, which in our setting must hold \textit{uniformly over the data-dependent} retained blocks.

\subsubsection{Refined change point localization (Step 3)}
\label{sec:theory-step3}

In Step~3, we apply the two-mean profile criterion to the filtered series $\{\widehat Y_t\}$ defined in \eqref{eq:single-filtered-series}. Unlike Step~1, the relevant jump after filtering is the structural intercept change
$\Delta_\mu$. Because the observed series splices two stationary segments, however, filtering the first $d$ post-change observations mixes pre- and post-change means. Specifically, the oracle-filtered series satisfies
\begin{equation}\label{eq:intro-filtered-mean}
Y_t=
\begin{cases}
\mu_1+\varepsilon_t, & d+1\le t\le\tau,\\
\mu_2+c_q+\varepsilon_t, & t=\tau+q,\quad 1\le q\le d,\\
\mu_2+\varepsilon_t, & \tau+d<t\le n,
\end{cases}
\qquad c_q:=\left(\sum_{\ell=q}^d A_\ell\right)\Delta_\eta .
\end{equation}
Note that filtering removes the \textit{stochastic temporal dependence}, but induces a \textit{deterministic transition layer} of length at most $d$.

To quantify the effect of estimating the filter, let
\begin{equation}\label{eq:single-ell2-rho-alpha}
r_n:=s\ell_{2,n},\qquad \rho_n:=\bar Q\gamma_p\sqrt{\frac{r_n}{n}},\qquad
b_n:=\bar Q^2\gamma_p^2\frac{r_n^2}{n}.
\end{equation}
For a sufficiently large constant $C_B$, define
\[
\mathcal D_n := \left\{D:\vecop(D)\in\mathcal C(S),\ \|D\|_F\le C_B\bar Q\sqrt{r_n/n} \right\},
\]
wherein $S:=\operatorname{supp}(\beta^\star)$, and $\mathcal C(S)$ denotes the cone defined in Proposition~\ref{thm:S2}. Under the conditions of Proposition~\ref{thm:S2}, one can see that $P\{\widehat B-B\in\mathcal D_n\}\to1$.  Step~3 combines this coefficient bound with uniform partial-sum bounds for the centered lag process to control the estimated filter over the full sample.

\begin{assumption}\label{SNR3}
The intrinsic filtered innovation fluctuations are negligible relative to the structural jump:
\begin{equation}\label{eq:single-S3-innovation-snr}
\frac{p\log n}{n\|\Delta_\mu\|_2^2}\to0.
\end{equation}
The target Step~3 localization radius is \(o(n)\):
\begin{equation}\label{eq:single-S3-radius-small}
\frac{(1+b_n)}{n\|\Delta_\mu\|_2^2}\to0.
\end{equation}
\end{assumption}

An additional condition is required for the transition layer. For $q=1,\cdots,d$, let $C_q:=\sum_{\ell=1}^q c_\ell$ (with $c_q$ defined in \eqref{eq:intro-filtered-mean}) and
$\nu_n:=\max_{1\le q\le d}\|C_q\|_2/
\{q\|\Delta_\mu\|_2\}$.

\begin{assumption}\label{assump:transition-alignment}
There exists a constant \(c_{\mathrm{al}}>0\) such that for every \(q=1,\ldots,d\),
\begin{equation}\label{eq:single-transition-alignment}
q\|\Delta_\mu\|_2^2
+
2\langle \Delta_\mu,C_q\rangle
\ge
c_{\mathrm{al}}q\|\Delta_\mu\|_2^2.
\end{equation}
Moreover, $\rho_n\to0$, $\rho_n\nu_n\to0,\ \nu_n^2/n\to 0$.
\end{assumption}

\begin{remark}
The condition $\rho_n\to0$ in
Assumption~\ref{assump:transition-alignment} connects the accuracy of Step~2 with the perturbation control required in Step~3.
Indeed, $\rho_n=\bar Q\gamma_p\sqrt{s\ell_{2,n}/n}$
is the Step~2 Frobenius-error rate in
Proposition~\ref{thm:S2} multiplied by the dependence factor
$\gamma_p$. When $\gamma_p=O(1)$, this condition reduces to
$\bar Q^2s\ell_{2,n}/n\to0$, the usual requirement that
the Step~2 estimation-error bound vanish.
This asymptotic requirement is not implied by the sufficient
RSC sample-size bound in \eqref{eq:n-regime} alone.
When $\gamma_p$ is allowed to diverge, the additional factor
$\gamma_p$ accounts for the amplification of the filter-estimation
error by temporal dependence.
\end{remark}

\begin{remark}
Assumption~\ref{assump:transition-alignment} ensures identification of the true change point under the filtered two-mean criterion. Moving a candidate split from $\tau$ to $\tau+q$ assigns the first $q$ transition observations to the pre-change segment, yielding a leading population loss increase of $q\|\Delta_\mu\|_2^2+2\langle\Delta_\mu,C_q\rangle$. The condition requires this increase to remain proportional to
$q\|\Delta_\mu\|_2^2$, without restricting the magnitude of the transition layer itself. The conditions involving $\nu_n$ ensure that this separation is preserved under filter estimation and long-segment averaging. Appendix~\ref{sec:remove-transition-alignment} provides examples of stable VAR models that violate alignment and develops a transition-corrected
refinement that explicitly estimates the transition layer and restores population identification without this condition
(Lemma~\ref{lem:tc-excess}).
\end{remark}

\begin{remark}
The signal conditions for Steps~1 and 3 involve different jumps. Step~1 operates on the raw series and therefore depends on the stationary-mean jump $\Delta_\eta$, whereas Step~3 operates on the filtered series and depends on the structural intercept jump $\Delta_\mu=A(1)\Delta_\eta$. If $A(1)$ is well conditioned in the direction of the change, so that
$\|\Delta_\mu\|_2\asymp\|\Delta_\eta\|_2$, the Step~3 signal condition differs from its Step~1 counterpart essentially by the logarithmic margin required for the candidate search. The quantities $b_n$ and $\rho_n$ account for estimation of the common VAR filter, while $\nu_n$ controls the effect of the finite transition layer.
\end{remark}

\begin{theorem}\label{prop:tau-rate}
Suppose the conditions of Proposition~\ref{thm:S2} and
Assumptions~\ref{SNR3} and
\ref{assump:transition-alignment} hold, and that $\tau$ belongs to the Step~3 candidate set. If $d$ is fixed and $0\le\lambda_3\le\bar\lambda_n$ for a deterministic $\bar\lambda_n\to0$, then
\begin{equation}\label{eq:refined-rate}
|\widehat\tau^{\,r}-\tau| = O_{\IP}\left(
\frac{1+b_n}{\|\Delta_\mu\|_2^2} \right).
\end{equation}
\end{theorem}

Two features distinguish the proof of Theorem~\ref{prop:tau-rate} from that of Proposition~\ref{prop:tau0}. First, filtering creates a deterministic transition layer of length at most $d$. Lemma~\ref{asf:lem:S3-det-drift-aligned} derives the corresponding population drift, which remains positive under the alignment
condition \eqref{eq:single-transition-alignment}. Second, the filter is estimated from the same observations, requiring uniform control
over the Step~2 coefficient-error set $\mathcal D_n$. Lemma~\ref{asf:lem:S3-deterministic-D} establishes that the population separation is preserved uniformly over this set.

The stochastic analysis combines Doob's maximal inequality for innovation partial sums, coordinatewise bounds and the Lasso cone for data-dependent filtering errors, and chaining for the centered quadratic innovation term
(Lemmas~\ref{asf:lem:S3-localized-EP} and
\ref{asf:lem:S3-quad-innov-shell}). Proposition~\ref{asf:prop:S3-fluct-new} combines these bounds, allowing for correlation between innovations and filtering errors. A dyadic-shell argument then yields the refined localization rate
\eqref{eq:refined-rate}; see Appendix~\ref{asf:sec:S3-final}.

Unlike Step~1, Step~3 requires $\lambda_3\to0$, since the alignment condition concerns the unregularized population criterion. Lemma~\ref{asf:lem:S3-det-drift-aligned} shows that this requirement, including $\lambda_3=0$, preserves positive population drift. The bound holds uniformly over $\lambda_3\in[0,\bar\lambda_n]$,
allowing data-driven tuning within this range.

\begin{remark}\label{rem:single-rate-comparison}
Comparing the established localization bounds, filtering removes the
explicit dependence factor $\gamma_p^2$, while estimation of the common
VAR dynamics contributes the additional term $b_n$. When $b_n=O(1)$,
the refined bound has the same order as its oracle-filtered counterpart;
when $b_n=o(1)$, the contribution from estimating the filter is
asymptotically negligible. More generally, refinement yields a
strictly smaller localization bound whenever
\[
\frac{1+b_n}{\|\Delta_\mu\|_2^2}
=o\left(
\frac{\gamma_p^2}{\|\Delta_\eta\|_2^2}
\right).
\]

For example, under the remaining theorem assumptions, sufficient
conditions are $\gamma_p\to\infty$,
$\|\Delta_\eta\|_2\lesssim\|\Delta_\mu\|_2$, and
$\bar Qs\ell_{2,n}=o(\sqrt n)$. Indeed, the definition of $b_n$ gives
$(1+b_n)/\gamma_p^2
=\gamma_p^{-2}+\bar Q^2s^2\ell_{2,n}^2/n=o(1)$.
These conditions require sufficient sample size relative to sparsity of the VAR dynamics and the spectral cost of filter estimation, and prevent $A(1)$ from
asymptotically attenuating the jump direction. They allow, for example, increasing persistence in directions outside the mean change.
With bounded dependence and comparable jump norms, filtering may
instead yield a constant-factor improvement.
\end{remark}

\begin{remark}
The numerical results provide complementary evidence for the precision gains achieved through filtering: as temporal dependence strengthens, measured by the spectral radius $\rho(\mathcal{A})$, the Post-to-Pre RMSE ratio decreases and remains consistently below unity, where \textnormal{Pre} and \textnormal{Post} denote the preliminary and refined change-point estimates, respectively, and \textnormal{RMSE} denotes root mean squared error. These finite-sample findings complement our theoretical results, where the magnitude of improvement depends on both temporal persistence and the mapping $\Delta_\mu = A(1) \Delta_\eta$.
\end{remark}

To illustrate the alignment condition and its relation to the VAR dynamics,
we conclude with the VAR$(1)$ case, where the transition layer consists of a
single observation.
\begin{remark}[Example: Alignment condition for VAR(1) process]
For a VAR$(1)$, the transition vector and its relative magnitude are
$c_1=A_1(\mathbf{I}_p-A_1)^{-1}\Delta_\mu,
\ \nu_n=
\frac{\|A_1(\mathbf{I}_p-A_1)^{-1}\Delta_\mu\|_2}
     {\|\Delta_\mu\|_2}$.
Assumption~\ref{assump:transition-alignment} reduces to
\[
\|\Delta_\mu\|_2^2
+2\langle\Delta_\mu,
A_1(\mathbf{I}_p-A_1)^{-1}\Delta_\mu\rangle
\ge c_{\mathrm{al}}\|\Delta_\mu\|_2^2.
\]
If $A_1$ is diagonal with entries $a_j^{(A)}$, the left-hand side becomes $\sum_{j=1}^p
\frac{1+a_j^{(A)}}{1-a_j^{(A)}}[(\Delta_\mu)_j]^2$, where $(\Delta_\mu)_j$ is the $j$th coordinate of $\Delta_\mu$, making the condition a jump-weighted alignment requirement. In particular, if $\max_j|a_j^{(A)}|\le\rho<1$, then it holds with
$c_{\mathrm{al}}=(1-\rho)/(1+\rho)$. In the scalar case, write $a$ for the autoregressive coefficient. For a scalar AR$(1)$, the alignment
condition reduces to $(1+a)/(1-a)\ge c_{\mathrm{al}}>0$, while the robustness
conditions become
\[
\bar Q\gamma_p\sqrt{\frac{r_n}{n}}
\left|\frac{a}{1-a}\right|\to0,
\qquad
\frac1n\left|\frac{a}{1-a}\right|^2\to0.
\]
\end{remark}

\section{Multiple Change-Point Detection}\label{sec:multiple}

Let $m_0$ denote the true number of change points, with ordered locations $0 = \tau_0^\star < \tau_1^\star < \cdots < \tau_{m_0}^\star < \tau_{m_0+1}^\star = n$. We extend the three-step procedure to the multiple-change model:
\begin{equation}\label{eq:mean-shift-1}
y_t = \sum_{k=1}^{m_0+1} \eta_k \one\{\tau_{k-1}^\star < t \le \tau_k^\star\} + e_t, \qquad t = 1, \ldots, n,
\end{equation}
where $\eta_k = A(1)^{-1}\mu_k$ denotes the stationary mean of regime $k$, $\mu_k \in \mathbb{R}^p$ is the regime-specific intercept, and $e_t$ is the zero-mean common stationary error process defined in \eqref{eq:common-centered}.

The three-step structure remains unchanged: Step~1 obtains preliminary change-point estimates, Step~2 estimates the VAR transition matrices common across regimes, and Step~3 filters the series and refines the estimated locations from Step 1. The main additional challenge lies in Step~1, which must now determine both the number and locations of the change points. To address this problem, we adapt the two-stage procedure of \citet{bai2023multiple}: a rolling-window step generates candidate change points, which are subsequently screened to estimate the number and locations of the changes. We detail Step 1 of this procedure in Section~\ref{ssc:multi-step1} alongside its theoretical guarantees, followed by Steps 2 and 3 in Sections~\ref{ssc:multi-step2} and~\ref{sec:multi-step3}, respectively.

\subsection{Step 1: Initial localization of multiple change points}\label{ssc:multi-step1}

\noindent\textit{Stage 1: Candidate generation.}
We apply the single-change procedure in \eqref{eq:Q-def}--\eqref{eq:tau0-def} across a sequence of $M_n$ overlapping local windows of fixed length $h$. The $i$th window $[b_i, e_i)$ contains observations $y_{b_i}, \ldots, y_{e_i-1}$. We set $b_1 = 1$ and $e_1 = 1+h$, advancing both endpoints by integer shift $\ell$ while $e_i \le n+1$, appending the full window $[n-h+1, n+1)$ if the final window ends prior to $n+1$ (in accordance with the interval notation in Appendix~\ref{app:numerical-implementation}).

For window $i$, let $L_i(\zeta)$ denote the profile loss in \eqref{eq:L-def} evaluated on that window at local split point $k=\zeta-b_i+1$. Thus, $\zeta=k+b_i-1$ is the global split point, and we define
\[
\mathcal{T}_i := \left\{ \zeta \in \mathbb{Z} : c^\ast h \le \zeta - b_i + 1 \le (1 - c^\ast) h \right\}.
\]
Collecting the local minimizers across all $M_n$ windows yields the candidate set $\widetilde{\mathcal{S}}$, containing $\widetilde{m}$ distinct candidate locations:
\[
\widetilde{\mathcal{S}} := \left\{ \arg\min_{\zeta \in \mathcal{T}_i} L_i(\zeta) : 1 \le i \le M_n \right\} = \left\{ \widetilde{\tau}_1 < \cdots < \widetilde{\tau}_{\widetilde{m}} \right\}.
\]
Duplicate candidate locations from overlapping windows are removed. Because every window yields a candidate, including those containing no true change point, $\widetilde{\mathcal{S}}$ typically contains spurious locations.

Let $\mathcal{S}^\star := \{\tau_1^\star, \ldots, \tau_{m_0}^\star\}$ denote the set of true change points, $\Delta_n := \min_{0 \le j \le m_0}(\tau_{j+1}^\star - \tau_j^\star)$ the minimum regime duration, and $v_{\min} := \min_{1 \le j \le m_0} \|\eta_{j+1} - \eta_j\|_2$ the minimum jump size in the stationary mean. For finite subsets $A, B \subset \mathbb{R}$, the \textit{directed Hausdorff distance} is defined as $d_H(A,B) := \max_{b \in B} \min_{a \in A} |b - a|$. Finally, let $\lambda_{1,i}$ denote the ridge regularization parameter in rolling window $i$, bounded above by a fixed constant $\bar{\lambda}_1$.
\begin{proposition}\label{prop:Haus}
Consider \eqref{eq:mean-shift-1} under Assumption~\ref{ass:innov}. Suppose Assumptions~\ref{assump:LJ} and \ref{assump:WS} hold and
$0\le\lambda_{1,i}\le\bar\lambda_1<\infty$ uniformly over the rolling
windows. Then,
\[
d_H(\widetilde{\mathcal S},\mathcal S^\star)
= O_{\IP}\left((1+\log m_0)\frac{\gamma_p^2}{v_{\min}^2}\right).
\]
\end{proposition}

The proof begins by establishing, for each true change point, the existence of a rolling window within which the change point is well-separated from the window boundaries. Assumption~\ref{assump:WS} makes this window contain exactly one change, at a fixed fraction from its endpoints, and Assumption~\ref{assump:LJ} gives the window-level signal condition, so the single-change theory of Section~\ref{sec:single} applies window by window. Because the rolling windows are fixed in advance, we apply the single-change tail bound in \ref{asf:eq:tau0-tail} directly within these windows. Its explicit exponential decay under serial dependence is the key ingredient, and combining it with this window construction is a technical contribution of the paper: with a tail level of order $1+\log m_0$, a union bound over the $m_0$ changes costs only a logarithmic factor, even when $m_0$ diverges, and the second-chaos part of the tail is absorbed because the window-level signal grows faster than $1+\log m_0$ under Assumption~\ref{assump:LJ}. This yields the stated localization rate in directed Hausdorff distance, so every true change has a nearby candidate, while spurious candidates are removed in the second stage; see Appendix~\ref{app:multiple-proofs}.

Let $\xi_n \downarrow 0$ be an auxiliary deterministic sequence satisfying
\[
\frac{(1+\log m_0)\gamma_p^2}{v_{\min}^2} \ll n\xi_n \ll \Delta_n,
\]
as specified in Assumption~\ref{assump:WS}. The neighborhood radius $n\xi_n$ strictly dominates the candidate localization rate while remaining negligible relative to the minimal spacing $\Delta_n$. Such a sequence exists under Assumption~\ref{assump:LJ}; for information criterion (IC) screening, it is additionally required to satisfy the penalty condition in Assumption~\ref{assump:pen}. Note that $\xi_n$ serves purely as an analytical device and is not an algorithmic tuning parameter.

We define the candidate-coverage event
\begin{equation}\label{eq:candidate-coverage-event}
\mathcal{C}_n := \left\{ d_H(\widetilde{\mathcal{S}}, \mathcal{S}^\star) \le n\xi_n \right\}.
\end{equation}
Under the conditions of Proposition~\ref{prop:Haus}, the lower bound on $n\xi_n$ ensures that $\mathbb{P}(\mathcal{C}_n) \to 1$ as $n \to \infty$. Because $n\xi_n = o(\Delta_n)$, the event $\mathcal{C}_n$ guarantees that, for all sufficiently large $n$, the candidate pool contains distinct candidates within an $n\xi_n$-neighborhood of every true change point. Stage 2 subsequently prunes redundant and spurious candidates from this pool.

\medskip
\noindent\textit{Stage 2: Candidate screening.}
For an ordered subset
$0<s_1<\cdots<s_m<n$ of $\widetilde{\mathcal S}$, let
$s_0:=0$, $s_{m+1}:=n$, and define the segment means and residual loss by
\[
\widehat\eta_i
:=
\frac{1}{s_i-s_{i-1}}
\sum_{t=s_{i-1}+1}^{s_i}y_t,
\qquad
\mathcal L_n(s_1,\ldots,s_m)
:=
\sum_{i=1}^{m+1}
\sum_{t=s_{i-1}+1}^{s_i}
\|y_t-\widehat\eta_i\|_2^2.
\]
We screen the candidates using
\begin{equation}\label{eq:Ln-def}
\mathrm{IC}_n(s_1,\ldots,s_m)
=
\mathcal L_n(s_1,\ldots,s_m)+m\omega_n,
\end{equation}
where $\omega_n$ controls the penalty for additional change points.

Starting from the full candidate set, we perform backward elimination. At
each iteration, we remove the candidate whose deletion produces the smallest
value of $\mathrm{IC}_n$ and continue as long as this deletion does not
increase the criterion. The procedure terminates at the screened set $\check{\mathcal S}=
\{\check\tau_1<\cdots<\check\tau_{\widehat m}\}$,
which provides the estimated number $\widehat m$ and the preliminary locations passed to Steps~2 and 3. The global IC selector, which the screening procedure approximates, is 
\begin{equation}\label{eq:IC-selection-clean}
(\widehat m,\check\tau_1,\ldots,\check\tau_{\widehat m})
\in
\arg\min_{\substack{0\le m\le\widetilde m\\
s_1<\cdots<s_m,\;s_i\in\widetilde{\mathcal S}}}
\mathrm{IC}_n(s_1,\ldots,s_m).
\end{equation}

\begin{algorithm}[htbp]
\caption{Backward elimination of spurious candidates in $\widetilde{\mathcal S}$}
\label{alg:backward-elimination}
Let the Step–1 pool be $\widetilde{\mathcal S}=\{\widetilde\tau_1<\cdots<\widetilde\tau_{\widetilde m}\}$ and recall
$\mathrm{IC}_n(\tau_1,\ldots,\tau_m)$ is the information criterion evaluated at a sorted tuple.

\begin{enumerate}
\item \textbf{Input:} data $\{y_t\}_{t=1}^n$; candidate set $\widetilde{\mathcal S}=\{\widetilde\tau_1<\cdots<\widetilde\tau_{\widetilde m}\}$.
\item \textbf{Initialize:} $\boldsymbol{\widetilde\tau}\leftarrow(\widetilde\tau_1,\ldots,\widetilde\tau_{\widetilde m})$ (sorted),\quad
      $m\leftarrow\widetilde m$;\quad $W_m\leftarrow \mathrm{IC}_n(\boldsymbol{\widetilde\tau})$.
\item \textbf{While} $m\ge 1$:
  \begin{enumerate}
  \item For each $j\in\{1,\ldots,m\}$, let
        \[
        W_{m-1,j}\ \defeq\ \mathrm{IC}_n\bigl(\boldsymbol{\widetilde\tau}\setminus\{\widetilde\tau_j\}\bigr).
        \]
        Choose $j^\star\in\arg\min_{1\le j\le m} W_{m-1,j}$ and set
        $W_{m-1}\leftarrow W_{m-1,j^\star}$.
  \item \textbf{If} $W_{m-1}>W_m$, \textbf{stop} and set
        $\check{\mathcal S}\leftarrow$ current $\boldsymbol{\widetilde\tau}$; go to Output.
  \item \textbf{Else} update simultaneously
        \[
        \boldsymbol{\widetilde\tau}\leftarrow \boldsymbol{\widetilde\tau}\setminus\{\widetilde\tau_{j^\star}\},\qquad
        W_{m-1}\leftarrow W_{m-1,j^\star},\qquad
        m\leftarrow m-1,
        \]
        and reindex $\boldsymbol{\widetilde\tau}$ to maintain increasing order.
  \end{enumerate}
\item \textbf{Output:} the final screened set
      $\check{\mathcal S}=\{\check\tau_1<\cdots<\check\tau_{\widehat m}\}$ with {$\widehat m\defeq|\check{\mathcal S}|$}. 
\end{enumerate}
\end{algorithm}

\noindent \textit{Stage 3: Local refinement.}
We implement the IC screening step using backward elimination, which
need not attain the global minimum in \eqref{eq:IC-selection-clean}.
The consistency and localization guarantees established below concern
the global selector and do not necessarily extend to this greedy
implementation. The computational complexity bound in Remark~\ref{rem:complexity} applies to backward elimination. 

For the theoretical results, write $\check\tau_1<\cdots<\check\tau_{\widehat m}$ for the locations returned by the global selector in \eqref{eq:IC-selection-clean}. The same local-refinement construction is used after backward elimination in the implementation.
 For each $j$, define
\[
(b_j,e_j)=
\left(
\left\lfloor\tfrac23\check\tau_{j-1}+\tfrac13\check\tau_j\right\rfloor,
\left\lceil\tfrac13\check\tau_j+\tfrac23\check\tau_{j+1}\right\rceil
\right),
\qquad
\check\tau_0=0,\quad \check\tau_{\widehat m+1}=n.
\]
With $c_{\mathrm{sp}}$ and $C_{\mathrm{sp}}$ as in Assumption~\ref{assump:local-geometry}, we set
$a_{\mathrm{loc}}:=1/\bigl(20(1+\max\{C_{\mathrm{sp}},c_{\mathrm{sp}}^{-1}\})\bigr)$. Local refinement is restricted to
\begin{equation}\label{eq:multi-trimmed-set}
\mathcal T(b,e):= \{k\in\mathbb Z:
b+a_{\mathrm{loc}}(e-b)\le k\le e-a_{\mathrm{loc}}(e-b)\}.
\end{equation}
Under the conditions of Proposition~\ref{thm:ic-consistency-rate}, with probability tending to one, each true change point lies in the interior of its local interval;
 see Lemma~\ref{lem:F-local-geometry} and \eqref{eq:F-geometry-margin}. The value of $a_{\mathrm{loc}}$ therefore only affects constants: any fixed trimming fraction in $(0,10a_{\mathrm{loc}})$ leads to the same localization rates. We apply the Step~1 ridge criterion
\eqref{eq:Q-def}--\eqref{eq:tau0-def} on each $(b_j,e_j]$ and obtain the locally refined estimates $\widehat\tau_j$. Throughout, $\check\tau_j$ denotes the screened IC location and $\widehat\tau_j$ its local refinement.

We next state conditions ensuring consistent estimation of the number of changes by the global IC selector and accurate localization by the subsequent refinement. Recall the minimal spacing $\Delta_n$, minimal jump size $v_{\min}$, and dependence measure $\gamma_p=\sum_{\ell\ge0}\|\Psi_\ell\|_2$. The ridge parameters may vary across rolling and local windows, but are
assumed nonnegative and uniformly bounded.

\begin{assumption}\label{assump:LJ}
The minimal spacing and jump size satisfy
\[
\frac{\Delta_n v_{\min}^2}
     {p\gamma_p^2(1+\log m_0)}
\longrightarrow\infty.
\]
\end{assumption}

Assumption~\ref{assump:LJ} is the multiple change analogue of Assumption~\ref{H1}, with the effective sample size given by the minimal spacing $\Delta_n$.  We comment on the additional factor $1+\log m_0$ arising from simultaneous
localization of the $m_0$ changes below in Remark \ref{rem:m0-bounded}.

\begin{assumption}\label{assump:WS}
There exists $\xi_n\downarrow0$ such that, for fixed
$c^\ast\in(0,1/2)$ and $c_{\mathrm{len}}\in(0,1)$,
\[
0<\ell\le(1-2c^\ast)h,\qquad
c_{\mathrm{len}}<
\liminf_{n\to\infty}\frac{h}{\Delta_n}
\le
\limsup_{n\to\infty}\frac{h}{\Delta_n}
<1,
\]
and the auxiliary sequence $\xi_n\downarrow0$  satisfies
\[
\frac{(1+\log m_0)\gamma_p^2}{v_{\min}^2}
\ll n\xi_n\ll\Delta_n,
\]
\end{assumption}

Under these conditions, each true change point is isolated in at least one rolling window and remains sufficiently far from its endpoints, while the candidate localization error is negligible relative to the spacing.

\begin{assumption}\label{assump:pen}
Let
$v_{\max}:=\max_j\|\eta_{j+1}-\eta_j\|_2$. The IC penalty satisfies
\begin{equation}\label{eq:pen-sandwich}
m_0\{n\xi_n v_{\max}^2+\gamma_p^2p\log(n\vee p)\}
\ll\omega_n\ll
\frac{\Delta_n v_{\min}^2}{m_0},
\end{equation}
\end{assumption}

\noindent In other words, a penalty sequence exists if
\begin{equation}\label{eq:pen-feasible}
m_0^2\{n\xi_n v_{\max}^2+\gamma_p^2p\log(n\vee p)\}
\ll\Delta_n v_{\min}^2.
\end{equation}

The lower bound in \eqref{eq:pen-sandwich} prevents over-segmentation by
dominating stochastic fluctuations and candidate misalignment, whereas the
upper bound prevents under-segmentation by remaining below the loss incurred
when a true change is omitted. 
The feasibility relation \eqref{eq:pen-feasible} expresses compatibility of these requirements.
 The factor $\log(n\vee p)$ arises
from uniform control over the candidate intervals entering the IC comparison.

\begin{assumption}\label{assump:local-geometry}
The minimal spacing satisfies $\Delta_n\to\infty$, and there are fixed constants $0<c_{\mathrm{sp}}\le C_{\mathrm{sp}}<\infty$ such that the regime lengths $l_j:=\tau_j^\star-\tau_{j-1}^\star$ satisfy
\begin{equation}\label{eq:local-geometry}
c_{\mathrm{sp}}\le\frac{l_{j+1}}{l_j}\le C_{\mathrm{sp}},
\qquad j=1,\ldots,m_0.
\end{equation}
\end{assumption}
Assumption~\ref{assump:local-geometry} requires adjacent regimes to have comparable lengths, while allowing substantial variation across the sample. By Lemma~\ref{lem:F-local-geometry}, this condition ensures that each true change point lies a fixed fraction away from the boundaries of its local-refinement interval and hence belongs to the interior of the trimmed set \eqref{eq:multi-trimmed-set}. A sufficient condition is $l_j\asymp n/(m_0+1)$ uniformly in $j$, which, for fixed $m_0$, requires each regime to occupy a nonvanishing fraction of the sample.

\begin{proposition}\label{thm:ic-consistency-rate}
Consider \eqref{eq:mean-shift-1} under Assumption~\ref{ass:innov}. Suppose
Assumptions~\ref{assump:LJ}--\ref{assump:local-geometry} hold and the ridge parameters are nonnegative and uniformly bounded over all rolling and local windows. 
Then there is a sufficiently large fixed constant $C_{\mathrm{pre}}>0$ such that the global IC selector in \eqref{eq:IC-selection-clean} and its subsequent local refinement satisfy
\[
P\left\{\widehat m=m_0,\quad
\max_{1\le j\le m_0}|\widehat\tau_j-\tau_j^\star|
\le C_{\mathrm{pre}}\frac{\gamma_p^2\log n}{v_{\min}^2}\right\}\longrightarrow1.
\]

\end{proposition}

 Note that, since Assumption \ref{assump:pen} implies $\gamma_p^2\log n/v_{\min}^2=o(\Delta_n)$, this joint bound sharpens the spacing-scale matching of the screened locations in Lemma~\ref{lem:IC-coarse-matching}.

The proof first establishes consistency of the estimated number of changes and coarse matching between the estimated and true locations (Theorem~\ref{thm:IC-consistency} and Lemma~\ref{lem:IC-coarse-matching}). Here the information criterion compares arbitrary candidate partitions, whose segments may contain several true changes; we control them all through uniform concentration of the error means over every subinterval (Lemma~\ref{lem:mean-block}) and an exact projection identity (Lemma~\ref{lem:UB-wellloc}). Lemma~\ref{lem:F-local-geometry} then validates the local-refinement intervals. Since these are selected from the data, we build the fluctuation event of Lemma~\ref{lem:uniform-local-raw} for the whole deterministic family of intervals, changes, and cuts, uniformly over the ridge parameter, before evaluating it at the selected ones; this union produces the $\log n$ factor. Handling serial dependence and data-selected intervals in this way is a technical contribution of the paper; see Appendix~\ref{app:proof-ic-consistency-rate}.

\begin{remark}\label{rem:m0-bounded}
The two localization rates emanating from the stages of Step 1 are similar. The stage 1 rate, based on the directed Hausdorff distance as stated in Proposition~\ref{prop:Haus}, incurs the factor $1+\log m_0$ from simultaneous control over the $m_0$-many true changes. On the other hand, the final stage 3 local-refinement rate from Proposition \ref{thm:ic-consistency-rate} incurs $\log n$ from uniform control over the
data-selected intervals.

Even though $m_0$ does not show up explicitly in this refined rate, the factor $(1+\log m_0)$ does appear in Assumption \ref{assump:LJ}. When $m_0$ is bounded, its explicit contributions to the rate and concentration conditions are absorbed into constants.
\end{remark}

\subsection{Step 2: Estimation of the common VAR transition matrices}\label{ssc:multi-step2}

Let $\widehat\tau_1<\cdots<\widehat\tau_{\widehat m}$ denote the preliminary locations from Proposition~\ref{thm:ic-consistency-rate}. Choose $K_n^{(m)}\to\infty$ and set
\[
w
=
\left\lceil
K_n^{(m)}\frac{\gamma_p^2\log n}{v_{\min}^2}
\right\rceil,
\qquad
w+d=o(\Delta_n).
\]
Define the purity event
\begin{equation}\label{eq:multi-purity-event}
\mathcal E_{\mathrm{pure}}^{(m)}
:=
\left\{
\widehat m=m_0,\ 
\max_{1\le j\le m_0}
|\widehat\tau_j-\tau_j^\star|\le w
\right\},
\end{equation}
which has probability tending to one by
Proposition~\ref{thm:ic-consistency-rate}. Let
$\widehat\tau_0^+=0$, $\widehat\tau_{\widehat m+1}^-=n$, and
$\widehat\tau_j^\pm=\widehat\tau_j\pm w$. We retain
\begin{equation}\label{eq:multi-retained-sets}
I_j
=
\{\widehat\tau_{j-1}^++d+1,\ldots,\widehat\tau_j^-\},
\qquad
j=1,\ldots,\widehat m+1.
\end{equation}

On $\mathcal E_{\mathrm{pure}}^{(m)}$, every retained response and its $d$ lags belong to the same true regime. For all sufficiently large $n$, the retained blocks are nonempty and, with $J=m_0+1$, their total size is
\[
N_m:=n-2m_0w-Jd\asymp n,
\]
since $(m_0+1)\Delta_n\le n$ and $(w+d)/\Delta_n\to0$.

Stack the retained observations into $Y_I$ and $X_I$, and let
$G_{\mathrm{blk}}^{(m)}:=\operatorname{diag}(G_{|I_1|},\ldots,G_{|I_{\widehat m+1}|})$ and
$\widehat N_m:=\sum_{j=1}^{\widehat m+1}|I_j|$.
The size-indexed notation $G_r=\mathbf{I}_r-r^{-1}\mathbf1_r\mathbf1_r^\top$ in the general block calculations of Appendix~\ref{asf:sec:S2-proof} denotes centering within a single block.
We estimate the common transition matrix by the block-profiled Lasso
\begin{equation}\label{eq:multi-profile-lasso}
\widehat B\in\arg\min_B
\left\{\widehat N_m^{-1}\|G_{\mathrm{blk}}^{(m)}(Y_I-X_IB)\|_F^2
+\widehat\lambda_{m,n}\|\operatorname{vec}B\|_1\right\},
\end{equation}
which profiles out a separate intercept for each retained regime. The corresponding Gram matrix and score are
$\widehat\Gamma_m=\widehat N_m^{-1}\mathbf{I}_p\otimes(X_I^\top G_{\mathrm{blk}}^{(m)}X_I)$ and
$\widehat\gamma_m=\widehat N_m^{-1}\operatorname{vec}(X_I^\top G_{\mathrm{blk}}^{(m)}Y_I)$.
Using the constants from the single-change analysis, define
\begin{equation}\label{eq:multi-log-scale}
\ell_{m,n}:=1+\log(dp^2)+2(m_0+1)\log n,\qquad
\lambda_{m,n}:=C_\lambda Q_{\mathrm{dev}}\sqrt{\ell_{m,n}/N_m}.
\end{equation}
The tuning $\widehat\lambda_{m,n}$ in \eqref{eq:multi-profile-lasso} is obtained from \eqref{eq:multi-log-scale} by replacing $m_0$ and $N_m$ with $\widehat m$ and the actual retained size $\widehat N_m$. On $\mathcal E_{\mathrm{pure}}^{(m)}$, these replacements leave the quantities unchanged: $\widehat N_m=N_m$ and $\widehat\lambda_{m,n}=\lambda_{m,n}$. The theoretical bounds below therefore use $N_m$ and $\ell_{m,n}$.\footnote{If the retained blocks are empty or degenerate, we set $\widehat{B}$, $\widehat{\Gamma}_m$, and $\widehat{\gamma}_m$ to zero. This convention merely serves to define the estimators on the complement of the high-probability event analyzed in the proofs.}

\begin{proposition}\label{thm:S2-multi}
Suppose the conditions of Proposition~\ref{thm:ic-consistency-rate} hold, and use its global IC selector and subsequent local refinement. Use the buffer defined above with $K_n^{(m)}\to\infty$ and $w+d=o(\Delta_n)$. Suppose $\beta^\star$ is $s$-sparse, $s\ge1$, $C_\lambda\ge4$ is fixed, and
\begin{equation}\label{eq:S2-multi-uniform}
N_m
\ge
C_{\mathrm{S2}}
\left[
\chi_x^2\{s(1+\log(dp))+(m_0+1)\log n\}
+(m_0+1)\chi_x+\ell_{m,n}
\right].
\end{equation}
Let $\mathcal E_{\mathrm{S2}}^{(m)}$ denote the intersection of the purity,
RSC, and score events, with $J=m_0+1$. Then,
\[
P\{(\mathcal E_{\mathrm{S2}}^{(m)})^c\}
\le
{P\{(\mathcal E_{\mathrm{pure}}^{(m)})^c\}+Cn^{-2}\longrightarrow0.}
\]
On $\mathcal E_{\mathrm{S2}}^{(m)}$, the cone, coefficient, prediction, and
block-demeaned residual bounds of Proposition~\ref{thm:S2} hold with
$(\lambda_n,n_{\mathrm{obs}},G_{\mathrm{blk}})$ replaced by
$(\lambda_{m,n},N_m,G_{\mathrm{blk}}^{(m)})$. Consequently,

\begin{equation}\label{eq:multi-filter-rate}
\|\widehat B-B\|_F
=O_{\IP}\left(
\bar Q\sqrt{\frac{s\ell_{m,n}}{N_m}}
\right).
\end{equation}
\end{proposition}

Proposition~\ref{thm:S2-multi} is the multiple-change analogue of Proposition~\ref{thm:S2}, and its proof follows the same three steps. The proof first verifies that the retained responses and their lags belong to the same regimes: the rate of Proposition~\ref{thm:ic-consistency-rate} and the diverging buffer place every retained response and its $d$ lags inside one regime, and the block lengths telescope to $N_m\asymp n$. Applying Proposition~\ref{prop:uniform-block-concentration} with $J = m_0 + 1$ blocks under the profiled regression representation in Lemma~\ref{asf:lem:S2-profile}, yields uniform concentration bounds across all $(n+1)^{2J}$ candidate endpoint configurations, where \eqref{eq:S2-multi-uniform} specifies the exact sample-size requirement. Proposition~\ref{asf:prop:S2-eventwise} then translates these bounds into the stated estimation bounds; see Appendix~\ref{ssc:step2-mcp}.

Relative to the single-change case, the additional cost arises from uniform concentration over $m_0+1$ data-dependent blocks, reflected in $\ell_{m,n}$. Block profiling also lowers the population curvature by at most $(m_0+1)L_x/N_m$ and adds a score-mean term of order $(m_0+1)\sigma\sqrt{L_x}/N_m$, and the concentration bound accounts for both.

\subsection{Step 3: Refined localization}
\label{sec:multi-step3}

Using the common Step~2 estimator, define the filtered series
\[
\widehat Y_t
=
y_t-\sum_{\ell=1}^d\widehat A_\ell y_{t-\ell},
\qquad t=d+1,\ldots,n.
\]
For each preliminary location $\widehat\tau_i$, let
\[
(b_i,e_i)
=
\left(
\left\lfloor\tfrac45\widehat\tau_{i-1}
+\tfrac15\widehat\tau_i\right\rfloor,
\left\lceil\tfrac15\widehat\tau_i
+\tfrac45\widehat\tau_{i+1}\right\rceil
\right),
\qquad
N_i:=e_i-b_i,
\]
with $\widehat\tau_0=0$ and $\widehat\tau_{\widehat m+1}=n$. By Proposition~\ref{thm:ic-consistency-rate} and
Assumption~\ref{assump:local-geometry}, with probability tending to one, each interval contains exactly one true change separated from its endpoints by at least $10a_{\mathrm{loc}}N_i$ (Lemma~\ref{lem:F-local-geometry} with $\theta=4/5$).

Let $\lambda_{3,i}\ge0$ denote the ridge parameter for the $i$th filtered refinement interval.
For $\zeta\in\mathcal T(b_i,e_i)$, we estimate the two local means by minimizing
\[
\frac1{N_i}
\left\{
\sum_{t=b_i+1}^{\zeta}\|\widehat Y_t-\nu_1\|_2^2
+
\sum_{t=\zeta+1}^{e_i}\|\widehat Y_t-\nu_2\|_2^2
\right\}
+
\lambda_{3,i}\|\nu_1-\nu_2\|_2^2.
\]
Let $\widetilde L_i(\zeta)$ denote the corresponding unpenalized profiled loss, as in the single-change refinement, and define
\[
\widehat\tau_i^{\,r}
\in
\arg\min_{\zeta\in\mathcal T(b_i,e_i)}
\widetilde L_i(\zeta).
\]

Let
$\Delta_{\eta,j}:=\eta_{j+1}-\eta_j$,
$\Delta_{\mu,j}:=\mu_{j+1}-\mu_j$, and
$u_\star:=\min_j\|\Delta_{\mu,j}\|_2$. To account for estimation of the common filter, define
\begin{equation}\label{eq:multi-filter-scales}
\rho_{m,n}
:=
\bar Q\gamma_p\sqrt{\frac{s\ell_{m,n}}{n}},
\qquad
b_{m,n}
:=
\bar Q^2\gamma_p^2
\frac{s^2\ell_{m,n}\ell_{2,n}}{n}.
\end{equation}
Here $\ell_{m,n}$ reflects estimation over the selected regime blocks, whereas $\ell_{2,n}$ arises from the coordinate and time maxima in the filtering error.

For a sufficiently large constant $C_B > 0$, analogous to the single-change analysis, we define the deterministic coefficient-error set
\[
\mathcal D_{m,n}:=\left\{D\in\mathbb R^{dp\times p}:\operatorname{vec}D\in\mathcal C(S),\quad
\|D\|_F\le C_B\bar Q\sqrt{s\ell_{m,n}/n}\right\}.
\]
\color{black}

\begin{assumption}\label{assump:S3SNR-multi}
The minimal spacing and structural jump satisfy
\[
\frac{p\log n}{\Delta_nu_\star^2}\to0,
\qquad
\frac{(1+b_{m,n})\log n}{\Delta_nu_\star^2}\to0.
\]
\end{assumption}

\begin{assumption}\label{assump:S3-filter-multi}
The filter-estimation scale satisfies $\rho_{m,n}\to0$.
\end{assumption}

For $q=1,\ldots,d$ and $j=1,\ldots,m_0$, define
\[
c_{q,j} :=
\left(\sum_{\ell=q}^dA_\ell\right)\Delta_{\eta,j}, \qquad C_{q,j}:=\sum_{\ell=1}^q c_{\ell,j},
\qquad \nu_{n,\star} := \max_{j,q}
\frac{\|C_{q,j}\|_2}{q\|\Delta_{\mu,j}\|_2}.
\]

\begin{assumption}\label{assump:S3-transition-multi}
There exists $c_{\mathrm{al}}>0$ such that, uniformly over $j$ and
$q=1,\ldots,d$,
\begin{equation}\label{eq:multi-transition-alignment}
q\|\Delta_{\mu,j}\|_2^2
+
2\langle\Delta_{\mu,j},C_{q,j}\rangle
\ge
c_{\mathrm{al}}q\|\Delta_{\mu,j}\|_2^2.
\end{equation}
Moreover,
$\rho_{m,n}\nu_{n,\star}\to0$ and
$\nu_{n,\star}^2/\Delta_n\to0$.
\end{assumption}

\begin{theorem}\label{prop:multi-local}
Suppose the conditions of Proposition~\ref{thm:S2-multi} and
Assumptions~\ref{assump:S3SNR-multi}--\ref{assump:S3-transition-multi} hold, with fixed $d$.
Let $\bar\lambda_n\downarrow0$ be deterministic and suppose
$0\le\lambda_{3,i}\le\bar\lambda_n$ for every selected refinement interval.
Then, for a sufficiently large fixed constant $C_{\mathrm{post}}>0$,
\begin{equation}\label{eq:multi-refined-rate}
P\left\{\widehat m=m_0,\quad
\max_{1\le j\le m_0}|\widehat\tau_j^{\,r}-\tau_j^\star|
\le C_{\mathrm{post}}\frac{\log n+b_{m,n}}{u_\star^2}\right\}\longrightarrow1.
\end{equation}

\end{theorem}

Theorem~\ref{prop:multi-local} is the multiple-change analogue of Theorem~\ref{prop:tau-rate}. The proof first uses Lemma~\ref{lem:F-local-geometry} to validate the
data-dependent refinement intervals and Proposition~\ref{thm:S2-multi}
to control the estimated common filter. Combining the population-drift
bound of Lemma~\ref{asf:lem:S3-deterministic-D}, now applied on each interval under the multiple-change alignment condition \eqref{eq:multi-transition-alignment}, with the uniform
fluctuation bound of Lemma~\ref{lem:uniform-local-filtered} yields
\eqref{eq:multi-refined-rate} simultaneously over all change points. The fluctuation bound holds on a single event, built for all admissible deterministic intervals, their changes and cuts, and all coefficient errors in $\mathcal D_{m,n}$ before the selected intervals and the estimated filter are substituted. Its fixed-direction terms involve only the $(d+1)m_0$ population jump and transition directions, and the two logarithms in $b_{m,n}$ enter separately: $\ell_{m,n}$ through estimation of the filter and $\ell_{2,n}$ through the coordinate maximum. See Appendix~\ref{app:multiple-proofs}.

\begin{remark}
As in the single-change case, filtering removes the explicit dependence
factor $\gamma_p^2$ from the localization bound, while $b_{m,n}$
accounts for estimation of the common VAR filter. Refinement yields
a strictly smaller simultaneous localization bound whenever
\[
\frac{\log n+b_{m,n}}{u_\star^2}
=o\left(\frac{\gamma_p^2\log n}{v_{\min}^2}\right).
\]

For example, under the remaining theorem assumptions, sufficient
conditions are $\gamma_p\to\infty$, $v_{\min}\lesssim u_\star$, and
$\bar Q^2s^2\ell_{m,n}\ell_{2,n}=o(n\log n)$. Indeed,
$(\log n+b_{m,n})/(\gamma_p^2\log n)
=\gamma_p^{-2}
+\bar Q^2s^2\ell_{m,n}\ell_{2,n}/(n\log n)=o(1)$.
The jump condition holds if $A(1)$ does not asymptotically attenuate
any of the jump directions, while the last condition requires
sufficient sample size relative to sparsity and the spectral cost
of filter estimation. These conditions allow increasing persistence
outside the directions of the mean changes. With bounded dependence
and comparable minimal jump norms, filtering may instead yield
a constant-factor improvement.

Unlike the single-change result, these bounds hold simultaneously over all change points, with the additional logarithmic factor
arising from the uniform control used in the analysis.
\end{remark}

\begin{remark}[Computational complexity of the 3-stage detection procedure]\label{rem:complexity}
We analyze the computational cost of the detection procedures conditional
on the lag order $d$ and penalty $\omega_n$, excluding their selection.
The ridge, Lasso, and window-length grids, as well as the number of
coordinate-descent passes, are fixed. Cumulative sums of $y_t$ and
$\|y_t\|_2^2$ require $O(np)$ preprocessing operations and allow the
profiled loss to be evaluated over all candidate splits of an interval
of length $N$ in $O(Np)$ operations per ridge value.

For a single change point, the searches in Steps~1 and~3 each cost
$O(np)$. Step~2 fits $p$ Lasso regressions, each with $pd$ predictors
and at most $n$ observations. Under the fixed iteration budget,
proximal coordinate descent costs $O(np^2d)$, as does filtering the series.
The total computational cost is therefore $O(np+np^2d)$.

For multiple change points, Stage~1 scans $M_n\le n/\ell+1$
overlapping windows of length $h$, at a cost of
$O(pM_nh)=O(nph/\ell)$. In the backward-elimination algorithm
(Algorithm~\ref{alg:backward-elimination}), removing a candidate merges
two adjacent segments. Using cumulative sums, each deletion criterion
can be evaluated in $O(p)$ operations, giving a total screening cost
of $O(p\widetilde m^2)$, where $\widetilde m\le M_n$.
When $\ell\asymp h$ and $h\gtrsim\sqrt n$, both candidate generation
and screening cost $O(np)$.

The local-refinement intervals have total length $O(n)$, so their
profile searches also cost $O(np)$. Estimating the common VAR from the
pooled retained observations and filtering the series each cost
$O(np^2d)$. Hence, under the stated window-size conditions and fixed
computational budget, the overall cost is $O(np+np^2d)$ for both
single and multiple change points, and is linear in $n$ for fixed
$p$ and $d$.
\end{remark}

\section{Performance Evaluation}\label{sec:simulations}
We evaluate the finite-sample performance of the proposed procedures for single and multiple change points using synthetic data. In the summaries, Pre and Post denote the preliminary Step~1 and refined Step~3 estimates, respectively. For multiple change points, Pre includes the local refinement in Stage~3 of Step~1.

\subsection{Single change point setting}\label{sec:single-CP-sims}

We first compare the preliminary estimator from Step1 and the refined estimator from Step3 with several methods for change-point detection in high-dimensional or panel time series: Inspect~\citep{WangSamworth2018}, sparsified binary segmentation (\textbf{sbs})\citep{cho2015multiple}, double CUSUM (\textbf{dc})\citep{cho2016change}, the scan statistic (\textbf{scanEH}) based on \citet{enikeeva2013high}, and the $\ell_2$- and $\ell_\infty$-CUSUM aggregation procedures (\textbf{agg2} and \textbf{aggoo}) based on \citet{jirak2015uniform}. These procedures operate broadly within a signal-plus-noise framework for panel or high-dimensional time series, with temporal dependence, when allowed, incorporated into the noise structure rather than explicitly estimated and removed through a common VAR filter. They therefore provide natural benchmarks for assessing the localization gains from exploiting the autoregressive structure.

The simulation design consists of 192 parameter configurations. Throughout,
there is a single change point at $\tau=\lceil n/2\rceil$, the VAR
transition matrices are common across regimes, and the innovations are
i.i.d.\ $N(0,\mathbf{I}_p)$. The parameters $c_k$, $c_A$, and $c_\vartheta$ control,
respectively, the fraction of coordinates affected by the mean change, the
density of the transition matrices, and the magnitude of the mean shift.
The design is the full factorial combination
\[
\begin{split}
n&\in\{100,250,500\},\qquad
p/n\in\{0.5,0.8\},\qquad
c_k\in\{0.5,1\},\\
c_A&\in\{0.5,1\},\qquad
\rho(\mathcal A)\in\{0.3,0.5,0.7,0.9\},\qquad
d\in\{1,2\},
\end{split}
\]
with $c_\vartheta=0.5$. We therefore have $p\in\{50,125,250\}$ when $p/n=0.5$ and
$p\in\{80,200,400\}$ when $p/n=0.8$.

The transition-matrix density is determined by
$q_A=\operatorname{round}(c_Ap)$: each lag coefficient matrix contains $q_A$ nonzero coefficients, with locations sampled uniformly without replacement and independent equiprobable signs. Prior to stability rescaling, their magnitudes are sampled from $[1.5,2.5]$ for $A_1$ and, when $d=2$, from $[0.5,1.5]$ for $A_2$. For a VAR$(1)$, $A_1$ is rescaled so that $\rho(A_1)=\rho(\mathcal A)$; for a VAR$(2)$, both lag matrices are rescaled
by a common factor so that the companion matrix has spectral radius $\rho(\mathcal A)$.

The mean shift has $k=\operatorname{round}(c_kp)$ active coordinates. Let
$\Delta_n:=\min\{\tau,n-\tau\}$ denote the single-change instance of the minimum-spacing notation, and set $\vartheta:=c_\vartheta\sqrt{p/\Delta_n}$, which determines the $\ell_2$-magnitude of the stationary-mean jump. Specifically,
\[
(\Delta_\eta)_r=
\begin{cases}
\vartheta w_rs_r, & r\le k,\\
0, & r>k,
\end{cases}
\qquad
w_r\propto r^{-1/2},\quad
\sum_{r=1}^k w_r^2=1,
\]
where $s_r$ are independent Rademacher variables. By construction, $\|\Delta_\eta\|_2=\vartheta$. The sign vector is drawn once for each parameter configuration and held fixed across repetitions, so that only the
innovations vary across repetitions.

Implementation details for Steps~1--3, including lag selection, buffered transition-matrix estimation, and ridge tuning, are provided in Appendix~\ref{app:numerical-implementation}. The true lag order is used only for data generation; estimation of the VAR lag order is data-driven.

For the Inspect algorithm, we follow the implementation and tuning guidelines of \citet{WangSamworth2018}. Specifically, we estimate the error variance for each series, normalize each series by its estimated standard deviation, and set the regularization parameter to $\sqrt{\frac{1}{2}\log(p\log n)}$.

For each design configuration and method, we compute the localization RMSE across 50 replicates. To assess the gain from refinement and compare it with Inspect, which proves to be the strongest of the competitors (the full comparison is reported in Appendix~\ref{app:single-cp-tables}), we report $\mathrm{RMSE}_{\mathrm{post}}/\mathrm{RMSE}_{\mathrm{pre}}$ and
$\mathrm{RMSE}_{\mathrm{Inspect}}/\mathrm{RMSE}_{\mathrm{pre}}$; ratios below one indicate lower localization error than the preliminary Step~1 estimator. The figures stratify these ratios by $c_A$, $\rho(\mathcal A)$, and $n$, while pooling over the remaining design parameters. The dashed horizontal line marks a ratio of one.

We report the VAR$(1)$ results for $p/n\in\{0.5,0.8\}$ in the main text,
with $d=1$ and $c_\vartheta=0.5$. The corresponding VAR$(2)$ results are
reported in Appendix~\ref{app:numerical-results}.

The complete single-change-point results are reported in
Appendix~\ref{app:single-cp-tables}, including comparisons with unregularized versions of Steps~1 and~3, denoted by Pre-NR and Post-NR (NR: no ridge regularization). Each version uses its own preliminary estimate and Step~2 VAR fit, with the Lasso penalty retained. These comparisons isolate the finite-sample contribution of ridge
regularization from the gains due to VAR filtering.
\begin{figure}[tbp]
  \centering
  \IfFileExists{numerical_results/panel_single_cp_rmse_ratios_pn0p50_lag1_cvartheta0p50.pdf}{    \includegraphics[width=\linewidth]{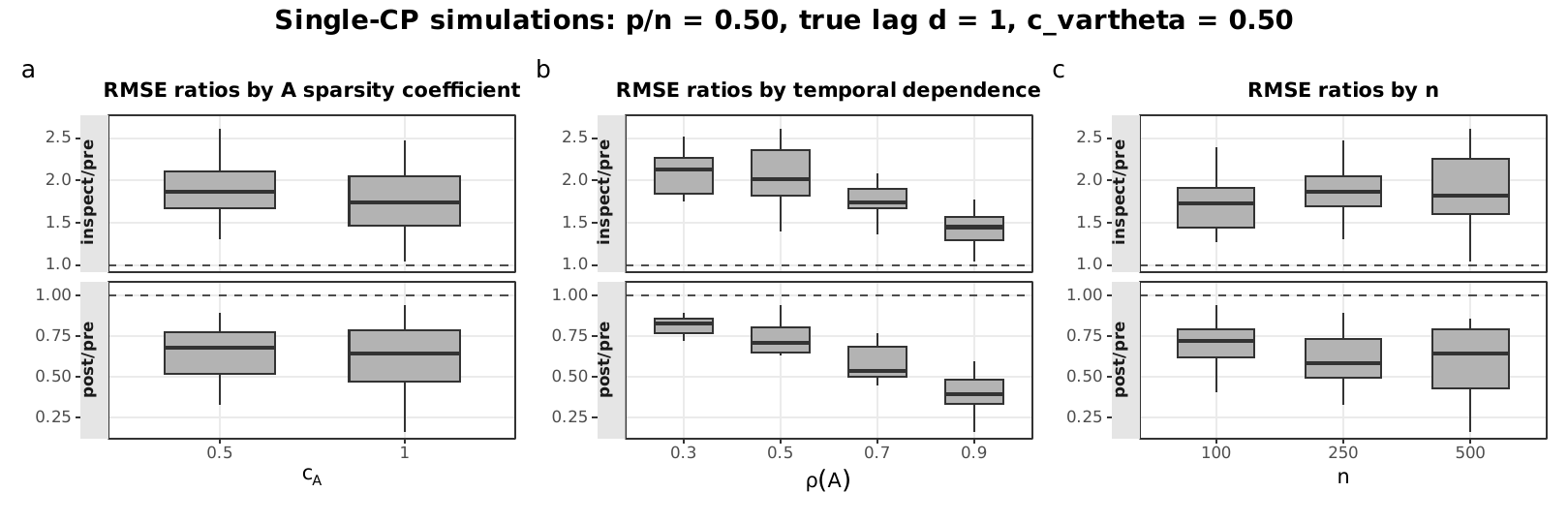}  }{    \fbox{\parbox[c][2.1in][c]{0.94\linewidth}{\centering
      Placeholder for \texttt{\detokenize{numerical_results/panel_single_cp_rmse_ratios_pn0p50_lag1_cvartheta0p50.pdf}}}}  }
  \caption{Single-change-point RMSE ratios for
  $(p/n,d,c_\vartheta)=(0.5,1,0.5)$. Columns stratify the results by
  $c_A$, $\rho(\mathcal A)$, and $n$; rows report Inspect/pre and post/pre.
  The horizontal line marks a ratio of one.}
  \label{fig:single-box-pn05-lag1}
\end{figure}

\begin{figure}[tbp]
  \centering
  \IfFileExists{numerical_results/panel_single_cp_rmse_ratios_pn0p80_lag1_cvartheta0p50.pdf}{    \includegraphics[width=\linewidth]{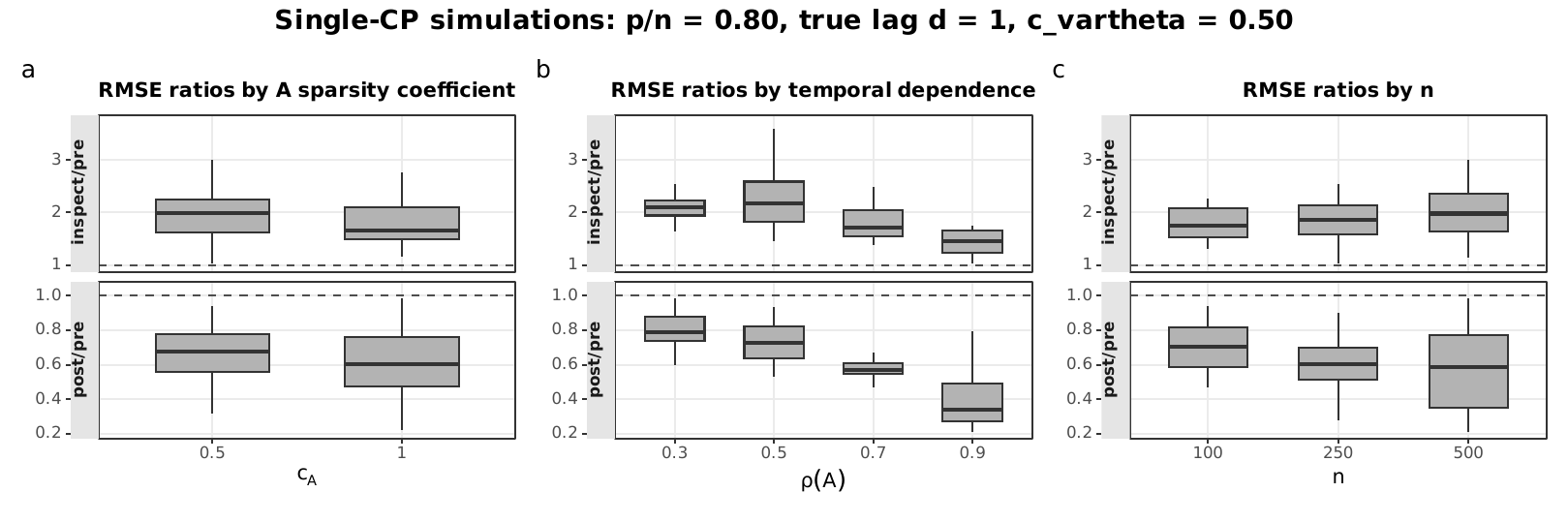}  }{    \fbox{\parbox[c][2.1in][c]{0.94\linewidth}{\centering
      Placeholder for \texttt{\detokenize{numerical_results/panel_single_cp_rmse_ratios_pn0p80_lag1_cvartheta0p50.pdf}}}}  }
  \caption{Single-change-point RMSE ratios for
  $(p/n,d,c_\vartheta)=(0.8,1,0.5)$. Columns stratify the results by
  $c_A$, $\rho(\mathcal A)$, and $n$; rows report Inspect/pre and post/pre.
  The horizontal line marks a ratio of one.}
  \label{fig:single-box-pn08-lag1}
\end{figure}

Figures~\ref{fig:single-box-pn05-lag1} and
\ref{fig:single-box-pn08-lag1} exhibit consistent gains from refinement. The median post/pre RMSE ratio is below one in every displayed stratum, with most
of each distribution also lying below one. These results show that Step~3 systematically improves
localization accuracy relative to the preliminary Step~1 estimator across the designs considered. By contrast, the Inspect/pre RMSE ratios are centered above one, indicating larger localization errors than Step~1 in these VAR$(1)$
experiments.

The gain from refinement becomes more pronounced as temporal dependence strengthens. In particular, the post/pre ratio decreases systematically with
$\rho(\mathcal A)$ in panel~(b). This pattern complements the theoretical
comparison in Remark~\ref{rem:single-rate-comparison}, where filtering removes
the direct dependence factor from the localization rate.

We find no comparable monotone pattern with respect to $c_A$ or $n$ in
panels~(a) and~(c). These are marginal summaries that pool over the remaining
design parameters. Moreover, varying $c_A$ changes both the number and
configuration of nonzero transition coefficients before stability rescaling,
while varying $n$ at fixed $p/n$ changes the sample size and dimension
simultaneously. These features interact with the signal structure, temporal
dependence, and estimation of the transition matrices. We therefore emphasize
the persistent reduction in RMSE after refinement rather than a monotone
comparative-static relationship with $c_A$ or $n$.
\subsection{Multiple change points}\label{sec:multi-CP-sims}

We next evaluate the multiple-change procedure, focusing both on the gains from the final refinement step and on robustness to departures from the common-VAR assumption. 
Since Inspect performed best overall among the competing
methods in the single-change-point experiments
(see Appendix~\ref{app:single-cp-tables}), we retain it as
the sole external benchmark in the multiple-change-point simulations to provide a focused comparison.  
For Inspect, we retain the standardization and the rule for choosing
the regularization parameter described in
Section~\ref{sec:single-CP-sims}. The stopping threshold for wild
binary segmentation is calibrated following Section~5.3 of
\citet{WangSamworth2018}. We consider 12 parameter configurations. Throughout, $n=2000$, $p=200$, $m_0=3$, $d=1$, $c_k=1$, $c_A=0.5$, and the innovations are
i.i.d.\ $N(0,\mathbf{I}_p)$. With these choices, all coordinates are affected by each mean change, and the baseline transition matrix contains 100 nonzero coefficients. Its
support, signs, and magnitudes are generated as in the single-change experiments and rescaled to attain the prescribed spectral radius.

The dependence and signal levels vary jointly over
\[
\bigl(\rho(\mathcal A),c_\vartheta\bigr)
\in\{(0.3,1),(0.5,1.1),(0.7,1.3),(0.9,2.5)\}.
\]
Each jump is normalized so that
$\|\Delta_{\eta,j}\|_2=\vartheta=c_\vartheta$, $j=1,\ldots,3$.
All jumps share the support $\{1,\ldots,200\}$ and the normalized profile $w_r\propto r^{-1/2}$. Their coordinate signs are drawn independently across
change points and held fixed across simulation repetitions, while successive regime means are constructed by cumulatively adding the three jumps.

We consider three designs, each evaluated at the four dependence--signal configurations:
\[
\begin{array}{lll}
\text{C--E:} & \text{common transition matrix,} & \text{equal spacing},\\
\text{L--E:} & \text{locally misspecified transition matrices,}
             & \text{equal spacing},\\
\text{C--U:} & \text{common transition matrix,} & \text{unequal spacing}.
\end{array}
\]
Under equal spacing, all four segments have length 500, with change points located at
$(500,1000,1500)$. Under unequal spacing, the segment lengths are
$(572,476,381,571)$, with change points
located at $(572,1048,1429)$. For this VAR$(1)$ design, write $A:=A_1$ for the baseline transition matrix. In the locally misspecified design, segment $j$ has transition matrix $A^{(j)}=A+D_j$, where $D_j$ contains approximately
one quarter as many nonzero coefficients as $A$, sampled from the support of $A$ with independent random signs and raw magnitudes drawn uniformly from $[0.5,1]$, and is scaled to satisfy $\|D_j\|_F/\|A\|_F=0.05$, subject to stability of every segment-specific VAR. To limit variations in temporal dependence across regimes, we require
$|\rho(A+D_j)-\rho(A)|\le 0.03$ for all $j$. This design assesses robustness when the common-dynamics assumption is only approximately satisfied.

Following the rolling-window strategy of \citet{bai2023multiple}, we consider window lengths of the $h(\upsilon)=\lfloor c_{\mathrm{win}}n^\upsilon\rfloor$,
$\upsilon\in\mathcal G_{\mathrm{win}}$, with rolling shift
$\ell(\upsilon)=\max\{1,\operatorname{round}(c_{\mathrm{shift}}h(\upsilon))\}$, wherein $c_{\mathrm{win}},c_{\mathrm{shift}}\in(0,1)$ are fixed constants and $\mathcal G_{\mathrm{win}}\subset[1/2,1]$ is a fixed finite grid; duplicate lengths and lengths below $\lceil\sqrt n\rceil$ are omitted.
The window length is selected from this grid using stability of the estimated number of change points across neighboring values of $h$. Importantly, neither
the true number of changes nor the minimum true spacing $\Delta_n$ enters the selection rule. Details of the stability rule, together with lag selection,
ridge and Lasso tuning, local refinement, and the data-driven choice of $\omega_n$, are provided in Appendix~\ref{app:numerical-implementation}.

Table~\ref{tab:mc-summary} summarizes the 12 configurations. The design labels
C--E, L--E, and C--U correspond to the three cases above. The estimated-count distribution is reported as frequencies over 50 repetitions. The localization measure $\mathrm{RMSE}_{\widehat\tau}$ is the square root of the mean squared localization error, averaged over change points and repetitions with $\widehat m=m_0$. The count measure $\mathrm{RMSE}_{\widehat m}$ is the square root of the mean squared error of $\widehat m$, computed over all valid repetitions. Localization RMSE is reported as NA when no repetition recovers the true number of changes.
Since estimated change points partition the observation sequence
into segments, we also assess segmentation accuracy using the
adjusted Rand index (ARI), which measures agreement between the
estimated and true partitions
\citep{rand1971objective,hubert1985comparing};
see Appendix~\ref{app:ari-definition} for its formal definition.
Larger ARI values indicate greater agreement. We report the
average ARI across repetitions and the percentage of repetitions
in which each method attains the highest ARI among Pre, Post,
and Inspect, with ties credited to all tied methods.

\IfFileExists{numerical_results/multi_cp_table.tex}{%
  \begin{table}[tbp]
\centering
\caption{Finite-sample performance of multiple change-point estimators (50 replications).}
\label{tab:mc-summary}
\tiny
\setlength{\tabcolsep}{2.0pt}
\resizebox{\textwidth}{!}{%
\begin{tabular}{c c c l r r r r r r r r r r r r r}
\toprule
\multirow{2}{*}{Design} & \multirow{2}{*}{$\rho(\mathcal A)$} & \multirow{2}{*}{$\vartheta$} & \multirow{2}{*}{Method} & \multicolumn{8}{c}{$\widehat m$ (frequency over 50 runs)} & \multirow{2}{*}{$\mathrm{RMSE}_{\widehat\tau}$} & \multirow{2}{*}{$\mathrm{RMSE}_{\widehat m}$} & \multirow{2}{*}{ARI} & \multirow{2}{*}{\shortstack{ARI \\ \% Best}} & \multirow{2}{*}{$d_H$} \\
\cmidrule(lr){5-12}
& & & & 0 & 1 & 2 & 3 & 4 & 5 & 6 & $\ge 7$ & & & & & \\
\midrule
\multirow{3}{*}{C--E} & \multirow{3}{*}{0.30} & \multirow{3}{*}{1.00} & inspect & 0 & 0 & 0 & 41 & 8 & 1 & 0 & 0 & 12.98 & 0.49 & 0.953 & 28.0 & 16.34 \\
& & & Pre & 0 & 1 & 1 & 48 & 0 & 0 & 0 & 0 & 14.25 & 0.32 & 0.951 & 26.0 & 35.60 \\
& & & Post & 0 & 1 & 1 & 48 & 0 & 0 & 0 & 0 & 11.36 & 0.32 & 0.960 & 62.0 & 33.14 \\
\cmidrule(lr){1-17}
\multirow{3}{*}{C--E} & \multirow{3}{*}{0.50} & \multirow{3}{*}{1.10} & inspect & 0 & 0 & 0 & 34 & 11 & 4 & 1 & 0 & 16.04 & 0.85 & 0.933 & 16.0 & 18.80 \\
& & & Pre & 0 & 0 & 0 & 50 & 0 & 0 & 0 & 0 & 15.71 & 0.00 & 0.966 & 24.0 & 17.76 \\
& & & Post & 0 & 0 & 0 & 50 & 0 & 0 & 0 & 0 & 7.21 & 0.00 & 0.984 & 68.0 & 8.30 \\
\cmidrule(lr){1-17}
\multirow{3}{*}{C--E} & \multirow{3}{*}{0.70} & \multirow{3}{*}{1.30} & inspect & 0 & 0 & 0 & 21 & 13 & 12 & 3 & 1 & 22.23 & 1.44 & 0.898 & 2.0 & 24.18 \\
& & & Pre & 0 & 0 & 0 & 50 & 0 & 0 & 0 & 0 & 10.30 & 0.00 & 0.977 & 24.0 & 11.72 \\
& & & Post & 0 & 0 & 0 & 50 & 0 & 0 & 0 & 0 & 3.27 & 0.00 & 0.992 & 82.0 & 4.28 \\
\cmidrule(lr){1-17}
\multirow{3}{*}{C--E} & \multirow{3}{*}{0.90} & \multirow{3}{*}{2.50} & inspect & 0 & 0 & 0 & 0 & 0 & 0 & 0 & 50 & NA & 49.71 & 0.156 & 0.0 & 12.54 \\
& & & Pre & 0 & 2 & 2 & 13 & 9 & 4 & 4 & 16 & 26.49 & 3.78 & 0.754 & 28.0 & 87.56 \\
& & & Post & 0 & 2 & 2 & 13 & 9 & 4 & 4 & 16 & 0.68 & 3.78 & 0.808 & 80.0 & 44.26 \\
\cmidrule(lr){1-17}
\addlinespace[2pt]
\multirow{3}{*}{L--E} & \multirow{3}{*}{0.30} & \multirow{3}{*}{1.00} & inspect & 0 & 0 & 0 & 37 & 12 & 1 & 0 & 0 & 18.62 & 0.57 & 0.938 & 22.0 & 23.98 \\
& & & Pre & 0 & 0 & 0 & 50 & 0 & 0 & 0 & 0 & 12.47 & 0.00 & 0.969 & 30.0 & 16.04 \\
& & & Post & 0 & 0 & 0 & 50 & 0 & 0 & 0 & 0 & 9.86 & 0.00 & 0.977 & 60.0 & 12.34 \\
\cmidrule(lr){1-17}
\multirow{3}{*}{L--E} & \multirow{3}{*}{0.50} & \multirow{3}{*}{1.10} & inspect & 0 & 0 & 0 & 23 & 16 & 8 & 3 & 0 & 21.17 & 1.22 & 0.910 & 28.0 & 23.90 \\
& & & Pre & 0 & 3 & 7 & 40 & 0 & 0 & 0 & 0 & 17.15 & 0.62 & 0.896 & 10.0 & 111.76 \\
& & & Post & 0 & 3 & 7 & 40 & 0 & 0 & 0 & 0 & 4.84 & 0.62 & 0.918 & 64.0 & 102.02 \\
\cmidrule(lr){1-17}
\multirow{3}{*}{L--E} & \multirow{3}{*}{0.70} & \multirow{3}{*}{1.30} & inspect & 0 & 0 & 0 & 0 & 0 & 0 & 0 & 50 & NA & 16.86 & 0.419 & 6.0 & 22.70 \\
& & & Pre & 0 & 8 & 12 & 19 & 9 & 2 & 0 & 0 & 24.01 & 1.10 & 0.791 & 20.0 & 219.44 \\
& & & Post & 0 & 8 & 12 & 19 & 9 & 2 & 0 & 0 & 3.43 & 1.10 & 0.814 & 78.0 & 212.98 \\
\cmidrule(lr){1-17}
\multirow{3}{*}{L--E} & \multirow{3}{*}{0.90} & \multirow{3}{*}{2.50} & inspect & 0 & 0 & 0 & 26 & 18 & 6 & 0 & 0 & 4.27 & 0.92 & 0.955 & 18.0 & 4.44 \\
& & & Pre & 0 & 0 & 0 & 50 & 0 & 0 & 0 & 0 & 1.53 & 0.00 & 0.997 & 42.0 & 1.86 \\
& & & Post & 0 & 0 & 0 & 50 & 0 & 0 & 0 & 0 & 0.77 & 0.00 & 0.998 & 78.0 & 0.94 \\
\cmidrule(lr){1-17}
\addlinespace[2pt]
\multirow{3}{*}{C--U} & \multirow{3}{*}{0.30} & \multirow{3}{*}{1.00} & inspect & 0 & 0 & 0 & 38 & 12 & 0 & 0 & 0 & 18.56 & 0.49 & 0.949 & 36.0 & 22.74 \\
& & & Pre & 0 & 1 & 8 & 41 & 0 & 0 & 0 & 0 & 15.36 & 0.49 & 0.923 & 22.0 & 80.70 \\
& & & Post & 0 & 1 & 8 & 41 & 0 & 0 & 0 & 0 & 11.55 & 0.49 & 0.929 & 52.0 & 77.56 \\
\cmidrule(lr){1-17}
\multirow{3}{*}{C--U} & \multirow{3}{*}{0.50} & \multirow{3}{*}{1.10} & inspect & 0 & 0 & 0 & 35 & 13 & 1 & 1 & 0 & 25.47 & 0.72 & 0.930 & 20.0 & 29.68 \\
& & & Pre & 0 & 0 & 4 & 46 & 0 & 0 & 0 & 0 & 13.86 & 0.28 & 0.955 & 30.0 & 42.76 \\
& & & Post & 0 & 0 & 4 & 46 & 0 & 0 & 0 & 0 & 9.12 & 0.28 & 0.965 & 62.0 & 36.70 \\
\cmidrule(lr){1-17}
\multirow{3}{*}{C--U} & \multirow{3}{*}{0.70} & \multirow{3}{*}{1.30} & inspect & 0 & 0 & 0 & 16 & 25 & 6 & 3 & 0 & 19.38 & 1.23 & 0.898 & 20.0 & 22.34 \\
& & & Pre & 0 & 0 & 10 & 39 & 1 & 0 & 0 & 0 & 20.66 & 0.47 & 0.917 & 12.0 & 91.88 \\
& & & Post & 0 & 0 & 10 & 39 & 1 & 0 & 0 & 0 & 3.61 & 0.47 & 0.945 & 70.0 & 78.74 \\
\cmidrule(lr){1-17}
\multirow{3}{*}{C--U} & \multirow{3}{*}{0.90} & \multirow{3}{*}{2.50} & inspect & 0 & 0 & 0 & 0 & 0 & 0 & 0 & 50 & NA & 42.57 & 0.174 & 0.0 & 10.40 \\
& & & Pre & 0 & 0 & 3 & 41 & 4 & 2 & 0 & 0 & 17.64 & 0.55 & 0.952 & 12.0 & 39.88 \\
& & & Post & 0 & 0 & 3 & 41 & 4 & 2 & 0 & 0 & 0.95 & 0.55 & 0.972 & 90.0 & 23.90 \\
\bottomrule
\end{tabular}%
}
\vspace{2pt}
\begin{minipage}{0.98\textwidth}
\footnotesize Design codes are C--E (common transition/equal spacing), L--E (locally misspecified transitions/equal spacing), and C--U (common transition/unequal spacing). Each integer heading is an estimated-count frequency. Localization $\mathrm{RMSE}_{\widehat\tau}$ averages squared errors over change points and replications with $\widehat m=m_0$, then takes the square root; count $\mathrm{RMSE}_{\widehat m}$ uses all valid replications. ARI is computed from the segment labels, and ARI \% Best includes ties. The directed $d_H$ is computed from the true set to the estimated set: $\max_{b\in\mathcal S^\star}\min_{a\in\widehat{\mathcal S}}|b-a|$.
\end{minipage}
\end{table}

}{%
  \begin{center}
  \fbox{\parbox[c][1.4in][c]{0.94\linewidth}{\centering
    Placeholder for the consolidated multi-change-point table:\\[4pt]
    \texttt{\detokenize{numerical_results/multi_cp_table.tex}}}}
  \end{center}
}

Table~\ref{tab:mc-summary} shows systematic gains from refinement. Post has lower conditional localization error and higher average ARI than Pre in all 12 configurations. Because Step~3 refines the retained change-point locations without reselecting their number, Pre and Post have identical
estimated-count distributions. Their comparison therefore isolates the effect of filtering and refinement: Step~3 reduces both
$\mathrm{RMSE}_{\widehat\tau}$ and the directed Hausdorff distance in every configuration. Post attains average ARI values between $0.808$ and $0.998$
and has the highest average ARI among the three methods in 11 out of the 12 configurations. For each dependence level, we compute the ratio of Post to Pre $\mathrm{RMSE}_{\widehat\tau}$ separately for C--E, L--E, and C--U, and take average of these three ratios. At $\rho(\mathcal A)=0.3$, $0.5$, $0.7$, and $0.9$, the resulting mean Post/Pre ratios are $0.780$, $0.466$, $0.212$, and $0.194$, respectively, corresponding to average relative reductions of approximately $22.0\%$, $53.4\%$, $78.8\%$, and $80.6\%$. These localization gains highlight the contribution of Steps~2--3 even when the preliminary procedure estimates the number of changes correctly. Thus, correct estimation of the number of changes does not remove the need for refinement: estimating the common dynamics in Step~2 and refining the locations on the filtered series in Step~3 yield substantial additional localization accuracy.

The results also indicate robustness to moderate departures from the common-dynamics assumption. Under local transition misspecification, Post continues to improve on Pre and retains high segmentation accuracy across most dependence levels. Within the L--E design, performance deteriorates primarily at
$\rho(\mathcal A)=0.7$, where the count RMSE reaches $1.10$, while the average ARI remains $0.814$. At this configuration, refinement reduces conditional localization RMSE from $24.01$ to $3.43$. Thus, the localization gains from refinement persist even when the VAR dynamics \textit{vary across regimes}.

The results further demonstrate robustness to
unequal change-point spacing. In the C--U design, the proposed procedure recovers the correct number of changes in 39--46 of the 50 replicates across all dependence levels, while Post maintains average ARI values between $0.929$ and $0.972$. The benefits of refinement remain substantial under stronger
temporal dependence: at $\rho(\mathcal A)=0.7$ and $0.9$, the conditional localization RMSE decreases from $20.66$ to $3.61$ and from $17.64$ to $0.95$, respectively. These findings indicate that the localization gains from refinement are not restricted to equally spaced change points.

The estimated number of change points further distinguishes the proposed procedure from Inspect. The proposed procedure recovers the true count in at least 39 of 50 replicates in ten of the twelve configurations, with the largest count errors occurring in the C--E design at $\rho(\mathcal A)=0.9$ and the L--E design at $\rho(\mathcal A)=0.7$. Inspect substantially overestimates the number of changes in several configurations. At $\rho(\mathcal A)=0.9$, all 50 Inspect replicates in the C--E and C--U designs produce at least seven estimated change points, with count RMSE ranging from $42.57$ to $49.71$ and average ARI from $0.156$ to $0.174$. This overestimation can also yield a smaller directed Hausdorff distance $d_H$: adding estimated change points cannot increase this measure, so a smaller value need not indicate better overall segmentation. Turning to localization accuracy, Post also has lower conditional $\mathrm{RMSE}_{\widehat\tau}$ than Inspect in every configuration for which both values are available. Averaging the design-specific ratios of Post to Inspect $\mathrm{RMSE}_{\widehat\tau}$ gives approximately $0.676$, $0.345$, $0.167$, and $0.180$ at $\rho(\mathcal A)=0.3$, $0.5$, $0.7$, and $0.9$, respectively.

\section{Application to Macroeconomic Indicators}
\label{sec:fred_md_empirical}

We apply the proposed multiple-change procedure to a panel of macroeconomic indicators extracted from the FRED-MD database \citep{StLouisFedFREDMD}. We use the May 2026 vintage and transform each series following the recommendations in \citet{McCrackenNg2016}. Restricting the transformed panel to common non-missing observations yields $p=126$ variables with $n=403$ consecutive monthly observations spanning March 1992 to September 2025. The variables used in the analysis, together with their descriptions, transformation codes, and economic group classifications, are documented in Appendix~\ref{app:fred-md-data-documentation}.

As a benchmark, we apply Inspect~\citep{WangSamworth2018}, which was the strongest competing procedure in our single-change simulations. Our method selects seven change-points, with final refined estimates (Post) located at:
\[
\text{1999-09},\quad
\text{2003-02},\quad
\text{2007-03},\quad
\text{2009-01},\quad
\text{2016-11},\quad
\text{2020-06},\quad
\text{2022-03}.
\]
The preliminary estimates (Pre), obtained after Stage~3 of Step-1, are located at 2000-02, 2003-07, 2007-03, 2009-07, 2017-01, 2020-04, and 2022-03. The final filtering and refinement step moves the first, second, fourth, and fifth locations earlier by five, five, six, and two months, respectively, and the sixth location two months later, while leaving the 2007-03 and 2022-03 estimates unchanged. The estimated lag order is $\widehat d=1$. In contrast, Inspect detects 162 change points and appears overly sensitive on this panel of macroeconomic indicators; therefore, we only use it as a descriptive benchmark.

For the selected VAR(1) specification, the Step~2 coefficient estimate $\widehat{A}_1$ contains 353 non-zero elements out of $126 \times 126 = 15,876$ parameters, corresponding to a density of $2.2\%$. Its spectral radius is $\rho(\widehat{A}_1) = 0.93 < 1$, confirming that the fitted common VAR process is stationary. This elevated spectral radius reflects substantial persistence in macroeconomic dynamics, underscoring the practical necessity of accounting for common temporal dependence prior to refining change-point locations.

\begin{figure}[!htbp]
\centering
\includegraphics[width=\textwidth,height=0.78\textheight,keepaspectratio]
{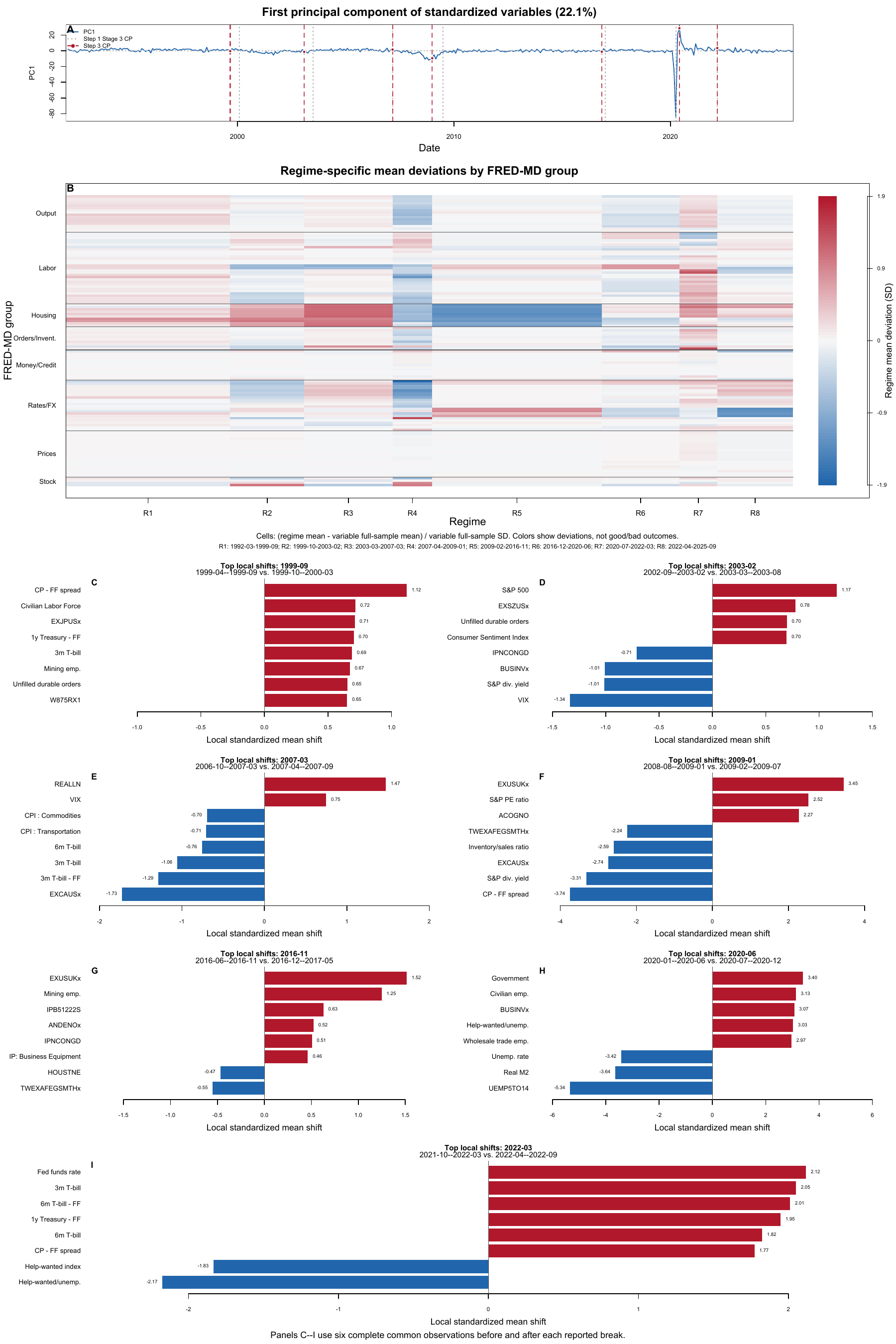}
\caption{Change-point estimates for the transformed FRED-MD panel. Panel~A plots the first principal component of the standardized panel (explaining 22.1\% of total variance), marking preliminary (Pre) and refined (Post) break dates. Panel~B reports standardized regime-mean deviations across the eight Post-defined regimes. Panels~C--I display the eight largest absolute standardized local mean shifts around each refined estimate, based on a six-observation window on either side.}
\label{fig:fred_md}
\end{figure}

Figure~\ref{fig:fred_md} summarizes the estimated segmentation. Panel~A plots the first principal component of the standardized panel (explaining 22.1\% of total variance), Panel~B displays standardized regime means across the eight estimated regimes, and Panels~C--I report variables with the largest standardized local mean shifts around each break point. The breaks at 2007-03 and 2009-01 bracket the Great Financial Crisis. Although 2007-03 precedes the official NBER recession onset (December 2007), it pinpoints initial subprime and credit distress—marked by late-February 2007 equity declines, Freddie Mac subprime loan restrictions, Greenspan's recession warning, and New Century Financial's April bankruptcy—concentrated mainly in financial, credit, and interest-rate series. Similarly, while 2009-01 leads the NBER trough (June 2009), it aligns with the shift toward policy-led stabilization around the passage of the $\$787$~billion American Recovery and Reinvestment Act (ARRA) in February 2009, reflected in realignments across exchange rates, equities, credit aggregates, and inventories.

The 2016-11 estimate coincides with the U.S. presidential election and the accompanying shift in policy expectations and financial conditions.\footnote{See Board of Governors of the Federal Reserve System, \emph{Monetary Policy Report}, February 14, 2017, Part~1: \url{https://www.federalreserve.gov/monetarypolicy/2017-02-mpr-part1.htm}.} The largest local mean shifts at this boundary occur in exchange rates, mining employment, industrial production, and new orders, reflecting a broader late-2016 macroeconomic realigning without implying direct election causality. The 2020-06 break is driven by shifts in labor-market indicators, business inventories, and real money balances. The six-observation window in Panel~H contrasts January--June 2020 with July--December 2020, capturing the transition from initial pandemic disruption to mid-2020 economic resumption rather than the recession's onset.\footnote{See U.S. Bureau of Labor Statistics, \emph{The Employment Situation---June 2020}, July 2, 2020: \url{https://www.bls.gov/news.release/archives/empsit_07022020.htm}.}

The 2022-03 break centers on short-term interest rates and labor-market tightness. Finally, the 1999-09 and 2003-02 estimates frame the early-2000s cycle: 1999-09 marks the end of the late-1990s expansion prior to the 2001 recession, while 2003-02 captures the initial employment-lagging recovery before the mid-2000s expansion. Importantly, these associations are descriptive, as the algorithm detects statistical mean shifts in the transformed panel rather than structural economic shocks.

\begin{figure}[!htbp]
\centering
\IfFileExists{numerical_results/01_normalized_loss_profiles_main_2panels.pdf}{%
  \includegraphics[width=\textwidth]{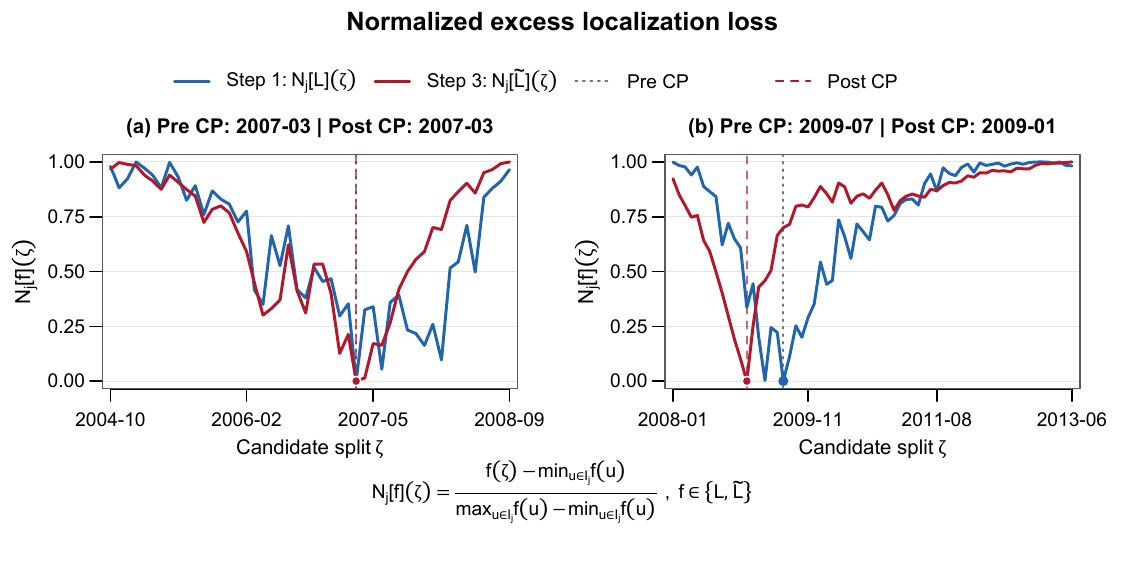}%
}{%
  \begin{minipage}[t]{0.48\textwidth}
    \centering
    \textcolor{blue}{\fbox{\parbox[c][1.65in][c]{0.90\linewidth}{\centering
    (a) Loss-profile placeholder\\[5pt]
    Pre CP: 2007-03\\Post CP: 2007-03}}}
  \end{minipage}\hfill
  \begin{minipage}[t]{0.48\textwidth}
    \centering
    \textcolor{blue}{\fbox{\parbox[c][1.65in][c]{0.90\linewidth}{\centering
    (b) Loss-profile placeholder\\[5pt]
    Pre CP: 2009-07\\Post CP: 2009-01}}}
  \end{minipage}%
}
\caption{Loss profiles evaluate step-specific data windows and ridge parameters, normalized by subtracting their minimum and dividing by their range. Vertical lines and dots denote estimated break locations and profile minima, respectively; vertical scales depict relative criterion contrast rather than absolute loss magnitudes.}
\label{fig:fred_md_loss_profiles}
\end{figure}

Figure~\ref{fig:fred_md_loss_profiles} compares normalized excess localization losses before (Pre) and after (Post) VAR filtering for the refined estimates corresponding to 2007-03 and 2009-01. Filtering sharpens candidate separation by suppressing competing low-loss regions: near both dates, the Post-filtering loss rises more sharply to the right of its global minimum, whereas Pre-filtering profiles display broader, more irregular low-loss basins. The remaining five profiles are provided in Figure~\ref{fig:fred_md_loss_profiles_appendix} of Appendix~\ref{app:fred-md-additional-loss-profiles}, which illustrate date shifts from 2017-01 to 2016-11 and from 2020-04 to 2020-06 (where a secondary local minimum persists near 2020-04). Overall, VAR filtering enhances localization contrast even when estimated break dates remain unchanged. Because each profile is individually normalized and ground-truth change points are unobserved, these curves demonstrate relative criterion sharpening rather than absolute loss magnitudes or direct estimation errors.

For comparison, \citet{HorvathLiuRiceZhao2022} apply CUSUM-based procedures to a transformed macroeconomic panel through June 2019, identifying August 2003 and multiple GFC-era breaks. Our 2003-02 estimate leads their 2003 break by six months, while our 2007-03 and 2009-01 estimates offer a distinct segmentation of the financial crisis. Given differences in methodology, sample coverage, and target parameters, these comparisons offer contextual alignment rather than direct validation. Overall, our procedure identifies economically interpretable structural shifts completely agnostic to NBER business-cycle dates or external event timing.

\section{Conclusion}\label{sec:conclusion}

We study the detection and localization of mean changes in high-dimensional VARs with common transition dynamics. Our procedure first obtains preliminary change-point estimates, uses buffered observations to estimate the common
sparse VAR, and then filters the series to refine the estimated locations. Our theory quantifies the resulting localization gain: filtering removes the direct dependence penalty from the oracle localization rate, while the feasible rate accounts explicitly for estimation of the VAR operator and the finite transition layer induced by filtering across a change point. We extend the procedure to an unknown number of changes using rolling-window candidate generation, information-criterion screening, and local refinement. Numerical experiments support the predicted localization gains and show robustness to moderate departures from common VAR dynamics, while the macroeconomic indicators application illustrates the method on a large macroeconomic panel.

The common-dynamics assumption isolates mean changes from changes in their propagation mechanism. Allowing the transition matrices and means to change jointly, possibly at different dates, is a natural extension and would require
joint identification and localization of the two forms of structural change. Another direction is a recursive intercept-change model, in which the
post-change mean adjusts dynamically rather than through the finite transition layer induced by the spliced formulation considered here.

\section*{AI use statement}

During the preparation of this work, the authors used ChatGPT and Codex (OpenAI) to assist with notation consistency and grammar checks, language refinement, reference formatting and the development and debugging of \texttt{R} code for simulations and real data analyses. After using these tools, the authors thoroughly reviewed and verified all outputs, revised and polished the manuscript and code as needed, and take full responsibility for the content of the current version.

\bibliographystyle{plainnat}
\bibliography{cpvar}

\newpage

\appendix
\begin{center}
    \begin{huge}
        Appendix
    \end{huge}
\end{center}
\begingroup
\noindent Appendix~\ref{sec:appendix} states the conventions used in the proofs.
Appendix~\ref{app:step1-proof} establishes the preliminary single-change localization bound.
Appendix~\ref{asf:sec:S2-proof} proves the transition-matrix estimation bound on data-dependent buffered blocks.
Appendix~\ref{asf:sec:S3-final} proves the localization bound for the filtered single-change refinement.
Appendix~\ref{app:multiple-proofs} extends the localization and filter-estimation arguments to multiple change points.
Appendix~\ref{sec:remove-transition-alignment} develops a transition-corrected refinement and presents examples that violate transition alignment.
Appendix~\ref{app:numerical-results} provides additional simulation results and implementation details.
Appendix~\ref{app:fred-md-data-documentation} lists the FRED-MD variables, transformations, and economic groups used in the empirical analysis.
\par\endgroup

\noindent For the reader’s convenience, we collect here the general notation used in the appendices.

\noindent \textbf{Notation for the appendices.} Write $\mathbb Z$ and $\mathbb R$ for the integers and real numbers. For a set $S$, $|S|$ denotes its cardinality and $S^c$ its complement. We use $\mathbf1\{\cdot\}$ for the indicator function, $\operatorname{supp}(v)$ for the support of a vector, and $v_S$ for its restriction to $S$. For vectors, $\|\cdot\|_q$ denotes the $\ell_q$ norm for $q\ge1$. For matrices, $\|\cdot\|_2$ denotes the operator norm and $\|\cdot\|_F$ the Frobenius norm. Unless otherwise specified, unsubscripted norms are Euclidean for vectors and operator norms for matrices. For a scalar random variable $X$, $\|X\|_{L_q}=(E|X|^q)^{1/q}$, and in moment bounds $\|X\|_2$ also denotes $\|X\|_{L_2}$; $\|X\|_{\psi_1}=\inf\{c>0:E\exp(|X|/c)\le2\}$ is the $\psi_1$-Orlicz norm. We write $\Var$ and $\Cov$ for variance and covariance. The symbols $H^\top$ and $H^*$ denote transpose and conjugate transpose, while $\operatorname{vec}$, $\operatorname{tr}$, $\otimes$, and $\langle\cdot,\cdot\rangle$ denote columnwise vectorization, trace, Kronecker product, and the Euclidean inner product, respectively. We write $\mathbf{I}_r$ and $\mathbf1_r$ for the identity matrix and the all-ones vector of dimension $r$, $\operatorname{diag}$ for a block-diagonal matrix, $\rho(H)$ for the spectral radius of a square matrix $H$, and $\succeq$ for the positive-semidefinite ordering. For real numbers, $a\vee b:=\max\{a,b\}$, while $\lfloor\cdot\rfloor$ and $\lceil\cdot\rceil$ denote the floor and ceiling functions. We use $O(\cdot)$, $o(\cdot)$, $O_{\mathbb P}(\cdot)$, and $o_{\mathbb P}(\cdot)$ in their usual senses. Finally, $\gtrsim$ and $\asymp$ denote comparisons up to constant factors, and $\gg$ denotes the reverse of little-o order.

Unless explicitly redefined locally, all model and estimator notation retains its meaning from the main text. We denote the set of natural numbers by $\mathbb{N} = \{1, 2, \ldots\}$, the zero vector in $\mathbb{R}^r$ by $\mathbf{0}_r$, and the principal submatrix indexed by set $T$ by $H_{TT}$. The Frobenius inner product is written as $\langle U, V \rangle_{\mathrm{F}}$, $H^{-\ast} := (H^{-1})^\ast$ denotes the conjugate transpose of the inverse matrix, and $\operatorname{blockdiag}(\cdot)$ denotes block-diagonal matrix assembly.

\section{Conventions for the proofs}\label{sec:appendix}

We use the spliced model \eqref{eq:single-cp} and the common centered process
\eqref{eq:common-centered}. The symbol $N$ denotes a sample size local to the argument: the total retained block size in Appendix~\ref{asf:sec:S2-proof}, $n-d$ in Appendix~\ref{asf:sec:S3-final}, and a local interval length in Appendix~\ref{app:multiple-proofs}. The scalar scales $b_n$ and $b_{m,n}$ measure filter-estimation error, whereas $b_\zeta$ in Appendix~\ref{app:step1-proof} is a scalar contrast weight; the transition-template vectors in Appendix~\ref{sec:remove-transition-alignment} are denoted by $\varphi_j$.
Constants $c,C$ may change between displays. They may depend on fixed trimming
constants, the fixed lag order, the fixed innovation variance, and stated bounds
on tuning parameters, but not on $n,p,s$. We display any dependence through $\gamma_p$ and
$\bar Q$ explicitly.

For the Step~1 calculations we rescale $y_t,\eta_j$ and $\varepsilon_t$ by
$\sigma$, so that the innovations have covariance $\mathbf{I}_p$. The minimizing split is invariant
under this common rescaling, and the fixed powers of $\sigma$ are
absorbed into constants on returning to the original units. The Step~2 formulas
retain $\sigma^2$ explicitly. Every probabilistic statement refers to the
original, unconditional law of the observations. In particular, whenever a
data-selected block is pure, we still treat its endpoints as random and never
condition on them.

\section{Single Changepoint: Step 1 initial localization}\label{app:step1-proof}

To make the appendix self-complete, we repeat the model and the estimators first. Consider the single-change model
\begin{equation}\label{asf:eq:model}
y_t=\eta_1\one\{t\le \tau\}+\eta_2\one\{t>\tau\}+e_t,\qquad t=1,\dots,n,
\end{equation}
where
\begin{equation}\label{asf:eq:error_rep}
e_t=\sum_{\ell=0}^\infty \Psi_\ell \varepsilon_{t-\ell},
\qquad
\varepsilon_t\stackrel{\mathrm{i.i.d.}}{\sim}N(0,\mathbf{I}_p),
\qquad
\gamma_p:=\sum_{\ell=0}^\infty \|\Psi_\ell\|_2.
\end{equation}
Since $\Psi_0=\mathbf{I}_p$, we have $\gamma_p\ge 1$.
In the VAR setting of Section~\ref{sec:single},
\[
\eta_j=(\mathbf{I}_p-A_1-\cdots-A_d)^{-1}\mu_j,\qquad j=1,2.
\]
Let
\[
\Delta_\eta:=\eta_2-\eta_1.
\]

\noindent For each candidate split \(\zeta\in\{1,\dots,n-1\}\), define
\begin{equation*}
\bar y_1(\zeta):=\frac1\zeta\sum_{t=1}^{\zeta}y_t,
\qquad
\bar y_2(\zeta):=\frac1{n-\zeta}\sum_{t=\zeta+1}^{n}y_t.
\end{equation*}

As in Step~1 of Section~\ref{ssc:singlecptau}, the criterion below is minimized over $\eta_1,\eta_2\in\mathbb R^p$, with ridge parameter $\lambda_1\ge0$.
\noindent For every \(\zeta\in\{1,\dots,n-1\}\), define
\begin{equation}\label{asf:eq:Q}
Q(\eta_1,\eta_2;\zeta)
=
\frac1n\Big(\sum_{t=1}^{\zeta}\|y_t-\eta_1\|_2^2+\sum_{t=\zeta+1}^{n}\|y_t-\eta_2\|_2^2\Big)
+\lambda_1\|\eta_1-\eta_2\|_2^2,
\end{equation}
and let the \textbf{fitted means} be 
\[
(\widehat\eta_1(\zeta),\widehat\eta_2(\zeta))
\in \arg\min_{\eta_1,\eta_2}Q(\eta_1,\eta_2;\zeta).
\]

\noindent Define the profile loss after plugging in the fitted means 
\begin{equation}\label{asf:eq:L1}
L(\zeta)
=
\frac1n\Big(\sum_{t=1}^{\zeta}\|y_t-\widehat\eta_1(\zeta)\|_2^2+\sum_{t=\zeta+1}^{n}\|y_t-\widehat\eta_2(\zeta)\|_2^2\Big),
\end{equation}
and we estimate \(\tau\) by
\begin{equation}\label{asf:eq:tauhat}
\widehat\tau=\arg\min_{\zeta\in\mathcal T_n}L(\zeta),
\qquad
\mathcal T_n:=\{\zeta\in\mathbb Z:\ c^\ast n\le \zeta\le (1-c^\ast)n\},
\end{equation}
where \(c^\ast\in(0,1/2)\) is fixed.

The contrast $T_\zeta$ and ridge weights $\Lambda_1(\zeta),b_\zeta$ are defined in Lemma~\ref{asf:lem:profile}; the centered contrast $G_\zeta$ and centered quadratic term $Q_\zeta$ are defined in Lemma~\ref{asf:lem:quadinc}.
Throughout this section we work under Assumptions~\ref{ass:innov}, \ref{H1} and
\ref{assump:tau-domain}, and we assume \(0\le\lambda_1\le\bar\lambda\) for a fixed \(\bar\lambda<\infty\); these are stated
in Section~\ref{sec:single}. Our goal in this section is to prove
Proposition~\ref{prop:tau0}. The ridge parameter enters only through the deterministic quantities $\Lambda_1(\zeta)$ and $b_\zeta$, whose bounds in Lemmas~\ref{asf:lem:weights}, \ref{asf:lem:signaldrift} and~\ref{asf:lem:noisedrift} depend on $\lambda_1$ only through $\bar\lambda$, while $T_\zeta$, $G_\zeta$ and $Q_\zeta$ do not involve $\lambda_1$. Bounding these random quantities once, as in the proof of Lemma~\ref{lem:uniform-local-raw}, we obtain Proposition~\ref{asf:prop:fluctuation} and the tail bound in ~\ref{asf:eq:tau0-tail} with the supremum over $\lambda_1\in[0,\bar\lambda]$ taken inside the expectation and the probability, while the drift bound of Proposition~\ref{asf:prop:drift} holds uniformly in $\lambda_1$. In particular, Proposition~\ref{prop:tau0} covers a data-driven $\lambda_1\in[0,\bar\lambda]$.

The proof proceeds in four main steps. First, Section~\ref{ssc:profilecriterion} derives an exact representation of the profiled criterion. Next, Section~\ref{ssc:localfluctuation} establishes a local fluctuation bound, while Section~\ref{ssc:drift} develops a deterministic drift inequality. Finally, Section~\ref{ssc:completingsinglecpproof} combines these bounds via a shell argument. Although this final step mirrors the structure of Theorem~3.2.5 in \cite{van1996weak} with the verification of its sufficient conditions driving much of the technical length, we present it independently for technical clarity.

\subsection{Profile criterion}\label{ssc:profilecriterion}

We use the following algebraic identity in both localization steps; it is
purely deterministic and holds for any data sequence.

\begin{lemma}\label{asf:lem:profile}
Let $y_1,\ldots,y_n\in\mathbb R^p$, $n\ge2$, and $\lambda_1\ge0$. For
$1\le\zeta<n$, let the fitted means minimize \eqref{asf:eq:Q}, and let
$L$ be the unpenalized loss \eqref{asf:eq:L1}. Define
\[
\Lambda_1(\zeta)=\frac{\lambda_1}{\lambda_1+\zeta(n-\zeta)/n^2}.
\]
The unique fitted means satisfy
\begin{equation}\label{asf:eq:step1A_matrix}
\begin{pmatrix}\widehat\eta_1\\\widehat\eta_2\end{pmatrix}
=\begin{pmatrix}
1-\frac{n-\zeta}{n}\Lambda_1&\frac{n-\zeta}{n}\Lambda_1\\
\frac{\zeta}{n}\Lambda_1&1-\frac{\zeta}{n}\Lambda_1
\end{pmatrix}
\begin{pmatrix}\bar y_1(\zeta)\\\bar y_2(\zeta)\end{pmatrix},
\end{equation}
where $\Lambda_1=\Lambda_1(\zeta)$. In particular,
\begin{equation}\label{asf:eq:step1A_diff}
\widehat\eta_2(\zeta)-\widehat\eta_1(\zeta)
=(1-\Lambda_1(\zeta))\{\bar y_2(\zeta)-\bar y_1(\zeta)\}.
\end{equation}
Writing $\bar y=n^{-1}\sum_t y_t$ and
$T_\zeta=\sum_{t\le\zeta}y_t-(\zeta/n)\sum_t y_t$, we have
\begin{equation}\label{asf:eq:L1profile}
L(\zeta)=\frac1n\left\{\sum_{t\le\zeta}\|y_t-\bar y_1\|_2^2+
\sum_{t>\zeta}\|y_t-\bar y_2\|_2^2\right\}
+\frac{\zeta(n-\zeta)}{n^2}\Lambda_1(\zeta)^2\|\bar y_1-\bar y_2\|_2^2,
\end{equation}
and
\begin{equation}\label{asf:eq:M1profile}
M_1(\zeta):=\frac1n\sum_t\|y_t\|_2^2-L(\zeta)
=\|\bar y\|_2^2+b_\zeta\|T_\zeta\|_2^2,
\qquad b_\zeta=\frac{1-\Lambda_1(\zeta)^2}{\zeta(n-\zeta)}.
\end{equation}
We see that minimizing $L$ is equivalent to maximizing $M_1$.
\end{lemma}
\begin{proof}
The first-order equations are
$(\zeta/n)(\widehat\eta_1-\bar y_1)+\lambda_1(\widehat\eta_1-\widehat\eta_2)=0$
and
$((n-\zeta)/n)(\widehat\eta_2-\bar y_2)+\lambda_1(\widehat\eta_2-\widehat\eta_1)=0$.
Their coefficient matrix is positive definite, and solving the two equations we obtain
\eqref{asf:eq:step1A_matrix}--\eqref{asf:eq:step1A_diff}.
Within each segment, the residuals about its sample mean sum to zero, so
substituting the fitted means we arrive at \eqref{asf:eq:L1profile}. Finally, the within-between sum-of-squares
identity, together with
$T_\zeta=\zeta(n-\zeta)(\bar y_1-\bar y_2)/n$, yields
\eqref{asf:eq:M1profile}.
\end{proof}

\subsection{Auxiliary lemmas for fluctuation control}

\begin{lemma}\label{asf:lem:weights}
Use the definitions of $T_\zeta$ and $b_\zeta$ in Lemma~\ref{asf:lem:profile}.
For $w_{\zeta,t}=\one\{t\le\zeta\}-\zeta/n$ and $1\le\zeta,\xi<n$,
\[
T_\zeta=\sum_t w_{\zeta,t}y_t,\qquad
\sum_t w_{\zeta,t}^2=\frac{\zeta(n-\zeta)}n,
\]
\[
\sum_t(w_{\zeta,t}-w_{\xi,t})^2
=|\zeta-\xi|-\frac{|\zeta-\xi|^2}{n},\qquad
\sum_t(w_{\zeta,t}+w_{\xi,t})^2\le n.
\]
If $0\le\lambda_1\le\bar\lambda<\infty$ and
$\zeta,\xi\in\mathcal T_n$ from \eqref{asf:eq:tauhat}, then
\[
0<b_\zeta\le Cn^{-2},\qquad
|b_\zeta-b_\xi|\le C|\zeta-\xi|n^{-3},
\]
where $C$ depends only on $c^\ast$ and $\bar\lambda$.
Under the mean structure \eqref{asf:eq:model} with $Ee_t=0$,
\[
ET_\zeta=-g_\zeta\Delta_\eta,\qquad
g_\zeta=\begin{cases}\zeta(n-\tau)/n,&\zeta\le\tau,\\
\tau(n-\zeta)/n,&\zeta>\tau,
\end{cases}
\]
and $0\le g_\zeta\le n$, $|g_\zeta-g_\xi|\le|\zeta-\xi|$.
\end{lemma}
\begin{proof}
We obtain the weight identities by separating the indices before, between, and
after the two candidate splits. For the last weight bound, we use
$(a+b)^2\le2a^2+2b^2$ and $\zeta(n-\zeta)\le n^2/4$.
For the coefficient bounds, we write
$b_\zeta=n^{-2}h(\zeta/n,\lambda_1)$, where
\[
h(r,\lambda)=\frac{1-[\lambda/\{\lambda+r(1-r)\}]^2}{r(1-r)}.
\]
Both $h$ and its derivative in $r$ are bounded on
$[c^\ast,1-c^\ast]\times[0,\bar\lambda]$, and $h$ is positive there.
Finally, substituting the two regime means into $ET_\zeta$, we see that the resulting
piecewise-linear function $g_\zeta$ has slopes bounded in absolute value by one.
\end{proof}

\begin{lemma}\label{asf:lem:projvar}
Let $e_t$ be the centered linear process in \eqref{asf:eq:error_rep}, with $\gamma_p<\infty$ and innovations of covariance $\mathbf{I}_p$ under the normalization in Appendix~\ref{sec:appendix}.
For any deterministic real numbers \(v_1,\dots,v_n\) and any \(u\in\mathbb R^p\),
\[
\Var\Big(\sum_{t=1}^n v_t\,u^\top e_t\Big)
\le
\gamma_p^2\|u\|_2^2\sum_{t=1}^n v_t^2.
\]

For $\Gamma_h=\operatorname{Cov}(e_t,e_{t+h})$, the same process also satisfies
\[
\sum_{h\in\mathbb Z}\|\Gamma_h\|_2\le\gamma_p^2,
\qquad \sum_{h\in\mathbb Z}|\operatorname{tr}(\Gamma_h)|\le p\gamma_p^2.
\]
\end{lemma}

\begin{proof}
From \eqref{asf:eq:error_rep},
\[
\sum_{t=1}^n v_t\,u^\top e_t
=
\sum_{s\in\mathbb Z}
\Big(\sum_{t=1}^n v_t\Psi_{t-s}^\top u\Big)^\top \varepsilon_s.
\]
By independence of the Gaussian innovations,
\[
\Var\Big(\sum_{t=1}^n v_t\,u^\top e_t\Big)
=
\sum_{s\in\mathbb Z}\Big\|\sum_{t=1}^n v_t\Psi_{t-s}^\top u\Big\|_2^2.
\]
Let \(a_\ell:=\|\Psi_\ell\|_2\) and \(\tilde v_t:=|v_t|\). Then
\[
\Big\|\sum_{t=1}^n v_t\Psi_{t-s}^\top u\Big\|_2
\le
\|u\|_2\sum_{t=1}^n \tilde v_t a_{t-s}.
\]
We therefore get
\[
\Var\Big(\sum_{t=1}^n v_t\,u^\top e_t\Big)
\le
\|u\|_2^2\sum_{s\in\mathbb Z}\Big(\sum_{t=1}^n \tilde v_t a_{t-s}\Big)^2.
\]
By Young's convolution inequality,
\[
\sum_{s\in\mathbb Z}\Big(\sum_{t=1}^n \tilde v_t a_{t-s}\Big)^2
\le
\Big(\sum_{\ell=0}^\infty a_\ell\Big)^2\sum_{t=1}^n v_t^2
=
\gamma_p^2\sum_{t=1}^n v_t^2,
\]

For $h\ge0$, $\Gamma_h=\sum_{j\ge0}\Psi_j\Psi_{j+h}^\top$;
negative lags give the transposes. Summing the operator-norm bounds over all
lags, we obtain $(\sum_{j\ge0}\|\Psi_j\|_2)^2=\gamma_p^2$.
The trace bound then follows from $|\operatorname{tr}(G)|\le p\|G\|_2$.
\end{proof}

\begin{lemma}\label{asf:lem:gaussmax}
Let $(Z_r)_{r=0}^m$ be a real process on a finite integer interval, $m\ge1$.
If $\|Z_r-Z_s\|_{\psi_1}\le C_{\mathrm{inc}}\sqrt{|r-s|}$ for all $r,s$ and some constant $C_{\mathrm{inc}}>0$, then there exists a $C>0$ such that, 
\[
\left\|\max_{0\le r\le m}|Z_r-Z_0|\right\|_{\psi_1}\le C C_{\mathrm{inc}}\sqrt m,
\qquad E\max_{0\le r\le m}|Z_r-Z_0|\le C C_{\mathrm{inc}}\sqrt m.
\]
In particular, the expectation bound holds for a centered Gaussian process
with $E(Z_r-Z_s)^2\le C_{\mathrm{inc}}^2|r-s|$.
\end{lemma}
\begin{proof}
At dyadic resolution $j$, we use a grid with at most $2^j+1$ points and mesh at
most $Cm2^{-j}$. Each increment from a grid point to its parent has
$\psi_1$ norm at most $C C_{\mathrm{inc}}\sqrt m\,2^{-j/2}$. The maximum of $N$ variables
with $\psi_1$ norms at most $a$ has $\psi_1$ norm at most $Ca\log(2N)$.
Applying the triangle inequality along the chains, we bound the norm of the
supremum by $C C_{\mathrm{inc}}\sqrt m\sum_{j\ge0}(j+1)2^{-j/2}$, and this series converges. Gaussian increments have $\psi_1$ norms bounded by
a universal constant times their standard deviations, which proves the final claim.
\end{proof}

\begin{lemma}\label{asf:lem:bilinear}
Let \((U,V)\) be a jointly centered Gaussian vector in \(\mathbb R^p\times\mathbb R^p\), and define
\[
\Sigma_{UU}:=\mathbb E(UU^\top),\qquad
\Sigma_{VV}:=\mathbb E(VV^\top),\qquad
\Sigma_{UV}:=\mathbb E(UV^\top).
\]
Then,
\[
\Var(\ip{U}{V})=\tr(\Sigma_{UU}\Sigma_{VV})+\tr(\Sigma_{UV}\Sigma_{UV}).
\]
In particular,
\[
\|\ip{U}{V}-E\ip{U}{V}\|_2
\le
\sqrt{2\,\tr(\Sigma_{UU})\,\|\Sigma_{VV}\|_2}.
\]
\end{lemma}

\begin{proof}
By Wick's formula,
\[
\mathbb E(U_iV_iU_jV_j)
=
\mathbb E(U_iU_j)\mathbb E(V_iV_j)+\mathbb E(U_iV_j)\mathbb E(V_iU_j)+\mathbb E(U_iV_i)\mathbb E(U_jV_j).
\]
Summing over \(i,j\) and subtracting \((E\ip{U}{V})^2\), we obtain the variance identity.

For the inequality, the first term satisfies
\(\tr(\Sigma_{UU}\Sigma_{VV})\le\tr(\Sigma_{UU})\|\Sigma_{VV}\|_2\), since
\(\tr(AB)\le\tr(A)\|B\|_2\) for positive semidefinite \(A,B\).  For the second
term we need a different argument, because \(\Sigma_{UV}\) need not be positive
semidefinite. Joint Gaussianity gives the representation
\(\Sigma_{UV}=\Sigma_{UU}^{1/2}K\Sigma_{VV}^{1/2}\) with \(\|K\|_2\le1\),
whence
\[
\bigl|\tr(\Sigma_{UV}\Sigma_{UV})\bigr|
\le\|\Sigma_{UV}\|_F^2
=\tr\bigl(\Sigma_{UU}^{1/2}K\Sigma_{VV}K^\top\Sigma_{UU}^{1/2}\bigr)
\le\|\Sigma_{VV}\|_2\,\tr(\Sigma_{UU}).
\]
Adding the two bounds, we arrive at the stated inequality.
\end{proof}

We also use the following consequence of Gaussian hypercontractivity. If
$W$ is centered and belongs to the second Wiener chaos, then
$\|W\|_{L_q}\le(q-1)\|W\|_{L_2}$ for $q\ge2$, and hence
\begin{equation}\label{eq:chaos-orlicz}
\|W\|_{\psi_1}\le C\|W\|_{L_2}.
\end{equation}
For the quadratic forms below one can also see this by diagonalizing the
form and applying the Gaussian quadratic-form moment bound.

\begin{lemma}\label{asf:lem:quadinc}
Let $y_t$ follow \eqref{asf:eq:model} with errors \eqref{asf:eq:error_rep}, $\gamma_p<\infty$, and let $T_\zeta$ and $\mathcal T_n$ be as in \eqref{asf:eq:tauhat} and Lemma~\ref{asf:lem:profile}. All suprema below are restricted to $\zeta\in\mathcal T_n$.
Let
\[
G_\zeta:=T_\zeta-E T_\zeta,
\qquad
Q_\zeta:=\|G_\zeta\|_2^2-E\|G_\zeta\|_2^2.
\]
Then there exists a constant \(C>0\) such that for all \(\zeta,\xi\in\mathcal T_n\),
\[
\|Q_\zeta-Q_\xi\|_2
\le
C\gamma_p^2\sqrt{pn}\sqrt{|\zeta-\xi|}.
\]
Moreover,
\[
\|Q_\zeta-Q_\xi\|_{\psi_1}
\le
C\gamma_p^2\sqrt{pn}\sqrt{|\zeta-\xi|},
\]
and therefore, for every integer \(m\ge 1\),
\[
E\sup_{|\zeta-\tau|\le m}|Q_\zeta-Q_\tau|
\le
C\gamma_p^2\sqrt{pn}\sqrt m.
\]
\end{lemma}

\begin{proof}
Write
\[
Q_\zeta-Q_\xi
=
\big(\|G_\zeta\|_2^2-\|G_\xi\|_2^2\big)-E\big(\|G_\zeta\|_2^2-\|G_\xi\|_2^2\big)
=
\ip{U_{\zeta,\xi}}{V_{\zeta,\xi}}-E\ip{U_{\zeta,\xi}}{V_{\zeta,\xi}},
\]
where $U_{\zeta,\xi}:=G_\zeta-G_\xi$ and $V_{\zeta,\xi}:=G_\zeta+G_\xi$.
For any unit vector \(u\in\mathbb R^p\), based on Lemmas~\ref{asf:lem:weights} and \ref{asf:lem:projvar}, we obtain
\[
\Var(u^\top U_{\zeta,\xi})
=
\Var\Big(\sum_{t=1}^n (w_{\zeta,t}-w_{\xi,t})u^\top e_t\Big)
\le
\gamma_p^2\sum_{t=1}^n (w_{\zeta,t}-w_{\xi,t})^2
\le
\gamma_p^2|\zeta-\xi|,
\]
and
\[
\Var(u^\top V_{\zeta,\xi})
=
\Var\Big(\sum_{t=1}^n (w_{\zeta,t}+w_{\xi,t})u^\top e_t\Big)
\le
\gamma_p^2\sum_{t=1}^n (w_{\zeta,t}+w_{\xi,t})^2
\le
\gamma_p^2 n.
\]
Therefore, we get

\[
\begin{gathered}
\|\Cov(U_{\zeta,\xi})\|_2\le \gamma_p^2|\zeta-\xi|,
\qquad
\|\Cov(V_{\zeta,\xi})\|_2\le \gamma_p^2 n,\\
\tr(\Cov(U_{\zeta,\xi}))
\le
p\,\|\Cov(U_{\zeta,\xi})\|_2
\le
p\gamma_p^2|\zeta-\xi|.
\end{gathered}
\]
By Lemma~\ref{asf:lem:bilinear}, we obtain
\[
\|Q_\zeta-Q_\xi\|_2
\le
\sqrt{2\,\tr(\Cov(U_{\zeta,\xi}))\,\|\Cov(V_{\zeta,\xi})\|_2}
\le
C\gamma_p^2\sqrt{pn}\sqrt{|\zeta-\xi|}.
\]
Since 
\(Q_\zeta-Q_\xi\) belongs to the second Gaussian Wiener chaos,  \eqref{eq:chaos-orlicz} yields
\[
\|Q_\zeta-Q_\xi\|_{\psi_1}
\le
C\|Q_\zeta-Q_\xi\|_2
\le
C\gamma_p^2\sqrt{pn}\sqrt{|\zeta-\xi|}.
\]
Finally, Lemma~\ref{asf:lem:gaussmax} gives the desired maximal bound.
\end{proof}

\subsection{Local fluctuation control}\label{ssc:localfluctuation}

\begin{proposition}\label{asf:prop:fluctuation}
Let $y_t$ follow \eqref{eq:single-cp} under Assumption~\ref{ass:innov}, and use the Step~1 estimator \eqref{eq:tau0-def}.
Suppose Assumptions~\ref{H1} and \ref{assump:tau-domain} hold, and let \(0\le\lambda_1=O(1)\).
Then there exist constants \(C>0\) and \(N\in\mathbb N\), depending only on \(c^\ast\), the eventual upper bound on \(\lambda_1\), and universal constants, such that for every \(n\ge N\) and every \(\delta>0\),
\begin{equation}\label{asf:eq:vdvw_fluct}
\E\Bigg[
\sup_{\zeta\in\mathcal V_{n,\delta}}
\Big|
\big(M_1(\zeta)-M_1(\tau)\big)-\E\big(M_1(\zeta)-M_1(\tau)\big)
\Big|
\Bigg]
\le
C\,\frac{\delta\,\gamma_p}{\sqrt n},
\end{equation}
where
\[
\mathcal V_{n,\delta}
:=
\Big\{\zeta\in\mathcal T_n:\ \|\eta_1-\eta_2\|_2\sqrt{\frac{|\zeta-\tau|}{n}}\le\delta\Big\}.
\]
\end{proposition}

\begin{proof}
Let $\Delta_\eta:=\eta_2-\eta_1$, and $\vartheta_n:=\frac{n\|\Delta_\eta\|_2^2}{p\gamma_p^2}$.
By Assumption~\ref{H1}, there exists \(N_0\) such that \(\vartheta_n\ge 1\) for all \(n\ge N_0\).

Recall from Lemma~\ref{asf:lem:profile} that
$M_1(\zeta)=\|\bar y\|_2^2+b_\zeta\|T_\zeta\|_2^2$ and define $Z_\zeta:=b_\zeta\big(\|T_\zeta\|_2^2-E\|T_\zeta\|_2^2\big)$. Based on this notation, we get
\[
(M_1(\zeta)-M_1(\tau))-\mathbb E(M_1(\zeta)-M_1(\tau))=Z_\zeta-Z_\tau.
\]
The bound is trivial when \(\mathcal V_{n,\delta}=\{\tau\}\), so we may assume that \(\mathcal V_{n,\delta}\neq\{\tau\}\) and define $
m_\delta:=\max_{\zeta\in\mathcal V_{n,\delta}}|\zeta-\tau|$.
By the definition of $\mathcal V_{n,\delta}$,
\begin{equation}\label{asf:eq:mdelta}
m_\delta\le \frac{n\delta^2}{\|\Delta_\eta\|_2^2}.
\end{equation}
Since \(m_\delta\ge 1\), we also have
\begin{equation*}
\delta\ge \frac{\|\Delta_\eta\|_2}{\sqrt n}.
\end{equation*}

We decompose $T_\zeta=E T_\zeta+G_\zeta$ and expanding its square, we obtain
\[
\|T_\zeta\|_2^2-E\|T_\zeta\|_2^2
=
2\ip{E T_\zeta}{G_\zeta}+Q_\zeta,
\]
which splits $Z_\zeta$ into a linear and a quadratic part,
\[
Z_\zeta=Z_\zeta^{\mathrm{lin}}+R_\zeta,
\qquad
Z_\zeta^{\mathrm{lin}}:=2b_\zeta\ip{E T_\zeta}{G_\zeta},
\qquad
R_\zeta:=b_\zeta Q_\zeta.
\]

\smallskip
\noindent We start with the linear part. By Lemma~\ref{asf:lem:weights}, $
E T_\zeta=-g_\zeta\Delta_\eta$
for some \(g_\zeta\in[0,n]\) with \(|g_\zeta-g_\tau|\le |\zeta-\tau|\).
We set
\[
Y_\zeta:=\Delta_\eta^\top G_\zeta,
\qquad
s_\zeta:=b_\zeta g_\zeta.
\]
We can then write $
Z_\zeta^{\mathrm{lin}}=-2s_\zeta Y_\zeta,
$
and therefore
$
Z_\zeta^{\mathrm{lin}}-Z_\tau^{\mathrm{lin}}=-2s_\zeta(Y_\zeta-Y_\tau)-2(s_\zeta-s_\tau)Y_\tau$.
Taking suprema, we get
\[
\sup_{\zeta\in\mathcal V_{n,\delta}}|Z_\zeta^{\mathrm{lin}}-Z_\tau^{\mathrm{lin}}|
\le
2\Big(\sup_{\zeta\in\mathcal V_{n,\delta}}|s_\zeta|\Big)\Big(\sup_{\zeta\in\mathcal V_{n,\delta}}|Y_\zeta-Y_\tau|\Big)
+
2\Big(\sup_{\zeta\in\mathcal V_{n,\delta}}|s_\zeta-s_\tau|\Big)|Y_\tau|.
\]

From Lemma~\ref{asf:lem:weights}, we get
\[
|s_\zeta|\le \frac{C}{n},
\qquad
|s_\zeta-s_\tau|
\le
|b_\zeta-b_\tau|\,|g_\zeta|+b_\tau|g_\zeta-g_\tau|
\le
C\frac{|\zeta-\tau|}{n^2}
\le
C\frac{m_\delta}{n^2}.
\]

Next, based on Lemmas~\ref{asf:lem:weights} and \ref{asf:lem:projvar} we obtain
\[
\Var(Y_\zeta-Y_\xi)
=
\Var\Big(\sum_{t=1}^n (w_{\zeta,t}-w_{\xi,t})\Delta_\eta^\top e_t\Big)
\le
\gamma_p^2\|\Delta_\eta\|_2^2\sum_{t=1}^n (w_{\zeta,t}-w_{\xi,t})^2
\le
\gamma_p^2\|\Delta_\eta\|_2^2|\zeta-\xi|.
\]
Applying Lemma~\ref{asf:lem:gaussmax} yields
\[
E\sup_{|\zeta-\tau|\le m_\delta}|Y_\zeta-Y_\tau|
\le
C\gamma_p\|\Delta_\eta\|_2\sqrt{m_\delta} \quad \text{and} \ 
\Var(Y_\tau)
\le
\gamma_p^2\|\Delta_\eta\|_2^2\sum_{t=1}^n w_{\tau,t}^2
\le
C\gamma_p^2\|\Delta_\eta\|_2^2 n,
\]
so that $
E|Y_\tau|\le C\gamma_p\|\Delta_\eta\|_2\sqrt n$.

Combining these bounds, we arrive at
\[
E\sup_{\zeta\in\mathcal V_{n,\delta}}|Z_\zeta^{\mathrm{lin}}-Z_\tau^{\mathrm{lin}}|
\le
C\left(
\frac{1}{n}\gamma_p\|\Delta_\eta\|_2\sqrt{m_\delta}
+
\frac{m_\delta}{n^2}\gamma_p\|\Delta_\eta\|_2\sqrt n
\right).
\]
Using \eqref{asf:eq:mdelta}, we get
\[
\frac{1}{n}\gamma_p\|\Delta_\eta\|_2\sqrt{m_\delta}
\le
\frac{\delta\gamma_p}{\sqrt n} \quad \text{and} \ \ 
\frac{m_\delta}{n^2}\gamma_p\|\Delta_\eta\|_2\sqrt n
\le
\frac{\gamma_p}{\sqrt n}\min\Big(\|\Delta_\eta\|_2,\frac{\delta^2}{\|\Delta_\eta\|_2}\Big)
\le
\frac{\delta\gamma_p}{\sqrt n}.
\]
Putting the two pieces together, we get
\begin{equation}\label{asf:eq:linearfinal}
E\sup_{\zeta\in\mathcal V_{n,\delta}}|Z_\zeta^{\mathrm{lin}}-Z_\tau^{\mathrm{lin}}|
\le
C\frac{\delta\gamma_p}{\sqrt n}.
\end{equation}

\smallskip
\noindent Next, we consider the quadratic part. We have
$
R_\zeta-R_\tau=b_\zeta(Q_\zeta-Q_\tau)+(b_\zeta-b_\tau)Q_\tau$.
For the first term, we get from Lemmas~\ref{asf:lem:weights} and \ref{asf:lem:quadinc} that
\[
E\sup_{\zeta\in\mathcal V_{n,\delta}}|b_\zeta(Q_\zeta-Q_\tau)|
\le
\frac{C}{n^2}\,\gamma_p^2\sqrt{pn}\sqrt{m_\delta}.
\]
Using \eqref{asf:eq:mdelta},
$
\sqrt{m_\delta}\le \min\Big(\sqrt n,\frac{\sqrt n\,\delta}{\|\Delta_\eta\|_2}\Big)$,
and consequently
\[
E\sup_{\zeta\in\mathcal V_{n,\delta}}|b_\zeta(Q_\zeta-Q_\tau)|
\le
C\frac{\gamma_p^2\sqrt p}{n}\min\Big(1,\frac{\delta}{\|\Delta_\eta\|_2}\Big).
\]
Since $\frac{\gamma_p^2\sqrt p}{n}=
\frac{\gamma_p\|\Delta_\eta\|_2}{\sqrt n}\cdot \frac{1}{\sqrt{\vartheta_n}}$,
we obtain
\[
E\sup_{\zeta\in\mathcal V_{n,\delta}}|b_\zeta(Q_\zeta-Q_\tau)|
\le
C\frac{\delta\gamma_p}{\sqrt n}\cdot \frac{1}{\sqrt{\vartheta_n}}
\le
C\frac{\delta\gamma_p}{\sqrt n}, \quad \text{for all} \  n\ge N_0.
\]
For the second term, we use \(E|Q_\tau|\le \|Q_\tau\|_2\). Since $Q_\tau=\|G_\tau\|_2^2-E\|G_\tau\|_2^2$, and for Gaussian \(Z\sim N(0,\Sigma)\), $\Var(\|Z\|_2^2)=2\tr(\Sigma^2)\le 2p\|\Sigma\|_2^2$
we find leveraging Lemmas~\ref{asf:lem:weights} and \ref{asf:lem:projvar} that $\|Q_\tau\|_2\le C\gamma_p^2 n\sqrt p$.
Combining the latter result with Lemma~\ref{asf:lem:weights}, we obtain
\[
E\sup_{\zeta\in\mathcal V_{n,\delta}}|(b_\zeta-b_\tau)Q_\tau|
\le
C\frac{m_\delta}{n^3}\cdot \gamma_p^2 n\sqrt p
=
C\gamma_p^2\sqrt p\,\frac{m_\delta}{n^2}.
\]
Again by \eqref{asf:eq:mdelta},
$\gamma_p^2\sqrt p\,\frac{m_\delta}{n^2}
\le
\frac{\gamma_p^2\sqrt p}{n}\min\Big(1,\frac{\delta^2}{\|\Delta_\eta\|_2^2}\Big).$

If \(\delta\le \|\Delta_\eta\|_2\), then
$
\frac{\gamma_p^2\sqrt p}{n}\cdot \frac{\delta^2}{\|\Delta_\eta\|_2^2}
=
\frac{\delta\gamma_p}{\sqrt n}\cdot \frac{\delta}{\|\Delta_\eta\|_2}\cdot \frac{1}{\sqrt{\vartheta_n}}
\le
\frac{\delta\gamma_p}{\sqrt n}\cdot \frac{1}{\sqrt{\vartheta_n}}.
$
If \(\delta>\|\Delta_\eta\|_2\), then $
\frac{\gamma_p^2\sqrt p}{n}
=
\frac{\gamma_p\|\Delta_\eta\|_2}{\sqrt n}\cdot \frac{1}{\sqrt{\vartheta_n}}
\le
\frac{\delta\gamma_p}{\sqrt n}\cdot \frac{1}{\sqrt{\vartheta_n}}.
$
In both cases,
\[
E\sup_{\zeta\in\mathcal V_{n,\delta}}|(b_\zeta-b_\tau)Q_\tau|
\le
C\frac{\delta\gamma_p}{\sqrt n}\cdot \frac{1}{\sqrt{\vartheta_n}}
\le
C\frac{\delta\gamma_p}{\sqrt n}, \quad \text{for all } \ n\ge N_0.
\]
Combining the two quadratic estimates, we conclude that
\begin{equation}\label{asf:eq:quadfinal}
E\sup_{\zeta\in\mathcal V_{n,\delta}}|R_\zeta-R_\tau|
\le
C\frac{\delta\gamma_p}{\sqrt n}.
\end{equation}

Finally, adding \eqref{asf:eq:linearfinal} and \eqref{asf:eq:quadfinal}, we obtain the desired result
\[
\E\Bigg[
\sup_{\zeta\in\mathcal V_{n,\delta}}
\Big|
\big(M_1(\zeta)-M_1(\tau)\big)-\E\big(M_1(\zeta)-M_1(\tau)\big)
\Big|
\Bigg]
\le
C\,\frac{\delta\,\gamma_p}{\sqrt n}.
\]
\end{proof}

\subsection{Deterministic drift of the population criterion}\label{ssc:drift}

\begin{lemma}\label{asf:lem:signaldrift}
Assume the mean structure \eqref{asf:eq:model}, $\tau\in\mathcal T_n$, and $0\le\lambda_1\le\bar\lambda<\infty$. Use $b_\zeta,T_\zeta$ from Lemma~\ref{asf:lem:profile}.
Define
\[
S_1(\zeta):=b_\zeta \|E T_\zeta\|_2^2.
\]
There exist constants \(c_{\mathrm{sig}}>0\) and \(N\in\mathbb N\), depending only on \(c^\ast\) and the eventual upper bound on \(\lambda_1\), such that for all \(n\ge N\) and all \(\zeta\in\mathcal T_n\),
\[
S_1(\tau)-S_1(\zeta)\ge c_{\mathrm{sig}}\frac{|\zeta-\tau|}{n}\,\|\Delta_\eta\|_2^2.
\]
\end{lemma}

\begin{proof}
Let $\rho:=\frac{\zeta}{n}$, and $\rho^\ast:=\frac{\tau}{n}$. We first consider the case \(\zeta\le \tau\), that is, \(\rho\le \rho^\ast\).
From Lemmas~\ref{asf:lem:profile} and \ref{asf:lem:weights},
we have 
\[
S_1(\zeta)
=
b_\zeta \frac{\zeta^2(n-\tau)^2}{n^2}\|\Delta_\eta\|_2^2
=
(1-\rho^\ast)^2 f_-(\rho)\|\Delta_\eta\|_2^2,
\]
where $
f_-(\rho):=
\frac{\rho^2\{\rho(1-\rho)+2\lambda_1\}}{\{\lambda_1+\rho(1-\rho)\}^2}$.
A direct differentiation yields 
$f_-'(\rho)=
\frac{\rho\big(4\lambda_1^2+3\lambda_1\rho+\rho^2(1-\rho)\big)}
{\{\lambda_1+\rho(1-\rho)\}^3}$.
For \(\rho\in[c^\ast,1-c^\ast]\) and \(0\le \lambda_1\le \bar\lambda\), the right-hand side is continuous and strictly positive,
so there exists \(m_1>0\) such that $
f_-'(\rho)\ge m_1$ for all $\rho\in[c^\ast,1-c^\ast]$, 
and all sufficiently large \(n\). By the mean value theorem,
we get $f_-(\rho^\ast)-f_-(\rho)\ge m_1(\rho^\ast-\rho)$.
Since $(1-\rho^\ast)^2\ge(c^\ast)^2$, we conclude that
\[
S_1(\tau)-S_1(\zeta)
\ge
(c^\ast)^2 m_1\,\frac{\tau-\zeta}{n}\,\|\Delta_\eta\|_2^2.
\]

Next, we consider the case \(\zeta\ge \tau\), that is, \(\rho\ge \rho^\ast\). In this case,
$S_1(\zeta)=(\rho^\ast)^2 f_+(\rho)\|\Delta_\eta\|_2^2$,
where $f_+(\rho):=
\frac{(1-\rho)^2\{\rho(1-\rho)+2\lambda_1\}}{\{\lambda_1+\rho(1-\rho)\}^2}$.
Again, direct differentiation yields
$f_+'(\rho)
=-\frac{(1-\rho)\big(4\lambda_1^2+3\lambda_1(1-\rho)+\rho(1-\rho)^2\big)} {\{\lambda_1+\rho(1-\rho)\}^3}$.
In particular, \(f_+'(\rho)<0\) on \([c^\ast,1-c^\ast]\).
By compactness there exists \(m_2>0\) such that
$-f_+'(\rho)\ge m_2$ for all $\rho\in[c^\ast,1-c^\ast]$,
and all sufficiently large \(n\). Another application of the  mean value theorem yields $
f_+(\rho^\ast)-f_+(\rho)\ge m_2(\rho-\rho^\ast)$,
and since $(\rho^\ast)^2\ge(c^\ast)^2$, we obtain
$
S_1(\tau)-S_1(\zeta)
\ge
(c^\ast)^2 m_2\,\frac{\zeta-\tau}{n}\,\|\Delta_\eta\|_2^2.
$

Combining the two cases, we obtain the result with
\[
c_{\mathrm{sig}}:=\min\{(c^\ast)^2m_1,(c^\ast)^2m_2\}=(c^\ast)^2\min\{m_1,m_2\}.
\]
\end{proof}

\begin{lemma}\label{asf:lem:noisedrift}
Let $e_t$ be the process \eqref{asf:eq:error_rep} with $\gamma_p<\infty$, let $G_\zeta=T_\zeta-ET_\zeta$, and assume $0\le\lambda_1\le\bar\lambda<\infty$ on the trimmed set \eqref{asf:eq:tauhat}.
Define
$
N_1(\zeta):=b_\zeta\,E\|G_\zeta\|_2^2$.
There exist constants \(C_{\mathrm{noi}}>0\) and \(N\in\mathbb N\), depending only on \(c^\ast\) and the eventual upper bound on \(\lambda_1\), such that for all \(n\ge N\) and all \(\zeta,\xi\in\mathcal T_n\),
\[
|N_1(\zeta)-N_1(\xi)|
\le
C_{\mathrm{noi}}\frac{p\gamma_p^2}{n^2}\,|\zeta-\xi|.
\]
\end{lemma}

\begin{proof}
Let $G_\zeta=\sum_{t=1}^n w_{\zeta,t}e_t$.
Expanding the square, we get
\[
E\|G_\zeta\|_2^2
=
\sum_{s,t=1}^n w_{\zeta,s}w_{\zeta,t}\,\tr(\Gamma_{t-s})
=
\sum_{h=-(n-1)}^{n-1}\tr(\Gamma_h)\,C_\zeta(h),
\]
where $C_\zeta(h):=\sum_{t=1}^{n-|h|}w_{\zeta,t}w_{\zeta,t+|h|}$.
Let \(\delta_t:=w_{\zeta,t}-w_{\xi,t}\). Then,
\[
C_\zeta(h)-C_\xi(h)
=
\sum_{t=1}^{n-|h|}\delta_t\,w_{\zeta,t+|h|}
+
\sum_{t=1}^{n-|h|}w_{\xi,t}\,\delta_{t+|h|}.
\]
Since \(|w_{\zeta,t}|\le 1\) and \(|w_{\xi,t}|\le 1\),
$
|C_\zeta(h)-C_\xi(h)|\le 2\sum_{t=1}^n |\delta_t|.
$
Then,
\[
\sum_{t=1}^n |\delta_t|
=
(n-|\zeta-\xi|)\frac{|\zeta-\xi|}{n}+|\zeta-\xi|\Big(1-\frac{|\zeta-\xi|}{n}\Big)
=
2|\zeta-\xi|-\frac{2|\zeta-\xi|^2}{n}
\le 2|\zeta-\xi|.
\]
Consequently, $
|C_\zeta(h)-C_\xi(h)|\le 4|\zeta-\xi|.$
By Lemma~\ref{asf:lem:projvar},
\[
|E\|G_\zeta\|_2^2-E\|G_\xi\|_2^2|
\le
\sum_{h=-(n-1)}^{n-1}|\tr(\Gamma_h)|\,|C_\zeta(h)-C_\xi(h)|
\le
4p\gamma_p^2|\zeta-\xi|.
\]

Next, by Lemma~\ref{asf:lem:weights},
$|b_\zeta-b_\xi|\le C\frac{|\zeta-\xi|}{n^3}, \
b_\xi\le \frac{C}{n^2}$ and  by Lemmas~\ref{asf:lem:weights} and \ref{asf:lem:projvar},
\[
E\|G_\zeta\|_2^2
=
\tr(\Cov(G_\zeta))
\le
p\|\Cov(G_\zeta)\|_2
\le
p\gamma_p^2\sum_{t=1}^n w_{\zeta,t}^2
\le
Cp\gamma_p^2 n.
\]
Putting these together, we obtain
\begin{align*}
|N_1(\zeta)-N_1(\xi)|
&\le
|b_\zeta-b_\xi|\,E\|G_\zeta\|_2^2+b_\xi\,|E\|G_\zeta\|_2^2-E\|G_\xi\|_2^2|\\
&\le
C\frac{|\zeta-\xi|}{n^3}\cdot (p\gamma_p^2 n)
+ \frac{C}{n^2}\cdot (p\gamma_p^2|\zeta-\xi|) \le C_{\mathrm{noi}}\frac{p\gamma_p^2}{n^2}\,|\zeta-\xi|.
\end{align*}
\end{proof}

\begin{proposition}\label{asf:prop:drift}
Let $y_t$ follow \eqref{eq:single-cp} under Assumption~\ref{ass:innov}, and use the Step~1 estimator \eqref{eq:tau0-def}.
Suppose Assumptions~\ref{H1} and \ref{assump:tau-domain} hold, and let \(0\le\lambda_1=O(1)\).
Then there exist constants \(c_0>0\) and \(N\in\mathbb N\), depending only on \(c^\ast\), the eventual upper bound on \(\lambda_1\), and universal constants, such that for all \(n\ge N\) and all \(\zeta\in\mathcal T_n\),
\begin{equation}\label{asf:eq:vdvw_drift}
E\{M_1(\tau)-M_1(\zeta)\}
\ge
c_0\,\frac{|\zeta-\tau|}{n}\,\|\eta_1-\eta_2\|_2^2.
\end{equation}
Equivalently, $
E\{L(\zeta)-L(\tau)\}
\ge
c_0\,\frac{|\zeta-\tau|}{n}\,\|\eta_1-\eta_2\|_2^2$.
\end{proposition}

\begin{proof}
From Lemma~\ref{asf:lem:profile}, we get
$
E M_1(\zeta)=E\|\bar y\|_2^2+S_1(\zeta)+N_1(\zeta),
$
where $S_1(\zeta)=b_\zeta\|E T_\zeta\|_2^2, \ 
N_1(\zeta)=b_\zeta E\|G_\zeta\|_2^2$.
We therefore have
\[
E\{M_1(\tau)-M_1(\zeta)\}
=
\big(S_1(\tau)-S_1(\zeta)\big)+\big(N_1(\tau)-N_1(\zeta)\big).
\]
By Lemma~\ref{asf:lem:signaldrift}, we obtain $
S_1(\tau)-S_1(\zeta)\ge c_{\mathrm{sig}}\frac{|\zeta-\tau|}{n}\,\|\eta_1-\eta_2\|_2^2$, whereas 
by Lemma~\ref{asf:lem:noisedrift}, we get
$
|N_1(\tau)-N_1(\zeta)|\le C_{\mathrm{noi}}\frac{p\gamma_p^2}{n^2}\,|\zeta-\tau|.$
Combining the two bounds, we get
\[
E\{M_1(\tau)-M_1(\zeta)\}
\ge
\frac{|\zeta-\tau|}{n}
\left(
c_{\mathrm{sig}}\|\eta_1-\eta_2\|_2^2-C_{\mathrm{noi}}\frac{p\gamma_p^2}{n}
\right).
\]
Assumption~\ref{H1} then yields
$\frac{n}{p\gamma_p^2}\|\eta_1-\eta_2\|_2^2\to\infty.$
We can therefore choose \(N\) such that for all \(n\ge N\),
$
C_{\mathrm{noi}}\frac{p\gamma_p^2}{n}\le \frac{c_{\mathrm{sig}}}{2}\|\eta_1-\eta_2\|_2^2$.
For such $n$,
\[
E\{M_1(\tau)-M_1(\zeta)\}
\ge
\frac{c_{\mathrm{sig}}}{2}\frac{|\zeta-\tau|}{n}\,\|\eta_1-\eta_2\|_2^2.
\]
We have proved the first inequality with \(c_0:=c_{\mathrm{sig}}/2\). Since
$
M_1(\tau)-M_1(\zeta)=L(\zeta)-L(\tau)$, 
the equivalent statement for \(L\) follows immediately.
\end{proof}

\subsection{Completing the proof of Proposition~\ref{prop:tau0}}\label{ssc:completingsinglecpproof}

\begin{proof}[Proof of Proposition \ref{prop:tau0}] As a first step, we prove the following generic tail bound \ref{asf:eq:tau0-tail} about Step~1 estimator \eqref{eq:tau0-def} under the current assumptions. In particular, let $y_t$ follow \eqref{eq:single-cp} under Assumption~\ref{ass:innov} and Assumptions~\ref{H1} and \ref{assump:tau-domain} hold. We show that if \(0\le\lambda_1=O(1)\), t here exist constants \(c,C>0\) and \(N\in\mathbb N\),
depending only on \(c^\ast\) and the constants of
Propositions~\ref{asf:prop:fluctuation} and \ref{asf:prop:drift}, such that for
every \(n\ge N\) and every \(K\ge1\),
\begin{equation}\label{asf:eq:tau0-tail}
P\left(|\widehat\tau-\tau|>K\frac{\gamma_p^2}{\|\eta_1-\eta_2\|_2^2}\right)
\;\le\;
C\Bigl[e^{-cK}+e^{-c\sqrt{K\vartheta_n}}\Bigr],
\qquad
\vartheta_n:=\frac{n\|\eta_1-\eta_2\|_2^2}{p\gamma_p^2}.
\end{equation}

\noindent From the bound \ref{asf:eq:tau0-tail} above, we see that
\[
\lim_{K\to\infty}\limsup_{n\to\infty}
P\left(|\widehat\tau-\tau|>K\gamma_p^2/\|\Delta_\eta\|_2^2\right)=0,
\]
because $\vartheta_n\to\infty$ by Assumption~\ref{H1}. This completes the proof of Proposition \ref{prop:tau0}. Substituting $\Delta_\eta=A(1)^{-1}\Delta_\mu$, we obtain its
equivalent expression in terms of the intercept jump.

The rest of the proof establishes \ref{asf:eq:tau0-tail}. Set \(m_n:=K\gamma_p^2/\|\eta_1-\eta_2\|_2^2\) and, for \(j\ge0\),
\[
\mathcal A_j:=\bigl\{\zeta\in\mathcal T_n:\ 2^jm_n<|\zeta-\tau|\le2^{j+1}m_n\bigr\}.
\]
Note that \(m_n\) are kept unrounded: the shells are defined by inequalities, so they remain well
posed for real \(m_n\), and their union covers \(\{|\zeta-\tau|>m_n\}\).  Since
\(\widehat\tau\) maximizes \(M_1\) over \(\mathcal T_n\) and \(\tau\in\mathcal T_n\),
\begin{equation}\label{asf:eq:tau0-shellsplit}
P\bigl(|\widehat\tau-\tau|>m_n\bigr)
\le
\sum_{j\ge0}P\Bigl(\sup_{\zeta\in\mathcal A_j}\bigl(M_1(\zeta)-M_1(\tau)\bigr)\ge0\Bigr).
\end{equation}

\noindent By Proposition~\ref{asf:prop:drift}, for \(n\) large and every
\(\zeta\in\mathcal A_j\),
\[
\E\bigl\{M_1(\tau)-M_1(\zeta)\bigr\}
\ \ge\ c_0\frac{2^jm_n\|\eta_1-\eta_2\|_2^2}{n}
\ =\ c_0K2^j\frac{\gamma_p^2}{n}
\ =:\ t_j ,
\]
the middle equality using \(m_n\|\eta_1-\eta_2\|_2^2=K\gamma_p^2\) exactly.
Consequently, each summand in \eqref{asf:eq:tau0-shellsplit} is at most
\(P(\sup_{\mathcal A_j}|W(\zeta)-W(\tau)|\ge t_j)\), where
\(W:=M_1-\E M_1\).

We split the centered increments as follows. By Lemma~\ref{asf:lem:profile},
\[
\begin{aligned}
W(\zeta)-W(\tau)
={}&Z_\zeta^{\mathrm{lin}}-Z_\tau^{\mathrm{lin}}
+R_\zeta-R_\tau,
\end{aligned}
\]
where \(Z_\zeta^{\mathrm{lin}}\) and \(R_\zeta\) are the Gaussian linear and
second-chaos components defined in the proof of Proposition~\ref{asf:prop:fluctuation}.
The common term \(\|\bar y\|_2^2-\E\|\bar y\|_2^2\) cancels.
It is therefore enough to bound each increment at level \(t_j/2\).

For the Gaussian part, the linear-part calculation in the proof of
Proposition~\ref{asf:prop:fluctuation} gives, with
\(\delta_j:=\|\eta_1-\eta_2\|_2\sqrt{2^{j+1}m_n/n}=\sqrt{2^{j+1}K}\,\gamma_p/\sqrt n\),
\[
\E\sup_{\zeta\in\mathcal A_j}\bigl|Z_\zeta^{\mathrm{lin}}-Z_\tau^{\mathrm{lin}}\bigr|
\le C\frac{\delta_j\gamma_p}{\sqrt n}
= C\sqrt{2^{j+1}K}\,\frac{\gamma_p^2}{n},
\]
and the same computation bounds the process variance by
\[
\sigma_j^2:=\sup_{\zeta\in\mathcal A_j}\Var\bigl(Z_\zeta^{\mathrm{lin}}-Z_\tau^{\mathrm{lin}}\bigr)
\le C\,2^{j}K\,\frac{\gamma_p^4}{n^2}.
\]
For \(K\) exceeding a fixed constant, \(\E\sup\le t_j/4\), so by the
Borell--TIS inequality
\[
P\Bigl(\sup_{\mathcal A_j}\bigl|Z_\zeta^{\mathrm{lin}}-Z_\tau^{\mathrm{lin}}\bigr|\ge \tfrac{t_j}{2}\Bigr)
\le
\exp\Bigl(-\frac{(t_j/4)^2}{2\sigma_j^2}\Bigr)
\le
\exp\bigl(-c\,K2^{j}\bigr).
\]

We are left with the final chaos term.  By Lemma~\ref{asf:lem:quadinc} the increments of
\(R_\zeta\) over \(\mathcal A_j\) have \(\psi_1\)-norm at most
\(U_j:=C\gamma_p^2\sqrt{pn}\sqrt{2^{j+1}m_n}/n^2\), and the quadratic-part calculation in the proof of
Proposition~\ref{asf:prop:fluctuation} bounds the corresponding expected
supremum by \(t_j/4\) for \(K\) large.  Applying the $\psi_1$ maximal bound in Lemma~\ref{asf:lem:gaussmax}
to the quadratic increments and to the coefficient-difference term, we obtain
$\left\|\sup_{\zeta\in\mathcal A_j}
|R_\zeta-R_\tau|\right\|_{\psi_1}
\le C U_j$.
Consequently,
\[
P\left(\sup_{\mathcal A_j}|R_\zeta-R_\tau|
\ge t_j/2\right)\le 2\exp(-c t_j/U_j).
\]
Substituting \(m_n=K\gamma_p^2/\|\eta_1-\eta_2\|_2^2\), we find
\[
\frac{t_j}{U_j}
=c\,\frac{K2^{j}\gamma_p^2/n}{\gamma_p^2\sqrt{pn}\sqrt{2^{j+1}K}\,\gamma_p/(n^2\|\eta_1-\eta_2\|_2)}
= c\,2^{j/2}\sqrt{K}\,
\frac{\sqrt n\,\|\eta_1-\eta_2\|_2}{\sqrt p\,\gamma_p}
= c\,2^{j/2}\sqrt{K\vartheta_n}.
\]

To conclude the proof, we combine the two displays with
\eqref{asf:eq:tau0-shellsplit} and obtain
\[
P\bigl(|\widehat\tau-\tau|>m_n\bigr)
\le\sum_{j\ge0}\Bigl[e^{-cK2^{j}}+e^{-c2^{j/2}\sqrt{K\vartheta_n}}\Bigr]
\le C\Bigl[e^{-cK}+e^{-c\sqrt{K\vartheta_n}}\Bigr],
\]
since \(\sum_{j\ge0}e^{-a2^{j}}\le Ce^{-a}\) and
\(\sum_{j\ge0}e^{-a2^{j/2}}\le Ce^{-a}\) for \(a\ge1\).  This is
exactly \eqref{asf:eq:tau0-tail}.
\end{proof}

We intentionally state the explicit bound \eqref{asf:eq:tau0-tail} for a single change point.  In the
multiple change-point setting of Section~\ref{sec:multiple},  we apply it on each
of \(m_0\) rolling windows and combine the results by a union bound; see the proof of
Proposition~\ref{prop:Haus}. For the data-selected local intervals we use the
separate uniform argument in Lemma~\ref{lem:uniform-local-raw}.

\section{Step 2: Transition matrix estimation on buffered blocks}\label{asf:sec:S2-proof}

In the proof, we separate the deterministic Lasso argument from the probability
bounds for the design and score, following the organization of
Section~4.2 of \citet{basu2015regularized}: their Proposition~4.1 gives the
deterministic estimation bounds, and Propositions~4.2--4.3 verify the RSC
and deviation conditions for a stable Gaussian VAR. In our setting, however, the retained rows
are selected using the data and a separate intercept is profiled out on
each block. We therefore verify the probability bounds for the profiled
regression, uniformly over possible block endpoints, and then apply the fixed-design
Lasso calculation on that simultaneous event.

Throughout this section, the innovation and stability conditions are those specified in Assumption~\ref{ass:innov}. We define $\widetilde{x}_t = (e_{t-1}^\top, \dots, e_{t-d}^\top)^\top \in \mathbb{R}^{dp}$. For a deterministic collection $I = (I_1, \dots, I_J)$ of ordered, disjoint, non-empty integer intervals, we set
\[
n_j = |I_j|, \qquad N_I = \sum_{j=1}^J n_j, \qquad G = \operatorname{blockdiag}(G_{n_1}, \dots, G_{n_J}), \qquad G_m = \mathbf{I}_m - m^{-1} \mathbf{1}_m \mathbf{1}_m^\top.
\]
Data matrices are stacked in chronological order. In summations indexed by $t \in I$, the set $I$ denotes the union $\bigcup_{j=1}^J I_j$. For any sequence $\{v_t\}$, we write $\bar{v}_j = n_j^{-1} \sum_{t \in I_j} v_t$ for its block average. In particular, $G$ is a symmetric projection matrix ($G = G^\top = G^2$) satisfying $\|G\|_2 \le 1$ and $\operatorname{tr}(G) = N_I - J$. These definitions accommodate arbitrary dependence across blocks, and all bounding constants below are independent of interval endpoints, individual block lengths $n_j$, and the total number of blocks $J$.

The proof of Step 2 proceeds in four main steps. First, Section~\ref{ssc:S2-profile} profiles out the block intercepts and derives the Gram matrix and score for the resulting regression. Second, Section~\ref{ssc:S2-covariance} establishes covariance bounds and quantifies the computational cost of profiling. Third, Section~\ref{ssc:S2-concentration} constructs a simultaneous concentration event for restricted strong convexity and the score vector. Finally, Section~\ref{ssc:S2-lasso} combines these components with a deterministic Lasso inequality applied to the selected blocks, completing the proof. Crucially, while the concentration event in Section~\ref{ssc:S2-concentration} is constructed uniformly over all deterministic block collections, the data-dependent blocks enter only in the final step when evaluated at the selected endpoints.

\subsection{Profiling the intercepts and identifying the score}\label{ssc:S2-profile}

\begin{lemma}\label{asf:lem:S2-profile}
Let $I_1,\ldots,I_J$ be as above. For every penalty $\lambda\ge0$,
minimizing the retained squared loss plus
$\lambda\|\operatorname{vec}B\|_1$ over $B$ and the $J$ unpenalized
intercepts has the same set of $B$-minimizers as
\[
N_I^{-1}\|G(Y_I-X_IB)\|_F^2+\lambda\|\operatorname{vec}B\|_1.
\]
If every response and its lags in each retained block belong to the same
true regime, then
\begin{equation}\label{asf:eq:S2-score}
\widehat\Gamma=N_I^{-1}\mathbf{I}_p\otimes(\widetilde X_I^\top G\widetilde X_I),
\qquad
\widehat\gamma-\widehat\Gamma\beta^\star
=N_I^{-1}\operatorname{vec}(\widetilde X_I^\top G E_I),
\end{equation}
where $\widetilde X_I$ has rows $\widetilde x_t^\top$ and $E_I$ has rows
$\varepsilon_t^\top$.
\end{lemma}

\begin{proof}
For fixed $B$, let $r_t=y_t-B^\top x_t$ and
$\bar r_j=n_j^{-1}\sum_{t\in I_j}r_t$. For any block intercept $\theta_j$,
\[
\sum_{t\in I_j}\|r_t-\theta_j\|_2^2
=\sum_{t\in I_j}\|r_t-\bar r_j\|_2^2
+n_j\|\theta_j-\bar r_j\|_2^2.
\]
The optimal intercept is therefore $\bar r_j=\bar y_j-B^\top\bar x_j$, and summing the
first term over blocks we obtain the displayed profiled objective. Since this identity
holds for every $B$, the two problems have the same sets of minimizers, whether or not
the penalized fit is unique.

On pure blocks the regression can be written
\[
Y_I=X_IB+U\Theta+E_I,\qquad X_I=\widetilde X_I+UM,
\]
where $U$ is the $N_I\times J$ block-indicator matrix and the rows of
$\Theta$ and $M$ contain the appropriate regime intercepts and lag means.
Because $GU=0$,
\[
GY_I=G\widetilde X_IB+GE_I,\qquad GX_I=G\widetilde X_I.
\]
With $\widetilde Z_I(G):=\mathbf{I}_p\otimes(GX_I)$, we see after vectorization that
$\widehat\Gamma=N_I^{-1}\widetilde Z_I(G)^\top\widetilde Z_I(G)$ and
$\widehat\gamma=N_I^{-1}\widetilde Z_I(G)^\top\operatorname{vec}(GY_I)$, and using
$G^\top G=G$ we arrive at \eqref{asf:eq:S2-score}.
\end{proof}

We need purity only to identify the observed regression
with this stationary-error regression. In the probability calculation below
we use $\widetilde X_I$ and $E_I$ on every deterministic collection, whether
or not that collection is pure for the observed mean sequence.

\subsection{Covariance bounds and the cost of profiling}\label{ssc:S2-covariance}

Set $L_x=2\pi M_x$, $\alpha=\sigma^2/(4\mu_{\max}(A))$, and
$\chi_x=L_x/\alpha$, as in the main text. We need two distinct covariance
bounds: a lower bound for the marginal lag covariance and an upper bound
for the time-stacked covariance.

Throughout the spectral formulas, $\mathrm{i}$ denotes the imaginary unit in the complex exponential $\mathrm{e}^{\mathrm{i}\omega}$. For the first, the spectral density of the stationary error satisfies
\[
f_e(\omega)=\frac{\sigma^2}{2\pi}
A(e^{-i\omega})^{-1}A(e^{-i\omega})^{-*}
\succeq\frac{\sigma^2}{2\pi\mu_{\max}(A)}\mathbf{I}_p.
\]
For $v=(v_1^\top,\ldots,v_d^\top)^\top$, we put
$v(\omega)=\sum_{j=1}^d v_j e^{ij\omega}$. By Parseval's identity,
\begin{align*}
v^\top\Gamma_xv
&=\int_{-\pi}^{\pi}v(\omega)^*f_e(\omega)v(\omega)\,d\omega
\\
&\ge\frac{\sigma^2}{2\pi\mu_{\max}(A)}
\int_{-\pi}^{\pi}\|v(\omega)\|_2^2\,d\omega
=4\alpha\|v\|_2^2,
\qquad \Gamma_x=E\widetilde x_t\widetilde x_t^\top.
\end{align*}
This is an integrated lower bound, so it holds even when the spectral density
of the lag-stacked process is singular at some frequencies.

For the upper bound, we see from the definition of $M_x$ that, for arbitrary
deterministic real weights $a_t$,
\begin{equation}\label{eq:D-time-covariance}
\operatorname{Var}\left(\sum_t a_t v^\top\widetilde x_t\right)
\le L_x\|v\|_2^2\sum_t a_t^2.
\end{equation}
To see this, we apply the spectral representation to the trigonometric polynomial
$\sum_t a_t e^{it\omega}$ and then Parseval's identity. By setting unused
weights to zero, we get the same bound for a collection with gaps. In
particular, $\Gamma_x\preceq L_x\mathbf{I}_{dp}$ and $\chi_x\ge4$.
For the stationary block average
$\overline{\widetilde x}_j=n_j^{-1}\sum_{t\in I_j}\widetilde x_t$,
\eqref{eq:D-time-covariance} also gives $
\operatorname{Cov}(\overline{\widetilde x}_j)\preceq L_x n_j^{-1}\mathbf{I}_{dp}.$

The profiled Gram matrix has the exact sample decomposition
$
\widetilde X_I^\top G\widetilde X_I
=\sum_{t\in I}\widetilde x_t\widetilde x_t^\top
-\sum_{j=1}^J n_j
\overline{\widetilde x}_j\overline{\widetilde x}_j^\top$.
Taking expectations, we obtain
\begin{equation}\label{eq:profile-population-lower}
E\{N_I^{-1}\widetilde X_I^\top G\widetilde X_I\}
=\Gamma_x-N_I^{-1}\sum_{j=1}^J n_j
\operatorname{Cov}(\overline{\widetilde x}_j)
\succeq(4\alpha-JL_x/N_I)\mathbf{I}_{dp}.
\end{equation}
We see that demeaning comes at a price: it can reduce the population curvature by
$JL_x/N_I$. Since the projection is block diagonal, only within-block covariances
enter this expectation. If $N_I\ge J\chi_x$, the
remaining population curvature is at least $3\alpha$.

\subsection{A simultaneous concentration event}\label{ssc:S2-concentration}

We first record the Gaussian quadratic bound that we use for both the Gram matrix
and the score; it allows the two factors of a score
entry to be correlated.

\begin{lemma}\label{lem:gaussian-quadratic-uniform}
If $g$ is centered Gaussian with covariance $\Sigma$ and $A$ is symmetric,
then, for $x\ge1$,
\begin{equation}\label{eq:gaussian-quadratic-tail}
P\left\{|g^\top Ag-\operatorname{tr}(A\Sigma)|>
C\bigl(\|\Sigma^{1/2}A\Sigma^{1/2}\|_F\sqrt x+
\|\Sigma^{1/2}A\Sigma^{1/2}\|_2x\bigr)\right\}
\le2e^{-x}.
\end{equation}
The covariance may be singular. Moreover, the $L_q$ norm of the centered
form is bounded by
$C\{\sqrt q\,\|\Sigma^{1/2}A\Sigma^{1/2}\|_F+
q\|\Sigma^{1/2}A\Sigma^{1/2}\|_2\}$ for $q\ge1$.
\end{lemma}

\begin{proof}
We write $g=\Sigma^{1/2}z$ and diagonalize $\Sigma^{1/2}A\Sigma^{1/2}$.
Let $a_i$ denote the eigenvalues of $\Sigma^{1/2}A\Sigma^{1/2}$.
The centered form has the law of $\sum_i a_i(z_i^2-1)$ with independent
standard normal $z_i$. For $2|t|\max_i|a_i|\le1/2$,
\[
\log E\exp\left\{t\sum_i a_i(z_i^2-1)\right\}
=\frac12\sum_i\{-\log(1-2ta_i)-2ta_i\}
\le Ct^2\sum_i a_i^2.
\]
We apply the exponential Markov inequality to each sign and take $|t|$ to be
the smaller of $c\sqrt{x/\sum_i a_i^2}$ and
$(4\max_i|a_i|)^{-1}$, which leads to
\eqref{eq:gaussian-quadratic-tail}; zero eigenvalues cause no difficulty.
Integrating the tail, we obtain the moment bound. The lemma is simply the Gaussian case of
the Hanson--Wright inequality; see also \citet{rudelson2013hanson}.
\end{proof}

\begin{proposition}\label{prop:uniform-block-concentration}
Under Assumption~\ref{ass:innov}, suppose that $\beta^\star$ is $s$-sparse with $s \ge 1$. For $n \ge 2$ and a deterministic integer $1 \le J \le n$ (which may depend on $n$), define
\[
\ell_J = 1 + \log(dp^2) + 2J \log n, \qquad \kappa = \frac{\alpha}{64s}.
\]
Let $N_0$ denote a deterministic lower bound on the total retained sample size. There are universal constants such that, if
\begin{equation}\label{eq:uniform-block-size}
N_0\ge C\left[\chi_x^2\{s\{1+\log(dp)\}+J\log n\}
+J\chi_x+\ell_J\right],
\end{equation}
then, with probability at least $1-Cn^{-2}$, simultaneously over all ordered
collections of $J$ disjoint nonempty blocks with total size $N_I\ge N_0$,
\begin{align*}
N_I^{-1}\mathbf{I}_p\otimes(\widetilde X_I^\top G\widetilde X_I)
\sim\mathrm{RSC}(\alpha,\kappa), \qquad
\left\|N_I^{-1}\operatorname{vec}(\widetilde X_I^\top GE_I)\right\|_\infty
\le Q_{\mathrm{dev}}\sqrt{\ell_J/N_I},
\end{align*}
provided the universal constants in \eqref{eq:Qdev-explicit} are sufficiently
large. The individual block lengths are unrestricted.
\end{proposition}

\begin{proof}
We start with a Gram bound in a fixed sparse direction. We fix a block collection, abbreviate $N_I$ by $N$, and write
\[
H=N^{-1}\widetilde X_I^\top G\widetilde X_I
-E(N^{-1}\widetilde X_I^\top G\widetilde X_I).
\]
For a unit vector $v\in\mathbb R^{dp}$, the time vector
$V=(v^\top\widetilde x_t)_{t\in I}$ is Gaussian with covariance
$K_v\preceq L_x\mathbf{I}_N$, by \eqref{eq:D-time-covariance}. Its quadratic-form
matrix after whitening is $K_v^{1/2}GK_v^{1/2}/N$, for which
\[
\|K_v^{1/2}GK_v^{1/2}/N\|_2\le L_x/N,
\qquad
\|K_v^{1/2}GK_v^{1/2}/N\|_F
\le L_x\sqrt{N-J}/N\le L_x/\sqrt N.
\]
Taking $x=cN/\chi_x^2$ in Lemma~\ref{lem:gaussian-quadratic-uniform},
with $c$ small, we obtain
\begin{equation*}
P\{|v^\top Hv|>\alpha/8\}\le2\exp(-cN/\chi_x^2).
\end{equation*}
Here $\chi_x\ge4$, and the assumed sample size ensures the tail parameter
is at least one after constants are adjusted.

Next we turn to sparse nets and extend the bound to each coordinate sphere. We put $k_0=\lceil64s\rceil$ and $r_0=\min(2k_0,dp)$.
On each coordinate sphere with support $T$ of size $r_0$, we choose a
$1/4$-net $\mathcal N_T$ with at most $9^{r_0}$ points. The number of net
points over all such supports is bounded by
\[
\binom{dp}{r_0}9^{r_0}
\le\exp\{Cs\{1+\log(dp)\}\}.
\]
When $r_0=dp$, there is only one support and $dp\le2k_0$, so the same
bound applies. Every smaller support is contained in one of these supports.

For completeness, let $M_T=\|H_{TT}\|_2$ and let $v$ be a
unit vector on $T$. We choose $u\in\mathcal N_T$ with $\|v-u\|_2\le1/4$.
By symmetry,
\[
|v^\top H_{TT}v-u^\top H_{TT}u|
\le\|(v-u)\|_2\|H_{TT}v\|_2
+\|u^\top H_{TT}\|_2\|v-u\|_2
\le M_T/2.
\]
Taking the supremum over $v$, we see that
$M_T\le2\max_{u\in\mathcal N_T}|u^\top Hu|$.
Consequently, except on an event of probability
\begin{equation}\label{eq:D-fixed-collection-failure}
2\exp\{Cs\{1+\log(dp)\}-cN/\chi_x^2\},
\end{equation}
every $2k_0$-sparse vector satisfies
$|v^\top Hv|\le(\alpha/4)\|v\|_2^2$. The same event bounds the operator
norm of $H$ restricted to the union of any two supports of size $k_0$.

We now use these sparse bounds to derive an RSC inequality. Suppose first that $2k_0<dp$. For arbitrary $v$, we arrange its coordinates in
decreasing absolute value and partition them into groups $S_1,S_2,\ldots$
of size $k_0$, except possibly the last. For $l\ge2$, every entry of
$v_{S_l}$ is at most $\|v_{S_{l-1}}\|_1/k_0$ in absolute value. Hence
\[
\sum_l\|v_{S_l}\|_2
\le\|v\|_2+k_0^{-1/2}\sum_{l\ge1}\|v_{S_l}\|_1
=\|v\|_2+k_0^{-1/2}\|v\|_1.
\]
On the sparse-net event, restricting $H$ to $S_i\cup S_j$ we get
$|v_{S_i}^\top Hv_{S_j}|\le(\alpha/4)
\|v_{S_i}\|_2\|v_{S_j}\|_2$, and summing over both indices we obtain
\begin{equation*}
|v^\top Hv|
\le\frac\alpha4\left(\sum_l\|v_{S_l}\|_2\right)^2
\le\frac\alpha2\|v\|_2^2+\frac\alpha{2k_0}\|v\|_1^2.
\end{equation*}
Combining this with \eqref{eq:profile-population-lower} and using
$N\ge J\chi_x$, we arrive at
\[
v^\top(N^{-1}\widetilde X_I^\top G\widetilde X_I)v
\ge\frac{5\alpha}{2}\|v\|_2^2-\frac\alpha{2k_0}\|v\|_1^2
\ge\alpha\|v\|_2^2-\kappa\|v\|_1^2.
\]
If $2k_0\ge dp$, we obtain directly from the full-sphere operator bound the stronger lower
bound $(3\alpha-\alpha/4)\|v\|_2^2$.

To pass to the vectorized regression, we write
$\theta=(\theta_1^\top,\ldots,\theta_p^\top)^\top$ with
$\theta_b\in\mathbb R^{dp}$. Summing the preceding inequalities, we get
\[
\theta^\top\{\mathbf{I}_p\otimes(N^{-1}\widetilde X_I^\top G\widetilde X_I)\}\theta
\ge\alpha\|\theta\|_2^2-\kappa\sum_{b=1}^p\|\theta_b\|_1^2
\ge\alpha\|\theta\|_2^2-\kappa\|\theta\|_1^2.
\]
This is the stated RSC inequality, and it holds for every
$\theta\in\mathbb R^{dp^2}$.

Next we turn to the centering of the score. We fix a lag coordinate $a\le dp$ and response coordinate $b\le p$.
Let $U=(\widetilde x_{t,a}/\sqrt{L_x})_{t\in I}$ and
$V=(\varepsilon_{t,b}/\sigma)_{t\in I}$. Each diagonal block of the
covariance of $(U^\top,V^\top)^\top$ is bounded by $\mathbf{I}_N$. For arbitrary
$u,v$, we then have by Cauchy--Schwarz
\[
\operatorname{Var}(u^\top U+v^\top V)
\le2\operatorname{Var}(u^\top U)+2\operatorname{Var}(v^\top V)
\le2(\|u\|_2^2+\|v\|_2^2).
\]
Thus the joint covariance operator is at most two, although the two time
vectors are generally dependent. The normalized score entry is
\[
\frac1N\sum_{t,u\in I}\widetilde x_{t,a}G_{tu}\varepsilon_{u,b}
=\sigma\sqrt{L_x}
\begin{pmatrix}U\\V\end{pmatrix}^{\!\top}
\frac1{2N}\begin{pmatrix}0&G\\G&0\end{pmatrix}
\begin{pmatrix}U\\V\end{pmatrix}.
\]
The displayed matrix has squared Frobenius norm
$(N-J)/(2N^2)$ and operator norm at most $1/(2N)$. By Lemma~\ref{lem:gaussian-quadratic-uniform}, its centered
tail is bounded by
\begin{equation}\label{eq:D-centered-score-tail}
C\sigma\sqrt{L_x}\{\sqrt{x/N}+x/N\}
\end{equation}
with failure probability at most $2e^{-x}$.

We have to center this quadratic form carefully. In the unprofiled score,
$E(\widetilde x_{t,a}\varepsilon_{t,b})=0$ because the lag vector uses only
past innovations. After profiling, however, the expectation becomes
\begin{align}\label{eq:D-profile-score-mean}
E\left\{\frac1N(\widetilde X_I^\top GE_I)_{ab}\right\}
&=-\frac1N\sum_{j=1}^J n_j
\operatorname{Cov}(\overline{\widetilde x}_{j,a},\bar\varepsilon_{j,b}),
\nonumber\\
\left|E\left\{\frac1N(\widetilde X_I^\top GE_I)_{ab}\right\}\right|
&\le\frac{J\sigma\sqrt{L_x}}N.
\end{align}
For the inequality, we use that the two block-average variances are at most $L_x/n_j$
and $\sigma^2/n_j$. The means may be correlated because an innovation in a
block enters later lagged observations in that block, and our argument keeps this
nonzero mean throughout.

We conclude the proof by counting endpoint configurations. Each interval is specified by two endpoints from $\{0,\ldots,n\}$.
Ignoring ordering and disjointness only enlarges the count, so there are
at most $(n+1)^{2J}$ admissible collections. Since $n\ge2$,
$\log(n+1)\le2\log n$, so their logarithmic count is at most $4J\log n$.
For the Gram bound, summing
\eqref{eq:D-fixed-collection-failure} over these collections, we get failure probability at most
\[
2\exp\{Cs\{1+\log(dp)\}+2J\log(n+1)-cN_0/\chi_x^2\}
\le Cn^{-2}
\]
under \eqref{eq:uniform-block-size}. The term $J\log n$ also pays for
the final $n^{-2}$ probability because $J\ge1$.

For the score, we take $x=K\ell_J$, with $K$ a sufficiently large universal
constant. The union over all $dp^2$ entries and block collections has
failure probability at most
\[
2dp^2(n+1)^{2J}e^{-K\ell_J}\le Cn^{-2}.
\]
Combining \eqref{eq:D-centered-score-tail} and
\eqref{eq:D-profile-score-mean}, we bound the score on this event by
\[
C\sigma\sqrt{L_x}\left\{
\sqrt{\ell_J/N}+\ell_J/N+J/N\right\}
\le Q_{\mathrm{dev}}\sqrt{\ell_J/N}.
\]
The last step uses $N\ge C\ell_J$, $J\le2J\log n\le\ell_J$ for $n\ge2$, and the constants in
\eqref{eq:Qdev-explicit}. The other sample-size term, $J\chi_x$, pays for
the population curvature correction. We see that \eqref{eq:uniform-block-size} plays three
separate roles: sparse concentration and endpoint
entropy, loss of curvature from profiling, and absorption of the score's
linear-tail and mean terms. Intersecting the two simultaneous
events, we obtain the proposition.
\end{proof}

\subsection{The Lasso inequality and application to selected blocks}\label{ssc:S2-lasso}

\begin{proposition}\label{asf:prop:S2-eventwise}
For the quadratic Lasso objective in \eqref{eq:quad-argmin}, let
$\lambda_n>0$ and $S=\operatorname{supp}(\beta^\star)$ with $|S|\le s$.
Suppose its Gram matrix is positive semidefinite,
$\widehat\Gamma\sim\mathrm{RSC}(\alpha,\kappa)$,
$s\kappa\le\alpha/32$, and
$\|\widehat\gamma-\widehat\Gamma\beta^\star\|_\infty\le\lambda_n/4$.
Then every minimizer satisfies
\begin{equation}\label{eq:lasso-oracle-repaired}
\begin{gathered}
\|\widehat\beta-\beta^\star\|_1\le12s\lambda_n/\alpha,
\qquad
\|\widehat\beta-\beta^\star\|_2\le3\sqrt{s}\lambda_n/\alpha,\\
(\widehat\beta-\beta^\star)^\top\widehat\Gamma
(\widehat\beta-\beta^\star)\le9s\lambda_n^2/(2\alpha),
\qquad \widehat\beta-\beta^\star\in\mathcal C(S).
\end{gathered}
\end{equation}
\end{proposition}

\begin{proof}
We let $v=\widehat\beta-\beta^\star$ and
$r=\widehat\gamma-\widehat\Gamma\beta^\star$. Comparing the objective
at $\widehat\beta$ and $\beta^\star$, we get
\[
v^\top\widehat\Gamma v
\le2\langle r,v\rangle
+\lambda_n(\|\beta^\star\|_1-\|\beta^\star+v\|_1).
\]
Because $\beta^\star_{S^c}=0$, we obtain from the triangle inequality and the score bound
\begin{equation}\label{eq:D-basic-inequality}
v^\top\widehat\Gamma v
\le\frac{\lambda_n}{2}\|v\|_1
+\lambda_n(\|v_S\|_1-\|v_{S^c}\|_1)
=\frac{3\lambda_n}{2}\|v_S\|_1
-\frac{\lambda_n}{2}\|v_{S^c}\|_1.
\end{equation}
Since $\widehat\Gamma$ is positive semidefinite and $\lambda_n>0$, we see that
$\|v_{S^c}\|_1\le3\|v_S\|_1$, so $v\in\mathcal C(S)$ and
$\|v\|_1\le4\sqrt{s}\|v\|_2$. On this cone the RSC inequality becomes
\[
v^\top\widehat\Gamma v
\ge(\alpha-16s\kappa)\|v\|_2^2
\ge\frac\alpha2\|v\|_2^2.
\]
Combining this with \eqref{eq:D-basic-inequality}, we find
$\|v\|_2\le3\sqrt{s}\lambda_n/\alpha$ (trivially also if $v=0$).
From the cone condition we then get the stated $\ell_1$ bound. Finally,
\[
v^\top\widehat\Gamma v
\le\frac32\lambda_n\sqrt{s}\|v\|_2
\le\frac{9s\lambda_n^2}{2\alpha}.
\]
All steps are deterministic and apply to every minimizer.
\end{proof}

\begin{proof}[Proof of Proposition~\ref{thm:S2}]
Let $\mathcal E_{\mathrm{unif}}$ denote the simultaneous event in
Proposition~\ref{prop:uniform-block-concentration}, with $J=2$ and
$N_0=n_{\mathrm{obs}}=n-2w-2d\asymp n$.
Its entropy factor is exactly $\ell_{2,n}$, and the sample-size condition
\eqref{eq:n-regime} in Proposition~\ref{thm:S2} implies \eqref{eq:uniform-block-size}, after constants are
enlarged. On $\mathcal E_{\mathrm{pure}}$, both selected blocks are pure
and have this total size, so Lemma~\ref{asf:lem:S2-profile}
identifies their observed Gram matrix and score with members of the simultaneous family. For these two retained blocks, setting $G=G_{\mathrm{blk}}$ in the design $\widetilde Z_I(G)$ defined in the proof of Lemma~\ref{asf:lem:S2-profile} gives $\widetilde Z_I(G_{\mathrm{blk}})=\widetilde Z_I$, with $\widetilde Z_I$ as defined in Step~2 of Section~\ref{ssc:singlecptau}.

We therefore have the deterministic inclusion of events
\[
\mathcal E_{\mathrm{pure}}\cap\mathcal E_{\mathrm{unif}}
\subseteq
\left\{\widehat\Gamma\sim\mathrm{RSC}(\alpha,\alpha/(64s)),\quad
\|\widehat\gamma-\widehat\Gamma\beta^\star\|_\infty
\le Q_{\mathrm{dev}}\sqrt{\ell_{2,n}/n_{\mathrm{obs}}}\right\}
\cap\mathcal E_{\mathrm{pure}}.
\]
This inclusion holds pathwise, so the selected endpoints remain random throughout.
Since $\lambda_n\ge4Q_{\mathrm{dev}}
\sqrt{\ell_{2,n}/n_{\mathrm{obs}}}>0$, we can apply the preceding proposition.
For $v=\operatorname{vec}(\widehat B-B)$, we also have after vectorization
\[
n_{\mathrm{obs}}^{-1}\|G_{\mathrm{blk}}X_I(\widehat B-B)\|_F^2
=v^\top\widehat\Gamma v.
\]
This is the claimed prediction bound, which concerns the retained, block-demeaned
design rather than an uncentered whole-sample residual.
\color{black}
By Proposition~\ref{prop:tau0}, the preliminary error is $O_{\IP}(\gamma_p^2/\|\Delta_\eta\|_2^2)$; since $w\ge K_n\gamma_p^2/\|\Delta_\eta\|_2^2$ and $K_n\to\infty$, we have $P(\mathcal E_{\mathrm{pure}}^c)\to0$. Consequently,
\color{black}
\[
P(\mathcal E_{\mathrm{S2}}^c)
\le P(\mathcal E_{\mathrm{pure}}^c)+P(\mathcal E_{\mathrm{unif}}^c)
\le P(\mathcal E_{\mathrm{pure}}^c)+Cn^{-2}\longrightarrow 0.
\]
The coefficient and prediction rates follow from
\eqref{eq:lasso-oracle-repaired} and the stated tuning.
\end{proof}

\section{Step 3: Refined localization on the filtered series}\label{asf:sec:S3-final}

Here we complete the proof of Theorem~\ref{prop:tau-rate} that establishes the refined localization rate for single change-point regime. It proceeds by comparing the criterion function at a candidate split with its value at the true change point. This comparison consists of two key components: first, showing that the deterministic contrast remains well-separated despite the finite transition layer and filter estimation error; second, uniformly controlling the stochastic increment over a domain containing this estimation error. Because the estimated filter is constructed from the same sample observations, this second bound is established uniformly over a deterministic set rather than by conditioning on the filter.

Throughout this section, the lag order $d$ is fixed, $N = n - d \asymp n$, and we reindex observations $d+1, \dots, n$ as $1, \dots, N$. Under this reindexing, the true change point, its estimate, and all candidate splits are shifted by $-d$. For notational convenience, we retain the symbols $\tau$, $\widehat{\tau}^r$, and $\mathcal{T}_n^{(3)}$ for these reindexed quantities and adjust all time-indexed sequences accordingly; localization errors remain unaffected. By the definition of the Step~3 candidate set $\mathcal{T}_n^{(3)}$ in Section~\ref{ssc:singlecptau} and the hypothesis $\tau \in \mathcal{T}_n^{(3)}$ of Theorem~\ref{prop:tau-rate} (which holds under Assumption~\ref{assump:tau-domain} since $c_{\mathrm{tr}} < c^\star$), there exists a constant $a > 0$—where any $a \in (0, c_{\mathrm{tr}})$ suffices for all sufficiently large $n$—such that both $\tau$ and all candidate splits lie within $[aN, (1-a)N]$. We restrict all optimizations over candidate splits below to this trimmed interval and set
\[
\delta_0 := \|\Delta_{\mu}\|_2 > 0.
\]
Recall that the deterministic set $\mathcal D_n$ and scales $\rho_n,b_n$ are defined
in \eqref{eq:single-ell2-rho-alpha} and the display following it. For
$D\in\mathcal D_n$, we obtain from the cone condition
\begin{equation}\label{eq:E-cone-radius}
\|\operatorname{vec}D\|_1\le4\sqrt{s}\|D\|_F
\le C\bar Qs\sqrt{\ell_{2,n}/n}=:a_n,
\qquad
a_n\gamma_p\sqrt{\ell_{2,n}}\le C\sqrt{b_n}.
\end{equation}
We use this entrywise bound for directions that depend on $D$, where a
fixed-direction Gaussian estimate is too weak.
The parameter set is compact and the sample criteria are continuous in
$D$, so the suprema used below are measurable.

The proof proceeds in four main steps. First, Section~\ref{ssc:S3-profile} derives an exact representation of the profiled criterion. Second, for the deterministic component, Section~\ref{ssc:S3-separation} establishes separation of the population contrast under a transition layer, and Section~\ref{ssc:S3-perturbation} demonstrates that this separation is preserved over a deterministic family of estimated-filter perturbations. Third, for the stochastic component, Section~\ref{ssc:S3-partialsums} bounds the contrast weights and local partial sums, Section~\ref{ssc:S3-coefficients} records the deterministic coefficients in the stochastic expansion, Section~\ref{ssc:S3-quadratic} controls the pure innovation quadratic term, and Section~\ref{ssc:S3-termbyterm} assembles these into a term-by-term bound for the filtered contrast. Finally, Section~\ref{ssc:S3-completing} combines the deterministic separation and stochastic bound via a shell argument. Notably, this final step parallels Section~\ref{ssc:completingsinglecpproof} for Step~1, while the transition layer and estimated filter account for the additional technical steps in this proof.

\subsection{The profiled criterion}\label{ssc:S3-profile}

\begin{lemma}\label{asf:lem:S3-profile-new}
For a vector sequence $V_1,\ldots,V_N$, define
\[
D_V(k)=\frac1{N-k}\sum_{t=k+1}^N V_t-\frac1k\sum_{t=1}^k V_t,
\qquad q_k=\frac{k(N-k)}{N^2}.
\]
Fit the two means using the normalized squared loss plus
$\lambda\|\nu_1-\nu_2\|_2^2$, where $\lambda\ge0$, and evaluate the
unpenalized residual loss at those fitted means. Its minimizers over $k$
are the maximizers of
\begin{equation}\label{asf:eq:S3-profile-new}
G_\lambda(V;k)=c_\lambda(k)\|D_V(k)\|_2^2,
\qquad c_\lambda(k)=q_k w_\lambda(q_k),
\qquad w_\lambda(q)=1-\left(\frac\lambda{q+\lambda}\right)^2.
\end{equation}
On the trimmed set, uniformly for $0\le\lambda\le\bar\lambda<\infty$,
$c_\lambda$ is bounded above and away from zero and
$|c_\lambda(k)-c_\lambda(l)|\le C|k-l|/N$.
For $0\le\lambda\le1$,
\begin{equation}\label{eq:ridge-weight-small}
|1-w_\lambda(q_k)|\le C\lambda^2,
\qquad
|w_\lambda(q_k)-w_\lambda(q_l)|\le C\lambda^2|k-l|/N.
\end{equation}
\end{lemma}

\begin{proof}
We let $\pi=k/N$, $\bar V_L=k^{-1}\sum_{t\le k}V_t$,
$\bar V_R=(N-k)^{-1}\sum_{t>k}V_t$, and
$\bar V=\pi\bar V_L+(1-\pi)\bar V_R$, and parameterize the fitted means by
their pooled mean $\bar\nu=\pi\nu_1+(1-\pi)\nu_2$ and difference
$\theta=\nu_2-\nu_1$. Then
$\nu_1=\bar\nu-(1-\pi)\theta$ and $\nu_2=\bar\nu+\pi\theta$. If $W(k)$ denotes
the normalized within-segment sum of squares at the sample means, the
penalized objective equals
\[
W(k)+\|\bar\nu-\bar V\|_2^2+q_k\|\theta-D_V(k)\|_2^2
+\lambda\|\theta\|_2^2.
\]
Minimizing over $\bar\nu$ and $\theta$, we find $\widehat{\bar\nu}=\bar V$ and
$\widehat\theta=q_k(q_k+\lambda)^{-1}D_V(k)$, and the unpenalized loss at
this fit is
\[
W(k)+q_k\left(\frac\lambda{q_k+\lambda}\right)^2\|D_V(k)\|_2^2
=\frac1N\sum_{t=1}^N\|V_t-\bar V\|_2^2-G_\lambda(V;k).
\]
The first term on the right is free of $k$, which proves the claim.
We emphasize that the coefficient here refers to the unpenalized loss evaluated
at ridge-fitted means, as opposed to the value of the penalized objective.

On the trimmed set $q_k\in[a(1-a),1/4]$ and
$|q_k-q_l|\le|k-l|/N$. The expression $q^2(q+2\lambda)/(q+\lambda)^2$, evaluated at $q=q_k$,
is positive and has a bounded derivative in $q$ on the stated compact domain. Finally, $
1-w_\lambda(q)=\frac{\lambda^2}{(q+\lambda)^2},\ 
w_\lambda'(q)=\frac{2\lambda^2}{(q+\lambda)^3}$, 
which proves \eqref{eq:ridge-weight-small}.
\end{proof}

\subsection{Separation in the presence of a transition layer}\label{ssc:S3-separation}

The oracle filtered mean need not jump immediately to its eventual
post-change value. After removing a common baseline, we write its first
$d$ post-change values as $\Delta_\mu+c_j$, with $c_j$ defined in
\eqref{eq:intro-filtered-mean}.

\begin{lemma}\label{asf:lem:S3-det-drift-aligned}
Let $u_t=0$ for $t\le\tau$, $u_{\tau+j}=\Delta_\mu+c_j$ for $1\le j\le d$,
and $u_t=\Delta_\mu$ subsequently, where $\Delta_\mu\ne0$ and $d$ is fixed. Set
\[
C_h=\sum_{j=1}^{\min(h,d)}c_j,\qquad
\nu=\max_{1\le h\le d}\frac{\|C_h\|_2}{h\delta_0}.
\]
Suppose, for a fixed $c_*>0$,
$h\delta_0^2+2\langle\Delta_\mu,C_h\rangle\ge c_*h\delta_0^2$
for $1\le h\le d$, $\nu^2/N\to0$, and $0\le\lambda_N\to0$.
Then, for all sufficiently large $N$, uniformly over trimmed splits,
\begin{equation}\label{asf:eq:S3-det-drift-aligned}
G_{\lambda_N}(u;\tau)-G_{\lambda_N}(u;k)
\ge c\frac{|k-\tau|}{N}\delta_0^2.
\end{equation}
The same conclusion holds after any common baseline is added to $u$.
\end{lemma}

\begin{proof}
We first compare unregularized contrasts. We put $l=\tau$ and $r=N-\tau$.
For $k=\tau+h$, the segment sums at the true split are $0$ and
$r\Delta_\mu+C_d$, whereas those at the candidate are
$h\Delta_\mu+C_h$ and $(r-h)\Delta_\mu+C_d-C_h$. By the between-segment identity,
\begin{align*}
N\{G_0(u;\tau)-G_0(u;\tau+h)\}
={}\frac{\|r\Delta_\mu+C_d\|_2^2}{r}
-\frac{\|h\Delta_\mu+C_h\|_2^2}{l+h}
-\frac{\|(r-h)\Delta_\mu+C_d-C_h\|_2^2}{r-h}.
\end{align*}
In this expression, the coefficient of $\delta_0^2$ is
$h-h^2/(l+h)=lh/(l+h)$, and the terms linear in $\Delta_\mu$ sum to
$2l\langle\Delta_\mu,C_h\rangle/(l+h)$. We therefore arrive at the exact identity
\begin{align}\label{eq:exact-right-drift}
N\{G_0(u;\tau)-G_0(u;\tau+h)\}
&=\frac{l}{l+h}\{h\delta_0^2+2\langle\Delta_\mu,C_h\rangle\}+R_h,\\
R_h&=\frac{\|C_d\|_2^2}{r}
-\frac{\|C_h\|_2^2}{l+h}
-\frac{\|C_d-C_h\|_2^2}{r-h}.\nonumber
\end{align}

For $1\le h\le d$, we keep the cancellation between the first and third
terms in $R_h$:
\begin{equation*}
R_h=-\frac{h\|C_d\|_2^2}{r(r-h)}
+\frac{2\langle C_d,C_h\rangle}{r-h}
-\left(\frac1{l+h}+\frac1{r-h}\right)\|C_h\|_2^2.
\end{equation*}
By trimming, all denominators are bounded below by a fixed multiple of $N$, and
$\|C_h\|_2\le h\nu\delta_0$, $\|C_d\|_2\le d\nu\delta_0$. We therefore get
\[
|R_h|\le C\nu^2\delta_0^2
\left(\frac h{N^2}+\frac hN+\frac{h^2}N\right)
\le C\frac{\nu^2h\delta_0^2}N,
\]
where the last constant uses fixed $d$. If $h>d$, then $C_h=C_d$ and
$R_h=\|C_d\|_2^2\{r^{-1}-(l+h)^{-1}\}$, which satisfies the same bound.
For such $h$, the leading bracket in \eqref{eq:exact-right-drift} is
\[
h\delta_0^2+2\langle\Delta_\mu,C_d\rangle
=(h-d)\delta_0^2+\{d\delta_0^2+2\langle\Delta_\mu,C_d\rangle\}
\ge\min(1,c_*)h\delta_0^2.
\]
Since $l/(l+h)\ge a$ and $\nu^2/N\to0$, the remainder is a vanishing
fraction of this leading term, uniformly in $h$.

For a candidate on the left, $k=\tau-h$, the left segment still has sum
zero and the right segment has sum $r\Delta_\mu+C_d$, and direct subtraction yields
\begin{equation*}
N\{G_0(u;\tau)-G_0(u;\tau-h)\}
=\frac{hr}{r+h}\left\|\Delta_\mu+\frac{C_d}{r}\right\|_2^2.
\end{equation*}
Here $\|C_d\|_2/(r\delta_0)\le C\nu/N=o(1)$, while $r/(r+h)\ge a$.
We thus obtain the same lower bound on the left.

It remains to account for the ridge. On the trimmed set, the segment-sum
formulas show that $\sup_kG_0(u;k)\le C\delta_0^2(1+\nu/N)^2$.
Using the ridge weights, we have the exact comparison
\begin{align*}
G_{\lambda_N}(u;\tau)-G_{\lambda_N}(u;k)
={}&_{\lambda_N}(q_\tau)\{G_0(u;\tau)-G_0(u;k)\}
+\{w_{\lambda_N}(q_\tau)-w_{\lambda_N}(q_k)\}G_0(u;k).
\end{align*}
By \eqref{eq:ridge-weight-small}, $w_{\lambda_N}(q_\tau)=1+O(\lambda_N^2)$ and the
absolute value of the second term is at most
$C\lambda_N^2|k-\tau|\delta_0^2/N$. Since $\lambda_N\to0$, we can absorb it into the
unregularized gap. Both corrections are
proportional to $|k-\tau|$, so the lower bound carries no additive $\delta_0^2/N$ term.
Finally, contrasts are invariant to a common baseline.
\end{proof}

\subsection{A deterministic family of estimated-filter perturbations}\label{ssc:S3-perturbation}

We first connect the transition sequence to the spliced model. Before the
change, the mean of the oracle filtered observation is
$\eta_1-\sum_{\ell=1}^dA_\ell\eta_1=\mu_1$. At the $j$th post-change
observation, $1\le j\le d$, the lags with $\ell<j$ have mean $\eta_2$
and those with $\ell\ge j$ still have mean $\eta_1$. We therefore get
\begin{align*}
EY_{\tau+j}
=\eta_2-\sum_{\ell<j}A_\ell\eta_2
-\sum_{\ell=j}^d A_\ell\eta_1
=\mu_2+\left(\sum_{\ell=j}^dA_\ell\right)\Delta_\eta
=\mu_1+\Delta_\mu+c_j.
\end{align*}
After these $d$ observations all lags have the post-change mean and the
filtered mean is $\mu_2$. At every time the centered part is exactly
$e_t-\sum_{\ell=1}^d A_\ell e_{t-\ell}=\varepsilon_t$, because the same
stationary error process continues through the splice. We see that the transition
affects only the mean; the error process and the innovation law continue unchanged.
This derivation also explains why its partial sums $C_h$, rather than just the
eventual jump $\Delta_\mu$, enter the right-sided drift.

For each deterministic $D\in\mathcal D_n$, let
$L_t=Ex_t-\mathbf1_d\otimes\eta_1$, using the corresponding original
time index in the lag vector. Each coordinate of $L_t$ is zero or a
coordinate of $\Delta_\eta$. Subtracting the common baseline
$\mu_1-D^\top(\mathbf1_d\otimes\eta_1)$ from the filtered sequence, we obtain
the pathwise identity
\begin{equation}\label{asf:eq:S3-YD-decomp}
Y_t(D)=u_t(D)+\varepsilon_t+Z_t(D),
\qquad u_t(D)=u_t-D^\top L_t,
\qquad Z_t(D)=-D^\top\widetilde x_t.
\end{equation}
This defines a family on the original probability space, indexed by the
deterministic parameter $D$. On the Step~2 event,
we recover the actual filtered observations, up to their irrelevant baseline,
by substituting $D=\widehat B-B$.

We control this perturbation in two different ways. We bound its deterministic
mean using the Frobenius radius
$
\sup_D\|D^\top L_t\|_2
\le C\bar Q\sqrt{s\ell_{2,n}/n}\,\gamma_p\delta_0
=C\rho_n\delta_0$.

For its random component we instead use the entrywise cone radius
and a coordinate maximum, as in \eqref{eq:E-cone-radius}. The first calculation explains the small-perturbation condition $\rho_n\to0$; the
second explains the possibly nonvanishing variance inflation $b_n$.
We use each argument where it is sharp; interchanging them would either miss the adaptive direction
or introduce an unnecessary dimension factor.

\begin{lemma}\label{asf:lem:S3-deterministic-D}
Under Assumption~
\ref{assump:transition-alignment}, uniformly over $D\in\mathcal D_n$,
$u(D)$ satisfies Lemma~\ref{asf:lem:S3-det-drift-aligned}, with $\Delta_\mu$ and $\delta_0$ replaced by $\delta_D:=\Delta_\mu-D^\top(\mathbf1_d\otimes\Delta_\eta)$ and $\|\delta_D\|_2$, respectively, with an alignment constant bounded away from zero. Consequently, if $0\le\lambda_3\to0$,
\begin{equation}\label{asf:eq:S3-uniform-drift}
G_{\lambda_3}(u(D);\tau)-G_{\lambda_3}(u(D);k)
\ge c|k-\tau|\delta_0^2/N
\end{equation}
for all sufficiently large $n$, uniformly in $D$ and trimmed $k$.
\end{lemma}

\begin{proof}
We put $L_\infty=\mathbf1_d\otimes\Delta_\eta$. The effective jump and
transition vectors of $u(D)$ are
\[
\delta_D=\Delta_\mu-D^\top L_\infty,
\qquad c_j(D)=c_j+D^\top(L_\infty-L_{\tau+j}).
\]
From the norm bound $\|\Delta_\eta\|_2\le\gamma_p\delta_0$, fixed $d$, and
the radius of $\mathcal D_n$, we get
\begin{equation*}
\|\delta_D-\Delta_\mu\|_2+
\max_{j\le d}\|c_j(D)-c_j\|_2\le C\rho_n\delta_0.
\end{equation*}
Let $e_D=\delta_D-\Delta_\mu$ and
$F_{h,D}=\sum_{j=1}^h\{c_j(D)-c_j\}$, for $h\le d$.
The difference between the effective and original alignment brackets is
\begin{align*}
&h\|\delta_D\|_2^2
+2\left\langle\delta_D,\sum_{j=1}^h c_j(D)\right\rangle
-h\delta_0^2-2\langle\Delta_\mu,C_h\rangle\\
&\quad=2h\langle\Delta_\mu,e_D\rangle+h\|e_D\|_2^2
+2\langle\Delta_\mu,F_{h,D}\rangle
+2\langle e_D,C_h\rangle+2\langle e_D,F_{h,D}\rangle.
\end{align*}
Since $\|F_{h,D}\|_2\le Ch\rho_n\delta_0$ and
$\|C_h\|_2\le h\nu_n\delta_0$, its absolute value is at most
\[
Ch\delta_0^2(\rho_n+\rho_n^2+\rho_n\nu_n)=o(h\delta_0^2).
\]
This is where $\rho_n\to0$ and $\rho_n\nu_n\to0$ are used.
Further,  $\|\delta_D\|_2/\delta_0=1+o(1)$ uniformly, and
\[
\max_{h\le d}
\frac{\|C_h+F_{h,D}\|_2}{h\|\delta_D\|_2}
\le C(\nu_n+\rho_n).
\]
Its square divided by $N$ tends to zero, and we can therefore apply the preceding deterministic
lemma with constants uniform over the entire parameter set.
\end{proof}

\subsection{Contrast weights and local partial sums}\label{ssc:S3-partialsums}

When a cut moves, we must keep track of the denominator changes in the two
sample means. For any sequence $V$, we write $D_V(k)=\sum_t w_{k,t}V_t$, where $
w_{k,t}=-k^{-1}\mathbf1\{t\le k\}
+(N-k)^{-1}\mathbf1\{t>k\}$.

Write $w_k=(w_{k,1},\ldots,w_{k,N})^\top$.
If $l<k$ and $h=k-l$, their difference is
\begin{equation*}
w_{k,t}-w_{l,t}=
\begin{cases}
h/(kl),&t\le l,\\
-1/k-1/(N-l),&l<t\le k,\\
h/\{(N-k)(N-l)\},&t>k.
\end{cases}
\end{equation*}
The middle part has $h$ entries of order $N^{-1}$, whereas the remaining
entries are of order $hN^{-2}$, and consequently
\begin{equation}\label{eq:E-weight-norms}
\begin{gathered}
\|w_k\|_2^2=1/k+1/(N-k)\le C/N,\qquad
\|w_k\|_\infty\le C/N,\\
\|w_k-w_l\|_1\le Ch/N,\qquad
\|w_k-w_l\|_2^2\le C(h/N^2+h^2/N^3)\le Ch/N^2.
\end{gathered}
\end{equation}

We also use a partial-sum form. We let
$S_L=\sum_{t=1}^{\tau}V_t$, $S_R=\sum_{t=\tau+1}^NV_t$,
$P_+(v)=\sum_{t=\tau+1}^{\tau+v}V_t$, and
$P_-(v)=\sum_{t=\tau-v+1}^{\tau}V_t$. Direct subtraction yields
\begin{align}\label{eq:contrast-denominator-identity}
D_V(\tau+v)-D_V(\tau)
={}&-\left(\frac1{\tau+v}+\frac1{N-\tau-v}\right)P_+(v)
\nonumber\\
&+\frac{v}{\tau(\tau+v)}S_L
+\frac{v}{(N-\tau)(N-\tau-v)}S_R,
\end{align}
and, on the left,
\begin{align}\label{eq:E-left-denominator-identity}
D_V(\tau-v)-D_V(\tau)
={}&\left(\frac1{\tau-v}+\frac1{N-\tau+v}\right)P_-(v)
\nonumber\\
&-\frac{v}{\tau(\tau-v)}S_L
-\frac{v}{(N-\tau)(N-\tau+v)}S_R.
\end{align}
We see that a partial-sum bound of size $A\sqrt h$, together with long-sum bounds
of size $A\sqrt N$, leads to a local contrast increment of size
\[
C\left(\frac{A\sqrt h}N+\frac{hA\sqrt N}{N^2}\right)
\le C\frac{A\sqrt h}N.
\]
We keep the second term; it is the smaller one because $h\le N$.

We use $\|W\|_{L_q}=(E|W|^q)^{1/q}$ also for nonnegative suprema.
We take partial sums only over available indices on their indicated side
of $\tau$; if $h$ is larger, the maximum stops at the available endpoint.

We abbreviate the contrast of $Z(D)$ by $D_Z(k):=D_{Z(D)}(k)$, suppressing its dependence on $D$.
\begin{lemma}\label{asf:lem:S3-localized-EP}
Under Assumption~\ref{ass:innov}, for $q=2,4$ and $1\le h\le N$,
\begin{align}\label{eq:uniform-filter-partials}
\left\|\sup_{D\in\mathcal D_n}\max_{1\le v\le h}
\left\|\sum_{t=\tau+1}^{\tau+v}Z_t(D)\right\|_2\right\|_{L_q}
&\le C\sqrt{b_nh},\\
\left\|\sup_D\|D_Z(\tau)\|_2\right\|_{L_q}
&\le C\sqrt{b_n/N},\nonumber\\
\left\|\sup_D\max_{|k-\tau|\le h}
\|D_Z(k)-D_Z(\tau)\|_2\right\|_{L_q}
&\le C\sqrt{b_nh}/N.\nonumber
\end{align}
The partial-sum bound also holds on the left and for a fixed block of
length at most $h$. The contrast bounds for innovations hold with $b_n$
replaced by $p$; for a fixed unit-vector projection, by one; and for the
maximum of coordinate projections, by $\ell_{2,n}$. The corresponding
partial-sum bounds use the same replacements.
\end{lemma}

\begin{proof}
For $S_v=\sum_{t=\tau+1}^{\tau+v}\widetilde x_t$, the deterministic
cone bound \eqref{eq:E-cone-radius} gives 
\[
\sup_D\max_{v\le h}\|D^\top S_v\|_2
\le a_n\max_{a\le dp,\,v\le h}|S_{v,a}|.
\]
To obtain the matrix inequality, we bound the Euclidean norm by the sum of
absolute output coordinates and then sum $|D_{ab}||S_{v,a}|$ over $a,b$.
By Lemma~\ref{asf:lem:projvar}, the variance of every $S_{v,a}$ is at most
$C\gamma_p^2h$ (the fixed innovation scale is included in $C$), and
a Gaussian union bound over at most $dpN$ entries yields
\[
P\left\{\max_{a,v}|S_{v,a}|>
C\gamma_p\sqrt{h\{\log(2dpN)+x\}}\right\}\le2e^{-x},
\qquad x\ge1.
\]
Integrating the tail and using $\log(2dpN)\le C\ell_{2,n}$, we obtain
an $L_q$ bound $C\gamma_p\sqrt{h\ell_{2,n}}$ for $q=2,4$, and multiplying by
$a_n$ we get the first assertion. The same variance
calculation applies to left partial sums and to a fixed block.

For innovations, the Euclidean norm of a partial sum is a nonnegative
submartingale, so by Doob's inequality and Gaussian moments,
\[
E\max_{v\le h}\left\|\sum_{t=\tau+1}^{\tau+v}\varepsilon_t\right\|_2^q
\le\left(\frac q{q-1}\right)^q
E\left\|\sum_{t=\tau+1}^{\tau+h}\varepsilon_t\right\|_2^q
\le C_{\mathrm{mom},q}(ph)^{q/2}.
\]
where $C_{\mathrm{mom},q}>0$ depends only on $q$ and the fixed innovation variance.
Applying the same argument to a fixed unit-vector projection, we get $C_{\mathrm{mom},q}h^{q/2}$,
and a coordinate Gaussian union bound bounds the $L_q$ norm of the coordinate maximum by
$C\sqrt{h\ell_{2,n}}$.
For the left-sided versions we simply reverse a finite list of independent innovations.

Finally, at $\tau$, each long sum is divided by a denominator of order
$N$, and we obtain the stated long-contrast bounds. Substituting into
\eqref{eq:contrast-denominator-identity} and
\eqref{eq:E-left-denominator-identity} and applying Minkowski's inequality,
we get all the increment bounds. We perform these operations after taking the
supremum over $D$, so they remain valid for any dependence between $D$
and the observations.
\end{proof}

For later products we also have, by the triangle inequality and $h\le N$,
\begin{equation}\label{eq:E-long-suprema}
\left\|\max_{|k-\tau|\le h}\|D_\varepsilon(k)\|_2\right\|_{L_q}
\le C\sqrt{p/N},\qquad
\left\|\sup_D\max_{|k-\tau|\le h}\|D_Z(k)\|_2\right\|_{L_q}
\le C\sqrt{b_n/N}.
\end{equation}
The projected and coordinate versions follow in the same way from the
corresponding statements of the lemma.

\subsection{Deterministic coefficients in the stochastic expansion}\label{ssc:S3-coefficients}

The oracle and perturbed mean contrasts satisfy $
D_{u(D)}(k)=D_u(k)-D^\top D_L(k)$.
Further, the oracle mean contrast has the explicit form
\begin{equation}\label{eq:E-mean-contrast-coefficients}
D_u(k)=f_k\Delta_\mu+\sum_{j=1}^d\beta_j(k)c_j,
\qquad
f_k=\begin{cases}(N-\tau)/(N-k),&k\le\tau,\\
\tau/k,&k\ge\tau,
\end{cases}
\qquad \beta_j(k)=w_{k,\tau+j}.
\end{equation}
Set $C_0=0$ for the transition partial sums, and note that
 $f_\tau=1$, $|f_k|\le C$, and
$|f_k-f_\tau|\le C|k-\tau|/N$. Since $c_j=C_j-C_{j-1}$,
$\|c_j\|_2\le(2j-1)\nu_n\delta_0$. The transition weights satisfy
\begin{equation}\label{eq:E-transition-weight-increments}
|\beta_j(k)|\le C/N,\qquad
|\beta_j(k)-\beta_j(\tau)|
\le C\left\{\frac h{N^2}
+\frac{\mathbf1\{k>\tau,\ j\le h\}}N\right\},
\quad h=|k-\tau|.
\end{equation}
The indicator records the observations crossed by a right-moving split, and
there are at most $\min(h,d)$ such crossings. This crossing term is genuinely of order
$1/N$, rather than $h/N^2$.

For the lag-mean contrasts, each $L_t$ satisfies
$\|L_t\|_2\le\sqrt d\,\|\Delta_\eta\|_2$ and
$\|L_t\|_\infty\le\|\Delta_\eta\|_2$. Applying the $\ell_1$ weight
bounds in \eqref{eq:E-weight-norms}, we get, for $r=2,\infty$,
\begin{equation}\label{eq:E-lag-mean-contrasts}
\|D_L(k)\|_r\le C\gamma_p\delta_0,
\qquad
\|D_L(k)-D_L(\tau)\|_r\le C\gamma_p\delta_0h/N.
\end{equation}
From \eqref{eq:E-mean-contrast-coefficients}--\eqref{eq:E-lag-mean-contrasts}
and the Frobenius radius of $\mathcal D_n$, we obtain
\begin{equation}\label{eq:E-deterministic-contrast-bounds}
\begin{split}
\sup_{D,k}\|D_{u(D)}(k)\|_2&\le C\delta_0(1+\nu_n/N+\rho_n)\le C\delta_0,\\
\sup_D\|D_{u(D)}(k)-D_{u(D)}(\tau)\|_2
&\le C\delta_0\{h+\nu_n\min(h,d)\}/N.
\end{split}
\end{equation}
For the second bound, the noncrossing transition weights contribute
$C\nu_n\delta_0h/N^2$, which is bounded by the displayed transition term
when $h\ge1$. The filtering part contributes $C\rho_n\delta_0h/N$.
These bounds use $\rho_n\to0$ and $\nu_n^2/N\to0$, and they allow
$\nu_n$ itself to be unbounded.

\subsection{The pure innovation quadratic term}\label{ssc:S3-quadratic}

We treat this term separately, since a product of two
Euclidean norms would lose its $\sqrt p$ fluctuation scale.

\begin{lemma}\label{asf:lem:S3-quad-innov-shell}
Under Assumption~\ref{ass:innov}, let
$Q_\varepsilon(k)=G_\lambda(\varepsilon;k)-G_\lambda(\varepsilon;\tau)$,
where $0\le\lambda\le\bar\lambda<\infty$ is deterministic. For
$1\le h\le N$,
\begin{equation*}
E\max_{|k-\tau|\le h}|Q_\varepsilon(k)|
\le C\left(\frac{\sqrt{ph}}{N^{3/2}}+\frac{ph}{N^2}\right).
\end{equation*}
\end{lemma}

\begin{proof}
Let $H_k=c_\lambda(k)w_kw_k^\top$. For any two trimmed cuts $k,l$,
\begin{align*}
H_k-H_l={}&c_\lambda(k)\{(w_k-w_l)w_k^\top
+w_l(w_k-w_l)^\top\}\\
&+\{c_\lambda(k)-c_\lambda(l)\}w_lw_l^\top.
\end{align*}
Since the Frobenius norm of $uv^\top$ is $\|u\|_2\|v\|_2$, we obtain from
\eqref{eq:E-weight-norms} and the coefficient Lipschitz bound
\begin{equation}\label{eq:E-quadratic-matrix-increments}
\|H_k-H_l\|_F
\le C\left\{\frac{\sqrt{|k-l|}}{N^{3/2}}+
\frac{|k-l|}{N^2}\right\}
\le C\frac{\sqrt{|k-l|}}{N^{3/2}}.
\end{equation}
The same bound holds for its operator norm.

We stack the innovations by coordinate; their covariance is $\sigma^2\mathbf{I}_{Np}$
and the relevant quadratic matrix is $\mathbf{I}_p\otimes(H_k-H_l)$, whose
Frobenius norm is $\sqrt p\,\|H_k-H_l\|_F$.
By the moment bound in Lemma~\ref{lem:gaussian-quadratic-uniform}, together
with the moment characterization of the $\psi_1$ norm, we get
\[
\|\{Q_\varepsilon(k)-EQ_\varepsilon(k)\}
-\{Q_\varepsilon(l)-EQ_\varepsilon(l)\}\|_{\psi_1}
\le C\sqrt p\,\|H_k-H_l\|_F
\le C\frac{\sqrt{p|k-l|}}{N^{3/2}}.
\]
We apply Lemma~\ref{asf:lem:gaussmax} separately on the two sides of $\tau$, and it
bounds the expectation of the centered maximum by
$C\sqrt{ph}/N^{3/2}$. The chaining argument thus avoids a union over individual cuts,
together with the corresponding $\log N$ factor, in this single-change bound.

In more detail, on a side of length at most $h$, a dyadic grid at resolution
$j$ has at most $2^j+1$ points. The increments to their parent grid points
have $\psi_1$ norm at most
$C\sqrt{ph}\,2^{-j/2}/N^{3/2}$. Taking a maximum at that resolution
costs at most a factor $C(j+1)$, and summing along the chains we obtain
\[
\frac{C\sqrt{ph}}{N^{3/2}}
\sum_{j\ge0}(j+1)2^{-j/2}
\le\frac{C\sqrt{ph}}{N^{3/2}}.
\]
Here we use bounds for increments between arbitrary pairs of cuts,
as proved in \eqref{eq:E-quadratic-matrix-increments}, rather than just
pointwise tails measured from $\tau$.

Finally, $q_k\|w_k\|_2^2=1/N$, so
\begin{equation}\label{eq:E-quadratic-mean}
EG_\lambda(\varepsilon;k)
=\sigma^2p\,c_\lambda(k)\|w_k\|_2^2
=\frac{\sigma^2p}{N}w_\lambda(q_k).
\end{equation}
Its increment is at most $Cph/N^2$ on the bounded ridge domain.
For $\lambda=0$, the mean is constant and this contribution is zero.
Adding the centered and mean bounds, we obtain the lemma.
\end{proof}

\subsection{A term-by-term bound for the filtered contrast}\label{ssc:S3-termbyterm}

\begin{proposition}\label{asf:prop:S3-fluct-new}
Under Assumption~\ref{ass:innov}, use \eqref{asf:eq:S3-YD-decomp} and
suppose $\rho_n\to0$, $\rho_n\nu_n\to0$, and $\nu_n^2/N\to0$.
Define
$\mathcal R_D(k)=G_{\lambda_3}(Y(D);k)-G_{\lambda_3}(u(D);k)$.
For bounded deterministic nonnegative $\lambda_3$ and $1\le h\le N$,
\begin{equation}\label{eq:master-local-fluctuation}
\begin{split}
E\sup_{D\in\mathcal D_n}\max_{|k-\tau|\le h}
|\mathcal R_D(k)-\mathcal R_D(\tau)|
\le C\bigg[&\frac{\delta_0\sqrt{(1+b_n)h}}N\\
&+\frac{(\sqrt p+\sqrt{pb_n}+b_n)\sqrt h}{N^{3/2}}
+\frac{ph}{N^2}\bigg].
\end{split}
\end{equation}
\end{proposition}

\begin{proof}
Using $D_{u(D)}(k)=D_u(k)-D^\top D_L(k)$, the expansion is exactly
\begin{equation}\label{eq:E-six-term-expansion}
\begin{aligned}
\mathcal R_D(k)&=\sum_{i=1}^6 T_i(D;k),\\
T_1(k)&=2c_{\lambda_3}(k)\langle D_u(k),D_\varepsilon(k)\rangle,
&T_2(D;k)&=-2c_{\lambda_3}(k)\langle D^\top D_L(k),D_\varepsilon(k)\rangle,\\
T_3(D;k)&=2c_{\lambda_3}(k)\langle D_{u(D)}(k),D_Z(k)\rangle,
&T_4(k)&=c_{\lambda_3}(k)\|D_\varepsilon(k)\|_2^2,\\
T_5(D;k)&=2c_{\lambda_3}(k)\langle D_\varepsilon(k),D_Z(k)\rangle,
&T_6(D;k)&=c_{\lambda_3}(k)\|D_Z(k)\|_2^2.
\end{aligned}
\end{equation}
Let $\Delta_kT=T(k)-T(\tau)$ and let $B_i(h)$ be the expectation of
$\sup_D\max_{|k-\tau|\le h}|\Delta_kT_i|$, where for $i=1,4$ the
supremum over $D$ is vacuous. We now bound the six quantities in turn.

We will repeatedly use the following decomposition of an increment of a bilinear term:
\begin{align}\label{eq:E-bilinear-increment}
\Delta_k\{c_{\lambda_3}(k)\langle a_k,b_k\rangle\}
={}&c_{\lambda_3}(k)\langle a_k,b_k-b_\tau\rangle
+c_{\lambda_3}(k)\langle a_k-a_\tau,b_\tau\rangle
\nonumber\\
&+(c_{\lambda_3}(k)-c_{\lambda_3}(\tau))\langle a_\tau,b_\tau\rangle.
\end{align}
This decomposition retains the changing denominator and coefficient terms. For products
of random quantities we use Cauchy--Schwarz and the $L_2$ bounds
already proved, which hold under arbitrary dependence.

We start with $T_1$. For the $f_k\Delta_\mu$ part of \eqref{eq:E-mean-contrast-coefficients},
we obtain from the fixed-direction innovation bounds and
\eqref{eq:E-bilinear-increment}
\[
C\delta_0\left(\frac{\sqrt h}N+\frac h{N^{3/2}}\right)
\le C\frac{\delta_0\sqrt h}N.
\]
For the transition part, we apply the same fixed-direction bounds separately
to each nonzero $c_j/\|c_j\|_2$ and bound the three contributions by
\begin{align*}
C\sum_{j=1}^d\|c_j\|_2
\bigg[\frac1N\frac{\sqrt h}N
&+\left\{\frac h{N^2}
+\frac{\mathbf1\{j\le h\}}N\right\}\frac1{\sqrt N}
+\frac hN\frac1N\frac1{\sqrt N}\bigg]\\
&\le C\nu_n\delta_0\left\{
\frac{\sqrt h}{N^2}+\frac h{N^{5/2}}
+\frac{\min(h,d)}{N^{3/2}}\right\}.
\end{align*}
The indicator is an upper bound for the right-sided crossing indicator
in \eqref{eq:E-transition-weight-increments}. Since
$\min(h,d)\le\sqrt d\sqrt h$ and $d$ is fixed,
\begin{equation}\label{eq:E-T1-bound}
B_1(h)\le C\frac{\delta_0\sqrt h}N(1+\nu_n/\sqrt N).
\end{equation}
Our calculation thus allows growing transition vectors under the stated
$\nu_n^2/N\to0$ condition.

Next we turn to $T_2$. The vector $D^\top D_L(k)$ varies with $D$, so we replace the preceding fixed-direction
argument by an entrywise bound. For example, pathwise,
\[
\sup_D|\langle D^\top D_L(k),D_\varepsilon(k)-D_\varepsilon(\tau)\rangle|
\le a_n\|D_L(k)\|_\infty\|D_\varepsilon(k)-D_\varepsilon(\tau)\|_\infty.
\]
The other two terms in \eqref{eq:E-bilinear-increment} use respectively
$D_L(k)-D_L(\tau)$ and $c_{\lambda_3}(k)-c_{\lambda_3}(\tau)$. From
\eqref{eq:E-lag-mean-contrasts} and the coordinate innovation bounds,
\begin{align*}
B_2(h)
\le Ca_n\gamma_p\delta_0
\left\{\frac{\sqrt{h\ell_{2,n}}}N
+\frac hN\sqrt{\frac{\ell_{2,n}}N}\right\}
\le C\frac{\delta_0\sqrt{b_nh}}N.
\end{align*}
The factor $a_n\gamma_p\sqrt{\ell_{2,n}}$ is precisely the one in
\eqref{eq:E-cone-radius}. Since the analysis proceeds pathwise, this inequality applies to all candidate choices of $D$ on the same realization, including the filter estimation error substituted in subsequent steps.

Next, we examine the mean-noise product term $T_3$. We use the uniform deterministic bounds
\eqref{eq:E-deterministic-contrast-bounds}, followed by
\eqref{eq:uniform-filter-partials}. The local increment, change in mean contrast, and coefficient increment give respectively
\begin{align*}
B_3(h) \le C\delta_0\left\{
\frac{\sqrt{b_nh}}N
+\frac{h+\nu_n\min(h,d)}N\sqrt{\frac{b_n}N}
+\frac hN\sqrt{\frac{b_n}N}\right\}
\le C\frac{\delta_0\sqrt{b_nh}}N(1+\nu_n/\sqrt N).
\end{align*}
We take the supremum over $D$ in both the deterministic bounds and the
partial-sum bounds before multiplying them.

Next we treat the innovation quadratic term $T_4$. Lemma~\ref{asf:lem:S3-quad-innov-shell} gives directly
\begin{equation*}
B_4(h)\le C\left\{\frac{\sqrt{ph}}{N^{3/2}}+
\frac{ph}{N^2}\right\}.
\end{equation*}
We note that the first term is $\sqrt p$, rather than $p$, times the
one-dimensional centered fluctuation scale.

For $T_5$, we use the alternative exact decomposition
\begin{align*}
\Delta_k\{c_{\lambda_3}(k)\langle D_\varepsilon(k),D_Z(k)\rangle\}
={}&c_{\lambda_3}(k)\langle D_\varepsilon(k)-D_\varepsilon(\tau),D_Z(k)\rangle\\
&+c_{\lambda_3}(k)\langle D_\varepsilon(\tau),D_Z(k)-D_Z(\tau)\rangle\\
&+(c_{\lambda_3}(k)-c_{\lambda_3}(\tau))\langle D_\varepsilon(\tau),D_Z(\tau)\rangle.
\end{align*}
For the first term, Cauchy--Schwarz bounds the expectation of its supremum
by the product of the $L_2$ norms of the innovation increment and long
filtering contrast. We treat the second and third terms with the other
two products, and \eqref{eq:uniform-filter-partials} and
\eqref{eq:E-long-suprema} then give
\begin{align}\label{eq:E-T5-bound}
B_5(h)\le C\left\{
\frac{\sqrt{ph}}N\sqrt{\frac{b_n}N}
+\sqrt{\frac pN}\frac{\sqrt{b_nh}}N
+\frac hN\sqrt{\frac pN}\sqrt{\frac{b_n}N}\right\}
\le C\frac{\sqrt{pb_nh}}{N^{3/2}}.
\end{align}
This argument allows the innovations and filtering noise to be correlated.

Finally we deal with the filtering quadratic term $T_6$. The identity
\[
\Delta_k\{c_{\lambda_3}(k)\|D_Z(k)\|_2^2\}
=c_{\lambda_3}(k)\langle D_Z(k)-D_Z(\tau),D_Z(k)+D_Z(\tau)\rangle
+(c_{\lambda_3}(k)-c_{\lambda_3}(\tau))\|D_Z(\tau)\|_2^2
\]
and the same $L_2$ product bounds yield
\begin{equation}\label{eq:E-T6-bound}
B_6(h)\le C\left\{
\frac{\sqrt{b_nh}}N\sqrt{\frac{b_n}N}
+\frac hN\frac{b_n}N\right\}
\le C\frac{b_n\sqrt h}{N^{3/2}}.
\end{equation}
We absorb the coefficient terms in \eqref{eq:E-T5-bound} and
\eqref{eq:E-T6-bound} using $h/N^2\le\sqrt h/N^{3/2}$.

It remains to sum \eqref{eq:E-T1-bound}--\eqref{eq:E-T6-bound}. Since
$1+\nu_n/\sqrt N$ is bounded for large $n$ and
$1+\sqrt{b_n}\le\sqrt{2(1+b_n)}$, we arrive at
\eqref{eq:master-local-fluctuation}. These six terms exhaust the square
in \eqref{eq:E-six-term-expansion}, so every mixed term has been accounted for.
\end{proof}

\subsection{Completing the proof of Theorem~\ref{prop:tau-rate}}\label{ssc:S3-completing}
Finally, we combine the results from previous subsections in Section \ref{asf:sec:S3-final} to conclude the proof of the refined localization rate. 
\begin{proof}[Proof of Theorem~\ref{prop:tau-rate}]
We let $R_n=(1+b_n)/\delta_0^2$, and let $\mathcal E_{\mathrm{S2}}$ be the
coefficient event of Proposition~\ref{thm:S2}. On that event,
$\widehat B-B\in\mathcal D_n$. By the profile identity, the refined
estimator maximizes $G_{\lambda_3}$ for the actual filtered sequence.
Hence, if it selects $k$ instead of $\tau$, the stochastic increment
at $D=\widehat B-B$ must offset the deterministic gap in
\eqref{asf:eq:S3-uniform-drift}, which holds by Lemma~\ref{asf:lem:S3-deterministic-D} under Assumption~\ref{assump:transition-alignment}. Replacing $D=\widehat B-B$ by the supremum over the deterministic
set $\mathcal D_n$ only enlarges that event.

For $K\ge1$, we put $t_j=2^jKR_n$ and consider the shells
\[
\mathcal S_j=\{k:t_j<|k-\tau|\le2t_j\},\qquad j=0,1,\ldots.
\]
We omit empty shells. For a nonempty shell we have $2t_j\ge1$ and
$H_j=\min(N,\lceil2t_j\rceil)\le4t_j$. Since its deterministic gap is at
least $ct_j\delta_0^2/N$, Markov's inequality yields
\begin{align}\label{eq:E-shell-probability}
P\{\widehat\tau^{\,r}\in\mathcal S_j,\mathcal E_{\mathrm{S2}}\}
&\le\frac{CN}{t_j\delta_0^2}
E\sup_D\max_{|k-\tau|\le H_j}
|\mathcal R_D(k)-\mathcal R_D(\tau)|.
\end{align}
We bound the expectation in \eqref{eq:E-shell-probability} by Proposition~\ref{asf:prop:S3-fluct-new}. Its conditions $\rho_n\to0$, $\rho_n\nu_n\to0$ and $\nu_n^2/N\to0$ hold by Assumption~\ref{assump:transition-alignment}, since $N=n-d\asymp n$.
The first term of \eqref{eq:master-local-fluctuation} contributes at most
\[
C\frac{\sqrt{1+b_n}}{\delta_0\sqrt{t_j}}
=\frac{C2^{-j/2}}{\sqrt K}.
\]
Summing this geometric series, we get $C/\sqrt K$.
The terms with denominator $N^{3/2}$ similarly sum to at most
\begin{align*}
\frac C{\sqrt K}
\frac{\sqrt p+\sqrt{pb_n}+b_n}
{\sqrt N\,\delta_0\sqrt{1+b_n}}
&\le\frac C{\sqrt K}\left\{
\sqrt{\frac p{N\delta_0^2}}+
\sqrt{\frac{b_n}{N\delta_0^2}}\right\}=o(1),
\end{align*}
by Assumption~\ref{SNR3}.

The remaining mean term contributes at most $Cp/(N\delta_0^2)$ per
nonempty shell. Every such shell has $1/2\le t_j<N$, so there are at most
$C\log N$ of them, even if $R_n$ tends to zero. Its total contribution is
therefore $Cp\log N/(N\delta_0^2)=o(1)$, again by
Assumption~\ref{SNR3}. Adding $P(\mathcal E_{\mathrm{S2}}^c)=o(1)$, we conclude that
\[
\limsup_{n\to\infty}
P\{|\widehat\tau^{\,r}-\tau|>KR_n\}\le C/\sqrt K.
\]
Letting $K\to\infty$ proves the stated $O_{\IP}$ rate. The argument includes
$R_n\to0$: rounding only enlarges the stochastic radius of a nonempty
shell by a constant factor, and the deterministic drift has no additive
one-observation remainder.
\end{proof}

We close this section by noting the separate roles of the assumptions. In Assumption~\ref{assump:transition-alignment}, the alignment condition \eqref{eq:single-transition-alignment} controls the
first $d$ right-sided comparisons; $\nu_n^2/N\to0$ controls the remaining
transition terms and their stochastic coefficients. The two conditions
$\rho_n\to0$ and $\rho_n\nu_n\to0$ preserve those comparisons when the
filter is estimated. Vanishing $\lambda_3$ controls the change in the ridge
weight, whereas Assumption~\ref{SNR3} makes the quadratic fluctuations and
their shell sum negligible relative to the jump. Finally, we use Step~2
only to put the coefficient error in $\mathcal D_n$ with high probability; this event is
the only input from Step~2, and the argument uses the unconditional law of the
innovations throughout.

\paragraph{Data-driven ridge parameters.} The ridge parameter enters the argument only through the coefficients $c_{\lambda_3}(k)$ and the ridge weights $w_{\lambda_3}(q_k)$. These are bounded, with increments of order $h/N$, uniformly over $\lambda_3\in[0,\bar\lambda_n]$, and the random quantities they multiply do not involve $\lambda_3$. Separating coefficients from random quantities as in the proof of Lemma~\ref{lem:uniform-local-raw}, we see that Proposition~\ref{asf:prop:S3-fluct-new} and the proof of Theorem~\ref{prop:tau-rate} hold with the supremum over $\lambda_3\in[0,\bar\lambda_n]$ inside the expectation and the probability, for any deterministic $\bar\lambda_n\to0$. In particular, Theorem~\ref{prop:tau-rate} covers a data-driven $\lambda_3$ in this range.

\section{Proofs for detection of multiple change points}\label{app:multiple-proofs}

This section provides the proofs for the main results in Section~\ref{sec:multiple}, proceeding in four steps that mirror the sequence of the estimation procedure. First, Section~\ref{ssc:mcp-candidate} proves candidate localization, establishing Proposition~\ref{prop:Haus} alongside the consistency of the information criterion. Second, Section~\ref{ssc:mcp-matching} addresses matching and local refinement, where the geometry of the selected local intervals is verified in Section~\ref{ssc:mcp-geometry} and a uniform raw-series comparison is established in Section~\ref{ssc:mcp-rawcomparison}, together yielding Proposition~\ref{thm:ic-consistency-rate}. Third, Section~\ref{ssc:step2-mcp} analyzes estimation across multiple retained blocks to prove Proposition~\ref{thm:S2-multi}. Finally, Section~\ref{ssc:mcp-filtered} establishes simultaneous filtered localization, completing the proof of Theorem~\ref{prop:multi-local}. Crucially, every probability bound in this section is first derived for a deterministic family of intervals before being evaluated at the data-selected endpoints. Throughout, the single-change point results from Appendices~\ref{app:step1-proof}, \ref{asf:sec:S2-proof}, and \ref{asf:sec:S3-final} are applied repeatedly, so only the additional multi-change arguments are detailed here.

\subsection{Candidate localization}\label{ssc:mcp-candidate}
\begin{proof}[Proof of Proposition~\ref{prop:Haus}]
We first establish that Assumption~\ref{assump:WS} provides a rolling window around each true change point $\tau_j^\star$ to which Proposition~\ref{prop:tau0} applies directly.
(i) Since \(\limsup_n h/\Delta_n<1\), every window has length \(h<\Delta_n\) and
so contains at most one true change point.  (ii) Consecutive windows are offset
by \(\ell\le(1-2c^\ast)h\), so the windows containing \(\tau_j^\star\) place it
at distance at least \(c^\ast h\) from both endpoints for at least one index
\(i\); on that window the window-level form of Assumption~\ref{assump:tau-domain}
holds with the same constant \(c^\ast\).  (iii) Since
\(\liminf_n h/\Delta_n>c_{\mathrm{len}}>0\), Assumption~\ref{assump:LJ} gives
\(h\|\eta_{j+1}-\eta_j\|_2^2/(p\gamma_p^2)\ge c_{\mathrm{len}}\Delta_nv_{\min}^2/(p\gamma_p^2)\to\infty\),
which is the window-level form of Assumption~\ref{H1}.  Fix such a window
\([b_i,e_i)\). The minimizer of the single-change criterion
on that window belongs to \(\widetilde{\mathcal S}\).
Because \(m_0\) may diverge, we use the tail bound rather than the \(O_{\IP}\)
statement.  We apply the generic tail bound in ~\ref{asf:eq:tau0-tail} on the window
\([b_i,e_i)\), whose length is \(h\asymp\Delta_n\), with
\(K:=C_{\mathrm{tail}}(1+\log m_0)\) for a constant \(C_{\mathrm{tail}}\) to be chosen.  Writing
\(\vartheta_n^{(w)}=h\|\eta_{j+1}-\eta_j\|_2^2/(p\gamma_p^2)\) for the
window-level version of \(\vartheta_n\), Assumption~\ref{assump:LJ} gives
\(\vartheta_n^{(w)}\gtrsim\Delta_nv_{\min}^2/(p\gamma_p^2)\gg(1+\log m_0)\), so
\[
\begin{aligned}
&P\Bigl(\min_{s\in\widetilde{\mathcal S}}|s-\tau_j^\star|
> C_{\mathrm{tail}}\,(1+\log m_0)\frac{\gamma_p^2}{v_{\min}^2}\Bigr)\\
&\qquad\le
C\bigl[e^{-cC_{\mathrm{tail}}(1+\log m_0)}+e^{-c\sqrt{C_{\mathrm{tail}}(1+\log m_0)\vartheta_n^{(w)}}}\bigr]\\
&\qquad\le
2C\,e^{-cC_{\mathrm{tail}}(1+\log m_0)}
=2C\,(em_0)^{-cC_{\mathrm{tail}}}.
\end{aligned}
\]
Taking a union bound over the $m_0$ changes, we get $2Cm_0(em_0)^{-cC_{\mathrm{tail}}}$. This tends to zero for
fixed $C_{\mathrm{tail}}>2/c$ when $m_0\to\infty$. For bounded $m_0$, sending $C_{\mathrm{tail}}\to\infty$
after $n\to\infty$ gives the stated $O_{\IP}$ bound.
\color{black}
\end{proof}

We state the next lemma \emph{uniformly}
over all subintervals of \(\{1,\dots,n\}\), because the information criterion
compares configurations drawn from \(\widetilde{\mathcal S}\) and we need the bounds to
hold simultaneously for every interval that can arise.

\begin{lemma}\label{lem:mean-block}
Let $e_t$ be the common Gaussian linear process in \eqref{eq:common-centered}, assume $\gamma_p<\infty$, and let $n\ge2$.
There is a constant \(C>0\) such that, with probability at least \(1-2n^{-1}\),
simultaneously for every interval \(I\subseteq\{1,\dots,n\}\) with \(|I|\ge1\),
\[
\|\bar\varepsilon_I\|_2^2\le C\,\frac{\sigma^2 p\,\log(n\vee p)}{|I|},
\qquad
\|\bar e_I\|_2^2\le C\,\frac{\gamma_p^2\sigma^2 p\,\log(n\vee p)}{|I|},
\]
where \(\bar\varepsilon_I:=|I|^{-1}\sum_{t\in I}\varepsilon_t\) and
\(\bar e_I:=|I|^{-1}\sum_{t\in I}e_t\).

Multiplying by $|I|^2$, we obtain the equivalent form
\[
\left\|\sum_{t\in I}\varepsilon_t\right\|_2^2\le C\sigma^2|I|p\log(n\vee p),
\qquad
\left\|\sum_{t\in I}e_t\right\|_2^2\le C\sigma^2\gamma_p^2|I|p\log(n\vee p).
\]
\end{lemma}

\begin{proof}[Proof of Lemma~\ref{lem:mean-block}]
The two statements are equivalent, since
\(\bar\varepsilon_I=|I|^{-1}\sum_{t\in I}\varepsilon_t\); we prove the partial-sum
form.  We fix an interval \(I\) and put \(S_I:=\sum_{t\in
I}\varepsilon_t\).  Under Assumption~\ref{ass:innov}, \(S_I\sim
N(0,\sigma^2|I|\,\mathbf{I}_p)\), so \(\|S_I\|_2^2/(\sigma^2|I|)\sim\chi^2_p\).  The
Laurent--Massart bound yields, for \(u>0\),
\[
P\Bigl(\|S_I\|_2^2>\sigma^2|I|\,(p+2\sqrt{pu}+2u)\Bigr)\le e^{-u}.
\]
We take \(u=3\log(n\vee p)\).  Since \(p+2\sqrt{3p\log(n\vee p)}+6\log(n\vee p)\le C p \log(n\vee p)\), we obtain
\(P(\|S_I\|_2^2>C\sigma^2|I|p\log(n\vee p))\le n^{-3}\).  There are at most \(n^2\)
subintervals of \(\{1,\dots,n\}\), so a union bound yields the claim
simultaneously over all of them with probability at least \(1-n^{-1}\).

For the linear process, we write \(\sum_{t\in I}e_t=\sum_{s}\bigl(\sum_{t\in
I}\Psi_{t-s}\bigr)\varepsilon_s\), a centered Gaussian vector whose covariance
has operator norm at most \(\sigma^2\gamma_p^2|I|\), by the same convolution
argument as in Lemma~\ref{asf:lem:projvar}.  Applying the display above to this
vector and taking a second union bound, we obtain the second pair of inequalities and
the stated probability \(1-2n^{-1}\).
\end{proof}

In the next lemma we treat arbitrary candidate partitions. An estimated segment
can contain no true change, one change, or several changes, and our comparison
allows all three possibilities.

Write $\eta_{(t)}=Ey_t$, and stack $\eta_{(t)}^\top$ and $e_t^\top$ as the rows of the $n\times p$ matrices $\eta$ and $e$, respectively.

\begin{lemma}\label{lem:UB-wellloc}
Consider \eqref{eq:mean-shift-1} under Assumption~\ref{ass:innov}, with
$m_0\ge1$, and work on the uniform event of Lemma~\ref{lem:mean-block}.
Put $V_n=\gamma_p^2p\log(n\vee p)$, absorbing the fixed innovation variance into the
constants. For an ordered partition $\boldsymbol{s}=(s_1,\ldots,s_m)$ into nonempty
segments, let $P_{\boldsymbol{s}}$ denote segmentwise averaging, applied coordinatewise to
$\eta=(\eta_{(t)})_{t=1}^n$. Then, simultaneously over such partitions,
\begin{equation}\label{eq:partition-lower}
\mathcal L_n(\boldsymbol{s})\ge\sum_{t=1}^n\|e_t\|_2^2
+\tfrac12\|(\mathbf{I}-P_{\boldsymbol{s}})\eta\|_F^2-C(m+m_0+1)V_n.
\end{equation}
In particular, if $m<m_0$, then
\begin{equation}\label{eq:LB-m-less}
\mathcal L_n(\boldsymbol{s})\ge\sum_{t=1}^n\|e_t\|_2^2
+c\Delta_nv_{\min}^2-C(m+m_0+1)V_n.
\end{equation}
If $\widehat T=(\widehat\tau_1,\ldots,\widehat\tau_{m_0})$ satisfies
$\max_j|\widehat\tau_j-\tau_j^\star|\le n\xi_n=o(\Delta_n)$,
with endpoints $0$ and $n$, then
\begin{equation}\label{eq:UB-wellloc}
\mathcal L_n(\widehat T)\le\sum_{t=1}^n\|e_t\|_2^2
+Cm_0(n\xi_n v_{\max}^2+V_n).
\end{equation}
\end{lemma}

\begin{proof}
For $r=(\mathbf{I}-P_{\boldsymbol{s}})\eta$, we have by the projection identity
\[
\mathcal L_n(\boldsymbol{s})=\|e\|_F^2+\|r\|_F^2
+2\langle r,e\rangle_F-\|P_{\boldsymbol{s}}e\|_F^2.
\]
The vector $r$ is constant on the common refinement of $\boldsymbol{s}$ and the true
partition, which has at most $m+m_0+1$ nonempty intervals. If $J$ is one
of these intervals and $r_J$ its constant value, then
\[
2|\langle r_J,\sum_{t\in J}e_t\rangle|
\le\tfrac12|J|\|r_J\|_2^2+2|J|\|\bar e_J\|_2^2.
\]
Summing over $J$, using Lemma~\ref{lem:mean-block}, and noting that
$\|P_{\boldsymbol{s}}e\|_F^2\le C(m+1)V_n$, we obtain
\eqref{eq:partition-lower}, whatever the number of true changes
in an estimated segment.

When $m<m_0$, the disjoint neighborhoods of radius $\Delta_n/4$ around
the true changes cannot all contain a candidate split. We choose a change with
an empty neighborhood. A subinterval with lengths of order $\Delta_n$
on either side of this change lies in one estimated segment and contains
no other true change. The best constant fit on that subinterval has squared
bias at least $c\Delta_nv_{\min}^2$, and \eqref{eq:LB-m-less} follows.

For the upper bound, we fit $\eta_j$ on
$(\widehat\tau_{j-1},\widehat\tau_j]$. The resulting deterministic error
$q_t$ is supported on at most $m_0$ disjoint intervals between the estimated
and true boundaries. Each has length at most $n\xi_n$, and
$\sum_t\|q_t\|_2^2\le m_0n\xi_n v_{\max}^2$. The least-squares loss is
at most the loss at these chosen constants. On each error interval,
\[
2|\langle q_J,\sum_{t\in J}e_t\rangle|
\le |J|\|q_J\|_2^2+|J|\|\bar e_J\|_2^2.
\]
Expanding $\sum_t\|e_t+q_t\|_2^2$ and applying the same uniform event, we obtain
\eqref{eq:UB-wellloc}.
\end{proof}

\begin{theorem}\label{thm:IC-consistency}
Consider \eqref{eq:mean-shift-1} with $m_0\ge1$. Suppose the conditions of Proposition~\ref{prop:Haus} and Assumption~\ref{assump:pen} hold. Let
$(\widehat m,\check\tau_1,\ldots,\check\tau_{\widehat m})$ be any global minimizer in \eqref{eq:IC-selection-clean}, and let $\mathcal C_n$ be the candidate-coverage event in \eqref{eq:candidate-coverage-event}. Then, for all sufficiently large $n$,
\[
P(\widehat m\ne m_0)\le P(\mathcal C_n^c)+2n^{-1}\longrightarrow0.
\]
\end{theorem}

\begin{proof}
We write $V_n=\gamma_p^2p\log(n\vee p)$ and
$H_n=n\xi_nv_{\max}^2+V_n$. 
Let $\mathcal G_n$ be the uniform event of Lemma~\ref{lem:mean-block}, so $P(\mathcal G_n^c)\le2n^{-1}$. Proposition~\ref{prop:Haus} and Assumption~\ref{assump:WS} imply $P(\mathcal C_n^c)\to0$ as explained after \eqref{eq:candidate-coverage-event}. On $\mathcal C_n\cap\mathcal G_n$, the pool contains a distinct ordered $m_0$-tuple $\widehat T$ within $n\xi_n$ of the true changes, and Lemma~\ref{lem:UB-wellloc} bounds its loss by $\|e\|_F^2+Cm_0H_n$.

For a candidate partition $\boldsymbol{s}$ with $m<m_0$, we obtain from \eqref{eq:LB-m-less}
\[
\mathrm{IC}_n(\boldsymbol{s})-\mathrm{IC}_n(\widehat T)
\ge c\Delta_nv_{\min}^2-Cm_0H_n-m_0\omega_n>0
\]
for all large $n$, by Assumption~\ref{assump:pen}.

For $m=m_0+k$ with $k\ge1$, we use \eqref{eq:partition-lower} and drop its
nonnegative signal term, which gives
\[
\mathrm{IC}_n(\boldsymbol{s})-\mathrm{IC}_n(\widehat T)
\ge k\omega_n-CkV_n-Cm_0H_n>0,
\]
uniformly in $k\ge1$, since $m_0H_n=o(\omega_n)$ and
$V_n\le H_n$. 
Every global minimizer therefore has $m_0$ splits on $\mathcal C_n\cap\mathcal G_n$ for all sufficiently large $n$. Taking complements and applying the union bound gives the stated probability inequality.
\end{proof}

\subsection{Matching and local refinement}\label{ssc:mcp-matching}
\label{app:proof-ic-consistency-rate}

\begin{lemma}\label{lem:IC-coarse-matching}
Under the conditions of Theorem~\ref{thm:IC-consistency}, write $\check\tau_j$ for the ordered global IC locations. For every fixed $\epsilon>0$,
\[
P\left\{\widehat m=m_0,\quad
\max_{1\le j\le m_0}|\check\tau_j-\tau_j^\star|\le\epsilon\Delta_n\right\}\longrightarrow1.
\]
Equivalently, count consistency holds and the maximum matching error is $o_{\IP}(\Delta_n)$ when defined to be zero off $\{\widehat m=m_0\}$.
\end{lemma}

\begin{proof}
Having established count consistency above, we now fix $\epsilon \in (0, 1/4)$. Recall that $\mathcal{G}_n$ denotes the uniform concentration event defined in Lemma~\ref{lem:mean-block}.
On $\{\widehat m=m_0\}$, write $\boldsymbol{s}=(\check\tau_1,\ldots,\check\tau_{m_0})$
for the selected partition and $P_{\boldsymbol{s}}$ for its segmentwise averaging operator.
If a true
change has no selected split within $\epsilon\Delta_n$, its two neighboring
intervals of length $\lfloor\epsilon\Delta_n\rfloor$ lie in one fitted
segment. The best constant fit on this subinterval leaves a squared deterministic error
of at least $c\epsilon\Delta_nv_{\min}^2$, so
$\|(\mathbf{I}-P_{\boldsymbol{s}})\eta\|_F^2$ has this lower bound. By
\eqref{eq:partition-lower}, the loss of such a partition with $m_0$ splits
exceeds $\|e\|_F^2+c\epsilon\Delta_nv_{\min}^2-Cm_0\gamma_p^2p\log(n\vee p)$.
The well-localized tuple has upper bound \eqref{eq:UB-wellloc}, and
its penalty is the same. By Assumption~\ref{assump:pen}, the lower bound is
larger for all sufficiently large $n$. Hence every true change has a selected
split within $\epsilon\Delta_n$. These neighborhoods are disjoint, and
there are exactly $m_0$ selected splits, which gives ordered matching. 
Since these comparisons hold on $\mathcal C_n\cap\mathcal G_n$ for each fixed $\epsilon\in(0,1/4)$ and all sufficiently large $n$, their joint probability tends to one. The conclusion for any larger $\epsilon$ follows by inclusion.
This result establishes candidate matching on the spacing scale $\Delta_n$, which the subsequent local refinement further sharpens.
\end{proof}

\subsubsection{Geometry of the selected local intervals}\label{ssc:mcp-geometry}

The local intervals have random endpoints, but their geometric properties
follow from a deterministic calculation. We give it for both endpoint
fractions used by the procedure.

\begin{lemma}\label{lem:F-local-geometry}
Suppose Assumption~\ref{assump:local-geometry} holds, and let $l_j=\tau_j^\star-\tau_{j-1}^\star\ge\Delta_n$. Consider ordered locations
$t_0=0<t_1<\cdots<t_{m_0}<t_{m_0+1}=n$ with
$\max_j|t_j-\tau_j^\star|\le r$, where $r+1\le\Delta_n/20$.
For $\theta\in\{2/3,4/5\}$, set
\[
b_j=\lfloor\theta t_{j-1}+(1-\theta)t_j\rfloor,
\qquad e_j=\lceil(1-\theta)t_j+\theta t_{j+1}\rceil,
\qquad N_j=e_j-b_j.
\]
Then $(b_j,e_j]$ contains only the true change $\tau_j^\star$,
$N_j\ge\Delta_n$, and
\begin{equation}\label{eq:F-geometry-margin}
\frac{\tau_j^\star-b_j}{N_j}\ge\frac1{2(1+C_{\mathrm{sp}})},
\qquad
\frac{e_j-\tau_j^\star}{N_j}
\ge\frac{c_{\mathrm{sp}}}{2(1+c_{\mathrm{sp}})};
\end{equation}
in particular, both ratios are at least $1/\{2(1+\max\{C_{\mathrm{sp}},c_{\mathrm{sp}}^{-1}\})\}=10a_{\mathrm{loc}}$, with $a_{\mathrm{loc}}$ as in \eqref{eq:multi-trimmed-set}.
Moreover, $b_j-\tau_{j-1}^\star\ge3\Delta_n/20$ and
$\tau_{j+1}^\star-e_j\ge3\Delta_n/20$. In particular the starting
point is more than $d$ observations beyond the preceding boundary whenever
$d<3\Delta_n/20$.
\end{lemma}

\begin{proof}
The errors in each convex combination are at most $r$. Since we round the left
endpoint downward and the right endpoint upward, we get
\begin{align}\label{eq:F-endpoint-bounds}
\theta l_j-r
&\le\tau_j^\star-b_j\le\theta l_j+r+1,\\
\theta l_{j+1}-r
&\le e_j-\tau_j^\star\le\theta l_{j+1}+r+1.\nonumber
\end{align}
Because $2/3\le\theta\le4/5$ and $r+1\le l_i/20$, we have
\[
\frac12l_j\le\tau_j^\star-b_j\le l_j,
\qquad
\frac12l_{j+1}\le e_j-\tau_j^\star\le l_{j+1},
\]
and hence
$\tfrac12(l_j+l_{j+1})\le N_j\le l_j+l_{j+1}$.
In particular, $N_j\ge\Delta_n$. Dividing the left-side lower
bound by the total upper bound and using $l_{j+1}\le C_{\mathrm{sp}}l_j$ from \eqref{eq:local-geometry}, we obtain
\[
\frac{\tau_j^\star-b_j}{N_j}
\ge\frac{l_j}{2(l_j+l_{j+1})}\ge\frac1{2(1+C_{\mathrm{sp}})};
\]
in the same way, $l_{j+1}\ge c_{\mathrm{sp}}l_j$ from \eqref{eq:local-geometry} gives
\[
\frac{e_j-\tau_j^\star}{N_j}
\ge\frac{l_{j+1}}{2(l_j+l_{j+1})}\ge\frac{c_{\mathrm{sp}}}{2(1+c_{\mathrm{sp}})}.
\]
Since $c_{\mathrm{sp}}/(1+c_{\mathrm{sp}})=1/(1+c_{\mathrm{sp}}^{-1})$, both bounds are at least $1/\{2(1+\max\{C_{\mathrm{sp}},c_{\mathrm{sp}}^{-1}\})\}=10a_{\mathrm{loc}}$, which proves \eqref{eq:F-geometry-margin}.
We see that the true split lies strictly inside the trimmed set, with a fixed additional
margin.

Finally, from the upper bounds in \eqref{eq:F-endpoint-bounds} we get
\[
b_j-\tau_{j-1}^\star\ge(1-\theta)l_j-r-1
\ge\frac3{20}l_j,
\qquad
\tau_{j+1}^\star-e_j\ge\frac3{20}l_{j+1}.
\]
These inequalities exclude the adjacent true changes and give the claimed
lag margin, including at the outer boundaries $0$ and $n$.
\end{proof}

In the probability comparisons we use only the lower bound $N_j\ge\Delta_n$.
Local lengths may therefore be much larger than $\Delta_n$; the adjacent-spacing
condition \eqref{eq:local-geometry} is all we assume.

\subsubsection{A uniform raw-series comparison}\label{ssc:mcp-rawcomparison}

Consider a deterministic interval $I = (b, e]$ of length $N = e - b$ containing a single true change point $\tau_j^\star$. Reindexing observations $b+1, \dots, e$ to $1, \dots, N$, we write $\tau = \tau_j^\star - b$ for the true change point and $k$ for a candidate split in local coordinates. All series in the local contrasts, including filtered observations, are restricted to $I$ and reindexed; adding $b$ restores original-time positions without altering localization errors. Both $\tau$ and $k$ are restricted to $[aN, (1-a)N]$ for fixed $a > 0$ (with $a = a_{\mathrm{loc}}$ from \eqref{eq:multi-trimmed-set} in Proposition~\ref{thm:ic-consistency-rate} and Theorem~\ref{prop:multi-local}), and we set $h = |k - \tau| \ge 1$. For filtered comparisons, we also assume $b$ exceeds the preceding true boundary by more than $d$ observations. High-probability events for this deterministic family are established prior to evaluating them on the selected interval realizations.

\begin{lemma}\label{lem:uniform-local-raw}
Under Assumption~\ref{ass:innov}, with probability at
least $1-Cn^{-2}$, simultaneously over all such intervals, changes, and
cuts, and uniformly for $0\le\lambda\le\bar\lambda<\infty$,
\begin{equation}\label{eq:uniform-raw-fluctuation}
\begin{split}
|\mathcal R_I(k)-\mathcal R_I(\tau)|\le C\bigg[
&\frac{\gamma_p\|\Delta_{\eta,j}\|_2\sqrt{h\log n}}N\\
&+\frac{\gamma_p^2(\sqrt{p\log n}+\log n)\sqrt h}{N^{3/2}}
+\frac{\gamma_p^2ph}{N^2}\bigg],
\end{split}
\end{equation}
where $\mathcal R_I(k)=G_\lambda(y;k)-G_\lambda(Ey;k)$ and the local
criterion is defined in \eqref{asf:eq:S3-profile-new}.
\end{lemma}

\begin{proof}
Where $j$ indexes the true change point contained in the interval, removing the baseline constant yields a clean two-piece mean sequence with jump size $g_j = \Delta_{\eta,j}$.
Its contrast is $f_kg_j$,
where $f_k$ is given by \eqref{eq:E-mean-contrast-coefficients}, and therefore
\[
\mathcal R_I(k)=2c_\lambda(k)f_k\langle g_j,D_e(k)\rangle
+c_\lambda(k)\|D_e(k)\|_2^2.
\]
The convolution representation of $e_t$ and Young's inequality give an
operator bound $\sigma^2\gamma_p^2$ for the covariance of a time-stacked
error vector. In particular, by \eqref{eq:E-weight-norms}, the projected
contrast increment has variance at most
$C\gamma_p^2\|g_j\|_2^2h/N^2$, and the projected contrast at $\tau$
has variance at most $C\gamma_p^2\|g_j\|_2^2/N$.
Since $c_\lambda(k)f_k$ is bounded and has increments $O(h/N)$, we bound the linear increment
through Gaussian tails by
$C\gamma_p\|g_j\|_2\sqrt{h\log n}/N$. The coefficient-increment term
has size $C\gamma_p\|g_j\|_2h\sqrt{\log n}/N^{3/2}$ and is no larger.

For the centered quadratic term, the contrast-weight matrix difference
has Frobenius norm at most $C\sqrt h/N^{3/2}$ by
\eqref{eq:E-quadratic-matrix-increments}. After tensoring with $\mathbf{I}_p$, its
Frobenius norm is at most $C\sqrt{ph}/N^{3/2}$ and its operator norm is
at most $C\sqrt h/N^{3/2}$. Multiplying by the covariance square roots
costs at most $\sigma^2\gamma_p^2$, and Lemma~\ref{lem:gaussian-quadratic-uniform}
then gives the centered bound
\[
C\gamma_p^2(\sqrt{p\log n}+\log n)\sqrt h/N^{3/2}.
\]

Since $e_t$ is dependent, we treat the expectation separately.
Its covariance sequence $\Gamma_e(v)=\operatorname{Cov}(e_t,e_{t-v})$ uses the opposite lag convention to $\Gamma_h$ in Lemma~\ref{asf:lem:projvar}. Accounting also for the innovation normalization in Appendix~\ref{sec:appendix}, the relation is $\Gamma_e(v)=\sigma^2\Gamma_{-v}=\sigma^2\Gamma_v^\top$.
It satisfies
\[
\sum_{v\in\mathbb Z}\|\Gamma_e(v)\|_2
\le\sigma^2\left(\sum_{r\ge0}\|\Psi_r\|_2\right)^2
=\sigma^2\gamma_p^2.
\]
Let $H(k)=E\|D_e(k)\|_2^2
=\sum_{t,s}w_{k,t}w_{k,s}\operatorname{tr}\Gamma_e(t-s)$.
Subtracting the two products of weights, we obtain
\begin{align*}
|H(k)-H(\tau)|
&\le p\sum_{t,s}|w_{k,t}-w_{\tau,t}|\,
\|\Gamma_e(t-s)\|_2
(|w_{k,s}|+|w_{\tau,s}|)\\
&\le C\gamma_p^2p\|w_k-w_\tau\|_1
\max_l\|w_l\|_\infty
\le C\gamma_p^2ph/N^2.
\end{align*}
Also $H(\tau)\le C\gamma_p^2p/N$. Multiplying by $c_\lambda(k)$ and including
$(c_\lambda(k)-c_\lambda(\tau))H(\tau)$, we get the same mean-increment bound.

Uniformity in $\lambda$ comes at no extra cost, and we never discretize the ridge
parameter. For the centered variables
$Q_k=\|D_e(k)\|_2^2-H(k)$, we use
\[
c_\lambda(k)Q_k-c_\lambda(\tau) Q_\tau=c_\lambda(k)(Q_k-Q_\tau)+(c_\lambda(k)-c_\lambda(\tau))Q_\tau.
\]
We bound the two random variables on the right independently of $\lambda$,
using respectively the weight-matrix difference and the baseline matrix
$w_\tau w_\tau^\top$. Their coefficients have the uniform bounds already
stated. The same separation applies to the linear term. There are at most
$n^4$ choices of two endpoints, a true change, and a candidate cut. Taking
$x=K\log n$ in the Gaussian and quadratic tails, with $K$ large, and
summing the failure probabilities, we obtain a single event of probability
at least $1-Cn^{-2}$ for the entire family.
\end{proof}

\begin{proof}[Proof of Proposition~\ref{thm:ic-consistency-rate}]
Proposition~\ref{prop:Haus} and the strengthened neighborhood condition in
Assumption~\ref{assump:WS} place an $m_0$-tuple of the candidate pool within
$n\xi_n$ of the true changes with probability tending to one. We then apply
Theorem~\ref{thm:IC-consistency} and Lemma~\ref{lem:IC-coarse-matching}.
On an event whose probability tends to one, the count is correct and the
screened errors are at most $\Delta_n/100$. Since $\Delta_n\to\infty$ by Assumption~\ref{assump:local-geometry}, we can apply
Lemma~\ref{lem:F-local-geometry} to $t_j=\check\tau_j$ with
$\theta=2/3$. Each selected interval belongs to the deterministic family with $a=a_{\mathrm{loc}}$,
has $N\ge\Delta_n$, and contains the true split in its trimmed candidate
set, at distance at least $10a_{\mathrm{loc}}N$ from both endpoints by \eqref{eq:F-geometry-margin}. In this way we verify the geometric conditions before comparing any local
criterion.

For a clean two-mean signal, we obtain from Lemma~\ref{asf:lem:signaldrift} a
contrast gap $ch\|\Delta_{\eta,j}\|_2^2/N$, uniformly for the bounded
nonnegative raw-series ridge parameters. On the simultaneous event of
Lemma~\ref{lem:uniform-local-raw}, we divide each fluctuation term by this
gap. If $h\ge K\gamma_p^2\log n/v_{\min}^2$, the linear ratio is at most
$C/\sqrt K$. The centered quadratic ratio is at most
\[
\frac C{\sqrt K}
\frac{\gamma_p(\sqrt p+\sqrt{\log n})}{\sqrt{\Delta_n}v_{\min}}=o(1),
\]
and the mean ratio is at most
$C\gamma_p^2p/(\Delta_nv_{\min}^2)=o(1)$.
Indeed, Assumption~\ref{assump:pen} implies
$\gamma_p^2p\log(n\vee p)/(\Delta_nv_{\min}^2)\to0$, while
$p\log(n\vee p)\ge c(p+\log n)$ for all large $n$.

We choose $K$ so that the first ratio is smaller than a fixed fraction of the
gap; the other two ratios then vanish uniformly. Consequently, every maximizer of the
local contrast lies within this radius, simultaneously for all changes on the same event. 
We have therefore proved the joint count-and-localization probability in Proposition~\ref{thm:ic-consistency-rate}; the random endpoints enter only through the deterministic family.
\end{proof}

\subsection{Estimation on multiple retained blocks}\label{ssc:step2-mcp}

\begin{proof}[Proof of Proposition~\ref{thm:S2-multi}]

Under the conditions of Proposition~\ref{thm:ic-consistency-rate} (included in the present proposition) and the diverging buffer multiplier, $\mathbb{P}(\mathcal{E}_{\mathrm{pure}}^{(m)}) \to 1$. On this event, exactly $J = m_0 + 1$ blocks are retained. Within block $j$, the earliest retained response at an interior left boundary occurs at time $\widehat{\tau}_{j-1} + w + d + 1$, so its furthest lag satisfies $\widehat{\tau}_{j-1} + w + 1 > \tau_{j-1}^\star$. Similarly, the final retained response satisfies $\widehat{\tau}_j - w \le \tau_j^\star$. By the outer-boundary conventions, this property applies analogously to the first ($j = 1$) and last ($j = J$) blocks. Consequently, every retained response and all its corresponding lags lie entirely within their respective pure regime and share its true mean.

The retained blocks are non-empty for all sufficiently large $n$, as interior block lengths are bounded below by $\Delta_n - 4w - d$, where $w + d = o(\Delta_n)$. Summing the block lengths yields, via a telescoping argument,
\[
N_m = n - 2m_0 w - Jd, \qquad \frac{2m_0 w + Jd}{n} \le C \frac{w + d}{\Delta_n} = o(1).
\]
Consequently, $N_m \asymp n$. We apply Proposition~\ref{prop:uniform-block-concentration} with deterministic parameter choices $J = m_0 + 1$ and $N_0 = N_m$. Assumption~\eqref{eq:S2-multi-uniform} matches its sample-size requirement exactly, while the simultaneous concentration event explicitly accounts for the $(n+1)^{2J}$ endpoint configurations, the curvature correction $J L_x / N_m$, and the profiled score mean $J \sigma \sqrt{L_x} / N_m$.

On the purity event $\mathcal{E}_{\mathrm{pure}}^{(m)}$, Lemma~\ref{asf:lem:S2-profile} identifies the empirical Gram matrix and score vector with their stationary-error counterparts. Consequently, the empirical tuning parameter $\widehat{\lambda}_{m,n}$ in \eqref{eq:multi-profile-lasso}—which constructs penalty levels using the estimated number of breaks and the retained sample size—coincides with $\lambda_{m,n} = C_{\lambda} Q_{\mathrm{dev}} \sqrt{\ell_{m,n} / N_m}$, where $C_{\lambda} \ge 4$. Applying Proposition~\ref{asf:prop:S2-eventwise} then yields all claimed estimation, cone condition, and prediction error bounds. The retained residual energy satisfies the quadratic identity
\[
N_m^{-1} \left\| G_{\mathrm{blk}}^{(m)} X_I (\widehat{B} - B) \right\|_{\mathrm{F}}^2 = \operatorname{vec}(\widehat{B} - B)^\top \widehat{\Gamma}_m \operatorname{vec}(\widehat{B} - B).
\]
Because the uniform concentration proposition explicitly accounts for the union bound over all candidate partitions, the overall failure probability is bounded by $\mathbb{P}\bigl( (\mathcal{E}_{\mathrm{pure}}^{(m)})^c \bigr) + C n^{-2} \to 0$.
\end{proof}

\subsection{Simultaneous filtered localization}\label{ssc:mcp-filtered}

Let
\[
\mathcal D_{m,n}=\left\{D:\operatorname{vec}D\in\mathcal C(S),\quad
\|D\|_F\le C_B\bar Q\sqrt{s\ell_{m,n}/n}\right\}.
\]
On the Step~2 event the estimated error belongs to this deterministic set.
Its cone radius is
\begin{equation*}
\sup_{D\in\mathcal D_{m,n}}\|\operatorname{vec}D\|_1
\le C\bar Qs\sqrt{\ell_{m,n}/n}=:a_{m,n},
\qquad a_{m,n}\gamma_p\sqrt{\ell_{2,n}}\le C\sqrt{b_{m,n}}.
\end{equation*}
The denominator here is the full sample size $n$, because
the common filter is estimated from all retained blocks.

On any admissible deterministic interval containing change $j$, the
decomposition \eqref{asf:eq:S3-YD-decomp} holds after subtraction of the
local pre-change baseline. Thanks to the lag margin, no transition from the
preceding change enters, and the only transition vectors are
$c_{q,j}$ defined in Assumption~\ref{assump:S3-transition-multi}.
We write
$\mathcal R_{I,D}(k)=G_\lambda(Y(D);k)-G_\lambda(u(D);k)$.

\begin{lemma}\label{lem:uniform-local-filtered}
Under Assumptions~\ref{ass:innov}, \ref{assump:S3-filter-multi}, and
\ref{assump:S3-transition-multi}, let $d$ be fixed. For all sufficiently
large $n$, with probability at least $1-Cn^{-2}$, simultaneously over all
admissible intervals with $N\ge c\Delta_n$, their true changes and trimmed
cuts, and $D\in\mathcal D_{m,n}$,
\begin{equation}\label{eq:uniform-filtered-fluctuation}
\begin{split}
|\mathcal R_{I,D}(k)-\mathcal R_{I,D}(\tau)|\le C\bigg[
&\frac{\delta_j(\sqrt{\log n}+\sqrt{b_{m,n}})\sqrt h}N\\
&+\frac{\{\sqrt{p\log n}+\log n+\sqrt{b_{m,n}(p+\log n)}+b_{m,n}\}\sqrt h}{N^{3/2}}
+\frac{ph}{N^2}\bigg],
\end{split}
\end{equation}
where $\delta_j=\|\Delta_{\mu,j}\|_2$,
and $h=|k-\tau|\ge1$. The bound is uniform for ridge parameters in any
fixed bounded nonnegative interval.
\end{lemma}

\begin{proof}

We let $\mathcal V$ contain the nonzero structural jumps and transition
vectors, each divided by its Euclidean norm. This is a deterministic set
of at most $(d+1)m_0\le(d+1)n$ directions, built from population quantities only.
From Gaussian tails we see that, simultaneously for every
integer interval $J$ of length $r\ge1$,
\begin{equation}\label{eq:F-uniform-interval-sums}
\begin{gathered}
\left\|\sum_{t\in J}\varepsilon_t\right\|_2
\le C\sqrt{r(p+\log n)},\qquad
\max_{v\in\mathcal V}\left|\sum_{t\in J}v^\top\varepsilon_t\right|
\le C\sqrt{r\log n},\\
\left\|\sum_{t\in J}\varepsilon_t\right\|_\infty
\le C\sqrt{r\ell_{2,n}},\qquad
\max_{a\le dp}\left|\sum_{t\in J}\widetilde x_{t,a}\right|
\le C\gamma_p\sqrt{r\ell_{2,n}},
\end{gathered}
\end{equation}
with joint failure probability at most $Cn^{-2}$.

Next, we provide the counts behind this event. There are at most $n^2$ intervals.
For the Euclidean innovation bound, the Gaussian norm tail at
$x=K\log n$ gives $C\sqrt{r(p+\log n)}$. For the directional bound there
are at most $(d+1)n^3$ interval-direction pairs, so the same logarithmic
scale suffices. The coordinate bounds involve at most $dpn^2$ scalar
sums. Their variances are at most $Cr$ for innovations and
$C\gamma_p^2r$ for the lag coordinates, by
Lemma~\ref{asf:lem:projvar}. Taking $x=K\ell_{2,n}$ pays for the coordinate
count and leaves failure probability $Cn^{-2}$. Only the logarithm
of the dimension enters here, so $p$ may grow faster than any polynomial in $n$.

On this event, the cone bound is pathwise:
\begin{equation*}
\sup_{D\in\mathcal D_{m,n}}
\left\|D^\top\sum_{t\in J}\widetilde x_t\right\|_2
\le a_{m,n}\max_a\left|\sum_{t\in J}\widetilde x_{t,a}\right|
\le C\sqrt{b_{m,n}r}.
\end{equation*}
We see that the two logarithms in $b_{m,n}$ enter at different steps:
$\ell_{m,n}$ through estimation of $D$, and $\ell_{2,n}$ through the
coordinate maximum in \eqref{eq:F-uniform-interval-sums}.

We use \eqref{eq:contrast-denominator-identity} and
\eqref{eq:E-left-denominator-identity} on every admissible interval.
Each local sum has length at most $h$ and coefficient $C/N$, and each
long sum has length at most $N$ and coefficient $Ch/N^2$. Consequently,
\begin{equation}\label{eq:F-contrast-event}
\begin{gathered}
\|D_\varepsilon(\tau)\|_2\le C\sqrt{(p+\log n)/N},\qquad
\|D_\varepsilon(k)-D_\varepsilon(\tau)\|_2\le C\sqrt{(p+\log n)h}/N,\\
\max_{v\in\mathcal V}|v^\top D_\varepsilon(\tau)|\le C\sqrt{\log n/N},\qquad
\max_{v\in\mathcal V}|v^\top(D_\varepsilon(k)-D_\varepsilon(\tau))|\le C\sqrt{h\log n}/N,\\
\|D_\varepsilon(\tau)\|_\infty\le C\sqrt{\ell_{2,n}/N},\qquad
\|D_\varepsilon(k)-D_\varepsilon(\tau)\|_\infty\le C\sqrt{\ell_{2,n}h}/N,\\
\sup_D\|D_Z(\tau)\|_2\le C\sqrt{b_{m,n}/N},\qquad
\sup_D\|D_Z(k)-D_Z(\tau)\|_2\le C\sqrt{b_{m,n}h}/N.
\end{gathered}
\end{equation}
The long bounds hold at $k$ as well, with another constant, because
$h\le N$. All of these are deterministic inequalities on a single event.

Next we settle the centered innovation quadratic event. For each interval, true change, and cut, we apply
Lemma~\ref{lem:gaussian-quadratic-uniform} to the matrix differences
$\mathbf{I}_p\otimes(w_kw_k^\top-w_\tau w_\tau^\top)$ and the baseline matrix
$\mathbf{I}_p\otimes w_\tau w_\tau^\top$. The difference has Frobenius norm
at most $C\sqrt{ph}/N^{3/2}$ and operator norm at most
$C\sqrt h/N^{3/2}$; the baseline norms are $C\sqrt p/N$ and $C/N$.
Taking $x=K \log n$ and a union over at most $n^4$ tuples, we obtain, with another
failure probability $Cn^{-2}$, a centered increment bound
\[
C(\sqrt{p\log n}+\log n)\sqrt h/N^{3/2}.
\]
As in Lemma~\ref{lem:uniform-local-raw}, the coefficient separation gives
uniformity in the ridge parameter. The baseline form is multiplied
by $|c_\lambda(k)-c_\lambda(\tau)|\le Ch/N$, and hence has no larger order. The exact
expectation formula \eqref{eq:E-quadratic-mean} adds at most $Cph/N^2$.
We intersect this event with the interval-sum event.

We next bound each of the six terms on the joint concentration event by applying the local counterpart of expansion \eqref{eq:E-six-term-expansion}, replacing $\lambda_3$ with the local ridge parameter $\lambda$. The deterministic relations \eqref{eq:E-mean-contrast-coefficients}--\eqref{eq:E-deterministic-contrast-bounds} hold upon replacing $\delta_0$ with $\delta_j$ and $\rho_n$ with $\rho_{m,n}$, while bounding $\nu_n$ by $\nu_{n,\star}$. Because $N \ge c \Delta_n$, it follows that $\nu_{n,\star}/\sqrt{N} = o(1)$ uniformly. Below, we explicitly detail the resulting stochastic increments to distinguish the fixed parameter directions from the adaptive direction.

For $T_1$, the fixed-direction rows of \eqref{eq:F-contrast-event},
together with the transition crossing count in
\eqref{eq:E-transition-weight-increments}, give
\[
|\Delta_kT_1|\le
C\frac{\delta_j\sqrt{h\log n}}N
\left(1+\frac{\nu_{n,\star}}{\sqrt N}\right).
\]
We bound the crossing part using
$\min(h,d)\le\sqrt d\sqrt h$, exactly as in \eqref{eq:E-T1-bound}.
For $T_2$, we use the entrywise inequality rather than a direction from
$\mathcal V$:
\[
\sup_D|\langle D^\top D_L(k),D_\varepsilon(k)-D_\varepsilon(\tau)\rangle|
\le a_{m,n}\|D_L(k)\|_\infty\|D_\varepsilon(k)-D_\varepsilon(\tau)\|_\infty.
\]
Including the changes in $D_L(k)$ and $c_\lambda(k)$, we get
\[
\sup_D|\Delta_kT_2|
\le Ca_{m,n}\gamma_p\delta_j\sqrt{\ell_{2,n}}
\left(\frac{\sqrt h}N+\frac h{N^{3/2}}\right)
\le C\frac{\delta_j\sqrt{b_{m,n}h}}N.
\]
For $T_3$, the mean bounds and the last row of
\eqref{eq:F-contrast-event} give
\begin{align*}
\sup_D|\Delta_kT_3|
&\le C\delta_j\left\{
\frac{\sqrt{b_{m,n}h}}N+
\frac{h+\nu_{n,\star}\min(h,d)}N\sqrt{\frac{b_{m,n}}N}
+\frac hN\sqrt{\frac{b_{m,n}}N}\right\}\\
&\le C\frac{\delta_j\sqrt{b_{m,n}h}}N
\left(1+\frac{\nu_{n,\star}}{\sqrt N}\right).
\end{align*}
The preceding quadratic event gives
\[
|\Delta_kT_4|\le C\left\{
\frac{(\sqrt{p\log n}+\log n)\sqrt h}{N^{3/2}}+\frac{ph}{N^2}\right\}.
\]
Finally, the bilinear and square identities from the proof of
Proposition~\ref{asf:prop:S3-fluct-new}, now applied to the pathwise
Euclidean bounds in \eqref{eq:F-contrast-event}, give
\begin{align*}
\sup_D|\Delta_kT_5|
&\le C\left\{
\frac{\sqrt{(p+\log n)h}}N\sqrt{\frac{b_{m,n}}N}
+\sqrt{\frac{p+\log n}N}\frac{\sqrt{b_{m,n}h}}N
+\frac hN\frac{\sqrt{b_{m,n}(p+\log n)}}N\right\}
\\
&\le C\frac{\sqrt{b_{m,n}(p+\log n)h}}{N^{3/2}},\\
\sup_D|\Delta_kT_6|
&\le C\left\{\frac{\sqrt{b_{m,n}h}}N\sqrt{\frac{b_{m,n}}N}
+\frac{b_{m,n}h}{N^2}\right\}
\le C\frac{b_{m,n}\sqrt h}{N^{3/2}}.
\end{align*}
All products are deterministic on the joint event, so they hold under arbitrary
dependence. Summing the six inequalities and absorbing the
bounded transition factors, we obtain \eqref{eq:uniform-filtered-fluctuation}.
We constructed the event before choosing either the interval or $D$.
\end{proof}

\begin{proof}[Proof of Theorem~\ref{prop:multi-local}]
We first assemble the events on which a local comparison is legitimate.
Count consistency and Proposition~\ref{thm:ic-consistency-rate} give
\[
\max_j|\widehat\tau_j-\tau_j^\star|
=O_{\IP}(\gamma_p^2\log n/v_{\min}^2)=o_{\IP}(\Delta_n).
\]
The last relation follows from Assumption~\ref{assump:pen}, as in the
raw-series proof. With probability tending to one these errors are at most
$\Delta_n/100$, so Lemma~\ref{lem:F-local-geometry}, with $\theta=4/5$,
places all selected filtered intervals in the admissible deterministic
family with $a=a_{\mathrm{loc}}$, by \eqref{eq:F-geometry-margin}. Their starts have the required lag margin for all large $n$,
because $d$ is fixed and $\Delta_n\to\infty$ by Assumption~\ref{assump:local-geometry}. In particular, each true
split is an available candidate. Proposition~\ref{thm:S2-multi} puts
$\widehat B-B$ in $\mathcal D_{m,n}$ on another event of probability
tending to one. We intersect these events with that of
Lemma~\ref{lem:uniform-local-filtered}.

For the deterministic gap, we use the calculation in
Lemma~\ref{asf:lem:S3-deterministic-D} on each interval, with the single-change alignment condition \eqref{eq:single-transition-alignment} replaced by its multiple-change version \eqref{eq:multi-transition-alignment} in Assumption~\ref{assump:S3-transition-multi}, and with $\rho_n$, $\nu_n$ and $\mathcal D_n$ replaced by $\rho_{m,n}$, $\nu_{n,\star}$ and $\mathcal D_{m,n}$. Its perturbation
of an alignment bracket is bounded by
\[
Ch\delta_j^2\{
\rho_{m,n}+\rho_{m,n}^2+\rho_{m,n}\nu_{n,\star}\}=o(h\delta_j^2)
\]
uniformly in $j$ and $D$, because $\rho_{m,n}\to0$ by Assumption~\ref{assump:S3-filter-multi} and $\rho_{m,n}\nu_{n,\star}\to0$ by Assumption~\ref{assump:S3-transition-multi}. Its transition remainder is a vanishing fraction
of the leading gap because
$\nu_{n,\star}^2/N\le C\nu_{n,\star}^2/\Delta_n\to0$; here $N\ge\Delta_n$ by Lemma~\ref{lem:F-local-geometry}, and $\nu_{n,\star}^2/\Delta_n\to0$ by Assumption~\ref{assump:S3-transition-multi}.
Finally, the ridge comparison costs at most
$C\bar\lambda_n^2h\delta_j^2/N=o(h\delta_j^2/N)$, uniformly over the allowed data-selected ridge parameters.
We conclude that \eqref{asf:eq:S3-uniform-drift} holds on every interval with the same
constant $c>0$ and gap $ch\delta_j^2/N$.

We put $R_{m,n}=(\log n+b_{m,n})/u_\star^2$. For any candidate with
$h\ge KR_{m,n}$, dividing \eqref{eq:uniform-filtered-fluctuation} by
this gap, we bound the linear ratio by $C/\sqrt K$, since
$(\sqrt{\log n}+\sqrt{b_{m,n}})/\sqrt{\log n+b_{m,n}}\le\sqrt2$ and $\delta_j\ge u_\star$.
The centered quadratic ratio is at most
\[
\frac C{\sqrt K}
\frac{\sqrt{p\log n}+\log n+\sqrt{b_{m,n}(p+\log n)}+b_{m,n}}
{\sqrt{\Delta_n}\,u_\star\sqrt{\log n+b_{m,n}}}
\le\frac C{\sqrt K}
\sqrt{\frac{p+\log n+b_{m,n}}{\Delta_nu_\star^2}}=o(1).
\]
The mean ratio is $Cp/(\Delta_nu_\star^2)=o(1)$.
Assumption~\ref{assump:S3SNR-multi} implies these limits and also
$R_{m,n}=o(\Delta_n)$: it controls $p\log n$, $\log n$, and $b_{m,n}\log n$ relative to
$\Delta_nu_\star^2$, with $\log n\ge1$.

We choose $K$ large enough that the linear ratio is smaller than a fixed
fraction of the gap. For all large $n$ the other ratios are smaller than
the remaining fraction, uniformly over changes, intervals, and filters.
Consequently, every maximizer of a filtered local contrast lies within the radius
$KR_{m,n}$. We may substitute the selected intervals and the estimated
filter on the joint event already constructed. 
Its probability tends to one, proving the joint count-and-localization conclusion \eqref{eq:multi-refined-rate} and hence the simultaneous $O_{\IP}$ rate.
Since we work on a single joint event throughout, the bound holds
simultaneously over all changes.
\end{proof}

\section{Removing the transition-alignment condition in the single-change case}
\label{sec:remove-transition-alignment}

This section proceeds in two main steps. First, Section~\ref{ssc:tcr-refinement} develops the transition-corrected refinement, where Lemma~\ref{lem:tc-excess} establishes that its population criterion separates the true change point without requiring the alignment condition~\eqref{eq:single-transition-alignment}. Second, to conclude the section, we introduce three stable VAR(2) designs in which Assumption~\ref{assump:transition-alignment} fails, accompanied by a corresponding numerical comparison.

\subsection{A transition-corrected single-change refinement}\label{ssc:tcr-refinement}

Our default filtered two-mean refinement requires
\eqref{eq:single-transition-alignment} because the first \(d\) post-change
filtered observations retain a deterministic transition layer.  The
transition-corrected refinement (TCR) instead fits that finite shape.  We write
\begin{align}
\Delta_\eta&:=\eta_2-\eta_1,\qquad
\Delta_\mu:=\left(\mathbf{I}_p-\sum_{i=1}^dA_i\right)\Delta_\eta,\nonumber\\
F_r&:=\mathbf{I}_p-\sum_{i=1}^rA_i,\quad F_0:=\mathbf{I}_p,
\label{eq:tc-Fr}\\
\varphi_j&:=F_{j-1}\Delta_\eta\ (j=1,\ldots,d),\qquad
\varphi_\infty:=F_d\Delta_\eta=\Delta_\mu.
\label{eq:tc-pop-template}
\end{align}
Relative to the pre-change level, the oracle filtered-mean template is therefore $
0,\varphi_1,\ldots,\varphi_d,\varphi_\infty$.
Using the buffered pure blocks \(I_1,I_2\) and the Step~2 estimates, we set
\begin{align*}
\widehat\Delta_\eta
&:=|I_2|^{-1}\sum_{t\in I_2}y_t
   -|I_1|^{-1}\sum_{t\in I_1}y_t,\\
\widehat F_r
&:=\mathbf{I}_p-\sum_{i=1}^r\widehat A_i,\quad \widehat F_0:=\mathbf{I}_p,\qquad
\widehat{\varphi}_j:=\widehat F_{j-1}\widehat\Delta_\eta,\quad
\widehat{\varphi}_\infty:=\widehat F_d\widehat\Delta_\eta.
\end{align*}
For \(k\in\mathcal T_n^{(3)}\), we shift the estimated template to \(k\):
\begin{equation}\label{eq:tc-shifted-template}
\widehat s_t(k):=
\begin{cases}
0, & d+1\le t\le k,\\
\widehat{\varphi}_{t-k}, & k+1\le t\le k+d,\\
\widehat{\varphi}_\infty, & k+d+1\le t\le n.
\end{cases}
\end{equation}
We profile out one common baseline and define
\begin{align}
\widehat\theta(k)
&:=\frac1{n-d}\sum_{t=d+1}^n
\{\widehat Y_t-\widehat s_t(k)\},
\label{eq:tc-theta}\\
L_{\mathrm{tc}}(k)
&:=\frac1{n-d}\sum_{t=d+1}^n
\|\widehat Y_t-\widehat\theta(k)-\widehat s_t(k)\|_2^2,
\label{eq:tc-loss}\\
\widehat\tau_{\mathrm{tc}}
&\in\arg\min_{k\in\mathcal T_n^{(3)}}L_{\mathrm{tc}}(k).
\label{eq:tc-tau}
\end{align}

For the default criterion, comparing \(\tau+q\) with \(\tau\) produces the
signed drift
\(q\|\Delta_\mu\|_2^2+2\langle\Delta_\mu,C_q\rangle\).
TCR instead compares the correctly located template with a shifted copy after
removing a common baseline, so the oracle excess is a nonnegative
squared-template mismatch.  Lemma~\ref{lem:tc-excess} shows that this excess is at least
\(c_{\mathrm{tr}}q\|\Delta_\mu\|_2^2/(d+1)\) for all large \(n\), without
\eqref{eq:single-transition-alignment}.

\begin{lemma}\label{lem:tc-excess}
Let \(s_t(k)\) be defined as in \eqref{eq:tc-shifted-template} with
\(\widehat\varphi_j,\widehat\varphi_\infty\) replaced by \(\varphi_j,\varphi_\infty\), so that the
oracle filtered series has mean \(\mu_1+s_t(\tau)\) for \(t\ge d+1\). For
\(k\in\mathcal T_n^{(3)}\), let \(D_t(k):=s_t(\tau)-s_t(k)\),
\(\bar D(k):=(n-d)^{-1}\sum_{t=d+1}^nD_t(k)\), and
\[
\mathcal E_{\mathrm{tc}}(k):=\sum_{t=d+1}^n\|D_t(k)-\bar D(k)\|_2^2 ,
\]

This is $(n-d)$ times the excess obtained by replacing $\widehat Y_t$ with the oracle filtered mean $\mu_1+s_t(\tau)$ and using the true template. It is also $(n-d)$ times the expected excess loss when the true filter and true template are used but the innovations are retained: the noise contribution has the same expectation for every $k$, and the cross term has expectation zero. This is a population separation statement, not an identity for the realized noisy excess.
If $\tau\in\mathcal T_n^{(3)}$ and \(d<c_{\mathrm{tr}}n\), then for every \(k\in\mathcal T_n^{(3)}\)
with \(q:=|k-\tau|\ge1\),
\[
\mathcal E_{\mathrm{tc}}(k)\ge\Bigl(1-\frac{q+d}{n-d}\Bigr)\frac{q\|\Delta_\mu\|_2^2}{d+1}.
\]
In particular, for fixed $d$, \(\mathcal E_{\mathrm{tc}}(k)\ge c_{\mathrm{tr}}q\|\Delta_\mu\|_2^2/(d+1)\) for all
large \(n\).
\end{lemma}

\begin{proof}
We set \(\varphi_j:=\Delta_\mu\) for \(j>d\), consistently with \(\varphi_\infty=\Delta_\mu\),
and
\[
U_{j'}:=\sum_{i=j'}^{\min(j'+q-1,d)}A_i\Delta_\eta,
\qquad 1\le j'\le d.
\]
Since
\(\varphi_j=F_{j-1}\Delta_\eta\) for \(j\le d\) and \(\Delta_\mu=F_d\Delta_\eta\), we have
\(\varphi_{j'}-\varphi_{j'+q}=U_{j'}\) for every \(1\le j'\le d\).

We first take \(k=\tau+q\). Then \(D_t(k)=0\) unless \(\tau<t\le\tau+q+d\); for
\(1\le j\le q\) we have \(D_{\tau+j}(k)=\varphi_j\), and for \(1\le j'\le d\) we have
\(D_{\tau+q+j'}(k)=\varphi_{q+j'}-\varphi_{j'}=-U_{j'}\). For \(k=\tau-q\) the same vectors
appear with the opposite sign, namely \(D_{\tau-q+j}(k)=-\varphi_j\) and
\(D_{\tau+j'}(k)=U_{j'}\). The condition \(d<c_{\mathrm{tr}}n\) keeps all these indices in
\(\{d+1,\ldots,n\}\). In both cases we obtain
\[
M:=\sum_{t=d+1}^n\|D_t(k)\|_2^2=\sum_{j=1}^q\|\varphi_j\|_2^2+\sum_{j'=1}^d\|U_{j'}\|_2^2 .
\]
For each \(1\le j\le\min(q,d)\), telescoping gives
\[
\Delta_\mu=\varphi_j-\sum_{c\ge0:\,j+cq\le d}U_{j+cq},
\]
a sum of at most \(1+\lceil d/q\rceil\le d+1\) vectors. By the Cauchy--Schwarz
inequality, \(\|\Delta_\mu\|_2^2\le(d+1)\{\|\varphi_j\|_2^2+\sum_c\|U_{j+cq}\|_2^2\}\).
Every \(j'\in\{1,\ldots,d\}\) is of the form \(j+cq\) for exactly one pair with
\(1\le j\le q\) and \(c\ge0\). If \(q\le d\), summing over \(1\le j\le q\) we get
\(q\|\Delta_\mu\|_2^2\le(d+1)M\). If \(q>d\), only \(c=0\) occurs, so
\(\|\Delta_\mu\|_2^2\le2(\|\varphi_j\|_2^2+\|U_j\|_2^2)\) for \(j\le d\), while
\(\varphi_j=\Delta_\mu\) for \(d<j\le q\); hence
\(M\ge(q-d)\|\Delta_\mu\|_2^2+(d/2)\|\Delta_\mu\|_2^2\ge q\|\Delta_\mu\|_2^2/2\). In both cases
\(M\ge q\|\Delta_\mu\|_2^2/(d+1)\).

Finally, at most \(q+d\) of the vectors \(D_t(k)\) are nonzero, so by the
Cauchy--Schwarz inequality
\((n-d)\|\bar D(k)\|_2^2=\|\sum_tD_t(k)\|_2^2/(n-d)\le(q+d)M/(n-d)\), and
\[
\mathcal E_{\mathrm{tc}}(k)=M-(n-d)\|\bar D(k)\|_2^2\ge\Bigl(1-\frac{q+d}{n-d}\Bigr)M .
\]
Since \(k\) and \(\tau\) both lie in \([c_{\mathrm{tr}}n,(1-c_{\mathrm{tr}})n]\), we have
\(q\le(1-2c_{\mathrm{tr}})n\), and therefore \(1-(q+d)/(n-d)\ge c_{\mathrm{tr}}\) for all
large \(n\).
\end{proof}

For the following VAR(2) examples, we write \(G:=A_1+A_2\).  Since
\(C_1=G\Delta_\eta\) and \(\Delta_\mu=(\mathbf{I}_p-G)\Delta_\eta\),
\[
D_1:=\|\Delta_\mu\|_2^2+2\langle\Delta_\mu,C_1\rangle
=\|\Delta_\eta\|_2^2-\|G\Delta_\eta\|_2^2.
\]
We see that \(D_1<0\) suffices to violate
Assumption~\ref{assump:transition-alignment} at \(q=1\).

We next consider three stable VAR(2) designs that violate
Assumption~\ref{assump:transition-alignment}, illustrating that VAR
stability does not imply transition alignment.
Tables~\ref{tab:tcr_var2_alignment_failure_rmse}--\ref{tab:tcr_var2_positive_diag_blocks_alignment_failure_rmse}
correspond to Examples~1--3, respectively, and report localization
RMSEs for the preliminary estimator (Pre), our filtered two-mean
refinement (Post), TCR, and the benchmark methods. These comparisons
assess whether the population identification provided by TCR
translates into improved finite-sample localization when the
alignment condition fails.

\paragraph{Example 1: a diagonal VAR(2) failure design.}
Let \(k=2\), \(\vartheta=0.8\), and
\[
A_1=\operatorname{diag}(-1.5\mathbf{I}_k,\,0.30\mathbf{I}_{p-k}),
\quad
A_2=\operatorname{diag}(-0.75\mathbf{I}_k,\,0.10\mathbf{I}_{p-k}),
\]
with $
\eta_1=\mathbf0_p,\ 
\eta_2=\Delta_\eta
=\frac{\vartheta}{\sqrt{k}}(\mathbf1_k^\top,\mathbf0_{p-k}^\top)^\top$.
The companion matrix is stable, with spectral radius \(\sqrt3/2\), but the active jump satisfies
\[
G\Delta_\eta=-\frac94\Delta_\eta,
\qquad
D_1=\left\{1-\left(\frac94\right)^2\right\}\vartheta^2
=-\frac{65}{16}\vartheta^2<0.
\]
Assumption~\ref{assump:transition-alignment} therefore fails even though the process is stable.
\begingroup
\scriptsize
\setlength{\tabcolsep}{2.5pt}
\renewcommand{\arraystretch}{0.95}

\begin{longtable}{rrrrrrrrrrrrrr}
\caption[Single change-point localization under a stable diagonal VAR(2) design]{Root mean squared errors of change-point localization under the stable diagonal VAR(2) design violating the transition-alignment condition, with $k=2$ active coordinates. Pre and Post denote preliminary and filtered localization, respectively, and TCR denotes transition-corrected refinement.}\label{tab:tcr_var2_alignment_failure_rmse}\\
\toprule
n & p & Pre & Post & TCR & Inspect & dc & sbs & \makecell[c]{scan\\EH} & agg2 & aggoo & \makecell[c]{post/\\pre} & \makecell[c]{TCR/\\post} & \makecell[c]{inspect/\\pre} \\
\midrule
\endfirsthead
\toprule
n & p & Pre & Post & TCR & Inspect & dc & sbs & \makecell[c]{scan\\EH} & agg2 & aggoo & \makecell[c]{post/\\pre} & \makecell[c]{TCR/\\post} & \makecell[c]{inspect/\\pre} \\
\midrule
\endhead
100 & 20 & 9.31 & \textbf{1.82} & 6.45 & 22.52 & 34.87 & 21.53 & 28.16 & 26.76 & 34.58 & 0.20 & 3.55 & 2.42 \\
250 & 50 & 14.17 & \textbf{1.54} & 7.89 & 54.56 & 93.55 & 57.53 & 67.70 & 60.77 & 93.96 & 0.11 & 5.12 & 3.85 \\
500 & 100 & 25.95 & 1.98 & \textbf{1.58} & 81.50 & 74.12 & 126.84 & 56.30 & 104.78 & 94.07 & 0.08 & 0.80 & 3.14 \\
\bottomrule
\end{longtable}

\endgroup

\paragraph{Example 2: a VAR(2) failure design with cross-sectional lag dependence.}
Let \(k=p/10\), let \(P_k^{\mathrm{path}}\) be the path-adjacency matrix, whose $(i,j)$ entry is $\mathbf1\{|i-j|=1\}$, and set
\[
A_1=\operatorname{blockdiag}
(-1.5\mathbf{I}_k+0.10P_k^{\mathrm{path}},\,0.30\mathbf{I}_{p-k}),
\qquad
A_2=\operatorname{blockdiag}
(-0.75\mathbf{I}_k,\,0.10\mathbf{I}_{p-k}),
\]
using the same equal-sign active jump
\(\Delta_\eta=\vartheta(\mathbf1_k^\top,\mathbf0_{p-k}^\top)^\top/\sqrt{k}\).
This process is stable: along the eigenvectors of \(P_k^{\mathrm{path}}\), the active block splits into scalar AR(2) recursions with first coefficient in \((-1.7,-1.3)\) and second coefficient \(-0.75\), so the companion spectral radius is \(\sqrt3/2\). It also contains off-diagonal lag dependence.  Writing
\(a=\vartheta/\sqrt{k}\), for \(k\ge2\) the active entries satisfy
\[
(G\Delta_\eta)_j=
\begin{cases}
-2.15a, & j=1,k,\\
-2.05a, & 2\le j\le k-1.
\end{cases}
\qquad
D_1\le\{1-(2.05)^2\}\vartheta^2<0.
\]
Again, Assumption~\ref{assump:transition-alignment} fails here as well.
\begingroup
\scriptsize
\setlength{\tabcolsep}{2.5pt}
\renewcommand{\arraystretch}{0.95}

\begin{longtable}{rrrrrrrrrrrrrr}
\caption[Single change-point localization under a stable VAR(2) design with cross-sectional lag dependence]{Root mean squared errors of change-point localization under the stable VAR(2) design with cross-sectional lag dependence violating the transition-alignment condition, with $k=p/10$ active coordinates. Pre and Post denote preliminary and filtered localization, respectively, and TCR denotes transition-corrected refinement.}\label{tab:tcr_var2_cross_sectional_alignment_failure_rmse}\\
\toprule
n & p & Pre & Post & TCR & Inspect & dc & sbs & \makecell[c]{scan\\EH} & agg2 & aggoo & \makecell[c]{post/\\pre} & \makecell[c]{TCR/\\post} & \makecell[c]{inspect/\\pre} \\
\midrule
\endfirsthead
\toprule
n & p & Pre & Post & TCR & Inspect & dc & sbs & \makecell[c]{scan\\EH} & agg2 & aggoo & \makecell[c]{post/\\pre} & \makecell[c]{TCR/\\post} & \makecell[c]{inspect/\\pre} \\
\midrule
\endhead
100 & 50 & 7.74 & \textbf{4.78} & 7.33 & 17.41 & 43.26 & 26.43 & 38.03 & 28.01 & 41.55 & 0.62 & 1.53 & 2.25 \\
100 & 100 & 8.66 & \textbf{6.49} & 14.35 & 17.88 & 47.21 & 30.31 & 46.80 & 33.72 & 46.25 & 0.75 & 2.21 & 2.07 \\
250 & 50 & 19.04 & \textbf{1.52} & 7.79 & 52.80 & 106.81 & 59.24 & 100.02 & 75.07 & 102.15 & 0.08 & 5.13 & 2.77 \\
250 & 100 & 17.39 & \textbf{4.41} & 17.44 & 39.68 & 118.16 & 75.90 & 113.91 & 80.70 & 115.93 & 0.25 & 3.96 & 2.28 \\
500 & 50 & 18.58 & 1.39 & \textbf{1.27} & 102.78 & 212.15 & 63.57 & 149.67 & 103.63 & 209.45 & 0.07 & 0.92 & 5.53 \\
500 & 100 & 25.86 & \textbf{2.27} & 2.29 & 95.52 & 246.41 & 120.14 & 232.75 & 147.98 & 242.11 & 0.09 & 1.01 & 3.69 \\
\bottomrule
\end{longtable}

\endgroup

From the first two examples we see that TCR generally becomes advantageous only
when \(n\) is sufficiently large.  Both, however, rely on dominant negative
diagonal coefficients; they are deliberately adverse rather than
especially natural VAR constructions.  We next impose positive diagonal
coefficients.

\paragraph{Example 3: a positive-diagonal block VAR(2) failure design.}
We partition the \(k=p/10\) active coordinates into blocks of size two or three. Let $m\in\{2,3\}$ denote the size of the current active block, indexed by $b$.
For each block, let \(U_m\) have ones on its first superdiagonal and zeros
elsewhere, and set
\[
A_{1,b}=0.25\mathbf{I}_m-0.95U_m,\qquad A_{2,b}=0.25\mathbf{I}_m.
\]
The inactive coordinates use \(A_{1,i}=0.30\mathbf{I}_{p-k}\) and
\(A_{2,i}=0.10\mathbf{I}_{p-k}\).  Write $G_b:=A_{1,b}+A_{2,b}$ and let $\Delta_{\eta,b}$ be the restriction of $\Delta_\eta$ to block $b$.
Within each active block,
\(\Delta_\eta\) alternates in sign and is scaled to
\(\|\Delta_\eta\|_2=\vartheta\).  All diagonal coefficients are positive and
the companion spectral radius is approximately \(0.64\).  Let \(a_b>0\) be
the common absolute jump size within block \(b\).  Alternating signs give
\[
(G_b\Delta_{\eta,b})_j=
\begin{cases}
1.45(\Delta_{\eta,b})_j, & j<m,\\
0.50(\Delta_{\eta,b})_m, & j=m.
\end{cases}
\]
Consequently, each block contributes
\[
\begin{aligned}
\|\Delta_{\eta,b}\|_2^2-\|G_b\Delta_{\eta,b}\|_2^2
&=\{m-(m-1)(1.45)^2-(0.50)^2\}a_b^2\\
&=
\begin{cases}
-0.3525a_b^2, & m=2,\\
-1.455a_b^2, & m=3.
\end{cases}
\end{aligned}
\]
Summing over blocks, we get \(D_1<0\), violating
Assumption~\ref{assump:transition-alignment}.  Even so, we find no systematic advantage for TCR
in the reported results.
\begingroup
\scriptsize
\setlength{\tabcolsep}{2.5pt}
\renewcommand{\arraystretch}{0.95}

\begin{longtable}{rrrrrrrrrrrrrr}
\caption[Single change-point localization under a stable positive-diagonal block VAR(2) design]{Root mean squared errors of change-point localization under the stable block VAR(2) design with positive diagonal coefficients violating the transition-alignment condition, with $k=p/10$ active coordinates. Pre and Post denote preliminary and filtered localization, respectively, and TCR denotes transition-corrected refinement.}\label{tab:tcr_var2_positive_diag_blocks_alignment_failure_rmse}\\
\toprule
n & p & Pre & Post & TCR & Inspect & dc & sbs & \makecell[c]{scan\\EH} & agg2 & aggoo & \makecell[c]{post/\\pre} & \makecell[c]{TCR/\\post} & \makecell[c]{inspect/\\pre} \\
\midrule
\endfirsthead
\toprule
n & p & Pre & Post & TCR & Inspect & dc & sbs & \makecell[c]{scan\\EH} & agg2 & aggoo & \makecell[c]{post/\\pre} & \makecell[c]{TCR/\\post} & \makecell[c]{inspect/\\pre} \\
\midrule
\endhead
100 & 50 & 12.90 & \textbf{9.47} & 19.49 & 18.77 & 26.71 & 23.35 & 26.19 & 20.97 & 26.68 & 0.73 & 2.06 & 1.45 \\
100 & 100 & 11.51 & \textbf{9.07} & 19.08 & 17.95 & 26.99 & 24.00 & 24.73 & 22.95 & 27.02 & 0.79 & 2.10 & 1.56 \\
250 & 50 & 38.63 & \textbf{25.66} & 53.74 & 56.66 & 69.54 & 66.81 & 69.89 & 65.45 & 69.37 & 0.66 & 2.09 & 1.47 \\
250 & 100 & 28.05 & \textbf{23.94} & 43.93 & 43.95 & 72.38 & 69.77 & 69.52 & 65.07 & 71.26 & 0.85 & 1.83 & 1.57 \\
500 & 50 & 81.99 & \textbf{41.09} & 111.93 & 124.13 & 164.04 & 145.68 & 157.02 & 150.80 & 163.48 & 0.50 & 2.72 & 1.51 \\
500 & 100 & 44.41 & \textbf{37.14} & 75.20 & 86.49 & 130.45 & 136.86 & 138.16 & 136.00 & 131.26 & 0.84 & 2.02 & 1.95 \\
\bottomrule
\end{longtable}

\endgroup

Taken together, these examples suggest that the signed alignment condition is
difficult to violate in natural stable constructions.  Even in designs built
to violate it, our main refinement performs at least as well as TCR in most
finite-sample cases unless \(n\) is sufficiently large.  For a concise presentation, we
therefore keep TCR out of the proposed
procedure and the reported empirical analysis.

\section{Supplementary simulation results and implementation details}
\label{app:numerical-results}

\subsection{Additional single-change-point box plots}
\label{app:single-cp-boxplots}
In this section, we present box plots summarizing single-change-point performance under the VAR(2) designs. The data-generating process (DGP), signal-to-noise construction, competing benchmark methods, and RMSE ratio definitions mirror those in Section~\ref{sec:single-CP-sims}, with the true VAR lag order increased from $d = 1$ to $d = 2$. As in the main-text figures, Panels~(a)--(c) stratify the design-point ratios by $c_A$, $\rho(\mathcal{A})$, and sample size $n$, respectively.
\begin{figure}[!htbp]
  \centering
  \IfFileExists{numerical_results/panel_single_cp_rmse_ratios_pn0p50_lag2_cvartheta0p50.pdf}{    \includegraphics[width=\linewidth]{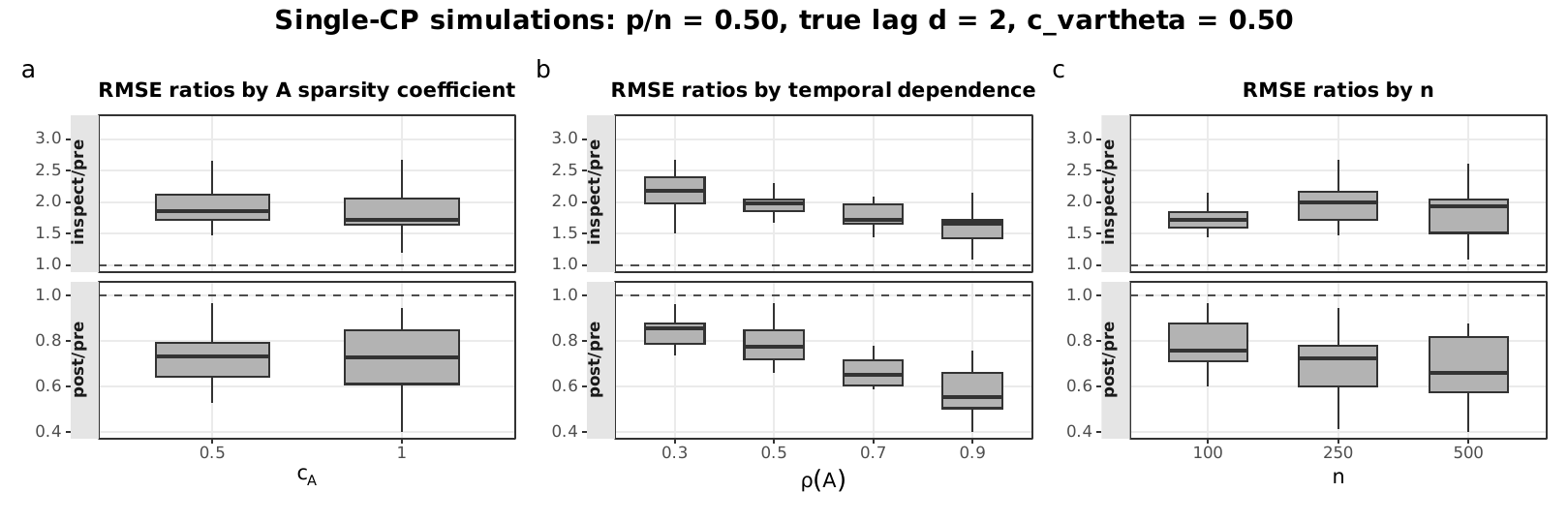}  }{    \fbox{\parbox[c][2.1in][c]{0.94\linewidth}{\centering
      Placeholder for \texttt{\detokenize{numerical_results/panel_single_cp_rmse_ratios_pn0p50_lag2_cvartheta0p50.pdf}}}}  }
  \caption{Single-change-point RMSE ratios for
  $(p/n,d,c_\vartheta)=(0.5,2,0.5)$.  The three columns stratify the
  design-point ratios by $c_A$, \(\rho(\mathcal A)\), and $n$, respectively;
  the stacked facets show Inspect/pre and post/pre.}
  \label{fig:single-box-pn05-lag2}
\end{figure}

\begin{figure}[!htbp]
  \centering
  \IfFileExists{numerical_results/panel_single_cp_rmse_ratios_pn0p80_lag2_cvartheta0p50.pdf}{    \includegraphics[width=\linewidth]{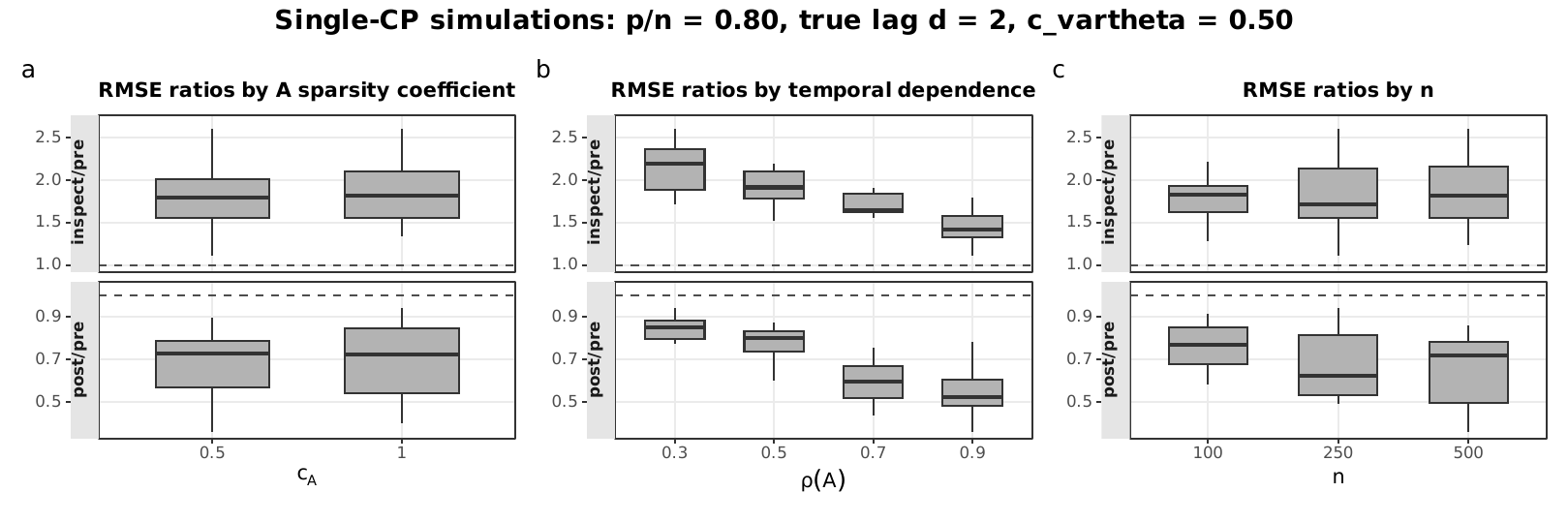}  }{    \fbox{\parbox[c][2.1in][c]{0.94\linewidth}{\centering
      Placeholder for \texttt{\detokenize{numerical_results/panel_single_cp_rmse_ratios_pn0p80_lag2_cvartheta0p50.pdf}}}}  }
  \caption{Single-change-point RMSE ratios for
  $(p/n,d,c_\vartheta)=(0.8,2,0.5)$.  The three columns stratify the
  design-point ratios by $c_A$, \(\rho(\mathcal A)\), and $n$, respectively;
  the stacked facets show Inspect/pre and post/pre.}
  \label{fig:single-box-pn08-lag2}
\end{figure}

\FloatBarrier
\subsection{Complete single-change-point results}\label{app:single-cp-tables}
The tables below present the full single-change-point RMSE comparisons across all competing estimators, including the no-ridge variants Pre-NR and Post-NR introduced in Section~\ref{sec:single-CP-sims}.

\IfFileExists{numerical_results/single_cp_table.tex}{  \begin{table}[htbp]
\centering
\caption{Single change-point localization: RMSE over 50 replications ($p/n=0.50$, $d=1$, and $c_{\vartheta}=0.50$).}
\label{tab:single_cp_pn_0p50_lag_1_cvartheta_0p50}
\tiny
\setlength{\tabcolsep}{2.4pt}
\resizebox{\textwidth}{!}{%
\begin{tabular}{r r r c r r r r r r r r r r r r}
\toprule
\multirow{2}{*}{$n$} & \multirow{2}{*}{$k$} & \multirow{2}{*}{$c_A$} & \multirow{2}{*}{$\rho(A)$} & \multicolumn{10}{c}{RMSE of estimated change-point location} & \multicolumn{2}{c}{RMSE ratio} \\
\cmidrule(lr){5-14}\cmidrule(lr){15-16}
& & & & Pre & Post & Pre-NR & Post-NR & inspect & dc & sbs & scanEH & agg2 & aggoo & post/pre & inspect/pre \\
\midrule
100 & 25 & 0.50 & 0.30 & 11.05 & \textbf{8.35} & 33.84 & 30.73 & 24.11 & 35.20 & 34.93 & 33.89 & 33.84 & 34.49 & 0.76 & 2.18 \\
250 & 62 & 0.50 & 0.30 & 25.24 & \textbf{20.88} & 81.52 & 78.99 & 47.55 & 80.82 & 79.79 & 83.68 & 81.52 & 79.15 & 0.83 & 1.88 \\
500 & 125 & 0.50 & 0.30 & 30.62 & \textbf{24.97} & 162.89 & 158.09 & 69.86 & 185.11 & 151.38 & 147.26 & 162.89 & 179.58 & 0.82 & 2.28 \\
100 & 50 & 0.50 & 0.30 & 10.54 & \textbf{8.13} & 34.06 & 29.93 & 21.88 & 34.33 & 33.11 & 34.70 & 34.06 & 34.10 & 0.77 & 2.07 \\
250 & 125 & 0.50 & 0.30 & 20.07 & \textbf{17.90} & 86.15 & 87.08 & 48.86 & 91.50 & 82.33 & 84.44 & 86.15 & 89.90 & 0.89 & 2.43 \\
500 & 250 & 0.50 & 0.30 & 27.95 & \textbf{23.61} & 131.51 & 130.39 & 70.43 & 181.15 & 148.09 & 142.70 & 131.51 & 180.45 & 0.84 & 2.52 \\
\cmidrule(lr){1-16}
100 & 25 & 0.50 & 0.50 & 11.51 & \textbf{8.45} & 26.77 & 21.11 & 19.35 & 30.04 & 28.38 & 26.19 & 26.77 & 29.98 & 0.73 & 1.68 \\
250 & 62 & 0.50 & 0.50 & 23.12 & \textbf{14.60} & 78.62 & 77.07 & 47.15 & 86.77 & 74.22 & 80.28 & 78.62 & 86.10 & 0.63 & 2.04 \\
500 & 125 & 0.50 & 0.50 & 34.70 & \textbf{29.49} & 154.29 & 145.84 & 90.66 & 175.16 & 148.61 & 159.12 & 154.29 & 178.69 & 0.85 & 2.61 \\
100 & 50 & 0.50 & 0.50 & 16.76 & \textbf{10.56} & 31.37 & 28.53 & 23.45 & 34.92 & 31.99 & 33.51 & 31.37 & 34.01 & 0.63 & 1.40 \\
250 & 125 & 0.50 & 0.50 & 20.79 & \textbf{14.23} & 84.83 & 71.79 & 39.70 & 87.78 & 84.90 & 91.85 & 84.83 & 89.99 & 0.68 & 1.91 \\
500 & 250 & 0.50 & 0.50 & 35.71 & \textbf{28.12} & 156.83 & 140.86 & 85.56 & 180.21 & 144.69 & 145.47 & 156.83 & 179.78 & 0.79 & 2.40 \\
\cmidrule(lr){1-16}
100 & 25 & 0.50 & 0.70 & 12.78 & \textbf{9.58} & 32.15 & 31.05 & 22.16 & 32.31 & 33.45 & 33.55 & 32.15 & 32.36 & 0.75 & 1.73 \\
250 & 62 & 0.50 & 0.70 & 33.57 & \textbf{16.81} & 82.50 & 73.73 & 70.02 & 81.43 & 88.71 & 82.04 & 82.50 & 81.43 & 0.50 & 2.09 \\
500 & 125 & 0.50 & 0.70 & 34.54 & \textbf{26.65} & 188.97 & 152.57 & 71.33 & 199.51 & 183.95 & 194.60 & 188.97 & 198.48 & 0.77 & 2.06 \\
100 & 50 & 0.50 & 0.70 & 11.23 & \textbf{7.64} & 35.80 & 28.45 & 20.78 & 38.03 & 33.06 & 36.71 & 35.80 & 37.40 & 0.68 & 1.85 \\
250 & 125 & 0.50 & 0.70 & 28.63 & \textbf{14.01} & 90.65 & 75.86 & 49.97 & 88.92 & 87.05 & 84.74 & 90.65 & 88.97 & 0.49 & 1.75 \\
500 & 250 & 0.50 & 0.70 & 48.72 & \textbf{25.60} & 172.41 & 145.92 & 88.52 & 191.54 & 161.35 & 187.92 & 172.41 & 190.06 & 0.53 & 1.82 \\
\cmidrule(lr){1-16}
100 & 25 & 0.50 & 0.90 & 16.66 & \textbf{8.97} & 40.14 & 29.08 & 25.64 & 42.27 & 36.70 & 40.26 & 40.14 & 42.12 & 0.54 & 1.54 \\
250 & 62 & 0.50 & 0.90 & 54.27 & \textbf{17.72} & 71.07 & 63.60 & 70.69 & 70.65 & 72.74 & 72.47 & 71.07 & 70.46 & 0.33 & 1.30 \\
500 & 125 & 0.50 & 0.90 & 55.72 & \textbf{21.52} & 208.49 & 118.57 & 80.10 & 208.85 & 190.40 & 211.12 & 208.49 & 209.08 & 0.39 & 1.44 \\
100 & 50 & 0.50 & 0.90 & 16.20 & \textbf{9.69} & 30.45 & 21.06 & 22.66 & 28.58 & 35.82 & 30.29 & 30.45 & 28.93 & 0.60 & 1.40 \\
250 & 125 & 0.50 & 0.90 & 33.21 & \textbf{16.97} & 93.48 & 79.12 & 54.79 & 102.12 & 92.61 & 98.13 & 93.48 & 102.24 & 0.51 & 1.65 \\
500 & 250 & 0.50 & 0.90 & 83.37 & \textbf{28.28} & 147.44 & 135.54 & 148.16 & 151.22 & 158.02 & 152.09 & 147.44 & 151.21 & 0.34 & 1.78 \\
\cmidrule(lr){1-16}
100 & 25 & 1.00 & 0.30 & 13.82 & \textbf{11.83} & 33.27 & 29.31 & 24.77 & 37.13 & 32.76 & 34.73 & 33.27 & 37.57 & 0.86 & 1.79 \\
250 & 62 & 1.00 & 0.30 & 25.12 & \textbf{20.97} & 80.62 & 72.23 & 43.95 & 83.76 & 75.93 & 81.83 & 80.62 & 83.94 & 0.83 & 1.75 \\
500 & 125 & 1.00 & 0.30 & 39.93 & \textbf{28.75} & 140.48 & 132.43 & 72.66 & 182.21 & 148.06 & 150.23 & 140.48 & 179.45 & 0.72 & 1.82 \\
100 & 50 & 1.00 & 0.30 & 11.15 & \textbf{9.99} & 33.78 & 30.24 & 20.55 & 31.09 & 34.40 & 33.95 & 33.78 & 30.29 & 0.90 & 1.84 \\
250 & 125 & 1.00 & 0.30 & 20.83 & \textbf{15.99} & 87.02 & 81.50 & 46.99 & 74.71 & 74.43 & 75.87 & 87.02 & 74.85 & 0.77 & 2.26 \\
500 & 250 & 1.00 & 0.30 & 30.96 & \textbf{26.52} & 156.26 & 131.58 & 69.51 & 165.07 & 150.68 & 147.43 & 156.26 & 165.24 & 0.86 & 2.25 \\
\cmidrule(lr){1-16}
100 & 25 & 1.00 & 0.50 & 11.41 & \textbf{10.71} & 33.09 & 31.23 & 26.76 & 34.28 & 34.43 & 32.41 & 33.09 & 33.11 & 0.94 & 2.35 \\
250 & 62 & 1.00 & 0.50 & 24.78 & \textbf{17.21} & 83.20 & 74.95 & 49.31 & 86.49 & 81.87 & 87.29 & 83.20 & 84.21 & 0.69 & 1.99 \\
500 & 125 & 1.00 & 0.50 & 43.57 & \textbf{28.75} & 163.57 & 128.63 & 74.97 & 167.59 & 149.14 & 164.40 & 163.57 & 167.64 & 0.66 & 1.72 \\
100 & 50 & 1.00 & 0.50 & 9.87 & \textbf{8.73} & 35.07 & 32.34 & 23.59 & 33.94 & 32.12 & 33.61 & 35.07 & 32.62 & 0.88 & 2.39 \\
250 & 125 & 1.00 & 0.50 & 26.09 & \textbf{18.78} & 92.34 & 79.77 & 48.19 & 93.27 & 80.95 & 88.30 & 92.34 & 92.37 & 0.72 & 1.85 \\
500 & 250 & 1.00 & 0.50 & 44.64 & \textbf{28.23} & 176.04 & 166.40 & 99.71 & 192.08 & 170.70 & 183.48 & 176.04 & 194.00 & 0.63 & 2.23 \\
\cmidrule(lr){1-16}
100 & 25 & 1.00 & 0.70 & 15.81 & \textbf{9.88} & 30.36 & 28.00 & 27.15 & 30.80 & 28.64 & 30.96 & 30.36 & 31.23 & 0.62 & 1.72 \\
250 & 62 & 1.00 & 0.70 & 29.39 & \textbf{15.85} & 87.88 & 78.14 & 72.52 & 89.51 & 86.74 & 79.64 & 87.88 & 88.98 & 0.54 & 2.47 \\
500 & 125 & 1.00 & 0.70 & 57.73 & \textbf{28.99} & 193.86 & 137.35 & 96.01 & 201.49 & 178.05 & 195.53 & 193.86 & 201.29 & 0.50 & 1.66 \\
100 & 50 & 1.00 & 0.70 & 15.13 & \textbf{10.75} & 30.92 & 30.35 & 20.55 & 31.61 & 29.59 & 30.74 & 30.92 & 31.42 & 0.71 & 1.36 \\
250 & 125 & 1.00 & 0.70 & 31.57 & \textbf{16.81} & 88.36 & 85.12 & 52.66 & 91.74 & 88.97 & 89.10 & 88.36 & 92.78 & 0.53 & 1.67 \\
500 & 250 & 1.00 & 0.70 & 57.04 & \textbf{25.46} & 155.86 & 144.34 & 79.44 & 184.99 & 168.16 & 182.68 & 155.86 & 187.28 & 0.45 & 1.39 \\
\cmidrule(lr){1-16}
100 & 25 & 1.00 & 0.90 & 21.11 & \textbf{8.53} & 41.20 & 32.02 & 30.85 & 42.74 & 41.39 & 42.01 & 41.20 & 42.40 & 0.40 & 1.46 \\
250 & 62 & 1.00 & 0.90 & 47.97 & \textbf{17.53} & 71.94 & 71.24 & 69.99 & 69.75 & 77.85 & 69.62 & 71.94 & 69.60 & 0.37 & 1.46 \\
500 & 125 & 1.00 & 0.90 & 104.97 & \textbf{16.94} & 172.27 & 115.53 & 118.63 & 178.99 & 176.17 & 169.98 & 172.27 & 183.78 & 0.16 & 1.13 \\
100 & 50 & 1.00 & 0.90 & 19.51 & \textbf{9.10} & 41.82 & 31.72 & 24.79 & 42.42 & 39.23 & 42.10 & 41.82 & 42.95 & 0.47 & 1.27 \\
250 & 125 & 1.00 & 0.90 & 33.69 & \textbf{16.07} & 87.93 & 66.14 & 57.21 & 88.36 & 80.97 & 85.54 & 87.93 & 90.29 & 0.48 & 1.70 \\
500 & 250 & 1.00 & 0.90 & 106.24 & \textbf{25.59} & 153.92 & 142.51 & 110.76 & 158.63 & 150.84 & 153.71 & 153.92 & 164.21 & 0.24 & 1.04 \\
\bottomrule
\end{tabular}%
}
\vspace{2pt}
\begin{minipage}{0.98\textwidth}
\footnotesize Pre and Post denote the preliminary Step-1 and refined Step-3 estimates. Pre-NR and Post-NR denote the corresponding estimates with zero ridge penalty in Steps 1 and 3; the lag and Step-2 Lasso are refitted from the no-ridge preliminary estimate. The ratios post/pre and inspect/pre are computed from the corresponding RMSE values. The signal magnitude is $\vartheta=c_{\vartheta}\sqrt{p/\Delta_n}$. The smallest method RMSE in each design row is boldfaced.
\end{minipage}
\end{table}

\begin{table}[htbp]
\centering
\caption{Single change-point localization: RMSE over 50 replications ($p/n=0.50$, $d=2$, and $c_{\vartheta}=0.50$).}
\label{tab:single_cp_pn_0p50_lag_2_cvartheta_0p50}
\tiny
\setlength{\tabcolsep}{2.4pt}
\resizebox{\textwidth}{!}{%
\begin{tabular}{r r r c r r r r r r r r r r r r}
\toprule
\multirow{2}{*}{$n$} & \multirow{2}{*}{$k$} & \multirow{2}{*}{$c_A$} & \multirow{2}{*}{$\rho(A)$} & \multicolumn{10}{c}{RMSE of estimated change-point location} & \multicolumn{2}{c}{RMSE ratio} \\
\cmidrule(lr){5-14}\cmidrule(lr){15-16}
& & & & Pre & Post & Pre-NR & Post-NR & inspect & dc & sbs & scanEH & agg2 & aggoo & post/pre & inspect/pre \\
\midrule
100 & 25 & 0.50 & 0.30 & 9.80 & \textbf{9.42} & 33.20 & 30.28 & 20.99 & 35.81 & 33.06 & 31.87 & 33.20 & 34.72 & 0.96 & 2.14 \\
250 & 62 & 0.50 & 0.30 & 16.62 & \textbf{13.08} & 77.13 & 72.47 & 44.08 & 82.09 & 80.74 & 81.79 & 77.13 & 80.45 & 0.79 & 2.65 \\
500 & 125 & 0.50 & 0.30 & 40.78 & \textbf{31.98} & 166.86 & 160.61 & 82.82 & 183.43 & 169.55 & 167.75 & 166.86 & 183.20 & 0.78 & 2.03 \\
100 & 50 & 0.50 & 0.30 & 11.47 & \textbf{9.83} & 34.59 & 31.18 & 25.79 & 35.63 & 35.71 & 36.37 & 34.59 & 37.37 & 0.86 & 2.25 \\
250 & 125 & 0.50 & 0.30 & 20.67 & \textbf{15.71} & 78.59 & 75.76 & 42.48 & 94.12 & 77.60 & 77.84 & 78.59 & 93.57 & 0.76 & 2.06 \\
500 & 250 & 0.50 & 0.30 & 30.16 & \textbf{26.48} & 141.76 & 136.23 & 55.72 & 168.99 & 139.99 & 135.73 & 141.76 & 165.41 & 0.88 & 1.85 \\
\cmidrule(lr){1-16}
100 & 25 & 0.50 & 0.50 & 9.77 & \textbf{8.54} & 31.06 & 29.15 & 20.68 & 33.93 & 32.26 & 31.30 & 31.06 & 34.51 & 0.87 & 2.12 \\
250 & 62 & 0.50 & 0.50 & 23.93 & \textbf{16.63} & 72.36 & 69.85 & 55.09 & 86.98 & 72.13 & 79.44 & 72.36 & 87.79 & 0.69 & 2.30 \\
500 & 125 & 0.50 & 0.50 & 39.84 & \textbf{32.31} & 181.38 & 175.08 & 78.82 & 190.89 & 174.23 & 168.16 & 181.38 & 192.59 & 0.81 & 1.98 \\
100 & 50 & 0.50 & 0.50 & 12.53 & \textbf{12.09} & 31.55 & 27.43 & 22.62 & 31.50 & 31.83 & 32.66 & 31.55 & 32.08 & 0.96 & 1.81 \\
250 & 125 & 0.50 & 0.50 & 24.40 & \textbf{18.17} & 84.71 & 84.47 & 47.07 & 93.14 & 81.29 & 87.61 & 84.71 & 91.88 & 0.75 & 1.93 \\
500 & 250 & 0.50 & 0.50 & 41.56 & \textbf{27.50} & 157.80 & 139.33 & 78.10 & 186.33 & 171.26 & 179.02 & 157.80 & 182.86 & 0.66 & 1.88 \\
\cmidrule(lr){1-16}
100 & 25 & 0.50 & 0.70 & 16.18 & \textbf{11.11} & 33.25 & 31.82 & 28.22 & 33.06 & 34.05 & 32.90 & 33.25 & 32.51 & 0.69 & 1.74 \\
250 & 62 & 0.50 & 0.70 & 28.20 & \textbf{21.95} & 83.55 & 76.12 & 48.86 & 93.62 & 83.11 & 86.04 & 83.55 & 92.18 & 0.78 & 1.73 \\
500 & 125 & 0.50 & 0.70 & 44.89 & \textbf{26.65} & 167.84 & 142.65 & 142.31 & 185.86 & 180.87 & 188.65 & 167.84 & 186.81 & 0.59 & 3.17 \\
100 & 50 & 0.50 & 0.70 & 15.14 & \textbf{11.46} & 32.77 & 31.06 & 23.45 & 35.63 & 33.08 & 34.63 & 32.77 & 34.12 & 0.76 & 1.55 \\
250 & 125 & 0.50 & 0.70 & 29.33 & \textbf{20.57} & 93.53 & 84.55 & 50.04 & 90.60 & 90.54 & 86.87 & 93.53 & 88.18 & 0.70 & 1.71 \\
500 & 250 & 0.50 & 0.70 & 49.65 & \textbf{29.08} & 164.97 & 135.45 & 75.13 & 200.23 & 161.82 & 187.86 & 164.97 & 202.76 & 0.59 & 1.51 \\
\cmidrule(lr){1-16}
100 & 25 & 0.50 & 0.90 & 14.47 & \textbf{10.43} & 33.87 & 30.72 & 25.73 & 35.91 & 32.17 & 35.38 & 33.87 & 35.41 & 0.72 & 1.78 \\
250 & 62 & 0.50 & 0.90 & 33.49 & \textbf{13.86} & 93.73 & 69.01 & 49.33 & 101.79 & 93.73 & 97.00 & 93.73 & 99.65 & 0.41 & 1.47 \\
500 & 125 & 0.50 & 0.90 & 111.12 & \textbf{58.80} & 173.58 & 153.17 & 120.65 & 188.83 & 171.13 & 184.46 & 173.58 & 187.61 & 0.53 & 1.09 \\
100 & 50 & 0.50 & 0.90 & 14.52 & \textbf{9.95} & 39.60 & 35.05 & 25.06 & 40.41 & 36.96 & 40.53 & 39.60 & 40.30 & 0.69 & 1.73 \\
250 & 125 & 0.50 & 0.90 & 33.56 & \textbf{17.78} & 76.26 & 69.83 & 72.06 & 79.55 & 75.22 & 78.58 & 76.26 & 79.65 & 0.53 & 2.15 \\
500 & 250 & 0.50 & 0.90 & 56.36 & \textbf{30.21} & 162.12 & 150.56 & 84.77 & 198.40 & 173.16 & 184.98 & 162.12 & 196.64 & 0.54 & 1.50 \\
\cmidrule(lr){1-16}
100 & 25 & 1.00 & 0.30 & 11.58 & \textbf{10.09} & 31.28 & 29.56 & 18.61 & 30.95 & 29.29 & 31.79 & 31.28 & 31.44 & 0.87 & 1.61 \\
250 & 62 & 1.00 & 0.30 & 20.39 & \textbf{17.21} & 80.12 & 77.85 & 45.45 & 86.08 & 85.83 & 79.19 & 80.12 & 83.92 & 0.84 & 2.23 \\
500 & 125 & 1.00 & 0.30 & 32.64 & \textbf{28.60} & 155.54 & 152.54 & 85.19 & 172.62 & 162.65 & 152.50 & 155.54 & 178.27 & 0.88 & 2.61 \\
100 & 50 & 1.00 & 0.30 & 14.28 & \textbf{10.51} & 33.35 & 31.79 & 21.47 & 33.03 & 33.90 & 31.83 & 33.35 & 33.10 & 0.74 & 1.50 \\
250 & 125 & 1.00 & 0.30 & 16.14 & \textbf{15.23} & 77.24 & 78.97 & 43.03 & 88.68 & 82.86 & 81.07 & 77.24 & 88.84 & 0.94 & 2.67 \\
500 & 250 & 1.00 & 0.30 & 31.94 & \textbf{27.35} & 160.19 & 155.17 & 74.17 & 174.52 & 172.05 & 163.80 & 160.19 & 174.73 & 0.86 & 2.32 \\
\cmidrule(lr){1-16}
100 & 25 & 1.00 & 0.50 & 12.52 & \textbf{11.08} & 32.68 & 29.60 & 24.80 & 31.87 & 30.91 & 30.87 & 32.68 & 30.95 & 0.88 & 1.98 \\
250 & 62 & 1.00 & 0.50 & 26.46 & \textbf{20.16} & 83.41 & 80.20 & 44.22 & 91.86 & 90.25 & 88.02 & 83.41 & 89.93 & 0.76 & 1.67 \\
500 & 125 & 1.00 & 0.50 & 34.54 & \textbf{28.96} & 158.17 & 140.88 & 70.28 & 180.22 & 166.63 & 171.79 & 158.17 & 177.71 & 0.84 & 2.03 \\
100 & 50 & 1.00 & 0.50 & 14.26 & \textbf{10.24} & 32.20 & 29.33 & 22.13 & 35.48 & 33.11 & 35.70 & 32.20 & 37.12 & 0.72 & 1.55 \\
250 & 125 & 1.00 & 0.50 & 23.18 & \textbf{18.33} & 81.53 & 78.72 & 48.17 & 93.91 & 82.27 & 86.76 & 81.53 & 91.71 & 0.79 & 2.08 \\
500 & 250 & 1.00 & 0.50 & 34.94 & \textbf{25.02} & 180.20 & 169.19 & 69.98 & 180.28 & 159.58 & 173.58 & 180.20 & 174.92 & 0.72 & 2.00 \\
\cmidrule(lr){1-16}
100 & 25 & 1.00 & 0.70 & 17.16 & \textbf{10.28} & 34.91 & 29.38 & 24.85 & 37.40 & 36.62 & 35.98 & 34.91 & 36.92 & 0.60 & 1.45 \\
250 & 62 & 1.00 & 0.70 & 33.45 & \textbf{21.39} & 87.15 & 72.84 & 64.41 & 95.30 & 90.42 & 86.86 & 87.15 & 94.30 & 0.64 & 1.93 \\
500 & 125 & 1.00 & 0.70 & 47.20 & \textbf{28.95} & 157.33 & 137.77 & 79.76 & 179.87 & 147.91 & 151.76 & 157.33 & 180.69 & 0.61 & 1.69 \\
100 & 50 & 1.00 & 0.70 & 13.79 & \textbf{12.99} & 32.42 & 27.34 & 23.48 & 33.56 & 29.49 & 31.16 & 32.42 & 33.10 & 0.94 & 1.70 \\
250 & 125 & 1.00 & 0.70 & 26.96 & \textbf{16.31} & 87.86 & 78.49 & 55.40 & 92.41 & 85.72 & 86.24 & 87.86 & 90.37 & 0.60 & 2.05 \\
500 & 250 & 1.00 & 0.70 & 46.49 & \textbf{30.80} & 182.46 & 170.22 & 96.74 & 186.44 & 181.91 & 182.66 & 182.46 & 186.06 & 0.66 & 2.08 \\
\cmidrule(lr){1-16}
100 & 25 & 1.00 & 0.90 & 16.85 & \textbf{12.80} & 35.79 & 28.83 & 27.75 & 41.08 & 35.80 & 39.00 & 35.79 & 40.73 & 0.76 & 1.65 \\
250 & 62 & 1.00 & 0.90 & 44.99 & \textbf{26.62} & 85.42 & 71.16 & 76.74 & 87.45 & 82.55 & 83.31 & 85.42 & 86.95 & 0.59 & 1.71 \\
500 & 125 & 1.00 & 0.90 & 107.46 & \textbf{42.93} & 166.65 & 144.43 & 139.01 & 177.36 & 166.73 & 174.26 & 166.65 & 177.64 & 0.40 & 1.29 \\
100 & 50 & 1.00 & 0.90 & 17.33 & \textbf{11.29} & 38.04 & 30.44 & 29.15 & 40.30 & 34.87 & 39.83 & 38.04 & 41.16 & 0.65 & 1.68 \\
250 & 125 & 1.00 & 0.90 & 45.57 & \textbf{26.12} & 88.50 & 76.02 & 78.25 & 89.07 & 89.16 & 89.62 & 88.50 & 86.46 & 0.57 & 1.72 \\
500 & 250 & 1.00 & 0.90 & 102.22 & \textbf{44.14} & 141.70 & 124.64 & 121.67 & 138.24 & 151.50 & 137.16 & 141.70 & 137.44 & 0.43 & 1.19 \\
\bottomrule
\end{tabular}%
}
\vspace{2pt}
\begin{minipage}{0.98\textwidth}
\footnotesize Pre and Post denote the preliminary Step-1 and refined Step-3 estimates. Pre-NR and Post-NR denote the corresponding estimates with zero ridge penalty in Steps 1 and 3; the lag and Step-2 Lasso are refitted from the no-ridge preliminary estimate. The ratios post/pre and inspect/pre are computed from the corresponding RMSE values. The signal magnitude is $\vartheta=c_{\vartheta}\sqrt{p/\Delta_n}$. The smallest method RMSE in each design row is boldfaced.
\end{minipage}
\end{table}

\begin{table}[htbp]
\centering
\caption{Single change-point localization: RMSE over 50 replications ($p/n=0.80$, $d=1$, and $c_{\vartheta}=0.50$).}
\label{tab:single_cp_pn_0p80_lag_1_cvartheta_0p50}
\tiny
\setlength{\tabcolsep}{2.4pt}
\resizebox{\textwidth}{!}{%
\begin{tabular}{r r r c r r r r r r r r r r r r}
\toprule
\multirow{2}{*}{$n$} & \multirow{2}{*}{$k$} & \multirow{2}{*}{$c_A$} & \multirow{2}{*}{$\rho(A)$} & \multicolumn{10}{c}{RMSE of estimated change-point location} & \multicolumn{2}{c}{RMSE ratio} \\
\cmidrule(lr){5-14}\cmidrule(lr){15-16}
& & & & Pre & Post & Pre-NR & Post-NR & inspect & dc & sbs & scanEH & agg2 & aggoo & post/pre & inspect/pre \\
\midrule
100 & 40 & 0.50 & 0.30 & 7.66 & \textbf{7.21} & 28.87 & 25.44 & 16.87 & 35.33 & 32.22 & 32.27 & 28.87 & 35.72 & 0.94 & 2.20 \\
250 & 100 & 0.50 & 0.30 & 20.59 & \textbf{15.95} & 78.07 & 78.91 & 42.51 & 81.98 & 76.65 & 74.59 & 78.07 & 82.16 & 0.77 & 2.07 \\
500 & 200 & 0.50 & 0.30 & 27.48 & \textbf{18.22} & 124.03 & 113.07 & 53.64 & 153.07 & 145.96 & 132.28 & 124.03 & 155.17 & 0.66 & 1.95 \\
100 & 80 & 0.50 & 0.30 & 8.99 & \textbf{7.84} & 33.17 & 30.07 & 17.21 & 30.91 & 29.40 & 29.56 & 33.17 & 31.11 & 0.87 & 1.91 \\
250 & 200 & 0.50 & 0.30 & 25.87 & \textbf{15.50} & 79.97 & 74.81 & 42.28 & 90.86 & 74.86 & 82.87 & 79.97 & 91.30 & 0.60 & 1.63 \\
500 & 400 & 0.50 & 0.30 & 23.11 & \textbf{17.91} & 140.13 & 116.86 & 58.68 & 142.67 & 129.11 & 137.40 & 140.13 & 142.55 & 0.77 & 2.54 \\
\cmidrule(lr){1-16}
100 & 40 & 0.50 & 0.50 & 9.78 & \textbf{9.13} & 34.93 & 30.14 & 20.25 & 34.93 & 30.96 & 34.24 & 34.93 & 34.99 & 0.93 & 2.07 \\
250 & 100 & 0.50 & 0.50 & 22.08 & \textbf{16.94} & 75.96 & 72.82 & 50.91 & 85.03 & 71.57 & 77.19 & 75.96 & 84.82 & 0.77 & 2.31 \\
500 & 200 & 0.50 & 0.50 & 28.54 & \textbf{21.90} & 145.75 & 136.76 & 85.84 & 164.28 & 147.66 & 165.83 & 145.75 & 166.75 & 0.77 & 3.01 \\
100 & 80 & 0.50 & 0.50 & 11.85 & \textbf{8.16} & 32.01 & 30.72 & 23.40 & 33.33 & 32.35 & 33.97 & 32.01 & 33.38 & 0.69 & 1.97 \\
250 & 200 & 0.50 & 0.50 & 20.92 & \textbf{14.18} & 84.14 & 74.37 & 52.89 & 91.96 & 86.23 & 89.30 & 84.14 & 94.59 & 0.68 & 2.53 \\
500 & 400 & 0.50 & 0.50 & 24.22 & \textbf{22.08} & 150.49 & 133.97 & 87.09 & 190.84 & 162.50 & 178.61 & 150.49 & 187.60 & 0.91 & 3.60 \\
\cmidrule(lr){1-16}
100 & 40 & 0.50 & 0.70 & 13.59 & \textbf{9.13} & 31.91 & 28.70 & 20.14 & 34.12 & 31.57 & 33.79 & 31.91 & 33.71 & 0.67 & 1.48 \\
250 & 100 & 0.50 & 0.70 & 24.55 & \textbf{13.97} & 71.37 & 69.33 & 61.17 & 77.69 & 73.05 & 71.06 & 71.37 & 76.11 & 0.57 & 2.49 \\
500 & 200 & 0.50 & 0.70 & 41.63 & \textbf{23.81} & 169.17 & 132.60 & 83.51 & 193.54 & 171.61 & 185.42 & 169.17 & 194.23 & 0.57 & 2.01 \\
100 & 80 & 0.50 & 0.70 & 11.78 & \textbf{8.46} & 34.56 & 30.81 & 23.99 & 33.15 & 35.72 & 34.72 & 34.56 & 33.51 & 0.72 & 2.04 \\
250 & 200 & 0.50 & 0.70 & 24.21 & \textbf{12.93} & 87.67 & 77.30 & 43.88 & 94.65 & 88.85 & 90.70 & 87.67 & 93.68 & 0.53 & 1.81 \\
500 & 400 & 0.50 & 0.70 & 32.52 & \textbf{19.48} & 167.45 & 134.50 & 72.18 & 175.31 & 165.42 & 182.78 & 167.45 & 179.17 & 0.60 & 2.22 \\
\cmidrule(lr){1-16}
100 & 40 & 0.50 & 0.90 & 13.00 & \textbf{10.34} & 32.56 & 27.67 & 20.65 & 35.91 & 33.65 & 35.93 & 32.56 & 36.11 & 0.80 & 1.59 \\
250 & 100 & 0.50 & 0.90 & 43.28 & \textbf{14.07} & 74.28 & 56.66 & 54.67 & 80.30 & 71.99 & 78.10 & 74.28 & 80.45 & 0.33 & 1.26 \\
500 & 200 & 0.50 & 0.90 & 89.31 & \textbf{18.66} & 147.65 & 94.44 & 150.78 & 157.73 & 154.76 & 155.23 & 147.65 & 156.27 & 0.21 & 1.69 \\
100 & 80 & 0.50 & 0.90 & 16.77 & \textbf{8.94} & 27.32 & 30.49 & 25.99 & 26.88 & 26.42 & 27.17 & 27.32 & 26.88 & 0.53 & 1.55 \\
250 & 200 & 0.50 & 0.90 & 51.06 & \textbf{16.19} & 72.72 & 64.72 & 52.33 & 78.51 & 71.60 & 74.56 & 72.72 & 78.89 & 0.32 & 1.02 \\
500 & 400 & 0.50 & 0.90 & 47.98 & \textbf{17.16} & 173.98 & 110.18 & 54.42 & 189.42 & 178.20 & 181.12 & 173.98 & 185.87 & 0.36 & 1.13 \\
\cmidrule(lr){1-16}
100 & 40 & 1.00 & 0.30 & 10.35 & \textbf{8.36} & 34.10 & 31.06 & 21.84 & 33.95 & 32.65 & 33.97 & 34.10 & 33.92 & 0.81 & 2.11 \\
250 & 100 & 1.00 & 0.30 & 16.94 & \textbf{15.22} & 78.83 & 76.17 & 43.11 & 86.41 & 80.04 & 87.25 & 78.83 & 87.25 & 0.90 & 2.55 \\
500 & 200 & 1.00 & 0.30 & 26.10 & \textbf{25.71} & 126.57 & 111.31 & 60.12 & 168.34 & 129.75 & 132.77 & 126.57 & 173.21 & 0.99 & 2.30 \\
100 & 80 & 1.00 & 0.30 & 8.71 & \textbf{6.52} & 35.44 & 30.59 & 18.78 & 32.83 & 29.61 & 32.19 & 35.44 & 31.39 & 0.75 & 2.16 \\
250 & 200 & 1.00 & 0.30 & 22.84 & \textbf{18.30} & 81.78 & 76.72 & 43.66 & 85.78 & 84.06 & 86.22 & 81.78 & 87.53 & 0.80 & 1.91 \\
500 & 400 & 1.00 & 0.30 & 31.73 & \textbf{22.63} & 153.50 & 137.41 & 66.46 & 179.79 & 149.01 & 169.68 & 153.50 & 184.56 & 0.71 & 2.09 \\
\cmidrule(lr){1-16}
100 & 40 & 1.00 & 0.50 & 12.49 & \textbf{7.45} & 31.61 & 29.18 & 19.53 & 34.75 & 31.83 & 33.31 & 31.61 & 35.43 & 0.60 & 1.56 \\
250 & 100 & 1.00 & 0.50 & 23.18 & \textbf{14.55} & 70.24 & 58.91 & 34.20 & 89.85 & 75.61 & 81.28 & 70.24 & 90.54 & 0.63 & 1.48 \\
500 & 200 & 1.00 & 0.50 & 28.63 & \textbf{23.38} & 147.20 & 124.00 & 79.08 & 154.66 & 148.98 & 152.10 & 147.20 & 155.11 & 0.82 & 2.76 \\
100 & 80 & 1.00 & 0.50 & 11.69 & \textbf{9.78} & 31.50 & 28.39 & 26.57 & 33.98 & 31.96 & 33.82 & 31.50 & 34.43 & 0.84 & 2.27 \\
250 & 200 & 1.00 & 0.50 & 23.02 & \textbf{14.73} & 81.35 & 72.97 & 43.95 & 94.75 & 77.29 & 79.69 & 81.35 & 93.37 & 0.64 & 1.91 \\
500 & 400 & 1.00 & 0.50 & 40.57 & \textbf{21.40} & 153.43 & 143.79 & 59.37 & 189.97 & 162.59 & 186.05 & 153.43 & 190.10 & 0.53 & 1.46 \\
\cmidrule(lr){1-16}
100 & 40 & 1.00 & 0.70 & 15.26 & \textbf{7.18} & 31.12 & 27.42 & 22.75 & 31.80 & 30.98 & 31.63 & 31.12 & 31.62 & 0.47 & 1.49 \\
250 & 100 & 1.00 & 0.70 & 28.91 & \textbf{16.50} & 83.35 & 59.19 & 59.85 & 94.21 & 76.12 & 89.88 & 83.35 & 93.21 & 0.57 & 2.07 \\
500 & 200 & 1.00 & 0.70 & 42.22 & \textbf{23.17} & 175.32 & 126.34 & 66.51 & 201.01 & 173.14 & 188.11 & 175.32 & 201.53 & 0.55 & 1.58 \\
100 & 80 & 1.00 & 0.70 & 15.08 & \textbf{9.18} & 35.46 & 24.97 & 24.12 & 36.54 & 35.27 & 37.23 & 35.46 & 37.63 & 0.61 & 1.60 \\
250 & 200 & 1.00 & 0.70 & 26.56 & \textbf{16.20} & 74.61 & 63.77 & 43.07 & 94.22 & 80.07 & 83.19 & 74.61 & 93.67 & 0.61 & 1.62 \\
500 & 400 & 1.00 & 0.70 & 61.31 & \textbf{20.24} & 169.98 & 108.82 & 84.66 & 189.36 & 165.08 & 176.60 & 169.98 & 194.50 & 0.33 & 1.38 \\
\cmidrule(lr){1-16}
100 & 40 & 1.00 & 0.90 & 20.20 & \textbf{9.60} & 31.17 & 28.85 & 26.47 & 30.28 & 31.42 & 29.79 & 31.17 & 30.19 & 0.48 & 1.31 \\
250 & 100 & 1.00 & 0.90 & 29.96 & \textbf{13.30} & 84.93 & 67.88 & 49.52 & 88.71 & 82.26 & 88.94 & 84.93 & 88.29 & 0.44 & 1.65 \\
500 & 200 & 1.00 & 0.90 & 83.62 & \textbf{18.57} & 161.40 & 98.56 & 145.96 & 202.22 & 150.51 & 192.04 & 161.40 & 201.86 & 0.22 & 1.75 \\
100 & 80 & 1.00 & 0.90 & 17.93 & \textbf{10.00} & 26.38 & 24.09 & 24.43 & 25.15 & 26.01 & 24.70 & 26.38 & 25.48 & 0.56 & 1.36 \\
250 & 200 & 1.00 & 0.90 & 49.70 & \textbf{13.80} & 83.94 & 62.24 & 57.42 & 96.46 & 78.21 & 87.00 & 83.94 & 97.58 & 0.28 & 1.16 \\
500 & 400 & 1.00 & 0.90 & 88.75 & \textbf{21.92} & 155.12 & 110.17 & 146.53 & 170.46 & 156.14 & 167.51 & 155.12 & 171.10 & 0.25 & 1.65 \\
\bottomrule
\end{tabular}%
}
\vspace{2pt}
\begin{minipage}{0.98\textwidth}
\footnotesize Pre and Post denote the preliminary Step-1 and refined Step-3 estimates. Pre-NR and Post-NR denote the corresponding estimates with zero ridge penalty in Steps 1 and 3; the lag and Step-2 Lasso are refitted from the no-ridge preliminary estimate. The ratios post/pre and inspect/pre are computed from the corresponding RMSE values. The signal magnitude is $\vartheta=c_{\vartheta}\sqrt{p/\Delta_n}$. The smallest method RMSE in each design row is boldfaced.
\end{minipage}
\end{table}

\begin{table}[htbp]
\centering
\caption{Single change-point localization: RMSE over 50 replications ($p/n=0.80$, $d=2$, and $c_{\vartheta}=0.50$).}
\label{tab:single_cp_pn_0p80_lag_2_cvartheta_0p50}
\tiny
\setlength{\tabcolsep}{2.4pt}
\resizebox{\textwidth}{!}{%
\begin{tabular}{r r r c r r r r r r r r r r r r}
\toprule
\multirow{2}{*}{$n$} & \multirow{2}{*}{$k$} & \multirow{2}{*}{$c_A$} & \multirow{2}{*}{$\rho(A)$} & \multicolumn{10}{c}{RMSE of estimated change-point location} & \multicolumn{2}{c}{RMSE ratio} \\
\cmidrule(lr){5-14}\cmidrule(lr){15-16}
& & & & Pre & Post & Pre-NR & Post-NR & inspect & dc & sbs & scanEH & agg2 & aggoo & post/pre & inspect/pre \\
\midrule
100 & 40 & 0.50 & 0.30 & 8.78 & \textbf{7.88} & 29.73 & 28.54 & 19.43 & 34.18 & 28.20 & 31.00 & 29.73 & 33.18 & 0.90 & 2.21 \\
250 & 100 & 0.50 & 0.30 & 21.70 & \textbf{14.00} & 85.29 & 82.26 & 40.98 & 84.14 & 77.97 & 79.23 & 85.29 & 80.88 & 0.65 & 1.89 \\
500 & 200 & 0.50 & 0.30 & 26.02 & \textbf{20.26} & 139.49 & 123.48 & 65.11 & 182.63 & 148.52 & 151.62 & 139.49 & 186.91 & 0.78 & 2.50 \\
100 & 80 & 0.50 & 0.30 & 10.46 & \textbf{9.17} & 32.09 & 30.38 & 20.96 & 32.60 & 32.17 & 32.57 & 32.09 & 31.61 & 0.88 & 2.00 \\
250 & 200 & 0.50 & 0.30 & 17.56 & \textbf{14.47} & 68.96 & 68.71 & 38.03 & 86.80 & 81.32 & 77.74 & 68.96 & 82.32 & 0.82 & 2.17 \\
500 & 400 & 0.50 & 0.30 & 34.57 & \textbf{26.72} & 118.52 & 111.98 & 64.32 & 164.56 & 147.52 & 137.95 & 118.52 & 174.20 & 0.77 & 1.86 \\
\cmidrule(lr){1-16}
100 & 40 & 0.50 & 0.50 & 10.58 & \textbf{8.72} & 31.85 & 30.33 & 21.50 & 34.64 & 32.60 & 30.81 & 31.85 & 34.08 & 0.82 & 2.03 \\
250 & 100 & 0.50 & 0.50 & 17.73 & \textbf{14.77} & 87.91 & 73.67 & 34.43 & 83.55 & 80.44 & 89.33 & 87.91 & 86.68 & 0.83 & 1.94 \\
500 & 200 & 0.50 & 0.50 & 33.87 & \textbf{24.73} & 142.06 & 122.53 & 61.93 & 185.94 & 152.96 & 162.58 & 142.06 & 186.49 & 0.73 & 1.83 \\
100 & 80 & 0.50 & 0.50 & 12.88 & \textbf{9.67} & 32.04 & 28.59 & 22.57 & 36.56 & 32.24 & 30.73 & 32.04 & 36.68 & 0.75 & 1.75 \\
250 & 200 & 0.50 & 0.50 & 25.02 & \textbf{15.08} & 84.61 & 80.26 & 44.74 & 88.41 & 83.56 & 81.31 & 84.61 & 88.61 & 0.60 & 1.79 \\
500 & 400 & 0.50 & 0.50 & 24.04 & \textbf{19.17} & 161.56 & 150.29 & 62.64 & 186.02 & 156.83 & 154.21 & 161.56 & 187.95 & 0.80 & 2.61 \\
\cmidrule(lr){1-16}
100 & 40 & 0.50 & 0.70 & 14.54 & \textbf{9.86} & 34.32 & 29.90 & 23.97 & 35.02 & 32.20 & 34.36 & 34.32 & 34.89 & 0.68 & 1.65 \\
250 & 100 & 0.50 & 0.70 & 32.86 & \textbf{17.60} & 85.26 & 78.30 & 53.80 & 94.19 & 87.69 & 89.73 & 85.26 & 96.05 & 0.54 & 1.64 \\
500 & 200 & 0.50 & 0.70 & 38.09 & \textbf{27.71} & 156.07 & 128.71 & 46.89 & 170.77 & 161.61 & 174.06 & 156.07 & 163.18 & 0.73 & 1.23 \\
100 & 80 & 0.50 & 0.70 & 10.07 & \textbf{7.59} & 32.38 & 28.15 & 18.17 & 34.86 & 31.09 & 33.87 & 32.38 & 34.90 & 0.75 & 1.80 \\
250 & 200 & 0.50 & 0.70 & 25.32 & \textbf{16.94} & 76.51 & 66.13 & 61.27 & 83.44 & 82.77 & 83.18 & 76.51 & 81.74 & 0.67 & 2.42 \\
500 & 400 & 0.50 & 0.70 & 51.61 & \textbf{22.70} & 167.63 & 148.00 & 81.32 & 191.12 & 169.05 & 174.21 & 167.63 & 189.00 & 0.44 & 1.58 \\
\cmidrule(lr){1-16}
100 & 40 & 0.50 & 0.90 & 15.61 & \textbf{9.09} & 46.65 & 28.97 & 19.97 & 46.79 & 43.92 & 47.88 & 46.65 & 46.79 & 0.58 & 1.28 \\
250 & 100 & 0.50 & 0.90 & 45.16 & \textbf{22.43} & 83.48 & 74.77 & 54.03 & 97.54 & 83.18 & 86.74 & 83.48 & 96.65 & 0.50 & 1.20 \\
500 & 200 & 0.50 & 0.90 & 47.08 & \textbf{24.37} & 170.17 & 130.58 & 81.57 & 187.93 & 155.93 & 177.47 & 170.17 & 184.23 & 0.52 & 1.73 \\
100 & 80 & 0.50 & 0.90 & 15.07 & \textbf{11.78} & 38.54 & 29.78 & 22.54 & 38.57 & 36.62 & 39.39 & 38.54 & 38.76 & 0.78 & 1.50 \\
250 & 200 & 0.50 & 0.90 & 36.44 & \textbf{18.01} & 87.09 & 69.24 & 40.27 & 101.62 & 79.82 & 99.14 & 87.09 & 102.24 & 0.49 & 1.11 \\
500 & 400 & 0.50 & 0.90 & 76.66 & \textbf{27.61} & 194.70 & 169.14 & 103.13 & 203.46 & 181.95 & 186.48 & 194.70 & 207.68 & 0.36 & 1.35 \\
\cmidrule(lr){1-16}
100 & 40 & 1.00 & 0.30 & 11.81 & \textbf{9.50} & 33.74 & 33.00 & 20.26 & 32.23 & 31.11 & 31.73 & 33.74 & 31.77 & 0.81 & 1.72 \\
250 & 100 & 1.00 & 0.30 & 16.45 & \textbf{14.05} & 70.75 & 71.44 & 36.87 & 83.83 & 71.12 & 73.08 & 70.75 & 83.83 & 0.85 & 2.24 \\
500 & 200 & 1.00 & 0.30 & 27.62 & \textbf{23.81} & 137.12 & 136.26 & 66.73 & 144.19 & 142.01 & 128.75 & 137.12 & 141.06 & 0.86 & 2.42 \\
100 & 80 & 1.00 & 0.30 & 9.29 & \textbf{8.50} & 31.61 & 27.35 & 17.57 & 34.53 & 31.39 & 26.20 & 31.61 & 34.02 & 0.92 & 1.89 \\
250 & 200 & 1.00 & 0.30 & 15.76 & \textbf{14.86} & 75.17 & 72.24 & 41.10 & 88.21 & 70.98 & 76.43 & 75.17 & 90.03 & 0.94 & 2.61 \\
500 & 400 & 1.00 & 0.30 & 28.62 & \textbf{24.21} & 128.85 & 129.85 & 67.17 & 169.44 & 144.76 & 137.77 & 128.85 & 168.88 & 0.85 & 2.35 \\
\cmidrule(lr){1-16}
100 & 40 & 1.00 & 0.50 & 10.69 & \textbf{9.32} & 34.31 & 31.97 & 23.41 & 34.61 & 31.93 & 32.92 & 34.31 & 34.96 & 0.87 & 2.19 \\
250 & 100 & 1.00 & 0.50 & 24.26 & \textbf{17.92} & 76.07 & 68.77 & 37.07 & 83.22 & 77.02 & 76.34 & 76.07 & 84.52 & 0.74 & 1.53 \\
500 & 200 & 1.00 & 0.50 & 35.82 & \textbf{25.55} & 153.09 & 133.42 & 54.31 & 179.42 & 145.85 & 142.23 & 153.09 & 182.43 & 0.71 & 1.52 \\
100 & 80 & 1.00 & 0.50 & 11.54 & \textbf{9.77} & 28.82 & 28.48 & 21.77 & 32.28 & 30.66 & 32.21 & 28.82 & 32.79 & 0.85 & 1.89 \\
250 & 200 & 1.00 & 0.50 & 21.58 & \textbf{17.46} & 75.42 & 69.66 & 45.91 & 80.15 & 77.85 & 78.62 & 75.42 & 82.97 & 0.81 & 2.13 \\
500 & 400 & 1.00 & 0.50 & 28.43 & \textbf{23.80} & 138.01 & 123.61 & 59.45 & 178.71 & 153.46 & 161.86 & 138.01 & 172.96 & 0.84 & 2.09 \\
\cmidrule(lr){1-16}
100 & 40 & 1.00 & 0.70 & 12.33 & \textbf{8.26} & 32.79 & 28.46 & 23.54 & 35.59 & 35.30 & 35.54 & 32.79 & 35.04 & 0.67 & 1.91 \\
250 & 100 & 1.00 & 0.70 & 30.18 & \textbf{15.39} & 70.60 & 60.78 & 49.58 & 78.70 & 80.06 & 79.93 & 70.60 & 79.51 & 0.51 & 1.64 \\
500 & 200 & 1.00 & 0.70 & 43.18 & \textbf{21.92} & 161.93 & 134.78 & 70.96 & 172.56 & 168.67 & 172.08 & 161.93 & 169.97 & 0.51 & 1.64 \\
100 & 80 & 1.00 & 0.70 & 13.14 & \textbf{8.54} & 32.30 & 28.86 & 24.22 & 35.11 & 33.15 & 34.92 & 32.30 & 35.08 & 0.65 & 1.84 \\
250 & 200 & 1.00 & 0.70 & 28.30 & \textbf{15.43} & 74.08 & 70.98 & 44.10 & 94.60 & 76.60 & 90.21 & 74.08 & 93.47 & 0.55 & 1.56 \\
500 & 400 & 1.00 & 0.70 & 42.01 & \textbf{22.06} & 175.62 & 161.54 & 77.63 & 194.65 & 174.74 & 188.02 & 175.62 & 195.69 & 0.53 & 1.85 \\
\cmidrule(lr){1-16}
100 & 40 & 1.00 & 0.90 & 18.20 & \textbf{13.49} & 32.06 & 25.91 & 27.17 & 36.18 & 31.28 & 33.71 & 32.06 & 35.82 & 0.74 & 1.49 \\
250 & 100 & 1.00 & 0.90 & 48.46 & \textbf{25.82} & 73.93 & 70.30 & 65.59 & 76.81 & 73.82 & 78.86 & 73.93 & 77.18 & 0.53 & 1.35 \\
500 & 200 & 1.00 & 0.90 & 75.40 & \textbf{30.35} & 170.08 & 136.02 & 100.88 & 170.75 & 172.55 & 165.67 & 170.08 & 171.57 & 0.40 & 1.34 \\
100 & 80 & 1.00 & 0.90 & 17.75 & \textbf{11.93} & 33.94 & 28.99 & 27.68 & 38.22 & 32.16 & 36.59 & 33.94 & 37.56 & 0.67 & 1.56 \\
250 & 200 & 1.00 & 0.90 & 44.03 & \textbf{24.81} & 73.06 & 70.85 & 71.55 & 76.34 & 71.65 & 74.06 & 73.06 & 74.89 & 0.56 & 1.63 \\
500 & 400 & 1.00 & 0.90 & 75.73 & \textbf{34.65} & 167.67 & 146.76 & 136.02 & 177.88 & 168.11 & 164.79 & 167.67 & 178.43 & 0.46 & 1.80 \\
\bottomrule
\end{tabular}%
}
\vspace{2pt}
\begin{minipage}{0.98\textwidth}
\footnotesize Pre and Post denote the preliminary Step-1 and refined Step-3 estimates. Pre-NR and Post-NR denote the corresponding estimates with zero ridge penalty in Steps 1 and 3; the lag and Step-2 Lasso are refitted from the no-ridge preliminary estimate. The ratios post/pre and inspect/pre are computed from the corresponding RMSE values. The signal magnitude is $\vartheta=c_{\vartheta}\sqrt{p/\Delta_n}$. The smallest method RMSE in each design row is boldfaced.
\end{minipage}
\end{table}

}{  \begin{center}
  \fbox{\parbox[c][1.4in][c]{0.94\linewidth}{\centering
    Placeholder for the complete single-change-point tables:\\[4pt]
    \texttt{\detokenize{numerical_results/single_cp_table.tex}}}}
  \end{center}
}
\FloatBarrier

\subsection{Segmentation evaluation: adjusted Rand index}
\label{app:ari-definition}

For multiple-change-point estimation, the estimated and true
change-point sets induce two partitions of the observation
sequence. Let
$\mathcal P^\star=\{\mathcal P_1^\star,\ldots,\mathcal P_K^\star\}$
and
$\widehat{\mathcal P}=\{\widehat{\mathcal P}_1,\ldots,
\widehat{\mathcal P}_{\widehat K}\}$
denote the true and estimated partitions, respectively, where
$K=m_0+1$ and $\widehat K=\widehat m+1$. Define
\[
n_{ij}=|\mathcal P_i^\star\cap\widehat{\mathcal P}_j|,
\qquad
a_i=\sum_{j=1}^{\widehat K}n_{ij},
\qquad
b_j=\sum_{i=1}^{K}n_{ij}.
\]
The adjusted Rand index (ARI)
\citep{rand1971objective,hubert1985comparing} is defined as
\begin{equation*}
\mathrm{ARI}
=
\frac{
\displaystyle
\sum_{i=1}^{K}\sum_{j=1}^{\widehat K}\binom{n_{ij}}{2}
-
\frac{
\displaystyle
\left(\sum_{i=1}^{K}\binom{a_i}{2}\right)
\left(\sum_{j=1}^{\widehat K}\binom{b_j}{2}\right)
}{\binom{n}{2}}
}{
\displaystyle
\frac12\left[
\sum_{i=1}^{K}\binom{a_i}{2}
+
\sum_{j=1}^{\widehat K}\binom{b_j}{2}
\right]
-
\frac{
\displaystyle
\left(\sum_{i=1}^{K}\binom{a_i}{2}\right)
\left(\sum_{j=1}^{\widehat K}\binom{b_j}{2}\right)
}{\binom{n}{2}}
}.
\end{equation*}

\subsection{Practical tuning parameters and implementation details}
\label{app:numerical-implementation}
This section details the finite-sample tuning rules used in our numerical experiments to ensure exact reproducibility. While the asymptotic rate conditions in Sections~\ref{sec:single} and~\ref{sec:multiple} govern the theoretical framework—requiring Step~1 ridge parameters to remain uniformly $O(1)$, Step~3 ridge parameters to vanish asymptotically, and $\omega_n$ to satisfy the sandwich condition—the rules specified below select concrete data-dependent values directly from the sample.

\subsubsection{A common normalized ridge scan in Steps 1 and 3}
The single- and multiple-change-point implementations use the same split
criterion.  Consider any interval $[b,e]$ of length $N=e-b+1$ and any
candidate $z\in\{b,\ldots,e-1\}$.  Let
\[
 m_1=z-b+1,\qquad m_2=e-z,\qquad
 q_{[b,e]}(z)=\frac{m_1m_2}{N^2},
\]
and denote the sample means on $[b,z]$ and $[z+1,e]$ by
$\bar y_1(z)$ and $\bar y_2(z)$, respectively.  For a dimensionless ridge
coefficient $\lambda>0$, the implemented profiled score is
\begin{equation}\label{eq:numerical-ridge-score}
\begin{split}
 \mathcal J_{[b,e]}(z;\lambda)
 ={}&\frac1N\left\{
 \sum_{t=b}^{z}\|y_t-\bar y_1(z)\|_2^2
 +\sum_{t=z+1}^{e}\|y_t-\bar y_2(z)\|_2^2
 \right\}\\
 &+\frac{q_{[b,e]}(z)}{(1+q_{[b,e]}(z)/\lambda)^2}
 \|\bar y_1(z)-\bar y_2(z)\|_2^2.
\end{split}
\end{equation}
We note that \eqref{eq:numerical-ridge-score} is the unpenalized residual sum of squares, normalized by the interval
length, evaluated at the two ridge-fitted means in
\eqref{eq:Q-def}.  Indeed, if
\[
 \Lambda_{[b,e]}(z;\lambda)=\frac{\lambda}{\lambda+q_{[b,e]}(z)},
\]
then $\Lambda_{[b,e]}(z;\lambda)$ is the fraction of the raw mean contrast removed by
shrinkage toward the pooled mean, and the last coefficient in
\eqref{eq:numerical-ridge-score} is $q_{[b,e]}(z)\Lambda_{[b,e]}(z;\lambda)^2$.
This formula is used with the ridge coefficient of the current scan: it is evaluated on the raw observations in Step~1 and on the filtered observations in Step~3. For raw observations on $[1,n]$, $\mathcal J_{[1,n]}(\zeta;\lambda_1)=L(\zeta)$ and $\Lambda_{[1,n]}(\zeta;\lambda_1)=\Lambda_1(\zeta)$. With $k=z-b+1$, the balance factor is $q_{[b,e]}(z)=q_k$ for a sequence of length $N=e-b+1$. In this closed-interval notation, the local interval $(b_i,e_i]$ in Section~\ref{sec:multi-step3} is represented by $[b_i+1,e_i]$.

\begingroup
Let $\mathcal C_1$ and $\mathcal C_3$ be fixed finite grids of positive
constants, with log-spaced points independent of $n$.
The single- and multiple-change-point implementations select the Step~1
and Step~3 ridge coefficients from
\begin{equation}\label{eq:numerical-ridge-grid}
 \lambda_1\in\mathcal C_1,
 \qquad
 \lambda_3\in\{c/\log n:c\in\mathcal C_3\}.
\end{equation}
The two constant grids may differ. Here $n$ is the original series length,
including for scans within rolling windows and local refinement intervals.
The Step~1 grid is also used for its raw-series local refinement.
\par\endgroup

For each candidate ridge value, we minimize \eqref{eq:numerical-ridge-score} over
$z$.  After ordering the grid increasingly, we identify consecutive plateaus on which the estimated split is unchanged.  We select the second grid point of the longest constant-split plateau; if several plateaus have the same length, we use the first one on the increasing grid.  If no
two adjacent grid points return the same split, the selected value is the
one with the smallest minimized score.  {\color{green!50!black}Write $\widehat\lambda$ for the ridge value selected in a scan.}
We apply the same rule separately
in every multiple-change rolling window and every local refinement interval.
The Step~1 grid in \eqref{eq:numerical-ridge-grid} satisfies
the theoretical $O(1)$ requirement. Proposition~\ref{prop:tau0} holds
uniformly over $\lambda_1\in[0,\max\mathcal C_1]$ and therefore covers
the data-driven choice from this grid. Log spacing resolves the
multiplicative scale on which $\Lambda_{[b,e]}(z;\lambda)$ changes.  For the multiple-change table,
the reported Step~1 and Step~3 tuning summaries are the averages of $\log\widehat\lambda$ over the corresponding scans within a repetition.

\begingroup
For Step~3, the largest grid value is $(\max\mathcal C_3)/\log n\to0$.
Thus, the entire grid satisfies the vanishing-ridge requirement in
Theorems~\ref{prop:tau-rate} and~\ref{prop:multi-local}.
Under the remaining assumptions of these theorems, uniformity over the
ridge parameter covers the data-driven selection from this shrinking grid.
\par\endgroup

The single-change Step~1 scan uses all observations.  After filtering, Step~3 begins at time $\widehat d+1$, the first time at which all estimated lags are available, so the starting time is governed by $\widehat d$ alone
rather than by the lag-search upper bound of five.

\subsubsection{Rolling-window length and shift}

\begingroup
We use the fixed constants $c_{\mathrm{win}}$, $c_{\mathrm{shift}}$ and exponent grid $\mathcal G_{\mathrm{win}}$ defined in Section~\ref{sec:multi-CP-sims}. After removing duplicate lengths and lengths below $\lceil\sqrt n\rceil$, we evaluate the remaining lengths in decreasing order. For each length $h$, all windows have length $h$: the first starts at observation 1, successive starts are separated by $\ell=\max\{1,\operatorname{round}(c_{\mathrm{shift}}h)\}$, and a final full window ending at $n$ is added when necessary. The count associated with $h$ is the count after backward elimination, before raw-series local refinement.

Among consecutive plateaus of at least two lengths with the same positive count, we select the longest plateau; ties favor the plateau with larger window lengths. Its middle length is selected, taking the smaller of the two middle lengths for an even plateau. If no eligible plateau exists, we use the largest length for which screening succeeds. Raw-series local refinement is performed only after the window length is selected, since it preserves the screened count. This selection uses no true change-point locations, true count, or true minimum spacing.
\par\endgroup

\subsubsection{Lag Estimation and Clean Block Construction}

The true lag order is used solely for data generation. Once a preliminary segmentation is established, we trim observations in the proximity of estimated boundaries before selecting the lag order or estimating the common autoregressive matrices. In our finite-sample implementation, each buffer is set to a fixed proportion of its respective preliminary segment length, balancing boundary contamination against the sample size retained for estimation. Specifically, for the single-change design, we define
\[
B_{\mathrm{L}} = \left\lceil \frac{\widehat{\tau}}{5} \right\rceil, \qquad B_{\mathrm{R}} = \left\lceil \frac{n - \widehat{\tau}}{5} \right\rceil.
\]
The resulting clean blocks used for lag order selection are $\{1, \dots, \widehat{\tau} - B_{\mathrm{L}}\}$ and $\{\widehat{\tau} + B_{\mathrm{R}} + 1, \dots, n\}$. Thus, both left and right buffers adhere to the same one-fifth trimming rule, while naturally accommodating sample size asymmetries across regimes.

For the multiple-change setting, we define boundary points $\widehat{\tau}_0 = 0$ and $\widehat{\tau}_{\widehat{m}+1} = n$, and denote the $q$-th preliminary segment by $\{\widehat{\tau}_{q-1} + 1, \dots, \widehat{\tau}_q\}$, containing $n_q$ observations. Setting $B_q = \lceil n_q / 5 \rceil$, we retain the clean sub-segment
\begin{equation*}
  \left\{
  \widehat{\tau}_{q-1} + 1 + \mathbf{1}\{q > 1\} B_q,
  \dots,
  \widehat{\tau}_q - \mathbf{1}\{q \le \widehat{m}\} B_q
  \right\}.
\end{equation*}
Thus, for outer boundary segments ($q = 1$ and $q = \widehat{m}+1$), trimming is applied exclusively to the boundary adjacent to an estimated change point, whereas interior segments are trimmed at both endpoints. In Step~2, we additionally omit the initial $\widehat{d}$ response vectors from each clean block to ensure that all retained responses and their corresponding lagged predictors remain strictly within that block.

Each clean block is first centered and standardized series-by-series (coordinatewise). We then fit a sparse VAR model using the \texttt{sparseVAR} function from the \texttt{bigtime} R package with an $\ell_1$ penalty, BIC model selection, and a maximum candidate lag of five. Letting $\widehat{A}^{(q)}_\ell$ ($\ell = 1, \dots, 5$) denote the fitted autoregressive coefficient matrices in clean segment $q$, we define the lag-block energy $E_{q\ell}$ and cumulative proportion $F_q(r)$ as
\[
  E_{q\ell} = \left\| \widehat{A}^{(q)}_\ell \right\|_{\mathrm{F}}^2,
  \qquad
  F_q(r) = \frac{\sum_{\ell=1}^r E_{q\ell}}{\sum_{\ell=1}^5 E_{q\ell}}.
\]
The segment-specific lag order $\widehat{d}_q$ and the global common lag order $\widehat{d}$ are determined via
\begin{equation*}
  \widehat{d}_q = \min \left\{ r \in \{1, \dots, 5\} : F_q(r) \ge 0.95 \right\},
  \qquad
  \widehat{d} = \min_q \widehat{d}_q.
\end{equation*}
This cumulative-energy rule prevents isolated, magnitude-negligible higher-lag coefficients from artificially inflating the estimated lag order. If all fitted coefficient matrices have zero energy, numerical optimization fails to converge, a clean block contains insufficient observations, or an invalid lag is returned, we default to $\widehat{d} = 1$. In the multiple-change pipeline, if Steps~2--3 fail to execute at a selected lag order $\widehat{d} > 1$, we re-estimate the system setting $\widehat{d} = 1$.

\subsubsection{Step~2 Lasso Path and AIC Selection}

Let $n_{\mathrm{obs}}$ denote the number of retained response periods across clean blocks: in the multiple-change setting, $n_{\mathrm{obs}}$ corresponds to the actual retained sample size $\widehat{N}_m$, whereas in the single-change case, $n_{\mathrm{obs}} = |I_1| + |I_2|$. Let $G$ denote the block-centering matrix for these retained observations, representing $G_{\mathrm{blk}}$ in the single-change setup and $G_{\mathrm{blk}}^{(m)}$ in the multiple-change setup. We define $\widetilde{\mathcal{Y}}_I = \operatorname{vec}(G Y_I)$ and $\widetilde{Z}_I(G) = \mathbf{I}_p \otimes (G X_I)$ as the vectorized, block-centered response vector and design matrix, constructed using the selected lag order $\widehat{d}$. Following within-block centering, the paper-scaled Step~2 objective function is
\[
  \frac{1}{n_{\mathrm{obs}}} \left\| \widetilde{\mathcal{Y}}_I - \widetilde{Z}_I(G) \beta \right\|_2^2 + \lambda_n \|\beta\|_1.
\]
We evaluate this objective over a 101-point decreasing grid of penalty parameters,
\begin{equation}\label{eq:numerical-step2-grid}
  \lambda_n(a) = 2^a \sqrt{\frac{\log \widehat{d} + 2 \log p}{n_{\mathrm{obs}}}}, \qquad a \in \{5, 4.9, \dots, -5\}.
\end{equation}
The tuning path $\lambda_n(a)$ governs practical data-driven regularization in both single- and multiple-change settings, distinguishing it from the theoretical choices $\lambda_n$ and $\lambda_{m,n}$. In our benchmark simulation designs where $p \asymp n$, the lag order $d$ is fixed, and the number of breaks $m_0$ is fixed, the theoretical rate terms $\ell_{2,n}$ and $\ell_{m,n}$ share the logarithmic order of $\log(d p^2)$. This provides empirical motivation for grid \eqref{eq:numerical-step2-grid}, where candidate penalty multipliers are selected via the Akaike Information Criterion (AIC) rather than specified using unobserved theoretical dependence constants.

Evaluating the grid in decreasing order enables warm starts from heavier to lighter regularization states. We stack the $p$ system equations into a single vectorized model prior to calling \texttt{glmnet}. Because \texttt{glmnet} normalizes its Gaussian loss by $2 N_{\mathrm{glmnet}} = 2 p n_{\mathrm{obs}}$, the parameter value supplied to the package is scaled as
\begin{equation*}
  \lambda_{\mathrm{glmnet}}(a) = \frac{2^a}{2p} \sqrt{\frac{\log\widehat{d} + 2\log p}{n_{\mathrm{obs}}}}.
\end{equation*}
We enforce \texttt{intercept = FALSE} and \texttt{standardize = FALSE}, as segment-specific intercepts are profiled out via prior within-block centering. For each candidate grid point $a$, we compute
\[
  \operatorname{AIC}(a) = -2\ell(a) + 2\operatorname{df}(a),
\]
where $\ell(a)$ is the fitted Gaussian log-likelihood and $\operatorname{df}(a)$ denotes the active model degrees of freedom, and select its minimizer $\widehat{a}$. The sparse block-Kronecker structure implemented in practice serves as an equivalent, memory-efficient representation of the global Lasso objective, preserving the exact specification of grid \eqref{eq:numerical-step2-grid} and its AIC selection rule.

\subsubsection{Local Intervals in the Multiple-Change Procedure}

For ordered preliminary break estimates $\widehat{\tau}_1 < \dots < \widehat{\tau}_{\widehat{m}}$, with boundary conventions $\widehat{\tau}_0 = 0$ and $\widehat{\tau}_{\widehat{m}+1} = n$, the local search interval $(b_j(q), e_j(q)]$ generated by a fixed fraction $q \in (1/2, 1)$ is defined by
\begin{equation*}
  b_j(q) = \left\lfloor q \widehat{\tau}_{j-1} + (1-q) \widehat{\tau}_j \right\rfloor,
  \qquad
  e_j(q) = \left\lceil (1-q) \widehat{\tau}_j + q \widehat{\tau}_{j+1} \right\rceil.
\end{equation*}
The raw local re-estimation at the end of Step~1 uses a fixed fraction $q_{\mathrm{raw}} \in (1/2, 1)$, whereas the final filtered Step~3 employs a fixed fraction $q_{\mathrm{filt}} \in (1/2, 1)$ and restricts its search strictly to where the filtered process is defined. These fractions need not coincide. Both choices guarantee that a fixed positive fraction of each adjacent regime gap is preserved on either side of the current estimate. Once the preliminary localization error satisfies $o_{\mathbb{P}}(\Delta_n)$, each interval contains exactly one true change point with probability tending to one; varying the choice of fixed fraction thus alters bounding constants without impacting the rate of localization. If independently refined break locations collide or violate strict ordering in finite samples, we retain the original pre-refinement estimates, thereby preserving the total number of estimated change points.

\subsubsection{Data-Driven Selection of the BEA Penalty \texorpdfstring{$\omega_n$}{omega}}

In the multiple-change setting, the penalty parameter $\omega_n$ is selected from the complete backward elimination algorithm (BEA) loss path, after which backward elimination is re-run using this fixed penalty. Let $\widetilde{\mathcal{S}}$ denote the rolling-window candidate set, and let $\mathcal{L}_n(\mathcal{S})$ denote the total unpenalized within-segment sum of squares for a candidate partition $\mathcal{S}$. The tuning path starts at $\mathcal{S}_0 = \widetilde{\mathcal{S}}$ and terminates when no candidates remain. At deletion step $r$, the removed candidate $s_r^\star$ and updated candidate set $\mathcal{S}_r$ are defined by
\[
  s_r^\star \in \arg\min_{s \in \mathcal{S}_{r-1}} \mathcal{L}_n(\mathcal{S}_{r-1} \setminus \{s\}), \qquad \mathcal{S}_r = \mathcal{S}_{r-1} \setminus \{s_r^\star\},
\]
with corresponding loss increment
\[
  v_r = \left| \mathcal{L}_n(\mathcal{S}_r) - \mathcal{L}_n(\mathcal{S}_{r-1}) \right|, \qquad V = \left\{ v_1, \dots, v_{|\widetilde{\mathcal{S}}|} \right\}.
\]
Small values of $v_r$ correspond to redundant candidate splits, whereas larger values signal genuine structural change points.

We allow for up to three jump clusters ($K \in \{2, 3\}$) rather than only two because, late in the elimination path, deleting one of the few remaining candidates can merge two long segments, generating an exceptionally large loss jump. Under a strict binary partition ($K = 2$), such late-stage merging jumps can inflate cluster variance and cause moderate structural changes to be misclassified alongside redundant ones. A three-cluster partition cleanly separates a small redundant scale $\mathcal{K}_1$, a moderate structural scale $\mathcal{K}_2$, and a large late-merge scale $\mathcal{K}_3$; both $\mathcal{K}_2$ and $\mathcal{K}_3$ are subsequently treated as structural.

Setting $\epsilon = 10^{-6}$, the data-driven selection rule proceeds as follows:
\begin{enumerate}
  \item Transform the deletion losses via $u_r = \log(v_r + \epsilon)$. Fit one-dimensional Hartigan--Wong $k$-means separately for $K = 2$ and, when feasible ($|\widetilde{\mathcal{S}}| \ge 3$), for $K = 3$.
  \item Compute the average silhouette width for each fit using absolute distances on the log scale, assigning a silhouette value of zero to singleton clusters. Select the cluster count $K$ that maximizes the average silhouette width; ties or infeasible $K = 3$ cases default to $K = 2$.
  \item Order the selected cluster centers such that $\mathcal{K}_1, \mathcal{K}_2$, and (if $K = 3$) $\mathcal{K}_3$ have strictly increasing centroids. Only cluster $\mathcal{K}_1$ is classified as redundant.
  \item Compute the separation ratio $R_{\mathrm{sep}} = \mathrm{BSS} / \mathrm{TSS}$ from the selected partition, where $\mathrm{BSS}$ and $\mathrm{TSS}$ denote the between-cluster and total sums of squares of the log-transformed deletion losses $\{u_r\}$. If $R_{\mathrm{sep}} \ge 0.20$ and $\mathcal{K}_2$ is non-empty, set
  \begin{equation*}
    \widehat{\omega}_n = \max \left\{ 0, \, \min_{r : u_r \in \mathcal{K}_2} v_r - \epsilon \right\}.
  \end{equation*}
  Subtracting $\epsilon$ preserves the boundary candidate, as the final BEA step deletes a candidate whenever its deletion loss increase does not exceed $\widehat{\omega}_n$.
  \item If the partition is weak ($R_{\mathrm{sep}} < 0.20$) or degenerate, set $\widehat{\omega}_n = \max V + \epsilon$, representing the conservative no-change fallback. If the initial candidate set $\widetilde{\mathcal{S}}$ is empty, set $\widehat{\omega}_n = 0$.
\end{enumerate}
Finally, re-initializing from $\widetilde{\mathcal{S}}$, the fixed-penalty BEA deletes a candidate whenever doing so does not increase $\mathcal{L}_n(\mathcal{S}) + |\mathcal{S}| \widehat{\omega}_n$. This tuning rule is entirely data-dependent and operates without knowledge of true change-point locations, total break count, regime spacings, or jump signs.

\subsection{Additional Loss Profiles for the Macroeconomic Application}
\label{app:fred-md-additional-loss-profiles}

\begin{figure}[!htbp]
\centering
\IfFileExists{numerical_results/01_normalized_loss_profiles_appendix_5panels.pdf}{%
  \includegraphics[width=\textwidth,height=0.72\textheight,keepaspectratio]{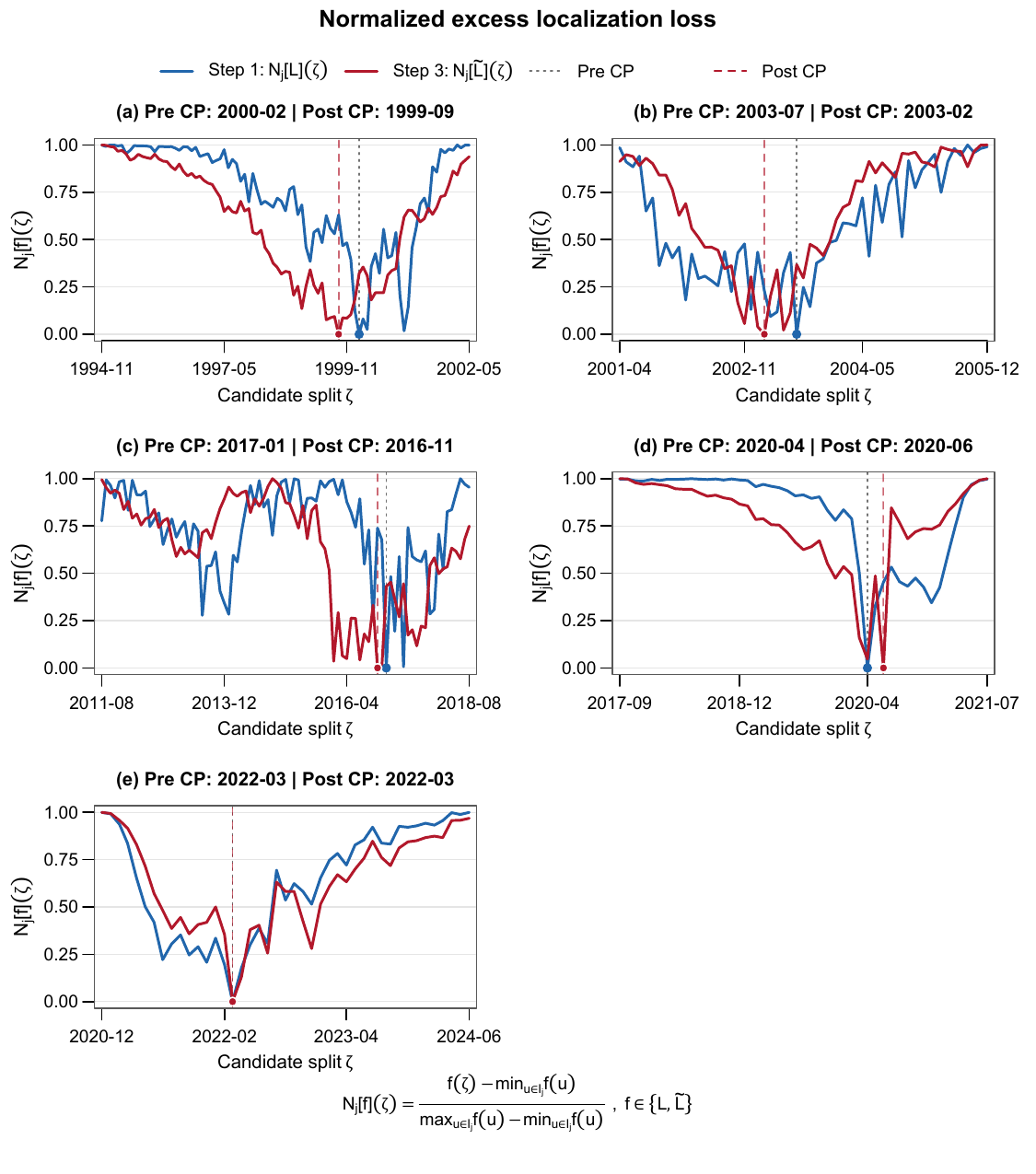}%
}{%
  \textcolor{blue}{\fbox{\parbox[c][3.0in][c]{0.94\linewidth}{\centering
  Placeholder for the remaining five loss-profile panels\\[8pt]
  Pre CP: 2000-02 \quad Post CP: 1999-09\\[4pt]
  Pre CP: 2003-07 \quad Post CP: 2003-02\\[4pt]
  Pre CP: 2017-01 \quad Post CP: 2016-11\\[4pt]
  Pre CP: 2020-04 \quad Post CP: 2020-06\\[4pt]
  Pre CP: 2022-03 \quad Post CP: 2022-03}}}%
}
\caption{Normalized excess localization losses before (Pre, blue) and after (Post, red) VAR filtering for the remaining five change points. Profiles evaluate step-specific data windows and ridge parameters, normalized by subtracting their minimum and dividing by their range. Vertical lines and dots mark estimated break locations and profile minima, respectively; vertical scales depict relative criterion contrast rather than absolute loss magnitudes.}
\label{fig:fred_md_loss_profiles_appendix}
\end{figure}

Figure~\ref{fig:fred_md_loss_profiles_appendix} displays the normalized excess localization losses for the remaining five refined estimates: 1999-09, 2003-02, 2016-11, 2020-06, and 2022-03. Together with Figure~\ref{fig:fred_md_loss_profiles}, these profiles encompass all seven estimated change points, adhering to the same normalization and graphical conventions.

\FloatBarrier

\begingroup

\section{FRED-MD variables and transformations}
\label{app:fred-md-data-documentation}

This appendix documents the 126 macroeconomic variables used in
Section~\ref{sec:fred_md_empirical}. Variable descriptions, transformation
codes, and economic group classifications are based on the FRED-MD
documentation \citep{McCrackenNg2016,StLouisFedFREDMD}.
Table~\ref{tab:fred-md-variables} presents this information for the
variables included in our analysis, in their analysis order.

The transformation code (T-code) follows \citet{McCrackenNg2016}:
1 denotes levels without transformation; 2, first differences;
3, second differences; 4, logarithms; 5, first differences of logarithms;
6, second differences of logarithms; and 7, first differences of the
simple growth rate. The economic group codes denote
1: output and income; 2: labor market; 3: housing;
4: consumption, orders, and inventories; 5: money and credit;
6: interest and exchange rates; 7: prices; and 8: stock market.

\small
\setlength{\tabcolsep}{4pt}
\renewcommand{\arraystretch}{1.05}
\begin{longtable}{@{}>{\raggedright\arraybackslash}p{0.25\linewidth}>{\raggedright\arraybackslash}p{\dimexpr0.59\linewidth-6\tabcolsep\relax}>{\centering\arraybackslash}p{0.08\linewidth}>{\centering\arraybackslash}p{0.08\linewidth}@{}}
\caption{Variables, transformations, and economic groups in the FRED-MD application.}
\label{tab:fred-md-variables}\\
\toprule
Variable & Description & T-code & Group \\
\midrule
\endfirsthead
\multicolumn{4}{@{}l}{\tablename~\thetable\ (continued)}\\
\toprule
Variable & Description & T-code & Group \\
\midrule
\endhead
\midrule
\multicolumn{4}{r@{}}{\footnotesize Continued on next page}\\
\endfoot
\bottomrule
\endlastfoot
\texttt{RPI} & Real Personal Income & 5 & 1 \\
\texttt{W875RX1} & Real personal income ex transfer receipts & 5 & 1 \\
\texttt{DPCERA3M086SBEA} & Real personal consumption expenditures & 5 & 4 \\
\texttt{CMRMTSPLx} & Real Manu.  and Trade Industries Sales & 5 & 4 \\
\texttt{RETAILx} & Retail and Food Services Sales & 5 & 4 \\
\texttt{INDPRO} & IP Index & 5 & 1 \\
\texttt{IPFPNSS} & IP: Final Products and Nonindustrial Supplies & 5 & 1 \\
\texttt{IPFINAL} & IP: Final Products (Market Group) & 5 & 1 \\
\texttt{IPCONGD} & IP: Consumer Goods & 5 & 1 \\
\texttt{IPDCONGD} & IP: Durable Consumer Goods & 5 & 1 \\
\texttt{IPNCONGD} & IP: Nondurable Consumer Goods & 5 & 1 \\
\texttt{IPBUSEQ} & IP: Business Equipment & 5 & 1 \\
\texttt{IPMAT} & IP: Materials & 5 & 1 \\
\texttt{IPDMAT} & IP: Durable Materials & 5 & 1 \\
\texttt{IPNMAT} & IP: Nondurable Materials & 5 & 1 \\
\texttt{IPMANSICS} & IP: Manufacturing (SIC) & 5 & 1 \\
\texttt{IPB51222S} & IP: Residential Utilities & 5 & 1 \\
\texttt{IPFUELS} & IP: Fuels & 5 & 1 \\
\texttt{CUMFNS} & Capacity Utilization:  Manufacturing & 2 & 1 \\
\texttt{HWI} & Help-Wanted Index for United States & 2 & 2 \\
\texttt{HWIURATIO} & Ratio of Help Wanted/No.  Unemployed & 2 & 2 \\
\texttt{CLF16OV} & Civilian Labor Force & 5 & 2 \\
\texttt{CE16OV} & Civilian Employment & 5 & 2 \\
\texttt{UNRATE} & Civilian Unemployment Rate & 2 & 2 \\
\texttt{UEMPMEAN} & Average Duration of Unemployment (Weeks) & 2 & 2 \\
\texttt{UEMPLT5} & Civilians Unemployed - Less Than 5 Weeks & 5 & 2 \\
\texttt{UEMP5TO14} & Civilians Unemployed for 5-14 Weeks & 5 & 2 \\
\texttt{UEMP15OV} & Civilians Unemployed - 15 Weeks \& Over & 5 & 2 \\
\texttt{UEMP15T26} & Civilians Unemployed for 15-26 Weeks & 5 & 2 \\
\texttt{UEMP27OV} & Civilians Unemployed for 27 Weeks and Over & 5 & 2 \\
\texttt{CLAIMSx} & Initial Claims & 5 & 2 \\
\texttt{PAYEMS} & All Employees:  Total nonfarm & 5 & 2 \\
\texttt{USGOOD} & All Employees:  Goods-Producing Industries & 5 & 2 \\
\texttt{CES1021000001} & All Employees:  Mining and Logging:  Mining & 5 & 2 \\
\texttt{USCONS} & All Employees:  Construction & 5 & 2 \\
\texttt{MANEMP} & All Employees:  Manufacturing & 5 & 2 \\
\texttt{DMANEMP} & All Employees:  Durable goods & 5 & 2 \\
\texttt{NDMANEMP} & All Employees:  Nondurable goods & 5 & 2 \\
\texttt{SRVPRD} & All Employees:  Service-Providing Industries & 5 & 2 \\
\texttt{USTPU} & All Employees:  Trade, Transportation \& Utilities & 5 & 2 \\
\texttt{USWTRADE} & All Employees:  Wholesale Trade & 5 & 2 \\
\texttt{USTRADE} & All Employees:  Retail Trade & 5 & 2 \\
\texttt{USFIRE} & All Employees:  Financial Activities & 5 & 2 \\
\texttt{USGOVT} & All Employees:  Government & 5 & 2 \\
\texttt{CES0600000007} & Avg Weekly Hours :  Goods-Producing & 1 & 2 \\
\texttt{AWOTMAN} & Avg Weekly Overtime Hours :  Manufacturing & 2 & 2 \\
\texttt{AWHMAN} & Avg Weekly Hours :  Manufacturing & 1 & 2 \\
\texttt{HOUST} & Housing Starts:  Total New Privately Owned & 4 & 3 \\
\texttt{HOUSTNE} & Housing Starts, Northeast & 4 & 3 \\
\texttt{HOUSTMW} & Housing Starts, Midwest & 4 & 3 \\
\texttt{HOUSTS} & Housing Starts, South & 4 & 3 \\
\texttt{HOUSTW} & Housing Starts, West & 4 & 3 \\
\texttt{PERMIT} & New Private Housing Permits (SAAR) & 4 & 3 \\
\texttt{PERMITNE} & New Private Housing Permits, Northeast (SAAR) & 4 & 3 \\
\texttt{PERMITMW} & New Private Housing Permits, Midwest (SAAR) & 4 & 3 \\
\texttt{PERMITS} & New Private Housing Permits, South (SAAR) & 4 & 3 \\
\texttt{PERMITW} & New Private Housing Permits, West (SAAR) & 4 & 3 \\
\texttt{ACOGNO} & New Orders for Consumer Goods & 5 & 4 \\
\texttt{AMDMNOx} & New Orders for Durable Goods & 5 & 4 \\
\texttt{ANDENOx} & New Orders for Nondefense Capital Goods & 5 & 4 \\
\texttt{AMDMUOx} & Unfilled Orders for Durable Goods & 5 & 4 \\
\texttt{BUSINVx} & Total Business Inventories & 5 & 4 \\
\texttt{ISRATIOx} & Total Business:  Inventories to Sales Ratio & 2 & 4 \\
\texttt{M1SL} & M1 Money Stock & 6 & 5 \\
\texttt{M2SL} & M2 Money Stock & 6 & 5 \\
\texttt{M2REAL} & Real M2 Money Stock & 5 & 5 \\
\texttt{BOGMBASE} & Monetary Base & 6 & 5 \\
\texttt{TOTRESNS} & Total Reserves of Depository Institutions & 6 & 5 \\
\texttt{NONBORRES} & Reserves Of Depository Institutions & 7 & 5 \\
\texttt{BUSLOANS} & Commercial and Industrial Loans & 6 & 5 \\
\texttt{REALLN} & Real Estate Loans at All Commercial Banks & 6 & 5 \\
\texttt{NONREVSL} & Total Nonrevolving Credit & 6 & 5 \\
\texttt{CONSPI} & Nonrevolving consumer credit to Personal Income & 2 & 5 \\
\texttt{S\&P 500} & S\&P’s Common Stock Price Index: Composite & 5 & 8 \\
\texttt{S\&P div yield} & S\&P’s Composite Common Stock: Dividend Yield & 2 & 8 \\
\texttt{S\&P PE ratio} & S\&P’s Composite Common Stock: Price-Earnings Ratio & 5 & 8 \\
\texttt{FEDFUNDS} & Effective Federal Funds Rate & 2 & 6 \\
\texttt{CP3Mx} & 3-Month AA Financial Commercial Paper Rate & 2 & 6 \\
\texttt{TB3MS} & 3-Month Treasury Bill: & 2 & 6 \\
\texttt{TB6MS} & 6-Month Treasury Bill: & 2 & 6 \\
\texttt{GS1} & 1-Year Treasury Rate & 2 & 6 \\
\texttt{GS5} & 5-Year Treasury Rate & 2 & 6 \\
\texttt{GS10} & 10-Year Treasury Rate & 2 & 6 \\
\texttt{AAA} & Moody’s Seasoned Aaa Corporate Bond Yield & 2 & 6 \\
\texttt{BAA} & Moody’s Seasoned Baa Corporate Bond Yield & 2 & 6 \\
\texttt{COMPAPFFx} & 3-Month Commercial Paper Minus FEDFUNDS & 1 & 6 \\
\texttt{TB3SMFFM} & 3-Month Treasury C Minus FEDFUNDS & 1 & 6 \\
\texttt{TB6SMFFM} & 6-Month Treasury C Minus FEDFUNDS & 1 & 6 \\
\texttt{T1YFFM} & 1-Year Treasury C Minus FEDFUNDS & 1 & 6 \\
\texttt{T5YFFM} & 5-Year Treasury C Minus FEDFUNDS & 1 & 6 \\
\texttt{T10YFFM} & 10-Year Treasury C Minus FEDFUNDS & 1 & 6 \\
\texttt{AAAFFM} & Moody’s Aaa Corporate Bond Minus FEDFUNDS & 1 & 6 \\
\texttt{BAAFFM} & Moody's Baa Corporate Bond Minus FEDFUNDS & 1 & 6 \\
\texttt{TWEXAFEGSMTHx} & Trade Weighted U.S. Dollar Index & 5 & 6 \\
\texttt{EXSZUSx} & Switzerland / U.S. Foreign Exchange Rate & 5 & 6 \\
\texttt{EXJPUSx} & Japan / U.S. Foreign Exchange Rate & 5 & 6 \\
\texttt{EXUSUKx} & U.S. / U.K. Foreign Exchange Rate & 5 & 6 \\
\texttt{EXCAUSx} & Canada / U.S. Foreign Exchange Rate & 5 & 6 \\
\texttt{WPSFD49207} & PPI: Finished Goods & 6 & 7 \\
\texttt{WPSFD49502} & PPI: Finished Consumer Goods & 6 & 7 \\
\texttt{WPSID61} & PPI: Intermediate Materials & 6 & 7 \\
\texttt{WPSID62} & PPI: Crude Materials & 6 & 7 \\
\texttt{OILPRICEx} & Crude Oil, spliced WTI and Cushing & 6 & 7 \\
\texttt{PPICMM} & PPI: Metals and metal products: & 6 & 7 \\
\texttt{CPIAUCSL} & CPI : All Items & 6 & 7 \\
\texttt{CPIAPPSL} & CPI : Apparel & 6 & 7 \\
\texttt{CPITRNSL} & CPI : Transportation & 6 & 7 \\
\texttt{CPIMEDSL} & CPI : Medical Care & 6 & 7 \\
\texttt{CUSR0000SAC} & CPI : Commodities & 6 & 7 \\
\texttt{CUSR0000SAD} & CPI : Durables & 6 & 7 \\
\texttt{CUSR0000SAS} & CPI : Services & 6 & 7 \\
\texttt{CPIULFSL} & CPI : All Items Less Food & 6 & 7 \\
\texttt{CUSR0000SA0L2} & CPI : All items less shelter & 6 & 7 \\
\texttt{CUSR0000SA0L5} & CPI : All items less medical care & 6 & 7 \\
\texttt{PCEPI} & Personal Cons.  Expend.:  Chain Index & 6 & 7 \\
\texttt{DDURRG3M086SBEA} & Personal Cons.  Exp:  Durable goods & 6 & 7 \\
\texttt{DNDGRG3M086SBEA} & Personal Cons.  Exp:  Nondurable goods & 6 & 7 \\
\texttt{DSERRG3M086SBEA} & Personal Cons.  Exp:  Services & 6 & 7 \\
\texttt{CES0600000008} & Avg Hourly Earnings :  Goods-Producing & 6 & 2 \\
\texttt{CES2000000008} & Avg Hourly Earnings :  Construction & 6 & 2 \\
\texttt{CES3000000008} & Avg Hourly Earnings :  Manufacturing & 6 & 2 \\
\texttt{UMCSENTx} & Consumer Sentiment Index & 2 & 4 \\
\texttt{DTCOLNVHFNM} & Consumer Motor Vehicle Loans Outstanding & 6 & 5 \\
\texttt{DTCTHFNM} & Total Consumer Loans and Leases Outstanding & 6 & 5 \\
\texttt{INVEST} & Securities in Bank Credit at All Commercial Banks & 6 & 5 \\
\texttt{VIXCLSx} & VIX & 1 & 8 \\
\end{longtable}
\endgroup

\end{document}